\documentclass[a4paper]{article}        
\usepackage{amsmath}
\usepackage{mathrsfs}
\usepackage{amssymb}
\usepackage{amsfonts}
\usepackage{latexsym}
\usepackage[mathscr]{eucal}    

\title{Finite Temperature QED:  
Non-Cancellation of Infrared Divergencies and Thermal Corrections to the 
Electron Magnetic Moment}
\author{}
\date{}

\begin{document}

\newcommand{\e}{\epsilon}
 
\newcommand{\wt}[1]{\widetilde{#1}}

\newcommand{\bel}[1]{\begin{equation}\label{eq:#1}}
\newcommand{\be}{\begin{equation}}
\newcommand{\ee}{\end{equation}}

\newcommand{\aerz}[2]{a^{#1}(\vec{#2},\beta)^{+}}
\newcommand{\aver}[2]{a^{#1}(\vec{#2},\beta)}
\newcommand{\aerzt}[2]{\tilde{a}^{#1}(\vec{#2},\beta)^{+}}
\newcommand{\avert}[2]{\tilde{a}^{#1}(\vec{#2},\beta)}

\newcommand{\berz}[2]{b_{#1}(\vec{#2},\beta)^{+}}
\newcommand{\bver}[2]{b_{#1}(\vec{#2},\beta)}
\newcommand{\berzt}[2]{\tilde{b}_{#1}(\vec{#2},\beta)^{+}}
\newcommand{\bvert}[2]{\tilde{b}_{#1}(\vec{#2},\beta)}

\newcommand{\derz}[2]{d_{#1}(\vec{#2},\beta)^{+}}
\newcommand{\dver}[2]{d_{#1}(\vec{#2},\beta)}
\newcommand{\derzt}[2]{\tilde{d}_{#1}(\vec{#2},\beta)^{+}}
\newcommand{\dvert}[2]{\tilde{d}_{#1}(\vec{#2},\beta)}

\newcommand{\unobar}[3]{u_{#1,#2}(\vec{#3})}
\newcommand{\vnobar}[3]{v_{#1,#2}(\vec{#3})}
\newcommand{\ubar}[3]{\bar{u}_{#1,#2}(\vec{#3})}
\newcommand{\vbar}[3]{\bar{v}_{#1,#2}(\vec{#3})}

\newcommand{\propD}[3]{\; {^{\beta}}\!D_{#1}^{#2}(#3)}
\newcommand{\propS}[4]{\; {^{\beta}}\!S_{#1}^{#2}(#3)_{#4}}

\newcommand{\pnoaD}[3]{\; {^{\beta}}\!D_{#1}^{#2}{#3}}
\newcommand{\pnoaS}[4]{\; {^{\beta}}\!S_{#1}^{#2}{#3}_{#4}}

\newcommand{\propDft}[3]{\; {^{\beta}}\!\widehat{D}_{#1}^{#2}(#3)}
\newcommand{\propSft}[4]{\; {^{\beta}}\!\widehat{S}_{#1}^{#2}(#3)_{#4}}

\newcommand{\propDtgl}[3]{\; {^{T=0}}\!D_{#1}^{#2}(#3)}
\newcommand{\propStgl}[4]{\; {^{T=0}}\!S_{#1}^{#2}(#3)_{#4}}

\newcommand{\propDtgr}[3]{\; {^{T>0}}\!D_{#1}^{#2}(#3)}
\newcommand{\propStgr}[4]{\; {^{T>0}}\!S_{#1}^{#2}(#3)_{#4}}

\newcommand{\propDfttgl}[3]{\; {^{T=0}}\!\widehat{D}_{#1}^{#2}(#3)}
\newcommand{\propSfttgl}[4]{\; {^{T=0}}\!\widehat{S}_{#1}^{#2}(#3)_{#4}}

\newcommand{\propDfttgr}[3]{\; {^{T>0}}\!\widehat{D}_{#1}^{#2}(#3)}
\newcommand{\propSfttgr}[4]{\; {^{T>0}}\!\widehat{S}_{#1}^{#2}(#3)_{#4}}

\newcommand{\propDoa}[3]{\; {^{\beta}}\!D_{#1}^{#2}#3}
\newcommand{\propSoa}[4]{\; {^{\beta}}\!S_{#1}^{#2}#3#4}

\newcommand{\slas}[1]{\slash \!\!\! #1}

\newcommand{\thermroot}[1]{<\!\Theta \sqrt{f_{\pm}(#1)}\!>}
\newcommand{\thermf}[1]{<\!\Theta f_{\pm}(#1)\!>}                

\newcommand{\wre}{\!\!:}
\newcommand{\wl}{:\!\!}

\newcommand{\sign}{\mathrm{sign}}
\newcommand{\tr}{\mathrm{tr}}
\newcommand{\supp}{\mathrm{supp}}

\newcommand{\F}{\mathrm{F}}
\newcommand{\ret}{\mathrm{ret}}
\newcommand{\av}{\mathrm{av}}

\newcommand{\ve}{\mathrm{V}}
\newcommand{\bs}{\mathrm{BS}}
\newcommand{\se}{\mathrm{SE}}

\renewcommand{\theequation}{\thesection.\arabic{equation}}

\maketitle
\thispagestyle{empty}

\vskip -0.2cm
\renewcommand{\thefootnote}{\fnsymbol{footnote}}
\centerline{Adrian M\"uller\footnote{\texttt{amueller@physik.unizh.ch}}}
\renewcommand{\thefootnote}{\arabic{footnote}}
\setcounter{footnote}{0}
\vskip 2mm

\centerline{\it{Institut f\"ur Theoretische Physik der Universit\"at
Z\"urich}}
\centerline{\it{Winterthurerstrasse 190, CH-8057 Z\"urich, Switzerland}}

\vskip 1.7cm

\begin{abstract}
\noindent
In this work quantum electrodynamics at $T\!>\!0$ is considered. 
For this purpose we use thermo field dynamics 
and the causal approach to quantum field theory 
according to Epstein and Glaser, the latter being a rigorous method to 
avoid the well-known ultraviolet divergencies of quantum field theory. 
It will be shown that the 
theory is infrared divergent if the usual scattering states are used. The 
same is true if we use more general mixed states. This is  
in contradiction to the results established in the literature, and 
we will point out why these earlier approaches fail to describe the 
infrared behaviour correctly. We also calculate the thermal corrections to the 
electron magnetic moment in the low temperature approximation 
$k_{{\scriptscriptstyle\mathrm{B}}}T\ll m_{\mathrm{e}^{-}}$. 
This is done by investigating 
the scattering of an electron on a C-number potential in third order 
in the limit of small momentum transfer $p\rightarrow q$. We reproduce one of 
the different results reported up to now in literature. In the low temperature 
approximation infrared finiteness is recovered in a very straightforward way:  
In contrast to the literature we do not have to introduce a 
thermal Dirac equation or thermal spinors.     
\vskip 2cm
\centerline{{\bf PACS number:} 11.10.Wx, 14.60.Cd, 12.20.Ds}
\vskip 2mm
\centerline{{\bf Keywords:} Thermal Field Theory, Infrared Divergencies}
\vskip 2mm
\centerline{{\bf Preprint:} ZU-TH 37/1999}
\vskip 7mm
\centerline{\large{\texttt{hep-th/9912240}}}
\end{abstract}
\newpage
                             
\thispagestyle{empty}
\tableofcontents
\newpage
                              
\section{Introduction}
\setcounter{equation}{0}
In this work we are concerned with quantum electrodynamics (QED) 
at finite temperatures. The main result 
is that this theory is infrared (IR) divergent already in the cross sections 
to fourth order, a result in contradiction to the literature where  
calculations are carried out for all temperatures (\cite{indum1}, 
\cite{ahmedsaleem}, \cite{PlaTa} and \cite{tyr} - \cite{keil}). At 
the end of Sec. \ref{subsubsec:disccan} we give a detailed review of the 
literature and point out where the shortcomings lie in those calculations.
 
If we restrict ourselves to small temperatures $k_{{\scriptscriptstyle
\mathrm{B}}}T\ll 
m_{\mathrm{e}^{-}}$ we reproduce the IR-finite results found in the 
literature (e.g. \cite{DHR}), 
but in a much more straightforward way. 

We also present a 
rigorous discussion of the thermal corrections to the magnetic moment of the 
electron in this low temperature approximation and reproduce the value 
of \cite{Spanier}. After having pointed out these 
results we now present some introductory remarks on the methods that will 
be applied. 
 
For the field theoretic aspect of QED at finite temperatures  
we utilize the causal approach to quantum field theory 
(QFT) because this provides us with 
a framework which is free of ultraviolet (UV) divergencies and allows a  
stringent investigation of the IR-behaviour (\cite{Ep-Gl}, 
\cite{ScharfBuch}). We will present an 
introduction to this theory in Sec. \ref{subsec:cauappQFT} where we 
will especially emphasize how the UV-divergencies and subsequent 
regularization and renormalization are avoided 
by carefully treating the involved mathematical objects 
as what they are, i.e. distributions and not functions: the origin of 
the UV-divergencies in the standard framework lies in  
ill-defined products of distributions. 

In addition, the use of the causal method avoids the so-called pinched 
singularities which arise through products of $\delta$-distributions. They  
stem from the Feynman-rules using thermal propagators. The causal method 
avoids such ill-defined products by construction. 

To include finite temperature we use the ideas of thermo field dynamics 
(TFD). The basics of this theory (\cite{LandsvWeert}, \cite{ojima}, 
\cite{TFDBuch}) will be presented in Sec. \ref{subsec:ftft}.

In Sec. \ref{sec:calc} we calculate all second order and 
some third order graphs we will need to investigate the IR-behaviour. 
The second order self energy contribution will be used to  
discuss the thermal corrections to the electron mass. There we reproduce the 
results of the literature (\cite{DHR}, \cite{Spanier}, etc.). 
The vertex graph will be the basis for our 
calculation of the corrections to the magnetic moment $\mu_{\mathrm{e}^{-}}$ 
of the electron. These corrections 
are unfortunately too small to be tested by current experimental techniques. 
The only experiments we are aware of that perhaps could detect finite 
temperature 
field theoretic effects in the not too distant future are measurements of 
the Casimir force (see \cite{cas1}, \cite{cas2} and references 
therein) and of energy level shifts in highly excitated atoms 
(Rydberg states), see \cite{ryd} (and more recent work on this topic). 

Finally we present all terms that have to be considered for the IR-behaviour 
of the theory in the adiabatic limit (which will be explained in Sec. 
\ref{subsubsec:caussm}) and investigate if their divergent parts cancel or 
not. This is done in Sec. \ref{subsubsec:crocrocro} and 
\ref{subsubsec:poscon} with the before-mentioned result.

A short discussion of this result and the relation to the literature can 
be found in Sec. \ref{subsubsec:disccan}. To conclude, we present some 
ideas that possibly could help to formulate an IR-finite thermal 
theory in Sec. \ref{sec:conclus}.
\\ \\
Throughout this paper we use units with $\hbar=1=c$ if not mentioned 
otherwise and define the Fourier transform as follows:
\bel{ftraf}
\hat{f}(p):=\frac{1}{(2\pi)^2}\int\!d^4xf(x)e^{ipx}.
\end{equation}
To keep this in mind is important for the correct factors $(2\pi)$.  
A remark on the notation: we often 
simplify the notation by the definition $m:=m_{\mathrm{e}^{-}}$ and thus 
write for the electron mass either $m_{\mathrm{e}^{-}}$ or just $m$. 

\section{Quantum Field Theory at Finite Temperatures}
\setcounter{equation}{0}
In this section we will present the formalism we are going to use to do  
QFT at finite temperatures. 
For the field theoretic aspect we utilize the 
so-called causal approach to QFT. To incorporate the temperature we make use 
of TFD. Both these formalisms will be introduced and 
explained in the following.
\subsection{The Causal Approach to QFT}\label{subsec:cauappQFT}
The traditional formalism of QFT is plagued with the well-known  
UV-divergencies one copes with by various regularization- and 
renormalization-procedures. 
The reason for these divergencies lies basically in the fact that one 
performs mathematically ill-defined operations by using Feynman rules or  
performing the usual time-ordering procedure that leads to them, respectively. 
There distributions (Feynman propagators or $\Theta$-functions and 
products of field operators, resp.) are multiplied as if they 
were ordinary functions, a procedure,   
which is in general not defined. This entails the appearance of 
the above-mentioned 
divergencies, which one has to get rid of by regularization and 
renormalization of physical parameters to extract 
useful information from the theory.

Instead of causing divergencies by wrong manipulations and having to cope 
with them afterwards, a better strategy is to avoid them from the 
very beginning, 
carefully paying attention to the mathematics that underlies the 
theory. 

This has been done by Epstein and Glaser in the early 
seventies \cite{Ep-Gl} and then further developed by Scharf et al. 
\cite{ScharfBuch}. 

This so-called causal (since causality plays a crucial r\^ole in the 
formalism we will soon present)  or finite 
(because there do not arise any 
UV-divergencies) QFT is a perturbative 
$S$-Matrix theory. Using some basic physical assumptions as 
causality and Poincar\'e-invariance the $S$-Matrix is constructed inductively 
order by order. In order to do that the basic fields of the theory 
(described by well-defined \emph{free} fields on Fock-space), 
their (anti-)commutation 
relations and how they couple in first order (which corresponds to 
the interaction  
Lagrangian in the standard formalism) have to be known.  
These free fields and the 
$S$-Matrix are operator-valued distributions. To extract physical 
information the relevant objects 
(e.g. $S$-matrix elements) have to be smeared out 
with test functions $g$, suitably falling off at infinity (e.g.\footnote{
$\mathbb{M}$ denotes the Minkowski space, i.e. the manifold $\mathbb{R}^4$ 
furnished with the Lorentz scalar product based on the metric tensor 
$g_{\mu\nu} = \mathrm{diag}(1,-1,-1,-1)$.} $g \in 
\mathcal{S}(\mathbb{M})$). These test functions cut off the long 
range part of the interactions. To end up with physically 
meaningful results the so called adiabatic limit 
$g \rightarrow 1$, which restores the total interaction has to be considered. 
This limit may not exist (since $1 \notin \mathcal{S}(\mathbb{M})$). 

If $g \in \mathcal{S}(\mathbb{M})$, $S(g)$ is well defined and free of 
UV- or IR-divergencies. The latter can show up in the 
limit $g \rightarrow 1$, e.g. in QED  
(because of the massless photon, which implies a 
long-range interaction like the Coulomb-Potential) for 
S-matrix elements. But even there (in QED at zero temperature) the adiabatic 
limit exists 
if the right physically measurable 
quantities (inclusive cross sections) are considered, making QED at $T\!=\!0$ 
a physically meaningful theory (in the cross sections at least up to fourth 
order - up to this level the existence and uniqueness of the adiabatic limit 
has been proven, see \cite{ScharfBuch}).

This last remark alludes to another aspect of the causal approach. One has 
to be modest; many results known in the standard 
theory are not yet rigorously verified. For example, it is difficult to 
prove non-perturbative statements. Likewise a rigorous proof of the 
cancellation of IR-singularities to all orders \`a la Grammar-Yennie 
\cite{Gra-Ye} is not yet known - to mention just two things. 
On the other side, by calculating on sound mathematical grounds one  
always knows exactly what is going on - and there is no other possibility to 
achieve this knowledge.  

\subsubsection{The Causal Construction of the \boldmath{$S$}-Matrix}
\label{subsubsec:caussm}
 
Here we want to present the causal formalism. For more details than we give 
here we refer the reader to \cite{ScharfBuch}, \cite{AsteDiss} and 
\cite{actapolon}. 

In quantum mechanical scattering 
theory the $S$-Matrix can be written under certain assumptions as a 
time-ordered exponential of the interaction Lagrangian (cf. e.g. Sec. 0.3. in 
\cite{ScharfBuch}). This and the work of St\"uckelberg and Bogoljubov, 
Shirkov\footnote
{For a short historical introduction and references we refer to 
the introductory remarks in \cite{ScharfBuch}.} motivated Epstein 
and Glaser to start in field theory with the following 
\emph{formal}\footnote{Throughout this paper we use the notation `formal' 
to denote infinite series without having control of convergence, 
which nevertheless can be treated 
in the well-defined setting of \cite{hardy}.} power series in the coupling 
constant:
\begin{equation}\label{eq:smatr}
S(g) = 1+\sum_{n=1}^{\infty}\frac{1}{n!}\int\!d^4x_{1}...d^4x_{n}
  T_{n}(x_{1},...,x_{n})g(x_{1})...g(x_{n}).
\end{equation}
We cannot make any statement about the convergence of this series - that is 
a main content of the expression \emph{formal}. But every $T_{n}$ will 
turn out to be  a well defined 
operator-valued distribution. After smearing it out with the test function 
$g \in \mathcal{S}(\mathbb{M})$, it is a well defined operator on Fock-space
\footnote{Furthermore, instead of treating the formulae to come in the 
context of formal power series, here it is always possible to take the 
sums up to a convenient large enough but finite value $m \in \mathbb{N}$ 
instead of $\infty$. (\ref{eq:invt}) would then be used as a definition, 
motivated by (\ref{eq:entwinvsmatr}); (\ref{eq:tcaus}) etc. can then be 
derived by calculations with polynomials.}. From the construction we will 
describe in the following, it is clear that $T_{n}$ is of the form 
\bel{tnnumeric}
T_{n}(x_{1},...,x_{n}) = \sum_{k}t^{k}_{n}(x_{1}-x_{n},...,x_{n-1}-x_{n})
  :\mathcal{O}_{k}(x_{1},...,x_{n}):, 
\end{equation}
where $:\mathcal{O}_{k}:$ is a normally ordered product of free 
field operators (a Wick monomial), and the numerical\footnote{Here `numerical' 
means that $d$ contains no 
field operators; but it does not mean that it is necessarily a scalar.} 
distribution $t^{k}_{n}$ 
depends only on relative coordinates as a consequence of translation 
invariance. 
 
The so-called $n$-point distributions or $n$-point functions 
$T_{n}$ will be constructed 
inductively order by order from $T_{1}$ 
which has to be known from the beginning and is given as a Wick-ordered 
product of free fields on 
Fock-space\footnote{E.g. for QED at $T\!=\!0$:
$T_{1}(x) := ie\wl\bar{\Psi}(x)\gamma^{\mu}\Psi(x)\wre A_{\mu}(x)$.}, i.e. 
expressed by well-defined objects only.
 
In standard field theory $T_{n}$ would be constructed as follows:
\begin{multline}\label{eq:Tord}
T_{n}(x_{1},...,x_{n}) = \mathcal{T}\big\{T_{1}(x_{1})...T_{1}(x_{n})\big\} \\ 
:= \sum_{\Pi}\Theta(x_{\Pi(1)}^0-x_{\Pi(2)}^0)\cdot...\cdot
  \Theta(x_{\Pi(n-1)}^0-x_{\Pi(n)}^0)T_{1}(x_{\Pi(1)})\cdot...\cdot
  T_{1}(x_{\Pi(n)}),
\end{multline}
where $\mathcal{T}$ denotes the time-ordering operator, $T_{1}$ the usual 
interaction Lagrangian and the sum runs over all 
$n!$ permutations $\Pi$ of the $n$ variables $x_{1},...,x_{n}$. This leads to 
the above-mentioned UV-divergencies; 
the reason is that the time-ordering by multiplication with 
$\Theta$-functions and subsequently the Feynman rules are usually 
not well defined procedures. 
The causal construction, which we will present now, is essentially a 
method to do this time-ordering in a well-defined way. 
 
The expansion of the $S$-Matrix in (\ref{eq:smatr}) can be formally 
inverted\footnote{Some remark 
concerning the typography: the literature on the causal method usually  
denotes $\overset{\frown}{T}_{n}$ by $\widetilde{T}_{n}$. Since we 
already want to use the `$\;\widetilde{\;\;}\;$'-symbol 
to denote a different 
object in the thermal field theory context in agreement with the literature 
of this latter area, we decided to change the notation in the causal 
context.}:
\begin{equation}\label{eq:invsmatr}
S(g)^{-1} = 1+\sum_{n=1}^{\infty}\frac{1}{n!}\int\!d^4x_{1}...d^4x_{n}
  \overset{\frown}{T}_{n}(x_{1},...,x_{n})g(x_{1})...g(x_{n}).
\end{equation}
By definition (c.f. (\ref{eq:smatr}) and (\ref{eq:invsmatr})), 
$T_{n}$ and $\overset{\frown}{T}_{n}$ are symmetric in 
$x_{1},...x_{n}$. For the disordered set of $n$ points in $\mathbb{M}$ we 
write 

\begin{equation}\label{eq:X}
X = \{x_{j} \in \mathbb{M} \;\vert\; j = 1,...n\}.
\end{equation}
If we write $S := 1+T$, then formally 
\begin{equation}\label{eq:entwinvsmatr}
S^{-1} = (1+T)^{-1} = 1+\sum_{r=1}^{\infty}(-T)^{r}.
\end{equation}
Now we collect all terms of the same order in the last sum and get  
\begin{equation}\label{eq:invt}
 \overset{\frown}{T}_{n}(X) = \sum_{j=1}^{n}(-1)^{j}
 \sum_{P_{j}}T_{n_{1}}(X_{1})...T_{n_{j}}(X_{j}),
\end{equation}
where the second sum runs over all partitions of $X$ into $j$ 
disjoint subsets

\begin{equation}\label{eq:subx}
X = X_{1}\cup ... \cup X_{j}, \;\; X_{h} \neq \emptyset,\;\; 
\lvert X_{h} \rvert  = n_{h}, \;\; h\in\{1,...,j\}\;.
\end{equation}
We want to emphasize that the products of distributions in (\ref{eq:invt}) 
are well-defined: they are \emph{direct} products of distributions, 
because the factors have disjoint sets of arguments.

Some other relations  between the $T_{n}$ and $\overset{\frown}{T}_{n}$, 
which for example will be used to obtain  
(\ref{eq:dsupp}) 
- (\ref{eq:asupp}), can 
now be derived with the help of $1 = S(g)S(g)^{-1}$, 
(\ref{eq:smatr}), (\ref{eq:invsmatr}) and collection of all terms of the 
same order. For that we refer the reader to \cite{ScharfBuch}. 
The consequences of unitarity, translation-invariance and Lorentz-covariance 
of the 
$S$-Matrix for the properties of $T_{n}$ and $\overset{\frown}{T}_{n}$ can be 
found there as well. 

We now turn to the most important property, that is causality. Consider two 
test functions $g_{1}, g_{2} \in \mathcal{S}(\mathbb{M})$, for which there is 
a space-like surface that separates their support. Then there exists a 
Lorentz-system, in which all points of $\supp(g_{1})$ are earlier\footnote{
Definition:  
Given $x,y \in \mathbb{M};\;\; x\;\;\textrm{is earlier then}\;\;y\;\;
\textrm{in a certain Lorentz-system}:
\Longleftrightarrow x_{0} < y_{0}$ in this system.}  then all 
points of $\supp(g_{2})$. In this case we write $\supp(g_{1}) < \supp(g_{2})$.
We now require for the $S$-Matrix 
\bel{scaus}
S(g_{1}+g_{2}) = S(g_{2})S(g_{1}) \;\;\forall g_{1}, g_{2} \in 
\mathcal{S}(\mathbb{M}) \;\;\textrm{if} \;\;\supp(g_{1}) < \supp(g_{2}).
\end{equation}
Combining this with (\ref{eq:smatr}) and (\ref{eq:invsmatr}), we get after 
permuting the integration variables in a suitable way 
(see \cite{ScharfBuch}) 
\bel{tcaus}
\begin{split}
T_{n}(x_{1},...,x_{n}) &= T_{m}(x_{1},...,x_{m})T_{n-m}(x_{m+1},...,x_{n}) 
 \;\; \textrm{for} \;\; \{x_{1},...,x_{m}\} > \{x_{m+1},...,x_{n}\} \;\; 
 \textrm{and}\\
 \overset{\frown}{T}_{n}(x_{1},...,x_{n}) &= 
 \overset{\frown}{T}_{m}(x_{1},...,x_{m})
 \overset{\frown}{T}_{n-m}(x_{m+1},...,x_{n}) \;\; 
 \textrm{for} \;\; \{x_{1},...,x_{m}\} < \{x_{m+1},...,x_{n}\}.
\end{split}
\end{equation} 
This property shows that the $T_{n}$ are time-ordered products; 
especially if all the arguments $x_{j}$ have different temporal components 
$x_{j}^0$, we can permute them to order them in time: 
$x_{1}^0 > x_{2}^0 > ... > x_{n}^0$. Repeated application 
of (\ref{eq:tcaus}) then yields 
\bel{timeord}
T_{n}(x_{1},...,x_{n}) = T_{1}(x_{1})\cdot...\cdot T_{1}(x_{n}).
\end{equation}

Now we turn to the general inductive step to construct $T_{n}$ on condition 
that we know all 
$T_{m}, \; 1 \leq m \leq n-1$, and that they have the 
properties mentioned above. 
Using (\ref{eq:invt}) we can calculate $\overset{\frown}{T}_{m}, \; 
1 \leq m \leq n-1$. Then we define the following distributions 
\begin{align}
A_{n}'(x_{1},...,x_{n}) &:= \sum_{P_{2}}\overset{\frown}{T}_{n_{1}}(X)
   T_{n-n_{1}}(Y,x_{n}), \label{eq:Aprime}\\
R_{n}'(x_{1},...,x_{n}) &:= \sum_{P_{2}}T_{n-n_{1}}(Y,x_{n}) 
   \overset{\frown}{T}_{n_{1}}(X) \label{eq:Rprime},  
\end{align}
where the sum runs over all partitions 
\bel{part2}
P_{2} : \{x_{1},...,x_{n-1}\} = X \cup Y, \;\; X \neq \emptyset,
\end{equation}
into two disjoint subsets with $\lvert X \rvert = n_{1}, \; \lvert Y \rvert 
 = n - 1 - n_{1}$. We also define 
\bel{dn}
D_{n}(x_{1},...,x_{n}) := R_{n}'(x_{1},...,x_{n}) - A_{n}'(x_{1},...,x_{n})\; .
\end{equation}
Then we define two other distributions, just extending the sums in 
(\ref{eq:Aprime}) and (\ref{eq:Rprime}) to all partitions $P_{2}^0$, including 
the empty set $X = \emptyset$:
\bel{An}
\begin{split}
A_{n}(x_{1},...,x_{n}) &:= \sum_{P_{2}^0}\overset{\frown}{T}_{n_{1}}(X)
   T_{n-n_{1}}(Y,x_{n}) \\
 & = A_{n}'(x_{1},...,x_{n}) + T_{n}(x_{1},...,x_{n}), 
\end{split}
\end{equation}
\bel{R}
\begin{split}
R_{n}(x_{1},...,x_{n}) &:= \sum_{P_{2}^0}T_{n-n_{1}}(Y,x_{n}) 
   \overset{\frown}{T}_{n_{1}}(X) \\
 & = R_{n}'(x_{1},...,x_{n}) + T_{n}(x_{1},...,x_{n}).
\end{split}
\end{equation}
The distributions $A_{n}, R_{n}$ are not known to us since they 
contain the unknown $T_{n}$. But we know the difference 
\bel{diffd}
D_{n} = R_{n}' - A_{n}' = R_{n} - A_{n}. 
\end{equation}
Now we will show, and this is the crucial step of the causal construction, 
how $R_{n}$ (or $A_{n}$) can be determined separately, thus making possible 
the construction of $T_{n}$ by 
\bel{t}
T_{n} = A_{n} - A_{n}' = R_{n} -R_{n}'.
\end{equation}
In order to achieve that, we investigate the support properties of 
$D_{n}, A_{n}$ and $R_{n}$. This is done in \cite{ScharfBuch}. The result is 
the following: Let 
\bel{fcx}
 \overline{V}^{+}(x) := \{y \in \mathbb{M} \;\vert (y-x)^2 \geq 0,\; 
 y^0 \geq x^0\}
\end{equation}
denote the closed forward cone of $x$ and 
\bel{bcx}
 \overline{V}^{-}(x) := \{y \in \mathbb{M} \;\vert (y-x)^2 \geq 0,\; 
 y^0 \leq x^0\}
\end{equation}
its closed backward cone. The $n$-dimensional generalizations are
\bel{ndimcone}
\overline{\Gamma}_{n}^{\pm}(x) := \{ (x_{1},...,x_{n}) \in \mathbb{M}^n \vert 
 x_{j} \in \overline{V}^{\pm}(x) \; \forall \; j = 1,...,n \}.
\end{equation}
Then it can be proven inductively that $\forall n \geq 3$ 
\begin{align}
\supp(D_{n}(x_{1},...,x_{n-1},x_{n})) & \subseteq 
  \overline{\Gamma}_{n}^{+}(x_{n}) \cup 
    \overline{\Gamma}_{n}^{-}(x_{n}) \label{eq:dsupp}\\
\supp(R_{n}(x_{1},...,x_{n-1},x_{n})) & \subseteq 
   \overline{\Gamma}_{n}^{+}(x_{n})
   \label{eq:rsupp}\\
\supp(A_{n}(x_{1},...,x_{n-1},x_{n})) & \subseteq 
  \overline{\Gamma}_{n}^{-}(x_{n})
   \label{eq:asupp}.
\end{align}

Therefore, we say that `$D_{n}$ has causal 
support'\footnote{Since $\supp(D_{n})$ lies 
in the $n$-dimensional generalization of the light cone with respect to 
$x_{n}$, i.e. each $x_{i},\;\;1\leq i\leq n$ lies in the region that can 
causally influence or be influenced by $x_{n}$.}. $A_{n}$ and $R_{n}$ 
are called the `advanced' and `retarded' distribution, respectively, 
because of 
their supports being located in the backward and forward light cone, 
respectively. It follows now 
from the equations (\ref{eq:dsupp}), (\ref{eq:rsupp}) and (\ref{eq:asupp}) 
that $A_{n}$ or $R_{n}$ and therefore $T_{n}$ 
(with (\ref{eq:t})) can be constructed 
if one knows $D_{n}$ and has split it into two parts according to these 
support properties. 

Because of the special r\^ole of $x_{n}$ in this procedure, $T_{n}$ as 
calculated in (\ref{eq:t}) in general has to be symmetrizised with respect 
to the arguments.

This splitting is the crucial point in the whole causal 
construction, and we will see that it has to be done 
very carefully and 
that 
just multiplying by $\Theta$-functions, which would be the most direct 
method, will in general fail. It is exactly this splitting-procedure that 
causes the UV-divergencies if done wrongly. We will present the splitting, 
its problems and how it is done correctly right after some other remarks.

Up to now we described the general inductive step from $n-1$ to $n$. 
To start the induction we have to know $T_{1}$, to construct $D_{2}$ 
according to (\ref{eq:dn}) and to prove explicitly that $D_{2}$ has causal 
support. Then $D_{2}$ has to be split into $R_{2}$ and $A_{2}$ according 
to the causal support properties. After that $T_{2}$ is constructed using  
(\ref{eq:t}). 

Since we will have to perform the adiabatic limit $g\rightarrow 1$ 
explicitly in our calculations, we present here how it is carried out 
in the most suitable way: we choose a fixed test function 
$g_{0}\in\mathcal{S}(\mathbb{M})$ with $g_{0}(0)=1$ and consider the limit 
\bel{adlim}
g(x):=g_{0}(\epsilon x),\;\;\textrm{with}\;\;\epsilon\rightarrow 0.
\end{equation} 
This so-called scaling limit describes a method to perform the adiabatic 
limit $g\rightarrow 1$ explicitly. Then we have in momentum space\footnote{
We caution the reader that we adopt the \emph{usual} definition of the 
Fourier transformation here, \emph{differently from} 
the corresponding formula 3.11.2 in \cite{ScharfBuch}.} 
\bel{adlimmom}
\hat{g}(k)=\frac{1}{(2\pi)^2}\int\!d^4xg(x)e^{ikx} = 
 \frac{1}{(2\pi)^2}\int\!d^4xg_{0}(\epsilon x)e^{ikx} = \frac{1}{\epsilon^4}
 \hat{g}_{0}\bigg(\frac{k}{\epsilon}\bigg)
\end{equation} 
and in the limit 
\bel{limitexpl}
\lim_{g \rightarrow 1} \hat{g}(k)= \lim_{\e\rightarrow 0}\frac{1}{\epsilon^4}
 \hat{g}_{0}\bigg(\frac{k}{\epsilon}\bigg)=(2\pi)^2\delta(k).
\end{equation}
{\footnotesize\hspace{1.5 cm} 
\textsc{Proof}: This is a distributional equation. Therefore we 
consider $\hat{g}\in \mathcal{S}(\mathbb{M})$ as a regular distribution 
$\hat{g}\in \mathcal{S}(\mathbb{M})'$. Then we have $\forall \varphi \in 
\mathcal{S}(\mathbb{M})$
\begin{equation*}
\begin{split}
\lim_{g \rightarrow 1}\langle\hat{g},\varphi\rangle=\lim_{g \rightarrow 1}
\langle g,\hat{\varphi}\rangle=
\lim_{g \rightarrow 1}\int\!d^4kg(k)\hat{\varphi}(k)=\lim_{\e \rightarrow 0}
\int\!d^4kg_{0}(\e k)\hat{\varphi}(k)=\int\!d^4k\hat{\varphi}(k)=\\=\langle 1,
\hat{\varphi}\rangle = \langle\hat{1},\varphi\rangle=(2\pi)^2\langle\delta, 
\varphi\rangle.\;\;
\textrm{qed}.
\end{split}
\end{equation*}
Here we used the theorem of Lebesgue in the fourth step.}\\
This leads to another useful formula: 
\bel{adlimintformel}
\int\!d^4k \hat{g}(k)=\int\!d^4k \frac{1}{\epsilon^4}\hat{g}_{0}
 \bigg(\frac{k}{\epsilon}\bigg) = \int\!d^4k(2\pi)^2\delta(k)=(2\pi)^2.
\end{equation}
In explicite calculations, which we will do mostly in momentum space, this is 
incorporated as follows. We are interested in expressions (e.g. a 
matrix element) of the form 
\bel{genadlim}
\int d^4k_{1}...d^4k_{n}\hat{g}(k_{1})...\hat{g}(k_{n})
\widehat{T}(k_{1},...,k_{n}),
\end{equation}
where $\widehat{T}(k_{1},...,k_{n})$ usually contains some normally ordered 
field operators and numerical distributions (cf. \ref{eq:tnnumeric}). 
Using (\ref{eq:adlimmom}) this equals
\bel{genadlim2}
\frac{1}{\epsilon^{4n}}\int d^4k_{1}...d^4k_{n}\hat{g}_{0}(\frac{k_{1}}
{\epsilon})...
\hat{g}_{0}(\frac{k_{n}}{\epsilon})\widehat{T}(k_{1},...,k_{n}).
\end{equation}
Now we perform the transformations $k_{i}\rightarrow\epsilon k_{i}\;\;
\textrm{for}\;\; i=1,...,n\;\;,$ and get
\bel{genadlim3}
\int d^4k_{1}...d^4k_{n}\hat{g}_{0}(k_{1})...\hat{g}_{0}(k_{n})
\widehat{T}(\epsilon k_{1},...,\epsilon k_{n}).
\end{equation}
In this way the adiabatic limit is usually performed: 
the variables $k_{i}$, in which the whole 
expression is smeared out are replaced by $\epsilon k_{i}$, and the 
test function $\hat{g}$ is replaced by $\hat{g}_{0}$. Then some integrations 
usually have to be performed, and the limit $\epsilon \rightarrow 0$ has to 
be taken in the 
end. However, sometimes the adiabatic limit can be performed `trivially' in 
some of the variables $k_{i}$. 
This means that all the involved expressions are of such a structure that 
$\epsilon$ can be set zero 
at the very beginning\footnote{Mathematically this means 
that we can apply the theorem of Lebesgue.}, and the remaining 
$k_{i}$-integration leaves us with factors $(2\pi)^2$ 
(use (\ref{eq:adlimintformel})). 
This amounts to setting the considered $k_{i}$ equal to zero in 
(\ref{eq:genadlim}) and 
replacing the corresponding integration $\int d^4k_{i}\hat{g}(k_{i})$ by 
$(2\pi)^2$. This will be used in many of the calculations in this work. 
Another often used trick is to set the involved $k_{i}$ equal to zero 
where this is allowed, 
although the adiabatic limit is not trivial in this variable. An example is 
$A^{\mu}(p-k_{1})$ with $p^2=m^2$: this can be replaced by $A^{\mu}(p)$. 
 
\subsubsection{Splitting of Causal Distributions}\label{subsubsec:split}
As mentioned in the preceeding section the crucial step of the causal 
construction of the $S$-matrix is the splitting\footnote{A detailed 
discussion of this splitting can be found in \cite{ScharfBuch} and 
\cite{AsteDiss}.} of $D_{n}$ into an advanced and a retarded part. 
 
By construction, $D_{n}$ has the following form (for an example see 
(\ref{eq:Dzwei}) where we calculate $D_{2}$ for QED at $T\!>\!0$)
\bel{dnnumeric}
D_{n}(x_{1},...,x_{n}) = \sum_{k}d^{k}_{n}(x_{1}-x_{n},...,x_{n-1}-x_{n})
  :\mathcal{O}_{k}(x_{1},...,x_{n}):, 
\end{equation}
where $:\mathcal{O}_{k}:$ is a normally ordered product of external free 
field operators (a Wick monomial), and the numerical 
distribution $d^{k}_{n}$ 
depends only on relative coordinates as a consequence of translation 
invariance. It is also the numerical distribution $d^{k}_{n}$, which 
contains the information on the support properties of $D_{n}$, that is, 
being concerned with the splitting problem, we can forget about the 
operator structure.

Given a tempered distribution $d \in \mathcal{S}'(\mathbb{M}^{\otimes n} \cong 
\mathbb{R}^m), \; m = 4n$, with $\supp(d) \subset 
\overline{\Gamma}_{n}^{+}(0) \cup \overline{\Gamma}_{n}^{-}(0)$, 
the splitting problem can be formulated as follows: 

\begin{itemize}
\item[]Is it possible to find a pair of
tempered distributions $\{r,a\} \subset \mathcal{S}'(\mathbb{M}^{\otimes n})$ 
with $\supp(r) \subset \overline{\Gamma}_{n}^{+}(0)$,  
$\supp(a) \subset \overline{\Gamma}_{n}^{-}(0)$ and $d = r-a$ ?
\item[] The answer is `yes' 
\footnote{Even the more general problem of decomposing a 
 not necessarily causal distribution into two parts with certain properties 
 is solvable under certain assumptions; see e.g. \cite{malgrange}.}. This 
will be shown in the following.  
\end{itemize} 
A remark on the uniqueness of the decomposition 
$d = r-a$: Assume there are two solutions of the splitting problem 
$\{r_{1},a_{1}\}$ and $\{r_{2},a_{2}\}$, $d=r_{1}-a_{1}=r_{2}-a_{2}$; 
then the difference $r_{1}-r_{2} = 
a_{1}-a_{2}$ has support $\{ 0 \} \subset \mathbb{M}^{\otimes n}$ and  
therefore it is of the following form:

\bel{pointsupp}
 \sum_{\lvert \alpha \rvert}c_{\alpha}\mathcal{D}^{\alpha}\delta(x), \; 
 x \in \mathbb{M}^{\otimes n}, \;\;
 \mathcal{D}^{\alpha} := \frac{\partial^{\alpha_{1}+...+\alpha_{4n}}}
 {\partial x_{1}^{\alpha_{1}}...\partial x_{4n}^{\alpha_{4n}}}, 
\end{equation}
where $\alpha$ is a multi-index and $x = (x_{1},...,x_{4n}) \in 
\mathbb{M}^{\otimes n}$.

So the splitting problem has an ambiguity in the solution at `$0$' only. 
We will see soon that it is exactly the behaviour of the distribution $d$ at 
this point 
`$0$', which is crucial for the splitting. Depending on how `singular' 
$d$ is in this region, the splitting can simply be done  by multiplication 
with $\Theta$-functions - or not; in this latter case $d$ has to be split  
more carefully and a multiplication with $\Theta$ would generate 
divergencies. 

To have a notion of how singular a causal distribution is near `$0$' we 
introduce the `singular order' of a causal distribution, which is a 
rigorous definition of the power counting degree of UV-divergence one knows 
from standars methods\footnote{An exemplary calculation, which shows how 
the singular order determined by sloppy power counting on the one hand and 
by the rigorous procedure presented here on the other can differ, can be found 
in \cite{ScharfAsteUrs}.}.

\begin{itemize}\item[]
\textsc{Definition:} The distribution $d \in \mathcal{S}'(\mathbb{R}^m)$ has 
the \emph{quasi-asymptotics}\footnote{We use the notation introduced by the 
authors who introduced this concept: \cite {tauber}.} 
$ d_{0}\in \mathcal{S}'(\mathbb{R}^m)$ at 
$x=0$ with respect to $\rho \in \mathcal{C}^{0}([0,\infty),[0,\infty))$ if the 
limit
\bel{quasias}
 \lim_{\delta \rightarrow 0}\rho(\delta)\delta^{m}d(\delta x) = d_{0}(x) \neq 0
\end{equation} 
exists in $\mathcal{S}'(\mathbb{R}^m)$.
\end{itemize}
The equivalent definition in momentum space reads as follows:
\begin{itemize}\item[]
\textsc{Definition:} The distribution $\hat{d} \in 
\mathcal{S}'(\mathbb{R}^m)$ has 
the \emph{quasi-asymptotics} $ \hat{d}_{0}\in \mathcal{S}'(\mathbb{R}^m)$ at 
$p=\infty$ with respect to $\rho \in \mathcal{C}^{0}([0,\infty),[0,\infty))$ 
if the limit 
\bel{quasiasimp}
 \lim_{\delta \rightarrow 0}\rho(\delta)\biggl\langle\hat{d}\biggl(
 \frac{p}{\delta}\biggr),\check{\varphi}(p)\biggr\rangle  = 
 \bigl <\hat{d}_{0},\check{\varphi}\bigr > \neq 
 \bigl <0,\check{\varphi}\bigr >
\end{equation} 
exists $\forall \check{\varphi} \in \mathcal{S}(\mathbb{R}^m)$. 
\end{itemize}
The condition `$\neq 0$' is necessary to have a unique definition 
of the term `singular order' as we will see soon\footnote{See footnote 
\ref{fn:unique}.}. The quasi-asymptotics 
probes the behaviour of $d$ and $\hat{d}$ in the vicinity of $x=0$ and  
$p=\infty$, respectively (cf.\cite{ScharfBuch}).
 
We derive by a scaling transformation that (cf. \cite{ScharfBuch})
\bel{omega}
\exists \omega \in \mathbb{R}:\;\; 
\lim_{\delta \rightarrow 0}\frac{\rho(a\delta)}{\rho(\delta)} = a^{\omega}\;\;
\forall a \in (0,\infty).
\end{equation}
Thus we call $\rho(\delta)$ the power-counting function and give the following
\begin{itemize}\item[]
\textsc{Definition:} The distribution $d \in \mathcal{S}'(\mathbb{R}^m)$ 
is said to have \emph{singular order} $\omega$ if it has a quasi-asymptotics 
$d_{0}$ at $x=0$ or its Fourier transform $\hat{d}$ has a quasi-asymptotics 
$\hat{d}_{0}$ at $p = \infty$, respectively, with power-counting function 
$\rho(\delta)$ satisfying (\ref{eq:omega})\footnote{If we do not require  
$d_{0} \neq 0$ in (\ref{eq:quasias}), and the corresponding condition in 
(\ref{eq:quasiasimp}), respectively, 
any $\omega' \geq \omega$ could be chosen as 
singular order of $d$. An exception is $\omega = -\infty$: then $d_{0}=0$ 
is possible and allowed. To have a definition where this case is incorporated 
as well the singular order has to be defined as an infimum of all 
the values for which the limits (\ref{eq:quasias}) and 
(\ref{eq:quasiasimp}), respectively, -  
leaving out the condition `$\neq 0$' - exist.
\label{fn:unique}}.
\end{itemize}
For examples and some remarks we refer the reader to \cite{ScharfBuch} and 
\cite{AsteDiss}.
 
This notion of \emph{singular order} $\omega$ is crucial for the splitting 
of the distribution $d=r-a$, because it can be shown that for $\omega < 0$ 
the splitting can be done essentially by multiplication with 
$\Theta$-functions and turns out to be unique\footnote{If  
$\omega(r) \leq \omega(d)\;\; \textrm{and}\;\; \omega(a) \leq \omega(d)$ is 
required \label{fn:req}.}, whereas 
for $\omega \geq 0$ one has to proceed more carefully and ends up with 
some ambiguities of the form of terms with support in $\{ 0 \}$. 
Now we turn to the splitting problem more explicitly:

\underline{\textsc{Case 1}}: $\omega < 0$. In this case we have
\bel{rhotoinf}
\rho(\delta) \rightarrow \infty \;\;\textrm{for}\;\; \delta \rightarrow 0, 
\end{equation}
which implies
\bel{deq0}
\biggl\langle d(x),\varphi\biggl(
 \frac{x}{\delta}\biggr)\biggr\rangle  \rightarrow 
 \frac{\bigl <d_{0},\varphi\bigr >}{\rho(\delta)} \rightarrow 0 \;\;
 \textrm{for} \;\; \delta \rightarrow 0.
\end{equation}
Now we choose an arbitrary but fixed vector $v = (v_{1},...,v_{n}) \in 
\overline{\Gamma}_{n}^{+}(0)$. Then we define a $(4n-1)$-dimensional 
hyperplane by
\bel{hypl}  
H := \{x \in \mathbb{M}^{\otimes n} \vert vx = \sum_{j=1}^{n}v_{j}x_{j} = 0\}.
\end{equation}
H splits the causal support of $d$: all products $v_{j}x_{j}$ are either 
$>0$ for $x\in \overline{\Gamma}_{n}^{+}(0)$ or $<0$ for 
$x\in \overline{\Gamma}_{n}^{-}(0)$. In addition we choose a monotonous 
$C^{\infty}$-function $\chi_{0}$ with
\begin{equation}
\chi_{0}(t) = 
\begin{cases}
 0 & \textrm{for} \;\;t\leq 0\\
  s \in \left[0,1\right)& \textrm{for} \;\;0<t<1 \;\;.\\
  1 & \textrm{for} \;\;t\geq 1
\end{cases}
\end{equation}
This function is sort of a smooth version of a $\Theta$-function. 
Now it can be shown (see \cite{ScharfBuch}) that the limit
\bel{splitkl0}
\lim_{\delta \rightarrow 0}\chi_{0}\biggl(\frac{vx}{\delta}\biggr)d(x) 
=:\Theta(vx)d(x) =:r(x)
\end{equation}
exists, defining the multiplication of $d$ by a $\Theta$-function 
unambiguously\footnote{Because of the claim in footnote 
\ref{fn:req}; adding distributions with point-support in $\{0\}$ would 
violate this claim.} and independent 
of $v$. Thus a solution of the splitting problem is given by 
\bel{splitted}
r \;\;\textrm{with}\;\; \supp(r) \subset \overline{\Gamma}_{n}^{+}(0) \;\;
\textrm{and}\;\; 
a := r-d \;\;\textrm{with}\;\; \supp(a) \subset \overline{\Gamma}_{n}^{-}(0).
\end{equation}
Many calculations in field theory are best done in momentum space, 
therefore we give a formula in the form of a dispersion integral for 
$\hat{r}$ (for details and derivation: \cite{ScharfBuch}, \cite{AsteDiss}):
\bel{disp}
\hat{r}(p) = \pm \frac{i}{2\pi}\int_{-\infty}^{\infty}dt\frac{\hat{d}(tp)}
 {1-t\pm i0}, \;\; \forall p \in \overline{\Gamma}^{\pm}.
\end{equation}

\underline{\textsc{Case 2}}: $\omega \geq 0$. Now we have 
\bel{pcomgr}
 \frac{\rho(\delta)}{\delta^{\omega+1}}\rightarrow \infty 
 \;\; \textrm{for}\;\; \delta \rightarrow 0.
\end{equation}
Proceeding as in Case 1 would result in ill-defined and UV-divergent 
expressions. Nevertheless it is possible to derive 
(see \cite{ScharfBuch}) a dispersion integral\footnote{Some intuitive 
remarks concerning this integral: The factor $\frac{1}{1-t\pm i0}$ is the 
Fourier transform of a $\Theta$-function and as a convolution in 
momentum space reflects the product with $\Theta$ in $x$-space. A factor 
$\frac{1}{t \mp i0}$ in momentum space is sort of integration in $x$-space 
(remember: differentiation in $x$-space becomes multiplication by $p$ in 
momentum space; therefore a division can be seen as an integration). Singular 
order $\omega \geq 0$ means that in momentum space $d \sim p^{\omega}$; 
if we divide this $(\omega +1)$ times by $p$, we get $\frac{1}{p}$,  
which has  $\omega = -1$ and can therefore be split trivially - thus 
$\frac{d}{p^\omega}$ becomes convoluted with the Fourier transform of 
$\Theta$. This may give some intuitive idea of (\ref{eq:dispomgr}). To derive 
this formula corresponding considerations for 
the test functions, with which $d$ is smeared out, have to be incorporated 
rigorously.} for $\hat{r}$  
in a more complicated way:
\bel{dispomgr}
\hat{r}_{0}(p) = \pm \frac{i}{2\pi}\int_{-\infty}^{\infty}dt\frac{\hat{d}(tp)}
 {(t \mp i0)^{\omega + 1}(1-t\pm i0)}, \;\; \forall p \in 
 \overline{\Gamma}^{\pm}.
\end{equation}
Here the subscript `$0$' refers to a special property of this solution of 
the splitting problem, namely 
\bel{css}
 \mathcal{D}^{\alpha}_{p}\hat{r}_{0}(p)\vert_{p=0} = 0, \;\; 
 \lvert \alpha \rvert \leq \omega.
\end{equation}
This means that $\hat{r}_{0}$ is in a certain way normalized at the origin 
in momentum space. Therefore this special solution $\hat{r}_{0}$ is called 
the `central splitting solution' with normalization point $0$. There is 
also a formula for a more 
general solution with an arbitrary normalization point 
$q \in \mathbb{M}^{\otimes n}$, but it is not very handy. 

The fact that for $\omega \geq 0$ the ambiguity 
(\ref{eq:pointsupp}) of the splitting solution cannot be removed 
and remains a priori undetermined is very important. 
In momentum space (\ref{eq:pointsupp}) 
takes the form of a polynomial:
\bel{mompointsupp}  
\sum_{\lvert \alpha \rvert = 0}^{\omega}c_{\alpha}p^{\alpha}.
\end{equation}
These terms are called normalization terms and the coefficients $c_{\alpha}$ 
normalization constants; they have to be determined by further physical 
conditions as Lorentz covariance, PCT-invariance, gauge-invariance, etc. 

For more remarks on the dispersion 
integrals, how analytic continuation is used to generalize (\ref{eq:dispomgr}) 
to arbitrary $p\in\mathbb{M}^{\otimes n}$,  
which consequences the existence of perhaps infinitely many 
normalization constants has for a theory etc., we refer 
to \cite{ScharfBuch} and \cite{AsteDiss}. 
\\
\textsc{Concluding remark:} In this section we have presented the basic 
formalism of the causal approach to QFT and pointed out the main features. 
The main point is that there are no UV-divergencies in the causal approach. 
This is related to the splitting of causal distributions. For singular order 
$\omega < 0$ it can be done trivially with a unique solution, and the results 
that are obtained for 
$T_{n}$ are the same as if they would be calculated using ordinary 
Feynman rules (if $T_{n}$ does not contain any subgraph with $\omega \geq 0$; 
an example of this is QED in the case of tree graphs). For $\omega \geq 0$ the 
trivial splitting yields UV-divergencies - just as they appear 
using Feynman rules, whereas with a more careful procedure as described 
above, we end up with UV-convergent and well defined quantities.   
But in this case the solution is not unique, there remain `local terms' 
of the form (\ref{eq:pointsupp}) resp. (\ref{eq:mompointsupp}) to be 
determined. 

Furthermore, the causal approach provides us with 
a well understood formalism to tackle the IR-problem,  
which amounts to inquiring into 
the adiabatic limit $g\rightarrow 1$. There is no  
need to introduce an unphysical photon mass or other regularization schemes. 
This will be crucial 
in our calculations for QED at $T\!>\!0$. In the literature this limit 
(or the equivalent 
procedure in the formalism used) is usually done at the 
very beginning of the calculations, whereas the correct proceeding is to do it 
at the end. This causes relevant differences in the 
results, as we will see later.

The causal formalism has up to now been used to investigate several 
theories (as QED in \cite{ScharfBuch} and \cite{actapolon}, 
Yang-Mills in \cite{YM1} - \cite{YM6}, 
the standard 
model in \cite{sm1} - \cite{sm3},  
quantum gravity in \cite{schorn1} - \cite{wellmann},
lower dimensional theories in \cite{ScharfAsteUrs},  
\cite{brazil1}, \cite{brazil2}, etc.) and   
fundamental questions (as the ghost structure in \cite{KraheGeist}, 
the convergence of the $S$-Matrix in \cite{ConvS}, the fundamental r\^ole 
of gauge invariance in QFT in \cite{ginv} and \cite{ginv2}, etc.). 
For an introduction 
to the theory we refer the reader to \cite{ScharfBuch},\cite{AsteDiss} and 
\cite{actapolon}.

\subsection{Field Theory at Finite Temperatures}\label{subsec:ftft}
In this section we make some remarks on field theory at finite temperatures 
and present the formalism we will use in our calculations.

The basic aim of thermal QFT is to describe particles that are embedded in 
some system that serves as a thermal background, e.g. an electron in a 
black body radiation background, within a quantum field theoretic setting. 
In this work we will only consider thermal QED, i.e. photons, electrons and 
positrons in a black body radiation background, if temperature is high 
enough including effects of $e^{-}e^{+}$-pairs generated in the bath. We 
will not consider the implication of our results for other things of 
great interest as the quark-gluon plasma, etc.   

There are several approaches to thermal QFT. A very complete review of these 
methods and their relation among each other is given in \cite{LandsvWeert}, 
where an extensive list of references up to 1987 can be found as well. 
  
Since we want to incorporate temperature in the causal approach to QFT we 
presented in the foregoing section, 
we decided to use the ideas of TFD, the latter being 
an operator formalism without path-integrals and by its structure allowing 
a quite straightforward combination 
with the former\footnote{Another reason to use 
TFD is the fact that it seems to be extendable to describe non-shell 
quasi-particles and non-equilibrium situations 
in a quite straightforward way (\cite{hen1}, \cite{hen2}). In 
\cite{LandsvWeert}, Sec. 2.5.6, 
it is shown that TFD and other formalisms are equivalent in some 
aspects as for example 
the perturbative calculation of thermal Green functions.}. 
We demand from TFD  
to provide us with field operators, their (anti-)commutators and $T_{1}$ to 
be able to use the causal formalism. In the following we will develop the 
ideas of TFD as far as we need them to achieve this goal. Then we will 
take this thermal $T_{1}$, in our case for QED at $T\!>\!0$, and 
start to calculate several things using the causal method. We compare the 
results with standard field theoretic calculations in QED at $T\!>\!0$ and 
establish several important differences, which will be traced back to some 
shortcomings in the standard calculations. 

We want to 
emphasize that these differences between our results and the literature 
do not depend on the fact that we use the causal method. 
The causal method differs 
from the standard calculation in the UV-region. 
But the problems in thermal QFT 
arise from the IR-behaviour, which can be treated using ordinary Feynman rules 
with an IR-regulator, 
since the involved graphs all have singular order $\omega < 0$, and graphs 
with ($\omega>0$)-subgraphs have a very transparent structure (at least the 
ones we are concerned with) - cf. the remark at the end of the introduction 
to Sec. \ref{sec:calc}. One just 
needs a formulation of the theory wherein the IR-problem, say the 
adiabatic limit, can be treated in a rigorous and controlled manner, which is 
the case in the causal method.

\subsubsection{Thermo Field Dynamics} \label{subsubsec:TFD}
We will present some short introduction to the ideas of TFD. This will be done 
in a rather loose manner, without any proofs or considerations concerning 
mathematical rigor, because we just 
need some basic features of TFD in an axiomatic way to build up a 
causal thermal QFT. All the things we will present concerning the 
algebraic background, etc. just serve to put the ideas we finally take 
over into some broader context. This part may be skipped without big  
consequences for the understanding of the calculations to come, since we 
will state the 
framework of our causal theory for $T\!>\!0$ in a rigorous way in section
\ref{subsubsec:causthermth}, quite independently of the motivating 
remarks that 
follow now.   
For more detailed introductions to TFD we 
refer the reader to \cite{LandsvWeert}, \cite{ojima} and \cite{TFDBuch}.

A basic idea of TFD is as follows: We write the statistical 
mechanical expectation value of some quantity $A$ in the mixed 
state $\rho$ as an expectation 
value on a Hilbert space as in ordinary quantum mechanics:
\bel{ew}
\rho(A) = \langle\Omega_{\rho} \lvert A \rvert \Omega_{\rho}\rangle. 
\end{equation}
More concrete: let $\left \{ \lvert n \rangle \right \} \subset \mathcal{H}$ 
be an orthonormal basis of eigenstates of a Hamiltonian on a Hilbert space 
$\mathcal{H}$:
\bel{bsp}
H\lvert n\rangle = E_{n}\lvert n\rangle, \;\; \langle m \vert n \rangle = 
 \delta_{mn}.
\end{equation}
Define a state $\lvert \Omega_{\beta} \rangle \in 
\mathcal{H}\otimes\mathcal{H}$ by\footnote{The `doubling 
of degrees of freedom' that shows up here by $\mathcal{H} \otimes 
\mathcal{H}$ will be explained later.} 
\bel{thermvak}
 \lvert \Omega_{\beta} \rangle := \frac{1}{Z_{\beta}}
 \sum_{n}e^{-\beta\frac{E_{n}}{2}}\lvert n 
 \rangle \otimes \lvert n \rangle \;\; \textrm{with} \;\;
 Z_{\beta}:= \tr(e^{-\beta H})= \sum_{n}e^{-\beta E_{n}}, \; 
 \beta := \frac{1}{k_{{\scriptscriptstyle\mathrm{B}}}T}.
\end{equation}
Then we have 
$\forall A \in \mathcal{L}(\mathcal{H},\mathcal{H})$
\bel{grundidee}
\langle \Omega_{\beta}\rvert A\otimes \mathbb{I} \lvert\Omega_{\beta}\rangle 
= \frac{\tr (e^{-\beta H}A)}{\tr (e^{-\beta H})} =: \langle A \rangle.
\end{equation}

(\ref{eq:grundidee}) can be considered in an algebraic setting: Compare  
quantum mechanics in the standard vs. the algebraic\footnote{See e.g. 
\cite{Haag}, \cite{Brat-Rob}.} formulation:
\begin{displaymath}\begin{array}{rcl}
 \textrm{standard:} &  & 
\textrm{algebraic:}\\& &\\
\textrm{state:} \;\;\;\Psi \in \mathcal{H} && 
\textrm{linear functional} \;\;\omega\\
\textrm{observable:} \;\;\;A \in \mathcal{L}(\mathcal{H},\mathcal{H})&
\longleftarrow \!\!\!\!\longrightarrow& \textrm{on a}\;\;  
C^{*}\textrm{-algebra of observables} \;\; \mathcal{A}\\
\textrm{expectation value:}\;\;\; \langle \Psi\lvert A \rvert \Psi\rangle && 
\omega (A).
\end{array}\end{displaymath}
Given a $C^{*}$-Algebra $\mathcal{A}$  and a state $\omega$ on $\mathcal{A}$,  
we can construct the following objects by the so-called 
GNS-construction (see \cite{Haag}):
\begin{displaymath}\begin{array}{rl}
\textrm{A representation} &\;\; \pi_{\omega} \;\; \textrm{of} \;\; \mathcal{A} 
\;\; \textrm{on a Hilbert space}\;\; \mathcal{H}_{\omega},\\
\textrm{wherein}&\;\; \exists \Omega \in \mathcal{H}_{\omega}: \Omega \;\; 
\textrm{is cyclic, and}\\
&\;\;\omega(\mathcal{A}) = \langle \Omega \lvert \pi_{\omega}(\mathcal{A})
\Omega \rangle \;\; \forall A \in \mathcal{A}.
\end{array}\end{displaymath}
If we now take an equilibrium state in the standard formulation, e.g. the 
grand-canonical ensemble, then we can define a linear functional 
on the algebra of observables $\mathcal{A}$ as follows:
\bel{equil}
\omega_{\beta}(A) := \tr (\rho A) \;\; \forall \; A \in \mathcal{A} \;\;
\textrm{with} \;\; \rho := \frac{e^{-\beta H}}{\tr (e^{-\beta H})}.
\end{equation}
Now we carry over the time evolution of the standard formalism to the 
algebraic one and get in 
this way a condition on $\omega_{\beta}$ restricting it 
to be an equilibrium state. This condition is the so-called KMS-condition 
(see \cite{Haag}) and reads as follows:
\bel{kms}
\omega_{\beta}(A_{t}B) = \omega_{\beta}(BA_{t+i\beta})\;\;\textrm{with} 
\;\; A_{t} := e^{itH}Ae^{-itH}.
\end{equation}
A state that obeys this condition is called KMS-state.

If we now carry out the GNS-construction for a KMS-state, we get a 
formalism whose constituent objects have all the special properties necessary 
to apply the results of Tomita and Takesaki on modular algebras (TT-theory) 
(see \cite{Haag}, \cite{LandsvWeert}, \cite{ojima}). By that we finally 
have a formalism to describe states with temperature $\neq 0$, which 
formally can be built up like an ordinary Fock space quantum field theoretic 
one. $\lvert\Omega_{\beta}\rangle$ serves as the vacuum state 
(the `thermal vacuum'), 
and we can define and construct emission and absorption operators, 
field operators, propagators, etc.
An important point is that the TT-theory for the GNS-construction on 
KMS-states demands a \emph{doubling of the degrees of freedom} to correctly 
incorporate the time-evolution in the formalism. Roughly speaking one has the 
ordinary objects of the standard theory and in addition a copy of them, often 
referred to as the `tilde-objects' (because they become marked by 
`$\widetilde{\;\;}$') and the time evolution is given by means of 
$H-\widetilde{H}$. 

Historically the whole story developed quite differently. The doubling of 
fields in TFD was established mostly on more or less stringent physical 
considerations (see e.g \cite{TFDBuch}). 
Beside that the doubling was necessary to get rid of certain 
singularities (the so-called `pinched singularities', see e.g. 
\cite{LandsvWeert}, p.205) that showed up in the theory in two- and more 
loop graphs as a consequence of products 
of $\delta$-distributions (as we will see later, temperature is incorporated 
in the Feynman propagators by terms $\sim \delta(p^2-m^2)$ -  
thus ill-defined products of $\delta$'s are generated using ordinary 
Feynman rules). This 
doubling could also be the solution of the problem the author of 
\cite{steinmann} sees with the existence of a perturbative thermal theory in 
general.   
Another argument for doubling the degrees of freedom is that a similar 
structure shows up in some other formalisms for thermal QFT. 

The connection between TFD and the algebraic results mentioned above has been  
established by Ojima in \cite{ojima}.

\subsubsection{TFD in the Causal Approach} \label{subsubsec:causthermth}
In this subsection we will build up the formalism to incorporate temperature 
in the causal QFT. This will give us a Fock space formalism as it is known 
from ordinary ($T\!=\!0$)-QFT, we just will have two copies of 
every field, etc., as motivated in the foregoing section. The interpretation 
of the physical content of these `tilde-fields' is somewhat unclear and 
controversial; in our opinion the calculations have to guide us 
in this respect. Thus we will for example see that the formalism urges us to 
interpret $\langle \tilde{a}^{\mu}(\vec{k},\beta)^{+}\Omega_{\beta}\rvert$ 
as a final state with one photon of momentum $\vec{k}$ absorbed from the 
thermal vacuum - so the argument that the `tilde-objects' are as unphysical 
as ghosts, as it is sometimes claimed in literature (e.g. in \cite{kosewe}), 
cannot provide us with the 
correct point of view. We will also see that `tilde-fields' can  
show up in external legs of graphs. All these things will become clear in the 
course of our concrete calculations in section \ref{sec:calc}. Some 
considerations concerning the physical significance of the `tilde-objects' 
can be found in \cite{TFDBuch}. 
\\
Now we will define our formalism for spinor QED at $T\!>\!0$:

\textsc{Bosons}: For the photons we take two copies of absorption and 
emission operators:
\bel{bosfock}
a^{\mu}(\vec{k},\beta),\;\; a^{\mu}(\vec{k},\beta)^{+} 
\qquad \textrm{and}\qquad
\tilde{a}^{\mu}(\vec{k},\beta),\;\;\tilde{a}^{\mu}(\vec{k},\beta)^{+},
\end{equation}
which obey the commutation relations:
\bel{bosvr}
[\aver{\mu}{k},\aerz{\nu}{k'}] = \delta(\vec{k} - \vec{k'})\delta^{\mu \nu} 
= [\avert{\mu}{k},\aerzt{\nu}{k'}], \;\;\;\; =0 \;\textrm{otherwise}.
\end{equation}
The common vacuum shall be denoted by $\lvert\Omega_{\beta}\rangle$:  
$\aver{\mu}{k}\lvert\Omega_{\beta}\rangle=0$ and 
$\avert{\mu}{k}\lvert\Omega_{\beta}\rangle=0$. 

Now we define the field operators:
\begin{eqnarray}
\label{eq:bosfeldop}
A^{\mu}(x) &:=&\frac{1}{(2\pi)^{\frac{3}{2}}}\int\!\frac{d^{3}k}{\sqrt{2
\lvert\vec{k}\rvert}}\Bigg\{\sqrt{1+f(k)}\aver{\mu}{k} e^{-ikx} + 
\sqrt{f(k)}\avert{\mu}{k} e^{ikx}\nonumber\\ 
&&\pm\sqrt{f(k)}\aerzt{\mu}{k} e^{-ikx} 
\pm\sqrt{1+f(k)}\aerz{\mu}{k} e^{ikx} \Bigg\}, \nonumber\\
\tilde{A}^{\mu}(x)&:=&\frac{1}{(2\pi)^{\frac{3}{2}}}
\int\!\frac{d^{3}k}{\sqrt{2\lvert\vec{k}\rvert}}
\Bigg\{\sqrt{1+f(k)}\avert{\mu}{k} e^{ikx} + 
\sqrt{f(k)}\aver{\mu}{k} e^{-ikx}\nonumber\\ 
&&\pm\sqrt{f(k)}\aerz{\mu}{k} e^{ikx} 
\pm\sqrt{1+f(k)}\aerzt{\mu}{k} e^{-ikx} \Bigg\}.
\end{eqnarray}
Here $f(k) := \frac{1}{e^{\beta \lvert ku\rvert} -1}$ is the Bose 
distribution function, $u \in \mathbb{M}$ denotes a time like normed (i.e. 
$u_{\mu}u^{\mu} =1$) vector
\footnote
{Incorporating this vector $u$ in the theory, 
we establish a covariant formalism. Cf. the remark at the end of this 
section.} that describes the four-velocity of the background 
heat bath\footnote{According to the ideas of TFD as presented in Sec. 
\ref{subsubsec:TFD}, the Fock space vacuum $\lvert\Omega_{\beta}\rangle$ 
may be interpreted as a thermal equilibrium heath bath of temperature 
$T = \frac{1}{k_{\mathrm{B}}\beta}$.} 
(thus we have $u = (1,\vec{0})$ in the rest frame of the bath). Furthermore,  
we define $\beta := \frac{1}{k_{{\scriptscriptstyle\mathrm{B}}}T}$, 
where $k_{{\scriptscriptstyle\mathrm{B}}}$ is  
the Boltzmann constant (which 
can be set to 1 in suitable units) and $T$ denotes the temperature (as a 
Lorentz scalar\footnote{For the different possibilities how the temperature 
could transform under Lorentz transformations see \cite{t1} and references 
therein and recent developments in this area, respectively. For the 
present calculations we adopt the most popular choice: the temperature is a 
Lorentz scalar. We did not investigate the consequences these other possible 
choices for the transformation properties of the temperature would 
have for thermal QFT.\label{fn:tscal}}). From   
`$\pm$' the `$+$'-sign is to be taken for the components\footnote{Compare 
with the quantization of the radiation field in \cite{ScharfBuch}, Sec. 2.11.} 
$\mu = 1,2,3$, `$-$' for $\mu = 0$. We emphasize that in (\ref{eq:bosfeldop}) 
we implicitly have $k^2=0$ and $k_{0}=\lvert\vec{k}\rvert$. 
Incorporating this covariantly (\ref{eq:bosfeldop}) reads as follows: 
\bel{Acov}
\begin{split}
A^{\mu}(x)=&\frac{1}{(2\pi)^{\frac{3}{2}}}\int\!d^{4}k\delta(k^2)\Theta(k_{0})
\Bigg\{\sqrt{1+f(k)}\aver{\mu}{k} e^{-ikx} + 
\sqrt{f(k)}\avert{\mu}{k} e^{ikx}\\
&\pm\sqrt{f(k)}\aerzt{\mu}{k} e^{-ikx} 
\pm\sqrt{1+f(k)}\aerz{\mu}{k} e^{ikx} \Bigg\}.
\end{split}
\end{equation}  
Commutation relations are given in App. \ref{sec:contrprop}.
 
\textsc{Fermions}: As for the photons we take two copies of absorption and 
emission operators for the electrons:
\bel{elfock}
b_{s}(\vec{p},\beta), \;\; b_{s}(\vec{p},\beta)^{+} \qquad \textrm{and} 
\qquad
\tilde{b}_{s}(\vec{p},\beta), \;\;\tilde{b}_{s}(\vec{p},\beta)^{+}, 
\end{equation}
and the same for the positrons: 
\bel{posfock}
d_{s}(\vec{p},\beta), \;\; d_{s}(\vec{p},\beta)^{+} \qquad \textrm{and}
\qquad\tilde{d}_{s}(\vec{p},\beta), \;\; \tilde{d}_{s}(\vec{p},\beta)^{+}.
\end{equation}
they shall obey the following anti-commutation relations:
\begin{align}\label{fermvr}
\{\bver{s}{p},\berz{s'}{p'}\} &= \delta(\vec{p} - \vec{p'})\delta_{ss'}  
=\{\bvert{s}{p},\berzt{s'}{p'}\},\;\;\;\; =0 \; \textrm{otherwise},\\
\{\dver{s}{p},\derz{s'}{p'}\} &= \delta(\vec{p} - \vec{p'})\delta_{ss'}  
=\{\dvert{s}{p},\derzt{s}{p'}\}, \;\;\;\; =0 \; \textrm{otherwise}.
\end{align}
Again we define field operators:
\bel{fermfeldop}
\begin{split}
\Psi_{a}(x):= &\frac{1}{(2\pi)^{\frac{3}{2}}}\sum_{s}\!\int\!d^{3}p
\Bigg\{\!\sqrt{1\!-\!f_{-}(p)}\bver{s}{p}\unobar{s}{a}{p}e^{\!-ipx} - 
i\sqrt{f_{-}(p)}\berzt{s}{p}\unobar{s}{a}{p}e^{\!-ipx} + \\
& +\sqrt{1-f_{+}(p)}\derz{s}{p}\vnobar{s}{a}{p}e^{ipx} + 
i\sqrt{f_{+}(p)}\dvert{s}{p}\vnobar{s}{a}{p}e^{ipx} \Bigg\},\\
\tilde{\Psi}_{a}(x):= &\frac{1}{(2\pi)^{\frac{3}{2}}}\sum_{s}\int\!d^{3}p
\Bigg\{\sqrt{1-f_{-}(p)}\bvert{s}{p}\ubar{s}{a}{p}e^{ipx} + 
i\sqrt{f_{-}(p)}\berz{s}{p}\ubar{s}{a}{p}e^{ipx} + \\
& +\sqrt{1-f_{+}(p)}\derzt{s}{p}\vbar{s}{a}{p}e^{-ipx} - 
i\sqrt{f_{+}(p)}\dver{s}{p}\vbar{s}{a}{p}e^{-ipx} \Bigg\},
\end{split}
\end{equation}
where $f_{\pm}(p) := \frac{1}{e^{\beta (\lvert pu \rvert \pm \mu)} +1}$ 
is the Fermi-Dirac distribution function with chemical potential $\pm\mu$, and 
$u, v$ are the ordinary four-spinors, solution of the Dirac equation, as used  
in the ($T\!=\!0$)-theory. The anti-commutation relations for the field 
operators can be found in App. \ref{sec:contrprop}.

In the definition of the field operators and in the following we adopt 
the viewpoint of describing the electrons, positrons, 
photons and their tilde-conjugates by 6 different sets of emission and 
absorption operators 
in the same Fock space (with vacuum $\lvert\Omega_{\beta}\rangle$) 
as it is ordinarily done in standard QED, 
with its three different `particles'.   

Here we can 
already see from (\ref{eq:bosfeldop}) that the incorporation of the 
temperature will cause additional IR-problems since $f(k)$ diverges 
as $\frac{1}{\lvert\vec{k}\rvert}$ for $\lvert\vec{k}\rvert\rightarrow 0$.  
 
Now we want to give some motivation for the formulae (\ref{eq:bosfeldop}) and 
(\ref{eq:fermfeldop}). In TFD the following relations between the ordinary 
emission and absorption operators 
of the ($T\!=\!0$)-theory and the ones for $T\!>\!0$ can 
be established:
\bel{bogolbos}
\begin{pmatrix}
\aver{\mu}{k}\\\aerzt{\mu}{k}
\end{pmatrix}=
\begin{pmatrix}
\mathrm{ch}&\mathrm{sh} \\ \mathrm{sh}&\mathrm{ch}
\end{pmatrix}
\begin{pmatrix}
a^{\mu}(\vec{k})\\\tilde{a}^{\mu}(\vec{k})^{+}
\end{pmatrix}
\longleftrightarrow
\begin{pmatrix}
a^{\mu}(\vec{k})\\\tilde{a}^{\mu}(\vec{k})^{+}
\end{pmatrix}=
\begin{pmatrix}
\mathrm{ch}&-\mathrm{sh} \\ -\mathrm{sh}&\mathrm{ch}
\end{pmatrix}
\begin{pmatrix}
\aver{\mu}{k}\\\aerzt{\mu}{k}
\end{pmatrix},
\end{equation}
 
\bel{bogolferm}
\begin{pmatrix}
\bver{s}{p}\\i\berzt{s}{p}
\end{pmatrix}=
\begin{pmatrix}
\mathrm{c}&\mathrm{s} \\ -\mathrm{s}&\mathrm{c}
\end{pmatrix}
\begin{pmatrix}
b_{s}(\vec{p})\\i\tilde{b}_{s}(\vec{p})^{+}
\end{pmatrix}
\longleftrightarrow
\begin{pmatrix}
b_{s}(\vec{p})\\i\tilde{b}_{s}(\vec{p})^{+}
\end{pmatrix}=
\begin{pmatrix}
\mathrm{c}&-\mathrm{s} \\ \mathrm{s}&\mathrm{c}
\end{pmatrix}
\begin{pmatrix}
\bver{s}{p}\\i\berzt{s}{p}
\end{pmatrix}.
\end{equation}
Here $(\mathrm{ch})^2 := 1+f(k), \;\; (\mathrm{sh})^2 := f(k)$ 
and $\mathrm{c}^2 := 1-f_{-}(p), \;\; 
\mathrm{s}^2 := f_{-}(p)$. `$\mathrm{ch}$', `$\mathrm{sh}$' 
can be written as a cosinus hyperbolicus and  
sinus hyperbolicus, resp., of a certain angle, `$\mathrm{c}$', `$\mathrm{s}$' 
as a cosinus and sinus, respectively. Such 
transformations as in (\ref{eq:bogolbos}), (\ref{eq:bogolferm}) are called 
Bogoliubov transformations. For the implications  
of this connection between $T\!=\!0$ and $T\!>\!0$, generalizations of them or 
further results related to them we refer the reader to the 
literature (see e.g. \cite{hen1}, \cite{TFDBuch}). 
We will not consider these things here because 
they are of no 
use to us. We just take these relations to give some more motivation for our 
formulation of the thermal QFT: Given (\ref{eq:bogolbos}) and 
(\ref{eq:bogolferm}), they can be inserted into the formulae for the 
field operators at $T\!=\!0$, 
which read as follows (see any QFT book; but we adopt 
here the normalization of \cite{ScharfBuch}, Sec. 2.2 and 2.11):
\bel{photstandard}
 A_{T=0}^{\mu}(x) := \frac{1}{(2\pi)^{\frac{3}{2}}}\int\!\frac{d^{3}k}{\sqrt{2
 \lvert\vec{k}\rvert}}\Bigg\{a^{\mu}(\vec{k})e^{-ikx}\pm a^{\mu}(\vec{k})^{+}
 e^{ikx} \Bigg\},
\end{equation}
\bel{fermstandard}
 \Psi^{T=0}_{a}(x):= \frac{1}{(2\pi)^{\frac{3}{2}}}\sum_{s}\int\!d^{3}p
 \Bigg\{ b_{s}(\vec{p})\unobar{s}{a}{p}e^{-ipx} 
 +d_{s}(\vec{p})^{+}\vnobar{s}{a}{p}e^{ipx} \Bigg\},
\end{equation}
and we get (\ref{eq:fermfeldop}) and an expression quite similar\footnote{
Since we want the `$\pm$' to be associated with the emission operator 
part of the 
field for $T >0$ as well (cf. the argumentation in \cite{ScharfBuch} 
Sec. 2.11) we have to adjust these signs in a suitable manner.} to 
(\ref{eq:bosfeldop}). Pay attention to the fact that unlike in the 
($T\!=\!0$)-theory here the `positive' and `negative frequency part', resp.' 
(`$e^{\pm ikx}$') 
is not anymore associated with the absorption and emission part, 
respectively. For 
the `tilde absorption' and `- emission' operators 
it is just the other way round. 
This will evoke quite different structures compared  
to $T\!=\!0$ in some calculations.

To use the causal method we now need to know the various contractions of 
the thermal field operators (because of the Wick theorem we will use) 
and $T_{1}$ for QED at $T\!>\!0$ to start the whole procedure. 
Beside that, we give some other 
useful formulae as contractions in momentum space, causal distributions, 
Feynman propagators and some relations among them. All this, except for 
$T_{1}$, which we want do discuss now, can be found 
in appendix \ref{sec:contrprop}. 

At the end of section \ref{subsubsec:TFD} we mentioned and shortly 
motivated the doubling of the degrees of freedom in thermal QFT, especially 
the new Hamiltonian $\widehat{H}:=H-\widetilde{H}$, which  
has to be used. Based on 
this consideration we define $T_{1}$ for QED at $T\!>\!0$ as follows\footnote{
This loosely could be written as `$T_{1}^{ T=0} - 
\widetilde{T_{1}}^{T=0}$', but we want to emphasize 
that it would rather be $T_{1}+\widetilde{T_{1}}$, using our definition for 
$T_{1}$ of QED at $T\!=\!0$, which 
we take over from \cite{ScharfBuch}, and the 
ordinary definition of a `tilde-operation' as given in the literature, 
because of the factor $i$ in our $T_{1}$ (the `tilde-operation' 
takes operators $\mathcal{O}$ to operators 
$\widetilde{\mathcal{O}}$ and numbers $c \in \mathbb{C}$ to $c^{*}$, 
the complex conjugate; it relates ordinary and `tilde-' 
objects. We will not define it in a rigorous way or discuss it because 
we will not need it).}: 
\bel{teins}
T_{1}(x) :\;= ie:\!\!\bar{\Psi}(x)\gamma^{\mu}\Psi(x)\!\!:A_{\mu}(x) -
               ie:\!\!\tilde{\Psi}(x)\gamma^{\mu}\bar{\tilde{\Psi}}(x)\!\!:
               \tilde{A}_{\mu}(x).
\end{equation}
From this equation we see that there are two types of vertices in the 
theory, one with `ordinary' fields, denoted by `type 1', the other with 
`tilde-fields', denoted by `type 2'.

A remark to avoid confusion: In our formalism of QED at $T\!>\!0$ 
\emph{we only have 
thermal objects}\footnote{Beside the spinors $u,v$. These are spinors as 
used in the ($T\!=\!0$)-theory. We emphasize this because of some calculations 
in the literature where so-called `thermal spinors' and a thermal Dirac 
equation are claimed to be necessary to do QED at $T\!>\!0$ (e.g. \cite{DHR}). 
This is not the case as we will see.}. 
We only have the thermal absorption and emission operators  
from (\ref{eq:bosfock}), (\ref{eq:elfock}) and (\ref{eq:posfock}), 
and the field operators 
built up of them as in (\ref{eq:bosfeldop}) and (\ref{eq:fermfeldop}). Any 
mentioning of ($T\!=\!0$)-quantities was only done to give some motivations for 
our formalism. Thus we emphasize especially that any normal ordering (e.g. in 
$T_{1}$) is done with respect to the absorption and emission operators from 
(\ref{eq:bosfock}),(\ref{eq:elfock}) and (\ref{eq:posfock}), and that 
expectation values have to be considered between 
states generated from the thermal 
vacuum $\lvert\Omega_{\beta}\rangle$.
 
\textsc{Remark:} In this section we have constructed a covariant formalism 
for QFT at $T\!>\!0$, here especially for QED. We see that covariance need 
not be lost in a thermal QFT as often claimed because of the preferred rest 
frame distinguished by the heat bath. Using a covariant 
formulation one 
usually has a better control of the calculations\footnote{Some people 
claim for example that the differences in the value of  
$\mu_{\mathrm{e}^{-}}$ in literature originate from problems with 
non-covariant calculations (\cite{Spanier}).}. In 
appendix \ref{sec:motiv} we give some short (not rigorous) 
calculations in this 
formalism, which may serve to strengthen the confidence in it.  
 
We want to emphasize that different authors in TFD sometimes use different 
conventions, which differ by 
factors $i, -1$ etc. in the formulae, e.g. for 
the Bogoljubov transformations.  

\section{Cross Sections to Fourth Order for QED at \boldmath{$T>0$}} 
\setcounter{equation}{0}
\label{sec:calc}
After having set up the framework of our theory in the foregoing section, 
we want to calculate all second order terms of QED at $T\!>\!0$, which will 
allow a discussion of the thermal mass correction (using the self energy of 
the electron). We also need them to calculate some third order contributions 
as the vertex, which will be needed to discuss the thermal corrections to 
the magnetic moment of the electron. Furthermore, we will gather all 
contributions from first to third order to the cross section 
(for scattering of an electron on a C-number potential) to fourth order 
to investigate the IR-behaviour of the theory. 

The main result is that thermal QED, as it is usually formulated,  
is IR-divergent already in cross sections to 
fourth order if it is \emph{not} restricted to small 
temperatures $k_{{\scriptscriptstyle\mathrm{B}}}T 
\ll m_{\mathrm{e}^{-}}$ (see Sec. 
\ref{subsubsec:poscon}): 
There is \emph{no complete cancellation} of 
IR-divergencies as it is claimed in literature 
(\cite{indum1}, \cite{ahmedsaleem}, etc.). Besides that 
we get a much simpler cancellation of IR-divergencies 
than the one in the literature (\cite{DHR}, etc.) for small $T$:   
we do not have to introduce thermal spinors and a Dirac-equation for $T\!>\!0$ 
in this approximation (see Sec. \ref{subsubsec:disccan}). 
 
Furthermore, we calculate the thermal 
corrections to the electron magnetic moment (see Sec. \ref{subsec:calcmue}) 
and the thermal mass correction 
(Sec. \ref{subsubsec:thermMass}). We calculate these two quantities 
$\mu_{\mathrm{e}^{-}}$ and $m_{\mathrm{e}^{-}}$ for 
$k_{{\scriptscriptstyle\mathrm{B}}}T\ll 
m_{\mathrm{e}^{-}}$ only, since the theory is not yet good for arbitrary 
temperatures because of the beforementioned IR-divergencies, and it does not 
make sense to extract much information before we have a theory that works 
for all temperatures at our disposal.

A rather complete version of the calculations can be found here and in the 
appendices since our results depend heavily on these detailed 
somewhat tedious derivations. So we want to provide a reader, who would 
possibly like or is forced to reproduce the results with as much material 
as possible. 
The calculations to come are done by the causal method as 
described in 
section \ref{subsubsec:caussm}, starting with $T_{1}$ from 
(\ref{eq:teins}). 
Doing the calculations we will see explicitly that ordinary Feynman 
rules could be used when $\omega \leq -1$ (cf. the remark at the end of Sec. 
\ref{subsec:cauappQFT}). This is for example the 
case in the second order calculations. In the third order calculations 
we see in addition 
that graphs with subgraphs with $\omega \geq 0$ have to be treated more 
carefully - as 
described in Sec. \ref{subsubsec:split}. Thus it is \emph{not} the case as 
often claimed in literature that the thermal contributions cause 
no additional UV-problems. Surely, all the subgraphs with $\omega \geq 0$ are 
temperature independent, but in combination with the rest of the graph they 
give $T$-dependent contributions. So the essential structure of possibly new 
UV-problems is already known from the ($T\!=\!0$)-theory, but how they contribute 
to matrix elements, etc. depends on the thermal structure as well.     

\subsection{Second Order QED at $\mathbf{T>0}$}
\label{subsec:secordQED}
\subsubsection{Calculation of \boldmath{$D_{2}$}}\label{subsubsec:calcdzwei}
According to (\ref{eq:dn}), (\ref{eq:Aprime}), (\ref{eq:Rprime}) and 
(\ref{eq:invt}) we have  
\bel{dezwei}
D_{2}(x,y) = R'_{2}(x,y) - A'_{2}(x,y) = T_{1}(y)\overset{\frown}{T}_{1}(x)-
               \overset{\frown}{T}_{1}(x)T_{1}(y) = 
               \underbrace{-T_{1}(y)T_{1}(x)}_{R'} + 
               \underbrace{T_{1}(x)T_{1}(y)}_{-A'} = 
\end{equation}
\bel{de2ausf}
\begin{split}
=\bigg[ -e^2\bigg(\;\;&+:\!\!\bar{\Psi}(x)\gamma^{\mu}\Psi(x)\!\!:A_{\mu}(x)
               :\!\!\bar{\Psi}(y)\gamma^{\nu}\Psi(y)\!\!:A_{\nu}(y)\\
              &-:\!\!\bar{\Psi}(x)\gamma^{\mu}\Psi(x)\!\!:A_{\mu}(x)
       :\!\!\tilde{\Psi}(y)\gamma^{\mu}\bar{\tilde{\Psi}}(y)\!\!:
            \tilde{A}_{\mu}(y)\\  
     &-:\!\!\tilde{\Psi}(x)\gamma^{\mu}\bar{\tilde{\Psi}}(x)\!\!:
             \tilde{A}_{\mu}(x)
                :\!\!\bar{\Psi}(y)\gamma^{\nu}\Psi(y)\!\!:A_{\nu}(y)\\
     &+:\!\!\tilde{\Psi}(x)\gamma^{\mu}\bar{\tilde{\Psi}}(x)\!\!:
        \tilde{A}_{\mu}(x)
       :\!\!\tilde{\Psi}(y)\gamma^{\mu}\bar{\tilde{\Psi}}(y)\!\!:
   \tilde{A}_{\mu}(y)\;\;\bigg)\bigg]\;\;\; 
  \underbrace{-\bigg[\textrm{dito with}\quad x \leftrightarrow y\bigg]}_{+R'}
  \;\;=
\end{split}
\end{equation}
\bel{de2ausf2}
\begin{split}
=\bigg[ -e^2\bigg(\;\;&+\gamma^{\mu}_{ab}\gamma^{\nu}_{cd}
                 :\!\!\bar{\Psi}_{a}(x)\Psi_{b}(x)\!\!:        
                 :\!\!\bar{\Psi}_{c}(y)\Psi_{d}(y)\!\!:A_{\mu}(x)A_{\nu}(y)\\
                  &-\gamma^{\mu}_{ab}\gamma^{\nu}_{cd}
                 :\!\!\bar{\Psi}_{a}(x)\Psi_{b}(x)\!\!:        
                 :\!\!\tilde{\Psi}_{c}(y)\bar{\tilde{\Psi}}_{d}(y)\!\!:
                 A_{\mu}(x)\tilde{A}_{\nu}(y)\\
                  &-\gamma^{\mu}_{ab}\gamma^{\nu}_{cd}
                 :\!\!\tilde{\Psi}_{a}(x)\bar{\tilde{\Psi}}_{b}(x)\!\!:        
           :\!\!\bar{\Psi}_{c}(y)\Psi_{d}(y)\!\!:\tilde{A}_{\mu}(x)A_{\nu}(y)\\
                  &+\gamma^{\mu}_{ab}\gamma^{\nu}_{cd}
                 :\!\!\tilde{\Psi}_{a}(x)\bar{\tilde{\Psi}}_{b}(x)\!\!:        
                 :\!\!\tilde{\Psi}_{c}(y)\bar{\tilde{\Psi}}_{d}(y)\!\!:
                 \tilde{A}_{\mu}(x)\tilde{A}_{\nu}(y) \;\;\bigg)\bigg]
  \;\;\;\underbrace{
  -\bigg[\textrm{dito with}\quad x \leftrightarrow y\bigg]}_{+R'}.
\end{split}
\end{equation}
Now we use $:\!\!\bar{\Psi}\Psi\!\!: = \bar{\Psi}\Psi - 
  C(\bar{\Psi}\Psi)$, where $C(\bar{\Psi}\Psi)
:=\{\bar{\Psi}^{(-)},\Psi^{(+)}\}$  
denotes the contraction, and the general Wick theorem
to get\footnote{For shortness we denote both $\bar{\Psi}_{a,c}$ and  
$\tilde{\Psi}_{a,c}$ by $\bar{\Psi}_{a,c}$, $\Psi_{b,d}$ and  
$\bar{\tilde{\Psi}}_{b,d}$ by $\Psi_{b,d}$ in (\ref{eq:wick}) etc.}
\bel{wick}
\begin{split}
:\!\!\bar{\Psi}_{a}\Psi_{b}\!\!::\!\!\bar{\Psi}_{c}\Psi_{d}\!\!:&= 
  \bar{\Psi}_{a}\Psi_{b}\bar{\Psi}_{c}\Psi_{d}-
  \bar{\Psi}_{a}\Psi_{b}C(\bar{\Psi}_{c}\Psi_{d})-
  C(\bar{\Psi}_{a}\Psi_{b})\bar{\Psi}_{c}\Psi_{d}+
  C(\bar{\Psi}_{a}\Psi_{b})
              C(\bar{\Psi}_{c}\Psi_{d})=\\
&=:\!\!\bar{\Psi}_{a}\Psi_{b}\bar{\Psi}_{c}\Psi_{d}\!\!:-
   C(\bar{\Psi}_{a}\bar{\Psi}_{c})
    :\!\!\Psi_{b}\Psi_{d}\!\!:+
  C(\bar{\Psi}_{a}\Psi_{d})
    :\!\!\Psi_{b}\bar{\Psi}_{c}\!\!:+
  C(\Psi_{b}\bar{\Psi}_{c})
    :\!\!\bar{\Psi}_{a}\Psi_{d}\!\!:+
  \\&+
  C(\Psi_{b}\Psi_{d})
    :\!\!\bar{\Psi}_{a}\bar{\Psi}_{c}\!\!:+
  C(\bar{\Psi}_{a}\Psi_{d})
              C(\Psi_{b}\bar{\Psi}_{c})+
  C(\bar{\Psi}_{a}\bar{\Psi}_{c})
              C(\Psi_{b}\Psi_{d}).
\end{split}
\end{equation}
Since $C(\bar{\Psi}\bar{\Psi})=0,\;\;
C(\Psi\Psi)=0$ etc., we finally have 
\bel{wickend}
:\!\!\bar{\Psi}_{a}\Psi_{b}\!\!::\!\!\bar{\Psi}_{c}\Psi_{d}\!\!:=
:\!\!\bar{\Psi}_{a}\Psi_{b}\bar{\Psi}_{c}\Psi_{d}\!\!:+
  C(\bar{\Psi}_{a}\Psi_{d})
     :\!\!\Psi_{b}\bar{\Psi}_{c}\!\!:+
  C(\Psi_{b}\bar{\Psi}_{c})
     :\!\!\bar{\Psi}_{a}\Psi_{d}\!\!:+
  C(\bar{\Psi}_{a}\Psi_{d})
              C(\Psi_{b}\bar{\Psi}_{c}).
\end{equation}
So we will need the following four equations:
\bel{wickferm}
\begin{split}
\wl\bar{\Psi}_{a}(x)\Psi_{b}(x)\wre &\wl\bar{\Psi}_{c}(y)\Psi_{d}(y)\wre 
   = 
   \wl\bar{\Psi}_{a}(x)\Psi_{b}(x)\bar{\Psi}_{c}(y)\Psi_{d}(y)\wre +
   \big\{ \bar{\Psi}_{a}^{(-)}(x),\Psi_{d}^{(+)}(y) \big\}
         \wl\Psi_{b}(x)\bar{\Psi}_{c}(y)\wre +\\
   &+ \big\{\Psi_{b}^{(-)}(x),\bar{\Psi}_{c}^{(+)}(y) \big\}
         \wl\bar{\Psi}_{a}(x)\Psi_{d}(y)\wre 
 +\big\{ \bar{\Psi}_{a}^{(-)}(x),\Psi_{d}^{(+)}(y) \big\}
     \big\{\Psi_{b}^{(-)}(x),\bar{\Psi}_{c}^{(+)}(y) \big\}\\
\wl\bar{\Psi}_{a}(x)\Psi_{b}(x)\wre&
  \wl\tilde{\Psi}_{c}(y)\bar{\tilde{\Psi}}_{d}(y)\wre= 
   \wl\bar{\Psi}_{a}(x)\Psi_{b}(x)\tilde{\Psi}_{c}(y)
              \bar{\tilde{\Psi}}_{d}(y)\wre+
   \big\{ \bar{\Psi}_{a}^{(-)}(x),\bar{\tilde{\Psi}}_{d}^{(+)}(y) \big\}
         \wl\Psi_{b}(x)\tilde{\Psi}_{c}(y)\wre +\\
  &+ \big\{\Psi_{b}^{(-)}(x),\tilde{\Psi}_{c}^{(+)}(y) \big\}
         \wl\bar{\Psi}_{a}(x)\bar{\tilde{\Psi}}_{d}(y)\wre 
 +\big\{ \bar{\Psi}_{a}^{(-)}(x),\bar{\tilde{\Psi}}_{d}^{(+)}(y) \big\}
     \big\{\Psi_{b}^{(-)}(x),\tilde{\Psi}_{c}^{(+)}(y) \big\}\\
\wl\tilde{\Psi}_{a}(x)\bar{\tilde{\Psi}}_{b}(x)\wre
       &\wl\bar{\Psi}_{c}(y)\Psi_{d}(y)\wre=
   \wl\tilde{\Psi}_{a}(x)\bar{\tilde{\Psi}}_{b}(x)
        \bar{\Psi}_{c}(y)\Psi_{d}(y)\wre+
   \big\{ \tilde{\Psi}_{a}^{(-)}(x),\Psi_{d}^{(+)}(y) \big\}
         \wl\bar{\tilde{\Psi}}_{b}(x)\bar{\Psi}_{c}(y)\wre +\\
  &+ \big\{\bar{\tilde{\Psi}}_{b}^{(-)}(x),\bar{\Psi}_{c}^{(+)}(y) \big\}
         \wl\tilde{\Psi}_{a}(x)\Psi_{d}(y)\wre 
 +\big\{ \tilde{\Psi}_{a}^{(-)}(x),\Psi_{d}^{(+)}(y) \big\}
     \big\{\bar{\tilde{\Psi}}_{b}^{(-)}(x),\bar{\Psi}_{c}^{(+)}(y) \big\}\\
\wl\tilde{\Psi}_{a}(x)\bar{\tilde{\Psi}}_{b}(x)\wre&
    \wl\tilde{\Psi}_{c}(y)\bar{\tilde{\Psi}}_{d}(y)\wre= 
   \wl\tilde{\Psi}_{a}(x)\bar{\tilde{\Psi}}_{b}(x)
   \tilde{\Psi}_{c}(y)\bar{\tilde{\Psi}}_{d}(y)\wre+
   \big\{ \tilde{\Psi}_{a}^{(-)}(x),\bar{\tilde{\Psi}}_{d}^{(+)}(y) \big\}
         \wl\bar{\tilde{\Psi}}_{b}(x)\tilde{\Psi}_{c}(y)\wre +\\
  &+ \big\{\bar{\tilde{\Psi}}_{b}^{(-)}(x),\tilde{\Psi}_{c}^{(+)}(y) \big\}
         \wl\tilde{\Psi}_{a}(x)\bar{\tilde{\Psi}}_{d}(y)\wre 
  +\big\{ \tilde{\Psi}_{a}^{(-)}(x),\bar{\tilde{\Psi}}_{d}^{(+)}(y) \big\}
     \big\{\bar{\tilde{\Psi}}_{b}^{(-)}(x),\tilde{\Psi}_{c}^{(+)}(y)\big\}.
\end{split}
\end{equation}
For the photons we need 
\bel{wickphot}
A_{\mu}(x)A_{\nu}(y) = \wl A_{\mu}(x)A_{\nu}(y)\wre + \big[A_{\mu}^{(-)}(x),
        A_{\nu}^{(+)}(y) \big],
\end{equation}
and the corresponding relations for 
$A\tilde{A}, \tilde{A}A, \tilde{A}\tilde{A}$.
Inserting (\ref{eq:wickferm}) and (\ref{eq:wickphot}) in  (\ref{eq:de2ausf2}) 
we finally get the expression (\ref{eq:Dzwei}), see App. \ref{subsec:dzwei}, 
for $D_{2}$.
 
Now we have to show that $D_{2}$, eq. (\ref{eq:Dzwei}), has causal support. 
This is done in App. \ref{subsec:caussupp}. Because the 
\emph{temperature independent} parts of $D_{2}$ have causal support 
separately  (see 
\cite{ScharfBuch}) the \emph{temperature dependent} parts have to have causal 
support as well. Thus we can split the $T$-dependent and -independent parts of 
$D_{2}$ separately. The $T$-independent parts are treated in 
\cite{ScharfBuch}, 
so we can restrict ourselves to the investigation of the $T$-dependent ones. 
This will turn out to be a great advantage because for these parts the 
splitting 
can always be done trivially; they all have singular order $\omega <0$ as 
we will see soon\footnote{This is true in the $n$-th order as well: As long 
as the temperature dependent part of $D_{n}$ contains no ($T\!=\!0$)-subgraphs 
with $\omega \geq 0$ it can be split trivially.}. 

\subsubsection{Calculation of \boldmath{$T_{2}$}}\label{subsubsec:tztzcalc}
In this section we calculate $T_{2}$. We will do that separately 
for every different process  
that is `contained' in $D_{2}$ (distinguished by different operator 
structures multiplying the numerical distributions). This is 
possible because each term that describes a certain process has 
causal support on its own (cf. the discussion of the causal support 
of $D_{2}$ on page \pageref{subsec:caussupp}). For the longer 
calculations we will refer to appendix \ref{subsec:tzwei}. A remark on the 
notation: we often handle the different parts of $D_{2}$ separately. The 
`$A_{kl},...,D_{kl}$' on the left indicates from which terms the 
expressions on the right origin (cf. eq. (\ref{eq:Dzwei})). 
We also want to emphasize that sometimes 
inessential factors or coefficients are omitted in intermediate steps (e.g. 
during the splitting procedure). Furthermore, we often take track of the 
$(T\!=\!0)$-parts just to have complete formulae. These parts are treated 
exactly in \cite{ScharfBuch} and are not involved here since the $T$-dependent 
and -independent parts can be treated separately as mentioned above at the 
end of Sec. \ref{subsubsec:calcdzwei}.
\subsubsection*{The `Trivial' Graph}
For $A_{11},B_{11},C_{11},D_{11}$ we have $D_{2}=0$ in all four cases; 
$\Longrightarrow T_{2} = R_{2}-R'_{2} = -R'_{2} \Longrightarrow$
\bel{trivialT2}
\begin{split}
{^{\beta}}\!T_{2}^{\textrm{trivial}}&=\\
A_{11}:\hspace{2.0 cm}&-e^2:\bar{\Psi}(y)\gamma^{\nu}\Psi(y)\bar{\Psi}(x)
     \gamma^{\mu}
              \Psi(x)::A_{\nu}(y)A_{\mu}(x):\\
B_{11}:\hspace{2.0 cm}&+e^2:
     \bar{\Psi}(y)\gamma^{\nu}\Psi(y)\tilde{\Psi}(x)\gamma^{\mu}
              \bar{\tilde{\Psi}}(x)::A_{\nu}(y)\tilde{A}_{\mu}(x):\\
C_{11}:\hspace{2.0 cm}&+e^2:\tilde{\Psi}(y)\gamma^{\nu}\bar{\tilde{\Psi}}(y)
             \bar{\Psi}(x)\gamma^{\mu}
              \Psi(x)::\tilde{A}_{\nu}(y)A_{\mu}(x):\\
D_{11}:\hspace{2.0 cm}&-e^2:\tilde{\Psi}(y)\gamma^{\nu}\bar{\tilde{\Psi}}(y)
             \tilde{\Psi}(x)\gamma^{\mu}\bar{\tilde{\Psi}}(x):
               :\tilde{A}_{\nu}(y)\tilde{A}_{\mu}(x):\;\;.
\end{split}
\end{equation}

\subsubsection*{Electron-Electron Scattering (M\o ller Scattering)}
\bel{d2moell}
\begin{split}
D_{2}^{\textrm{M\o ller}}&=\\
A_{12}^{\alpha}+A_{12}^{\beta}:\hspace{1.3 cm}&
            -e^2:\bar{\Psi}(x)\gamma^{\mu}\Psi(x)
            \bar{\Psi}(y)\gamma_{\mu}\Psi(y):
            i\underbrace{(\propD{11}{(+)}{x-y}-\propD{11}{(+)}{y-x})}_
               {={^{T=0}}\!D(x-y)}\\
B_{12}^{\alpha}+C_{12}^{\beta}:\hspace{1.3 cm}&
           +e^2:\bar{\Psi}(x)\gamma^{\mu}\Psi(x)
            \tilde{\Psi}(y)\gamma_{\mu}\bar{\tilde{\Psi}}(y):
            i\underbrace{(\propD{12}{(+)}{x-y}-\propD{12}{(+)}{y-x})}_
               {=0}\\
C_{12}^{\alpha}+B_{12}^{\beta}:\hspace{1.3 cm}&+e^2:\tilde{\Psi}(x)\gamma^{\mu}
            \bar{\tilde{\Psi}}(x)
            \bar{\Psi}(y)\gamma_{\mu}\Psi(y):
            i\underbrace{(\propD{12}{(+)}{x-y}-\propD{12}{(+)}{y-x})}_
               {=0}\\
D_{12}^{\alpha}+D_{12}^{\beta}:\hspace{1.3 cm}&-e^2:\tilde{\Psi}(x)\gamma^{\mu}
            \bar{\tilde{\Psi}}(x)
            \tilde{\Psi}(y)\gamma_{\mu}\bar{\tilde{\Psi}}(y):
            i\underbrace{(\propD{22}{(+)}{x-y}-\propD{22}{(+)}{y-x})}_
               {={^{T=0}}\!D(y-x)}\;\;.
\end{split}
\end{equation}
Since this $D_{2}$ just contains the causal Distribution $D$, which appears 
in the $(T\!=\!0)$-theory and which 
has singular order $\omega = -2$, cf. \cite{ScharfBuch}, the splitting can be 
done trivially. Thus we have \\
$T_{2}^{\textrm{M\o ller}}=R-R'\sim $
\bel{t2moell}
\begin{split}
A_{12}^{\alpha}+A_{12}^{\beta}&: \bigg[\;{^{T=0}}\!D(x-y)\bigg]^{\ret}
      -(-\propD{11}{(+)}{y-x})=\\
&\hspace{0.5 cm}
  =\propD{11}{\ret}{x-y}-\propD{11}{(-)}{x-y}=\propD{11}{\F}{x-y},\\
B_{12}^{\alpha}+C_{12}^{\beta}&: 0-(-\propD{12}{(+)}{y-x})
         =\propD{12}{\F}{y-x},\\
C_{12}^{\alpha}+B_{12}^{\beta}&: 0-(-\propD{12}{(+)}{y-x})
          =\propD{12}{\F}{y-x},\\
D_{12}^{\alpha}+D_{12}^{\beta}&: \bigg[\propD{22}{}{x-y}\bigg]^{\ret}
      -(-\propD{22}{(+)}{y-x})=\propD{22}{\F}{x-y}.
\end{split}
\end{equation} 
With all the factors this gives finally:
\bel{t2moelltotal}
\begin{split}
T_{2}^{\textrm{M\o ller}}&=\\
A_{12}^{\alpha}+A_{12}^{\beta}:\hspace{2.0 cm}
           &-ie^2:\bar{\Psi}(x)\gamma^{\mu}\Psi(x)
            \bar{\Psi}(y)\gamma_{\mu}\Psi(y):\propD{11}{\F}{x-y}\\
B_{12}^{\alpha}+C_{12}^{\beta}:\hspace{2.0 cm}
                 &+ie^2:\bar{\Psi}(x)\gamma^{\mu}\Psi(x)
            \tilde{\Psi}(y)\gamma_{\mu}\bar{\tilde{\Psi}}(y):
             \propD{12}{\F}{y-x}\\
C_{12}^{\alpha}+B_{12}^{\beta}:\hspace{2.0 cm}
                          &+ie^2:\tilde{\Psi}(x)\gamma^{\mu}
            \bar{\tilde{\Psi}}(x)
            \bar{\Psi}(y)\gamma_{\mu}\Psi(y):
            \propD{12}{\F}{y-x}\\
D_{12}^{\alpha}+D_{12}^{\beta}:\hspace{2.0 cm}
                     &-ie^2:\tilde{\Psi}(x)\gamma^{\mu}
            \bar{\tilde{\Psi}}(x)
            \tilde{\Psi}(y)\gamma_{\mu}\bar{\tilde{\Psi}}(y):
            \propD{22}{\F}{x-y}.
\end{split}
\end{equation}

\subsubsection*{Compton Scattering}
\bel{d2comp}
\begin{split}
D_{2}^{\mathrm{C}}&=\\
A_{21}^{\alpha}+A_{31}^{\beta}:\hspace{1.5 cm}
    &+e^2\frac{1}{i}:\bar{\Psi}(y)\gamma^{\nu}
              \propS{11}{}{y-x}{}\gamma^{\mu}\Psi(x)::A_{\mu}(x)A_{\nu}(y):\\
A_{31}^{\alpha}+A_{21}^{\beta}:\hspace{1.5 cm}
                     &-e^2\frac{1}{i}:\bar{\Psi}(x)\gamma^{\mu}
              \propS{11}{}{x-y}{}\gamma^{\nu}\Psi(y)::A_{\mu}(x)A_{\nu}(y):\\
B_{21}^{\alpha}+C_{31}^{\beta}:\hspace{1.5 cm}
                      &+e^2i:\tilde{\Psi}(y)\gamma^{\nu}
              \underbrace{\propS{21}{}{y-x}{}}_{=0}\gamma^{\mu}\Psi(x):
              :A_{\mu}(x)\tilde{A}_{\nu}(y):\\
C_{31}^{\alpha}+B_{21}^{\beta}:\hspace{1.5 cm}
             &-e^2i:\tilde{\Psi}(x)\gamma^{\mu}
              \underbrace{\propS{21}{}{x-y}{}}_{=0}\gamma^{\nu}\Psi(y):
              :\tilde{A}_{\mu}(x)A_{\nu}(y):\\
B_{31}^{\alpha}+C_{21}^{\beta}:\hspace{1.5 cm}
                   &+e^2i:\bar{\Psi}(x)\gamma^{\mu}
              \underbrace{\propS{12}{}{x-y}{}}_{=0}\gamma^{\nu}
              \bar{\tilde{\Psi}}(y):
              :A_{\mu}(x)\tilde{A}_{\nu}(y):\\
C_{21}^{\alpha}+B_{31}^{\beta}:\hspace{1.5 cm}
                 &-e^2i:\bar{\Psi}(y)\gamma^{\nu}
              \underbrace{\propS{12}{}{y-x}{}}_{=0}\gamma^{\mu}
              \bar{\tilde{\Psi}}(x):
              :\tilde{A}_{\mu}(x)A_{\nu}(y):\\
D_{21}^{\alpha}+D_{31}^{\beta}:\hspace{1.5 cm}
                      &+e^2\frac{1}{i}:\tilde{\Psi}(y)\gamma^{\nu}
              \propS{22}{}{y-x}{}\gamma^{\mu}\bar{\tilde{\Psi}}(x):
              :\tilde{A}_{\mu}(x)\tilde{A}_{\nu}(y):\\
D_{31}^{\alpha}+D_{21}^{\beta}:\hspace{1.5 cm}
               &-e^2\frac{1}{i}:\tilde{\Psi}(x)\gamma^{\mu}
              \propS{22}{}{x-y}{}\gamma^{\nu}\bar{\tilde{\Psi}}(y):
              :\tilde{A}_{\mu}(x)\tilde{A}_{\nu}(y):\;\;.
\end{split}
\end{equation}
Since $\propS{11/12}{}{z}{}=\propStgl{}{}{z}{}$ 
(see App. \ref{subsec:rel}) and 
the latter has singular order $\omega = -2$ (see \cite {ScharfBuch}) 
the splitting can be done trivially as well. For $T_{2}^{\mathrm{C}}$ this 
leads to (here we choose $x_{0}>y_{0}$ without\footnote{Since we are dealing 
with distributions it does not make sense to consider their value for a 
certain argument. Thus we can avoid $x=y$. By Lorentz transformations of 
these covariant expressions we can avoid $x_{0}=y_{0}$ as well.} 
restriction of the general validity of the expressions):
\bel{t2comp}
\begin{split}
A_{21}^{\alpha}+A_{31}^{\beta}\hspace{1.7 cm}&: (\propS{11}{}{y-x}{})^{\ret}
                -\propS{11}{(+)}{y-x}{}=\\=&-\propS{11}{\av}{y-x}{}
                -\propS{11}{(+)}{y-x}{}=\propS{11}{\F}{y-x}{},\\
A_{31}^{\alpha}+A_{21}^{\beta}\hspace{1.7 cm}&:(\propS{11}{}{x-y}{})^{\ret}
                -\propS{11}{(-)}{x-y}{}=-\propS{11}{\F}{x-y}{},\\
B_{21}^{\alpha}+C_{31}^{\beta}\hspace{1.7 cm}&:0-\propS{21}{(+)}{y-x}{}
                  =\propS{21}{\F}{y-x}{},\\
C_{31}^{\alpha}+B_{21}^{\beta}\hspace{1.7 cm}&:0-\propS{21}{(-)}{x-y}{}
              =-\propS{21}{\F}{x-y}{},\\
B_{31}^{\alpha}+C_{21}^{\beta}\hspace{1.7 cm}&:0-\propS{12}{(-)}{x-y}{}
              =-\propS{12}{\F}{x-y}{},\\
C_{21}^{\alpha}+B_{31}^{\beta}\hspace{1.7 cm}&:0-\propS{12}{(+)}{y-x}{}=
                      \propS{12}{\F}{y-x}{},\\
D_{21}^{\alpha}+D_{31}^{\beta}\hspace{1.7 cm}&: (\propS{22}{}{y-x}{})^{\ret}
                -\propS{22}{(+)}{y-x}{}=\\=&-\propS{22}{\av}{y-x}{}
                -\propS{22}{(+)}{y-x}{}=\propS{22}{\F}{y-x}{},\\
D_{31}^{\alpha}+D_{21}^{\beta}\hspace{1.7 cm}&:(\propS{22}{}{x-y}{})^{\ret}
                -\propS{22}{(-)}{x-y}{}=-\propS{22}{\F}{x-y}{}.
\end{split}
\end{equation}
With all the factors we finally have 
\bel{t2comptotal}
\begin{split}
T_{2}^{\mathrm{C}}&=\\
A_{21}^{\alpha}+A_{31}^{\beta}:\hspace{1.5 cm}
                  &-ie^2:\bar{\Psi}(y)\gamma^{\nu}
              \propS{11}{\F}{y-x}{}\gamma^{\mu}\Psi(x)::A_{\mu}(x)A_{\nu}(y):\\
A_{31}^{\alpha}+A_{21}^{\beta}:\hspace{1.5 cm}&+ie^2:\bar{\Psi}(x)\gamma^{\mu}
           (-1)\propS{11}{\F}{x-y}{}\gamma^{\nu}\Psi(y)::A_{\mu}(x)A_{\nu}(y):\\
B_{21}^{\alpha}+C_{31}^{\beta}:\hspace{1.5 cm}
                    &+ie^2:\tilde{\Psi}(y)\gamma^{\nu}
              \propS{21}{\F}{y-x}{}\gamma^{\mu}\Psi(x):
              :A_{\mu}(x)\tilde{A}_{\nu}(y):\\
C_{31}^{\alpha}+B_{21}^{\beta}:\hspace{1.5 cm}
                 &-ie^2:\tilde{\Psi}(x)\gamma^{\mu}
              (-1)\propS{21}{\F}{x-y}{}\gamma^{\nu}\Psi(y):
              :\tilde{A}_{\mu}(x)A_{\nu}(y):\\
B_{31}^{\alpha}+C_{21}^{\beta}:
               \hspace{1.5 cm}&+ie^2:\bar{\Psi}(x)\gamma^{\mu}
              (-1)\propS{12}{\F}{x-y}{}\gamma^{\nu}
              \bar{\tilde{\Psi}}(y):
              :A_{\mu}(x)\tilde{A}_{\nu}(y):\\
C_{21}^{\alpha}+B_{31}^{\beta}:\hspace{1.5 cm}
              &-ie^2:\bar{\Psi}(y)\gamma^{\nu}
              \propS{12}{\F}{y-x}{}\gamma^{\mu}
              \bar{\tilde{\Psi}}(x):
              :\tilde{A}_{\mu}(x)A_{\nu}(y):\\
D_{21}^{\alpha}+D_{31}^{\beta}:\hspace{1.5 cm}
              &-ie^2:\tilde{\Psi}(y)\gamma^{\nu}
              \propS{22}{\F}{y-x}{}\gamma^{\mu}\bar{\tilde{\Psi}}(x):
              :\tilde{A}_{\mu}(x)\tilde{A}_{\nu}(y):\\
D_{31}^{\alpha}+D_{21}^{\beta}:\hspace{1.5 cm}
              &+ie^2:\tilde{\Psi}(x)\gamma^{\mu}
              (-1)\propS{22}{\F}{x-y}{}\gamma^{\nu}\bar{\tilde{\Psi}}(y):
              :\tilde{A}_{\mu}(x)\tilde{A}_{\nu}(y):\;\;.
\end{split}
\end{equation} 

\subsubsection*{Vacuum Polarization}\label{subsec:VP}
\bel{d2vacpol}
\begin{split}
D_{2}^{\mathrm{VP}}&=\\
A_{41}^{\alpha}+A_{41}^{\beta}:\hspace{1.7 cm}&+e^2\tr\bigg[\gamma^{\nu}
      \propS{11}{(-)}{y-x}{}
           \gamma^{\mu}\propS{11}{(+)}{x-y}{}-\\&\hspace{0.5 cm}-
           \gamma^{\nu}\propS{11}{(+)}{y-x}{}\gamma^{\mu}\propS{11}{(-)}{x-y}{}
           \bigg]:A_{\mu}(x)A_{\nu}(y):\\
B_{41}^{\alpha}+C_{41}^{\beta}:\hspace{1.7 cm}&+e^2\tr\bigg[\gamma^{\nu}
       \propS{21}{(-)}{y-x}{}
           \gamma^{\mu}\propS{12}{(+)}{x-y}{}-\\&\hspace{0.5 cm}-
           \gamma^{\nu}\propS{21}{(+)}{y-x}{}\gamma^{\mu}\propS{12}{(-)}{x-y}{}
           \bigg]:A_{\mu}(x)\tilde{A}_{\nu}(y):\\
C_{41}^{\alpha}+B_{41}^{\beta}:\hspace{1.7 cm}&+e^2\tr\bigg[\gamma^{\nu}
       \propS{21}{(-)}{y-x}{}
           \gamma^{\mu}\propS{21}{(+)}{x-y}{}-\\&\hspace{0.5 cm}-
           \gamma^{\nu}\propS{12}{(+)}{y-x}{}\gamma^{\mu}\propS{21}{(-)}{x-y}{}
           \bigg]:\tilde{A}_{\mu}(x)A_{\nu}(y):\\
D_{41}^{\alpha}+D_{41}^{\beta}:\hspace{1.7 cm}&+e^2\tr\bigg[\gamma^{\nu}
      \propS{22}{(-)}{y-x}{}
           \gamma^{\mu}\propS{22}{(+)}{x-y}{}-\\&\hspace{0.5 cm}-
           \gamma^{\nu}\propS{22}{(+)}{y-x}{}\gamma^{\mu}\propS{22}{(-)}{x-y}{}
           \bigg]:\tilde{A}_{\mu}(x)\tilde{A}_{\nu}(y):\;\;.
\end{split}
\end{equation}
In appendix \ref{subsec:tzwei} we calculate ${\omega^{\mathrm{VP}}}$ with the 
following result: $\omega^{\mathrm{VP}} = -\infty$ for the temperature 
dependent 
parts. Thus we can split them trivially and arrive at:   
$T_{2} = R-R' \sim $ (for the terms $A_{41}^{\alpha}+A_{41}^{\beta}$):
\bel{t2vacpolprop}
\begin{split}
\big[(T=0)\textrm{-part}\big]^{\ret}& 
     + \propStgr{11}{(-)}{y-x}{}\gamma^{\mu}\propStgl{11}{\ret}{x-y}{} +\\
    & + \propStgl{11}{\av}{y-x}{}\gamma^{\mu}\propStgr{11}{(-)}{x-y}{} +
      \propS{11}{(+)}{y-x}{}\gamma^{\mu}\propS{11}{(-)}{x-y}{}=\\
=\big[(T=0)&\textrm{-part}\big]^{\ret} +
     \propStgl{11}{(+)}{y-x}{}\gamma^{\mu}\propStgl{11}{(-)}{x-y}{}+\\
     + &\propStgr{11}{(-)}{y-x}{}\gamma^{\mu}\propStgl{11}{\ret}{x-y}{} +
     \propStgr{11}{(+)}{y-x}{}\gamma^{\mu}\propStgl{11}{(-)}{x-y}{}+\\
    + &\propStgl{11}{\av}{y-x}{}\gamma^{\mu}\propStgr{11}{(-)}{x-y}{} 
     +\propStgl{11}{(+)}{y-x}{}\gamma^{\mu}\propStgr{11}{(-)}{x-y}{}+\\
   & +  \propStgr{11}{(+)}{y-x}{}\gamma^{\mu}\propStgr{11}{(-)}{x-y}{}=\\
= [(T=0)\textrm{-part}]& 
     - \propStgr{11}{\F}{y-x}{}\gamma^{\mu}\propStgl{11}{\F}{x-y}{} -\\
    & - \propStgl{11}{\F}{y-x}{}\gamma^{\mu}\propStgr{11}{\F}{x-y}{} -
     \propStgr{11}{\F}{y-x}{}\gamma^{\mu}\propStgr{11}{\F}{x-y}{}.
\end{split}
\end{equation}
For the other terms we get
\bel{t2vacpolproport}
\begin{split}
B_{41}^{\alpha}+C_{41}^{\beta}&: D_{2} =0 \longrightarrow T_{2} = -R'_{2}\sim\\
           &\sim\propS{21}{(+)}{y-x}{}\gamma^{\mu}\propS{12}{(-)}{x-y}{} = 
            -\propS{12}{\F}{y-x}{}\gamma^{\mu}\propS{12}{\F}{x-y}{}, \\
C_{41}^{\alpha}+B_{41}^{\beta}&: D_{2} =0 \longrightarrow T_{2} = -R'_{2}\sim\\
            &\sim\propS{12}{(+)}{y-x}{}\gamma^{\mu}\propS{21}{(-)}{x-y}{} = 
            -\propS{12}{\F}{y-x}{}\gamma^{\mu}\propS{12}{\F}{x-y}{}, \\
D_{41}^{\alpha}+D_{41}^{\beta}&: \textrm{calculation as for the $A$-part 
  gives}\\
                       T_{2}  &\sim [(T=0)\textrm{-part}] 
     - \propStgr{22}{\F}{y-x}{}\gamma^{\mu}\propStgl{22}{\F}{x-y}{} -\\
    & - \propStgl{22}{\F}{y-x}{}\gamma^{\mu}\propStgr{22}{\F}{x-y}{} -
     \propStgr{22}{\F}{y-x}{}\gamma^{\mu}\propStgr{22}{\F}{x-y}{}.
\end{split}
\end{equation}
With all factors we thus have
\bel{t2vacpol}
\begin{split}
T_{2}^{\mathrm{VP}}&=\\
A_{41}^{\alpha}+A_{41}^{\beta}:\hspace{1.7 cm}&+e^2\tr\bigg[\gamma^{\nu}\big(
          [(T=0)\textrm{-part}] 
     - \propStgr{11}{\F}{y-x}{}\gamma^{\mu}\propStgl{11}{\F}{x-y}{}-\\ 
     - \propStgl{11}{\F}{y-x}{}\gamma^{\mu}&\propStgr{11}{\F}{x-y}{} 
     -\propStgr{11}{\F}{y-x}{}\gamma^{\mu}\propStgr{11}{\F}{x-y}{}\big)\bigg]
      :A_{\mu}(x)A_{\nu}(y):\\
B_{41}^{\alpha}+C_{41}^{\beta}:\hspace{1.7 cm}&+e^2\tr\bigg[\gamma^{\nu}\big(
    -\propS{12}{\F}{y-x}{}\gamma^{\mu}\propS{12}{\F}{x-y}{}\big)\bigg]
      :A_{\mu}(x)\tilde{A}_{\nu}(y): \\
C_{41}^{\alpha}+B_{41}^{\beta}:\hspace{1.7 cm}&+e^2\tr\bigg[\gamma^{\nu}\big(
    -\propS{12}{\F}{y-x}{}\gamma^{\mu}\propS{12}{\F}{x-y}{}\big)\bigg]
       :\tilde{A}_{\mu}(x)A_{\nu}(y): \\
D_{41}^{\alpha}+D_{41}^{\beta}:\hspace{1.7 cm}&+e^2\tr\bigg[\gamma^{\nu}\big(
         [(T=0)\textrm{-part}] 
     - \propStgr{22}{\F}{y-x}{}\gamma^{\mu}\propStgl{22}{\F}{x-y}{} -\\
     - \propStgl{22}{\F}{y-x}{}\gamma^{\mu}&\propStgr{22}{\F}{x-y}{} 
   \!-\!\propStgr{22}{\F}{y-x}{}\gamma^{\mu}\propStgr{22}{\F}{x-y}{}\big)\bigg]
      \!\wl\tilde{A}_{\mu}(x)\tilde{A}_{\nu}(y)\wre\;.
\end{split}
\end{equation}

\subsubsection*{Vacuum Graph}
\bel{d2vgr}
\begin{split}
D_{2}^{\mathrm{VG}}&=\\
A_{42}^{\alpha}+A_{42}^{\beta}:\hspace{2.0 cm}& +ie^2\tr\bigg[
        \gamma^{\nu}\propS{11}{(-)}{y-x}{}\gamma_{\nu}\propS{11}{(+)}{x-y}{}
                       \propD{11}{(+)}{x-y}-\\&\hspace{1.2 cm}
         -\gamma^{\nu}\propS{11}{(+)}{y-x}{}\gamma_{\nu}\propS{11}{(-)}{x-y}{}
                       \propD{11}{(+)}{y-x}\bigg]\\
B_{42}^{\alpha}+C_{42}^{\beta}:\hspace{2.0 cm}&+ie^2\tr\bigg[
        \gamma^{\nu}\propS{21}{(-)}{y-x}{}\gamma_{\nu}\propS{12}{(+)}{x-y}{}
                       \propD{12}{(+)}{x-y}-\\&\hspace{1.2 cm}
         -\gamma^{\nu}\propS{21}{(+)}{y-x}{}\gamma_{\nu}\propS{12}{(-)}{x-y}{}
                       \propD{12}{(+)}{y-x}\bigg]\\
C_{42}^{\alpha}+B_{42}^{\beta}:\hspace{2.0 cm}&+ie^2\tr\bigg[
        \gamma^{\nu}\propS{12}{(-)}{y-x}{}\gamma_{\nu}\propS{21}{(+)}{x-y}{}
                       \propD{12}{(+)}{x-y}-\\&\hspace{1.2 cm}
         -\gamma^{\nu}\propS{12}{(+)}{y-x}{}\gamma_{\nu}\propS{21}{(-)}{x-y}{}
                       \propD{12}{(+)}{y-x}\bigg]\\
D_{42}^{\alpha}+D_{42}^{\beta}:\hspace{2.0 cm}&+ie^2\tr\bigg[
        \gamma^{\nu}\propS{22}{(-)}{y-x}{}\gamma_{\nu}\propS{22}{(+)}{x-y}{}
                       \propD{22}{(+)}{x-y}-\\&\hspace{1.2 cm}
         -\gamma^{\nu}\propS{22}{(+)}{y-x}{}\gamma_{\nu}\propS{22}{(-)}{x-y}{}
                       \propD{22}{(+)}{y-x}\bigg].
\end{split}   
\end{equation}
We use the relations from appendix \ref{subsec:rel} to rewrite the 
propagators in the $D$-term in a suitable way and thus prove  
$D_{42}^{\alpha}+D_{42}^{\beta}+A_{42}^{\alpha}+A_{42}^{\beta}=0$. 
Analogously we prove  
$B_{42}^{\alpha}+C_{42}^{\beta}=0$ and $C_{42}^{\alpha}+B_{42}^{\beta}=0$. 
This gives 
\bel{vgt2}
\begin{split}
T^{\mathrm{VG}}_{2} = -R'_{2}&=\\
A_{42}^{\alpha}+A_{42}^{\beta}+D_{42}^{\alpha}+D_{42}^{\beta}:\hspace{1.5 cm}
                &+ie^2\tr\bigg[\gamma^{\nu}\propS{11}{(+)}{y-x}{}
                \gamma_{\nu}\propS{11}{(-)}{x-y}{}\propD{11}{(+)}{y-x}\bigg]\\
                &+ie^2\tr\bigg[\gamma^{\nu}\propS{22}{(+)}{y-x}{}
                \gamma_{\nu}\propS{22}{(-)}{x-y}{}\propD{22}{(+)}{y-x}\bigg]\\
B_{42}^{\alpha}+C_{42}^{\beta}:\hspace{1.5 cm}
                       &+ie^2\tr\bigg[\gamma^{\nu}\propS{21}{(+)}{y-x}{}
                \gamma_{\nu}\propS{12}{(-)}{x-y}{}\propD{12}{(+)}{y-x}\bigg]\\
C_{42}^{\alpha}+B_{42}^{\beta}:\hspace{1.5 cm}
                      &+ie^2\tr\bigg[\gamma^{\nu}\propS{12}{(+)}{y-x}{}
                \gamma_{\nu}\propS{21}{(-)}{x-y}{}\propD{12}{(+)}{y-x}\bigg].
\end{split}
\end{equation}
In appendix \ref{subsec:tzwei} we show that the adiabatic limit in the vacuum 
expectation value of the vacuum graph is of the form 
$\frac{1}{\epsilon^{4}}c$, with $c=\textrm{const.}\neq 0$. 
Thus the adiabatic limit 
does not exist. But we have yet some undefined 
normalization constants at hand from the $(\omega=4)$-splitting in the 
$(T\!=\!0)$-case (see \cite{ScharfBuch}, 4.1.27): we can split 
different parts of a distribution separately if they have causal support each 
- to do that we each time use the singular order of the part considered. This 
provides us with the retarded and advanced parts of these different parts. 
But the normalization freedom is given by the maximal singular order of all 
these parts, since the philosophy is to split all parts with the same operator 
structure together - that is with the maximal singular order of these 
parts\footnote{Because of the remarks on the uniqueness of the retarded 
and advanced parts preceeding eq. (\ref{eq:pointsupp}) 
this does not contradict the statements above.}. 
If we now choose $C_{0}=0=C_{2}$ and $C_{4}=-c$ the 
limit exists. Since $c=0$ for $T\!=\!0$ 
(as can be seen from the expressions in 
App. \ref{subsec:tzwei}) the result for $T\!=\!0$ is not changed.  

\subsubsection*{Self Energy}\label{subsec:SE}
\bel{d2se}
\begin{split}
D_{2}^{\mathrm{SE}}&=\\
A_{22}^{\alpha}+A_{32}^{\beta}:\hspace{1.0 cm}
                  +e^2:\bar{\Psi}(y)\gamma^{\nu}\bigg[
                   \propS{11}{(-)}{y-x}{}&\propD{11}{(+)}{x-y}+\\&+
                   \propS{11}{(+)}{y-x}{}\propD{11}{(+)}{y-x}\bigg]\gamma_{\nu}
                   \Psi(x):\\
A_{32}^{\alpha}+A_{22}^{\beta}:\hspace{1.0 cm}
                  -e^2:\bar{\Psi}(x)\gamma^{\nu}\bigg[
                   \propS{11}{(+)}{x-y}{}&\propD{11}{(+)}{x-y}+\\&+
                   \propS{11}{(-)}{x-y}{}\propD{11}{(+)}{y-x}\bigg]\gamma_{\nu}
                   \Psi(y):\\
B_{22}^{\alpha}+C_{32}^{\beta}:\hspace{1.0 cm}
                    -e^2:\tilde{\Psi}(y)\gamma^{\nu}\bigg[
                   \propS{21}{(-)}{y-x}{}&\propD{12}{(+)}{x-y}+\\&+
                   \propS{21}{(+)}{y-x}{}\propD{12}{(+)}{y-x}\bigg]\gamma_{\nu}
                   \Psi(x):\\
C_{32}^{\alpha}+B_{22}^{\beta}:\hspace{1.0 cm}
                       +e^2:\tilde{\Psi}(x)\gamma^{\nu}\bigg[
                   \propS{21}{(+)}{x-y}{}&\propD{12}{(+)}{x-y}+\\&+
                   \propS{21}{(-)}{x-y}{}\propD{12}{(+)}{y-x}\bigg]\gamma_{\nu}
                   \Psi(y):\\
B_{32}^{\alpha}+C_{22}^{\beta}:\hspace{1.0 cm}
                   -e^2:\bar{\Psi}(x)\gamma^{\nu}\bigg[
                   \propS{12}{(+)}{x-y}{}&\propD{12}{(+)}{x-y}+\\&+
                   \propS{12}{(-)}{x-y}{}\propD{12}{(+)}{y-x}\bigg]\gamma_{\nu}
                   \bar{\tilde{\Psi}}(y):\\
C_{22}^{\alpha}+B_{32}^{\beta}:\hspace{1.0 cm}
                    +e^2:\bar{\Psi}(y)\gamma^{\nu}\bigg[
                   \propS{12}{(-)}{y-x}{}&\propD{12}{(+)}{x-y}+\\&+
                   \propS{12}{(+)}{y-x}{}\propD{12}{(+)}{y-x}\bigg]\gamma_{\nu}
                   \bar{\tilde{\Psi}}(x):\\
D_{22}^{\alpha}+D_{32}^{\beta}:\hspace{1.0 cm}
                     +e^2:\tilde{\Psi}(y)\gamma^{\nu}\bigg[
                   \propS{22}{(-)}{y-x}{}&\propD{22}{(+)}{x-y}+\\&+
                   \propS{22}{(+)}{y-x}{}\propD{22}{(+)}{y-x}\bigg]\gamma_{\nu}
                   \bar{\tilde{\Psi}}(x):\\
D_{32}^{\alpha}+D_{22}^{\beta}:\hspace{1.0 cm}
                   -e^2:\tilde{\Psi}(x)\gamma^{\nu}\bigg[
                   \propS{22}{(+)}{x-y}{}&\propD{22}{(+)}{x-y}+\\&=
                   \propS{22}{(-)}{x-y}{}\propD{22}{(+)}{y-x}\bigg]\gamma_{\nu}
                   \bar{\tilde{\Psi}}(y):\;\;.
\end{split}
\end{equation}
The calculation of the singular order of the self energy graph (see 
appendix \ref{subsec:tzwei}) shows that $\omega^{\mathrm{SE}} = -\infty$. 
Thus we can  
again split trivially and arrive after proceeding as in section 
\ref{subsec:VP} 
at the following expression for  $T_{2} = R - R'$:  
\bel{t2seprop}
\begin{split}
\big[(T=0)\textrm{-part}\big]+\propStgr{11}{\F}{y-x}{}\propDtgl{11}{\F}{x-y}+
             \\  +\propStgl{11}{\F}{y-x}{}\propDtgr{11}{\F}{x-y}
               +\propStgr{11}{\F}{y-x}{}\propDtgr{11}{\F}{x-y}.
\end{split}
\end{equation}
This are the terms from $A_{22}^{\alpha}+A_{32}^{\beta}$. The other terms of 
$A$ and $D$ are calculated analogously and the terms of $B$ and $C$ are easy 
to calculate since  $D_{2}=0$. Finally we arrive at  

\bel{t2se}
\begin{split}
T_{2}^{\mathrm{SE}}&=\\
A_{22}^{\alpha}+A_{32}^{\beta}:\hspace{1.0 cm}
                      &+e^2:\bar{\Psi}(y)\gamma^{\nu}\bigg[
                   ((T=0)\textrm{-part})+
                   \propStgl{11}{\F}{y-x}{}\propDtgr{11}{\F}{y-x}+\\
                   &\hspace{-0.5 cm}
           +\propStgr{11}{\F}{y-x}{}\propDtgl{11}{\F}{y-x}
                   +\propStgr{11}{\F}{y-x}{}\propDtgr{11}{\F}{y-x}
                   \bigg]\gamma_{\nu}\Psi(x):\\
A_{32}^{\alpha}+A_{22}^{\beta}:\hspace{1.0 cm}
                         &-e^2:\bar{\Psi}(x)\gamma^{\nu}\bigg[
                   ((T=0)\textrm{-part})-
                   \propStgl{11}{\F}{x-y}{}\propDtgr{11}{\F}{x-y}-\\
                   &\hspace{-0.5 cm}
           -\propStgr{11}{\F}{x-y}{}\propDtgl{11}{\F}{x-y}-
                   \propStgr{11}{\F}{x-y}{}\propDtgr{11}{\F}{x-y}
                   \bigg]\gamma_{\nu}\Psi(y):\\
B_{22}^{\alpha}+C_{32}^{\beta}:\hspace{1.0 cm}
                    &-e^2:\tilde{\Psi}(y)\gamma^{\nu}\bigg[
                   \propS{21}{\F}{y-x}{}\propD{12}{\F}{y-x}
                   \bigg]\gamma_{\nu}
                   \Psi(x):\\
C_{32}^{\alpha}+B_{22}^{\beta}:\hspace{1.0 cm}
                       &+e^2:\tilde{\Psi}(x)\gamma^{\nu}\bigg[
                   -\propS{21}{\F}{x-y}{}\propD{12}{\F}{y-x}
                  \bigg]\gamma_{\nu}
                   \Psi(y):\\
B_{32}^{\alpha}+C_{22}^{\beta}:\hspace{1.0 cm}
                     &-e^2:\bar{\Psi}(x)\gamma^{\nu}\bigg[
                   -\propS{12}{\F}{x-y}{}\propD{12}{\F}{y-x}
                   \bigg]\gamma_{\nu}
                   \bar{\tilde{\Psi}}(y):\\
C_{22}^{\alpha}+B_{32}^{\beta}:\hspace{1.0 cm}
                     &+e^2:\bar{\Psi}(y)\gamma^{\nu}\bigg[
                   \propS{12}{\F}{y-x}{}\propD{12}{\F}{y-x}
                   \bigg]\gamma_{\nu}
                   \bar{\tilde{\Psi}}(x):\\
D_{22}^{\alpha}+D_{32}^{\beta}:\hspace{1.0 cm}
                   &+e^2:\tilde{\Psi}(y)\gamma^{\nu}\bigg[
                   ((T=0)\textrm{-part})+
                   \propStgl{22}{\F}{y-x}{}\propDtgr{22}{\F}{y-x}+\\
                   &\hspace{-0.5 cm}
                +\propStgr{22}{\F}{y-x}{}\propDtgl{22}{\F}{y-x}
                   +\propStgr{22}{\F}{y-x}{}\propDtgr{22}{\F}{y-x}
                    \bigg]\gamma_{\nu}
                   \bar{\tilde{\Psi}}(x):\\
D_{32}^{\alpha}+D_{22}^{\beta}:\hspace{1.0 cm}
                       &-e^2:\tilde{\Psi}(x)\gamma^{\nu}\bigg[
                   ((T=0)\textrm{-part})-
                   \propStgl{22}{\F}{x-y}{}\propDtgr{22}{\F}{x-y}-\\
                   &\hspace{-0.5 cm}
              -\propStgr{22}{\F}{x-y}{}\propDtgl{22}{\F}{x-y}
                   -\propStgr{22}{\F}{x-y}{}\propDtgr{22}{\F}{x-y}
                   \bigg]\gamma_{\nu}
                   \bar{\tilde{\Psi}}(y):\;.
\end{split}
\end{equation}

\textsc{Remark:} Up to now we have calculated all contributions to $T_{2}$. It 
can be verified quite straightforward that all these contributions are 
symmetric under the transformation $x\rightarrow y, y\rightarrow x$ (use the 
relations from App. \ref{subsec:rel}). 
We see that their temperature dependent parts are the same as if they had 
been calculated using ordinary Feynman rules (they all have 
$\omega \leq -1$), whereas the $(T\!=\!0)$-parts have to be treated more 
carefully as it was done in \cite{ScharfBuch}. To get some insight in 
the consequences of these formulae, to extract physical quantities etc. it 
remains to perform the involved integrals. We will only do that 
in the case of the self energy under the assumption that 
$k_{B}T\ll m_{\mathrm{e}^{-}}$. This will allow the discussion of a possible 
thermal mass correction and is done in the next subsection.

\subsubsection{Self Energy and Thermal Corrections to 
\boldmath{$m_{\mathrm{e}^{-}}$}}\label{subsubsec:thermMass} 
Here we want to discuss a possible thermal mass correction by the self energy. 
This is usually looked at as follows: we consider the infinite summation of 
self energy graphs 
\bel{sesum}
\begin{split}
S^{\mathrm{F}}&(x-y) + \int\!d^4x_{1}d^4x_{2}
S^{\mathrm{F}}(x-x_{1})\Sigma(x_{1}-x_{2}) 
S^{\mathrm{F}}(x_{2}-y) + \\&+\int\!d^4x_{1}d^4x_{2}d^4x_{3}d^4x_{4}
S^{\mathrm{F}}(x-x_{1})\Sigma(x_{1}-x_{2}) S^{\mathrm{F}}(x_{2}-x_{3})
\Sigma(x_{3}-x_{4}) S^{\mathrm{F}}(x_{4}-y) + ... =:\\&=: 
S^{\mathrm{tot}}(x-y).
\end{split}
\end{equation}   
This can be interpreted as sort of a total propagator. Taking the Fourier 
transform of this gives
\bel{sesumimp}
\begin{split}
\frac{1}{(2\pi)^2}\hat{S}^{\mathrm{tot}}(p)=&\\=
\frac{1}{(2\pi)^2}\hat{S}^{\mathrm{F}}(p)+&
(2\pi)^2\hat{S}^{\mathrm{F}}(p)\hat{\Sigma}(p)
\hat{S}^{\mathrm{F}}(p)+(2\pi)^6\hat{S}^{\mathrm{F}}(p)\hat{\Sigma}(p)
\hat{S}^{\mathrm{F}}(p)\hat{\Sigma}(p)\hat{S}^{\mathrm{F}}(p)+...\;\;.
\end{split}
\end{equation}
Since this is a geometric series we can sum it and get 
\bel{totprop}
S^{\mathrm{tot}}=\frac{S^{\mathrm{F}}}{1-(2\pi)^4\Sigma S^{\mathrm{F}}}= 
\frac{1}{(S^{\mathrm{F}})^{-1}-(2\pi^4)\Sigma}=\frac{1}
{(2\pi)^2(\slas{p}-m+i0-(2\pi)^2\Sigma)}.
\end{equation}
These calculations are not rigorous, e.g. we set $g=1$ at the beginning. 
If we do it with $g\neq 1$, such a summation is impossible (just consider 
the Fourier transform of $S^{\F}(x_{1}-x_{2})g(x_{1})g(x_{2}) + 
S^{\F}(x_{1}-y_{1})\Sigma(y_{1}-y_{2})S^{\F}(y_{2}-x_{2})g(x_{1})g(x_{2})
g(y_{1})g(y_{2})+... $). 

Another difficulty origins in the fact that we have two types of vertices. 
In the considerations here we investigated the sum of terms with non-tilde 
vertices only. For a complete discussion we should somehow 
incorporate the terms with tilde vertices as well or give arguments why this 
is not necessary. 

However, we will not discuss this and if (\ref{eq:totprop}) 
can be derived rigorously without performing the adiabatic limit too early - 
to do that we should first have an IR-finite theory at hand, and this is 
not the case now as already mentioned (see Sec. \ref{subsubsec:poscon}). So  
we just want to discuss some consequences of this formula to compare our 
results with the literature. 
First we see that $(2\pi)^2\Sigma$ 
`appears as a mass' in the propagator - this 
point of view was also fruitfully adopted in the ($T\!=\!0$)-theory, e.g. 
to determine normalization constants in the causal approach 
(see \cite{ScharfBuch}, p.213). But as we will see soon, 
$\Sigma$ for $T\!>\!0$ has -  
beside the scalar component - terms proportional to $\slas{p}$ and 
$\slas{u}$, which perhaps should be interpreted as a thermal change of 
momentum of the particle. This idea finally leads to the 
definition and consideration of thermal spinors and a thermal Dirac equation 
as introduced in \cite{DH}, \cite{DHR}. 

Here we just calculate $\Sigma$ and evaluate it especially for 
$k_{{\scriptscriptstyle\mathrm{B}}}T \ll m_{\mathrm{e}^{-}}$,  
which practically means that we can set 
the Fermi-Dirac distribution functions $f_{\pm}$ to zero (i.e. we disregard 
terms $\mathcal{O}(e^{-\beta m_{\mathrm{e}^{-}}})$). Then we derive 
a thermal mass, which is in agreement with the results in the 
literature (e.g. \cite{Spanier}, \cite{DHR}).

To achieve this goal we have to investigate the various terms of 
(\ref{eq:t2se}). There are four terms to be considered, the others can be 
derived from them (e.g. by using the relations from app.\ref{subsec:rel}):
\bel{verschtermethermmasse}
\begin{split}
\mathrm{I}:\;\;\;\gamma^{\mu}\propStgl{11}{\F}{y-x}{}\propDtgr{11}{\F}{y-x}
        \gamma_{\mu}\\
\mathrm{II}:\;\;\;\gamma^{\mu}\propStgr{11}{\F}{y-x}{}\propDtgl{11}{\F}{y-x}
   \gamma_{\mu}\\
\mathrm{III}:\;\;\;\gamma^{\mu}\propStgr{11}{\F}{y-x}{}\propDtgr{11}{\F}{y-x}
     \gamma_{\mu}\\
\mathrm{IV}:\;\;\;\gamma^{\mu}\propStgr{12}{\F}{y-x}{}\propDtgr{12}{\F}{y-x}
      \gamma_{\mu}\\
\end{split}
\end{equation}
First we turn to $\mathrm{I}$.
Using the formula for the Fourier transform of a product:
\bel{ftprod}
 A(x)B(x) \longrightarrow \widehat{AB}(p) = \frac{1}{(2\pi)^{2}}\int\!d^4k
 \hat{A}(p-k)\hat{B}(k),
\end{equation}
we get the following:
\bel{Irech}
\begin{split}
\frac{1}{(2\pi)^2}\int\!d^4k \gamma^{\mu}&\propSfttgl{11}{\F}{p-k}{}
 \propDfttgr{11}{\F}{k}\gamma_{\mu} =\\&= \frac{i}{(2\pi)^5}\int\!d^4k 
 \frac{\gamma^{\mu}(\slas{p}-\slas{k}+m)\gamma_{\mu}}{(p-k)^2-m^2+i0}
 \delta(k^2)\frac{1}{e^{\beta\lvert ku \rvert} -1}.
\end{split}
\end{equation}
This has to be a covariant expression and therefore is of the following form:
\bel{covexprthermmass}
C_{0}+C_{p}\slas{p}+C_{u}\slas{u},\;\;\textrm{with} \;\;C_{0,p,u}\;\;
 \textrm{Lorentz scalars, depending on}\;\;p^2\;\;\textrm{and}\;\;pu.
\end{equation}
To evaluate (\ref{eq:Irech}) we use $\gamma^{\mu}(\slas{p}-\slas{k}+m)
\gamma_{\mu} = -2\slas{p}+2\slas{k}+4m$, and the fact that $C_{0}$ origins 
from the part proportional to $m$, $C_{p}$ and $C_{u}$ from the one 
proportional to $p^{\mu}+u^{\mu}$ (after omitting the $\gamma_{\mu}$ from 
$\slas{p}+\slas{u}$). If we multiply these terms $\sim(p^{\mu}+u^{\mu})$ 
by $p_{\mu}$ and $u_{\mu}$, respectively, we get the 
following three equations: 
\bel{thermmassbestimmgl}
\begin{split}
C_{0} = \frac{i}{(2\pi)^5}\int\!d^4k 
 \frac{4m}{(p-k)^2-m^2+i0}
 \delta(k^2)\frac{1}{e^{\beta\lvert ku \rvert} -1},\\
C_{p}(pu)+C_{u} = \frac{i}{(2\pi)^5}\int\!d^4k 
 \frac{-2pu+2pk}{(p-k)^2-m^2+i0}
 \delta(k^2)\frac{1}{e^{\beta\lvert ku \rvert} -1},\\
C_{p}p^2+C_{u}(pu) = \frac{i}{(2\pi)^5}\int\!d^4k 
 \frac{-2p^2 + 2pk}{(p-k)^2-m^2+i0}
 \delta(k^2)\frac{1}{e^{\beta\lvert ku \rvert} -1}.
\end{split}
\end{equation}
The calculation of $C_{0,p,u}$ is done in App. \ref{subsec:tzwei}.4. 
The full result (in the approximation 
\mbox{$k_{{\scriptscriptstyle\mathrm{B}}}T \ll m_{\mathrm{e}^{-}}$} 
and including the factor $e^2$) on the mass shell $p^2 = m^2$ is
\bel{resultthermmass}
\begin{split}
C_{0}=&0 \\
C_{p}=&\frac{ie^2}{48\beta^{2}\pi^{2}}\frac{1}{m^2-(pu)^2}
 \bigg(-1 + \frac{pu}{2\sqrt{(pu)^2-m^2}}\ln\bigg\lvert
  \frac{pu+\sqrt{(pu)^2-m^2}}
 {pu-\sqrt{(pu)^2-m^2}} \bigg\rvert\bigg)\\
C_{u}=&\frac{ie^2}{48\beta^{2}\pi^{2}}\frac{1}{m^2-(pu)^2}
 \bigg(pu -\frac{m^2}{2\sqrt{(pu)^2-m^2}}\ln\bigg\lvert
  \frac{pu+\sqrt{(pu)^2-m^2}}
 {pu-\sqrt{(pu)^2-m^2}} \bigg\rvert\bigg).  
\end{split}
\end{equation}
The terms $\mathrm{II - IV}$ of (\ref{eq:verschtermethermmasse}) are 
calculated in the same way, making an ansatz \mbox{$C_{0}\!+\!C_{p}\slas{p}+
C_{u}\slas{u}$} and determining these coefficients as above. However, in the 
approximation $k_{{\scriptscriptstyle\mathrm{B}}}T 
\ll m_{\mathrm{e}^{-}}$ only the term 
$\mathrm{I}$ is relevant, therefore we will not investigate the other 
terms anymore.

The result (\ref{eq:resultthermmass}), after inclusion of the factor 
$(2\pi)^2$ from $(2\pi)^2\Sigma$ in (\ref{eq:totprop}), is in 
agreement with \cite{Spanier} 
(see their reference [16] for their calculations). How to extract a 
possible mass correction due to the temperature from this result is not that 
clear as often claimed. One way would be to proceed in the usual way (e.g. 
\cite{ahmedsaleem}, \cite{DHR}) by determination of the pole of the total  
propagator (\ref{eq:totprop}) for $\lvert\vec{p}\rvert\rightarrow 0$. 
This leads to the result known from literature,
\bel{deltam}
\Delta m_{\beta}=\frac{e^2}{12m^2\beta^2},
\end{equation}

Another viewpoint could be the more direct one of taking the term $C_{0}$ 
as a mass correction and the `slash'-terms $C_{p}\slas{p}+C_{u}\slas{u}$ as 
some sort of momentum change; then there would be no mass correction since 
$C_{0}=0$ (see (\ref{eq:resultthermmass})). We emphasize that the 
contribution $C_{p}\slas{p}+C_{u}\slas{u}$ cannot be absorbed in 
normalization constants since they are not of a polynomial structure. 
We will not start to discuss this 
here any further since we take the viewpoint that we should first have a 
consistent thermal field theory for all temperatures. The determination of 
a mass and momentum correction certainly involves a discussion of the 
Dirac equation and the spinors in the thermal setting as it was done in 
\cite{DH}. As already mentioned it is not necessary to be concerned with this 
in the approximation $k_{{\scriptscriptstyle\mathrm{B}}}T
\ll m_{\mathrm{e}^{-}}$ - 
but it may has to be 
pursued to formulate a consistent thermal QFT for all temperatures (cf. Sec. 
\ref{sec:conclus}).

\subsubsection{Matrix Elements for Bremsstrahlung}\label{subsubsec:BS}
To discuss the IR-limit in the cross sections to fourth order (see Sec. 
\ref{subsec:IRlimit}) we need the contributions from the one photon 
bremsstrahlung processes. Because they are given essentially by Compton-graphs 
we calculate the matrix elements already in this section on second order 
terms. The process is described by Compton graphs with an external 
classical `non-tilde' photon leg (a C-number potential) and an absorbed or 
emitted soft photon. According to the theory we have to build states 
using the emission operators in the thermal Fock space. We start with the 
most simple case and discuss general coherent and incoherent superpositions 
at the end of this paper. Thus we have the following matrix elements:

Emission:
\bel{emphot}
\begin{split}
S_{fi}^{\mathrm{em}}=\langle b_{s}(\vec{p},\beta)^{+}
 \epsilon^{\mu}(\vec{k})a^{\mu}(\vec{k},\beta)^{+}\Omega 
 \lvert T_{2}^{\mathrm{Compton}} 
  \rvert b_{\sigma}(\vec{q},\beta)^{+} \Omega\rangle,
\end{split}
\end{equation}

Absorption:
\bel{absphot}
\begin{split}
S_{fi}^{\mathrm{abs}}=\langle b_{s}(\vec{p},\beta)^{+}
  \epsilon^{\mu}(-\vec{k})\tilde{a}^{\mu}(\vec{k},\beta)^{+}
 \Omega\lvert T_{2}^{\mathrm{Compton}} 
 \rvert b_{\sigma}(\vec{q},\beta)^{+} \Omega
 \rangle.
\end{split}
\end{equation}
We emphasize that there is no sum over the indices $\mu$, therefore we write 
both of them as superscripts. A remark on $\e$: Since physical photon states 
are transversal it is convenient to introduce real polarization vectors 
$\e_{\mu}$ with $\e_{\mu}=(0,\vec{\e}),\;\;\vec{k}\cdot\vec{\e}(\vec{k})=0, 
\;{\vec{\e}}^2=1$. Photon states are then given by $\lvert\e_{\mu}a_{\mu}
(\vec{k})^{+}\Omega\rangle$. For convenience we choose $\e(-\vec{k})$ for 
the absorption. Further remarks on this can be found in 
\cite{ScharfBuch} around (3.5.9) and (2.11.32). To describe the emission 
of a photon by the final state\footnote{
In the following discussion we omit the irrelevant polarization vectors and 
electron emission operators in the states.} $\langle\aerz{\mu}{k}
\Omega_{\beta}\rvert$ with initial state $\lvert\Omega_{\beta}\rangle$ is 
quite natural. The other graph has to be considered since it also gives a    
contribution to the investigated process where we start with 
$\lvert\Omega_{\beta}\rangle$ as the initial state and end up with one 
ordinary or one tilde photon. These are the only two possibilities to have 
$\langle\Omega_{\beta}\rvert$ with a one particle change in the photon sector 
as the final state. Thus it is quite natural to interpret this second 
contribution as an absorption from the photon bath since this is the only 
physical process besides emission, which takes place and changes the 
heath bath by 
one particle in the photon sector. It was already argued in \cite{TFDBuch}, 
p.120, that the second process describes 
the generation of a `hole', i.e. absorption of a photon. 
  
Now we consider the following contributions to BS: in $x_{2}$ we have a 
classical (thus non-tilde) external potential $A_{\nu}(x_{2})^{\mathrm{ext}}$ 
and an incoming field $\Psi$, in $x_{1}$ an outgoing field $\bar{\Psi}$ 
and a field $A_{\mu}$ or $\tilde{\Psi}$ and $\tilde{A}_{\mu}$. Or we have 
$A_{\nu}(x_{2})^{\mathrm{ext}}$ and an outgoing field $\bar{\Psi}$ in 
$x_{2}$, an incoming field $\Psi$ and $A_{\mu}$ or $\tilde{\Psi}$ and 
$\tilde{A}_{\mu}$ in $x_{1}$. Thus there are four terms from 
$T_{2}^{\mathrm{C}}$ (\ref{eq:t2comptotal}) that contribute: 
$A_{21}^{\alpha}+A_{31}^{\beta}, A_{31}^{\alpha}+A_{21}^{\beta}, 
C_{31}^{\alpha}+B_{21}^{\beta}$ and $C_{21}^{\alpha}+B_{31}^{\beta}$. In 
momentum space, the 
contributions with $x_{1}$ and $x_{2}$ interchanged give the same result, 
but this 
resulting factor two is canceled by the factor $\frac{1}{2!}=\frac{1}{2}$ 
from the $S$-matrix expansion (\ref{eq:smatr}). This gives  
\bel{BScompterm}
\begin{split}
S_{fi}^{\mathrm{em}}=
\int\!&d^4x_{1}d^4x_{2}g(x_{1})g(x_{2})
\frac{\sqrt{1-f_{-}(p)}}{(2\pi)^{\frac{3}{2}}}
 \bar{u}_{s}(p)e^{ipx_{2}}(-ie^2\gamma^{\nu}\propS{11}{F}{x_{2}-x_{1}}{}
 \gamma^{\mu})\cdot\\\cdot&\frac{\sqrt{1-f_{-}(q)}}{(2\pi)^{\frac{3}{2}}}
 u_{\sigma}(q)e^{-iqx_{1}}
 \frac{1}{(2\pi)^{\frac{3}{2}}}\int\!dk_{0}\delta(k^2)\Theta(k_{0})
\sqrt{1+f(k)}e^{ikx_{1}}\epsilon_{\mu}(\vec{k})A_{\nu}^{\mathrm{ext}}(x_{2})+\\
&+\textrm{the three other terms.}
\end{split}
\end{equation} 
Here we used the covariant expression from \ref{eq:Acov} for 
$A_{\mu}(x)$ in order to keep track in a rigorous way of the conditions 
$k^2=0$ and $k_{0}=\lvert\vec{k}\rvert$. The absorption part gives the same 
except that we have $e^{-ikx_{1}}$ instead of $e^{ikx_{1}}$ and $\sqrt{f}$ 
instead of $\sqrt{1+f}$. 
Now we use the following Fourier transformed expressions: 
\bel{ft}
\begin{split}
g(x_{i})=\frac{1}{(2\pi)^2}\int\!d^4k_{i}\hat{g}(&k_{i})e^{-ik_{i}x_{i}},\;\;
 S(x_{2}-x_{1})=\frac{1}{(2\pi)^2}\int\!d^4Q\hat{S}(Q)e^{-iQ(x_{2}-x_{1})},\\
&A(x_{2})=\frac{1}{(2\pi)^2}\int\!d^4r\hat{A}(r)e^{-irx_{2}},
\end{split}
\end{equation}
then the $x_{i}$-integrations give $(2\pi)^4\delta(-k_{1}+Q-q+k)
 (2\pi)^4\delta(-k_{2}+p-Q-r)$, which allows to calculate the $Q$- and $r$-
integrals (in the other three terms we proceed in an analogous way). 
In each term one $k_{i}$-integration can be done trivially leading to a factor 
$(2\pi)^2$ (cf. end of Sec. \ref{subsubsec:caussm}). 
With analogous calculations for the absorption we thus get the 
following matrix elements for emission and absorption:
\bel{matrelembs}
\begin{split}
S_{fi}&^{\mathrm{BS,em}}=
\frac{e^2}{(2\pi)^\frac{5}{2}}
\int\!d^4k_{1}\hat{g}(k_{1})
  \int\!dk_{0}\delta(k^2)\Theta(k_{0})\bar{u}_{s}(p)\bigg[\\
&+(-i)\slas{\hat{A}}^{\mathrm{ext}}(p-q+k-k_{1})\propSft{11}{\F}{q+k_{1}-k}{}
  \slas{\epsilon}(\vec{k})\sqrt{(1-f_{-}(p))(1-f_{-}(q))}\sqrt{1+f(k)}+\\  
&+(-i)\slas{\epsilon}(\vec{k})\propSft{11}{\F}{p-k_{1}+k}{}
   \slas{\hat{A}}^{\mathrm{ext}}
  (p-q+k-k_{1})\sqrt{(1-f_{-}(p))(1-f_{-}(q))}\sqrt{1+f(k)}-\\
&-\slas{\epsilon}(\vec{k})\propSft{21}{\F}{p-k_{1}+k}{}
  \slas{\hat{A}}^{\mathrm{ext}}
  (p-q+k-k_{1})\sqrt{f_{-}(p)(1-f_{-}(q))}\sqrt{f(k)}-\\
&-\slas{\hat{A}}^{\mathrm{ext}}(p-q+k-k_{1})\propSft{12}{\F}{q+k_{1}-k}{}
  \slas{\epsilon}(\vec{k})\sqrt{(1-f_{-}(p))f_{-}(q)}\sqrt{f(k)}
  \bigg]u_{\sigma}(q),\\
S_{fi}&^{\mathrm{BS,abs}}=
\frac{e^2}{(2\pi)^\frac{5}{2}}
\int\!d^4k_{1}\hat{g}(k_{1})
  \int\!dk_{0}\delta(k^2)\Theta(k_{0})\bar{u}_{s}(p)\bigg[\\
&+(-i)\slas{\hat{A}}^{\mathrm{ext}}(p-q-k-k_{1})\propSft{11}{\F}{q+k_{1}+k}{}
  \slas{\epsilon}(-\vec{k})\sqrt{(1-f_{-}(p))(1-f_{-}(q))}\sqrt{f(k)}+\\  
&+(-i)\slas{\epsilon}(-\vec{k})\propSft{11}{\F}{p-k_{1}-k}{}
   \slas{\hat{A}}^{\mathrm{ext}}
  (p-q-k-k_{1})\sqrt{(1-f_{-}(p))(1-f_{-}(q))}\sqrt{f(k)}-\\
&-\slas{\epsilon}(-\vec{k})\propSft{21}{\F}{p-k_{1}-k}{}
  \slas{\hat{A}}^{\mathrm{ext}}
  (p-q-k-k_{1})\sqrt{f_{-}(p)(1-f_{-}(q))}\sqrt{1+f(k)}-\\
&-\slas{\hat{A}}^{\mathrm{ext}}(p-q-k-k_{1})\propSft{12}{\F}{q+k_{1}+k}{}
  \slas{\epsilon}(-\vec{k})\sqrt{(1-f_{-}(p))f_{-}(q)}\sqrt{1+f(k)}
  \bigg]u_{\sigma}(q).
\end{split}
\end{equation}
Thus we see that the absorption is the same as the emission with $k$ 
replaced by $-k$ (except in $\Theta(k_{0})$), $1+f$ by $f$. 
 
Inserting the expressions for the propagators from App. 
\ref{subsec:covnotmomspre} in 
(\ref{eq:matrelembs}) - the definition of $<...>$ is given in 
(\ref{eq:defquant}) - using  
$\slas{p}\slas{\e}=-\slas{\e}\slas{p}+
2p\e$, etc. and the Dirac equation we get the following 2 times 6 terms:
\bel{matrelmbsexpl}
\begin{split}
S_{fi}^{\mathrm{BS,em}}=&
\frac{e^2}{(2\pi)^\frac{5}{2}}\int\!
   d^4k_{1}\hat{g}(k_{1})
   \int\!dk_{0}\delta(k^2)\Theta(k_{0})\bar{u}_{s}(p)
   \slas{\hat{A}}^{ext}(p-q+k-k_{1})u_{\sigma}(q)\bigg[\\
   &+\frac{(2p\epsilon-\slas{\e}\slas{k_{1}}+\slas{\e}\slas{k})
  (-i)\sqrt{(1-f_{-}(p))(1-f_{-}(q))}\sqrt{1+f(k)}}
    {(2\pi)^2((p-k_{1}+k)^2-m^2+i0)}+\\&+\frac{2p\epsilon-
  \slas{\e}\slas{k_{1}}+\slas{\e}\slas{k}}{2\pi}
    \sqrt{(1-f_{-}(p))(1-f_{-}(q))}\sqrt{1+f(k)}\cdot\\\hspace{2.0 cm}&\cdot
   \delta((p-k_{1}+k)^2-m^2)
   \thermf{p-k_{1}+k}+\\
&+(2p\epsilon-\slas{\e}\slas{k_{1}}+\slas{\e}\slas{k})
   (-1)\sqrt{f_{-}(p)(1-f_{-}(q))}\sqrt{f(k)}(\frac{-1}{2\pi})
   \cdot\\\hspace{2.0 cm}&\cdot
    \delta((p-k_{1}+k)^2-m^2)\thermroot{p-k_{1}+k} +\\
&+\frac{(2q\epsilon+\slas{k_{1}}\slas{\e}-\slas{k}\slas{\e})(-i)
  \sqrt{(1-f_{-}(p))(1-f_{-}(q))}
   \sqrt{1+f(k)}}
    {(2\pi)^2((q+k_{1}-k)^2-m^2+i0)}+\\&
 +\frac{2q\epsilon+\slas{k_{1}}\slas{\e}-\slas{k}\slas{\e}}{2\pi}
    \sqrt{(1-f_{-}(p))(1-f_{-}(q))}\sqrt{1+f(k)}
   \cdot\\\hspace{2.0 cm}\cdot&\delta((q+k_{1}-k)^2-m^2)
   \thermf{q+k_{1}-k}+\\
&+(2q\epsilon+\slas{k_{1}}\slas{\e}-\slas{k}\slas{\e})
   (-1)\sqrt{(1-f_{-}(p))f_{-}(q)}\sqrt{f(k)}(\frac{-1}{2\pi})
   \cdot\\\hspace{2.0 cm}&\cdot
   \delta((q+k_{1}-k)^2-m^2)\thermroot{q+k_{1}-k}\bigg].
\end{split}
\end{equation}
The matrix-element for the absorption
$S_{fi}^{\mathrm{BS,abs}}$ is the same, but with $k \leftrightarrow 
-k$ (except in $\Theta(k_{0})$) and $f\leftrightarrow 1+f$.

\subsection{Calculation of some Third Order Graphs}\label{subsec:drittordcalc}
In this section we calculate the third order graphs that play or could 
play a r\^ole in the discussion of the IR-behaviour 
of the theory in cross sections to fourth order. 
Thus we will calculate $T_{3}$ for the 
Vertex and the third order self energy graph. These have to be considered 
in investigating the IR-problem since they will turn out to be infinite in the 
adiabatic limit $g\rightarrow 1$. 
We also discuss the third order vacuum polarization and the graphs with 
two bremsstrahlung-photons - they are finite in the limit $g\rightarrow 1$,  
therefore we will not consider them further. To caution the reader we want 
to emphasize that in the following calculations 
we do not consider $(T\!=\!0)$- 
and $(T\!>\!0)$-parts separately. Therefore, all formulae have to be taken 
literally for the $(T\!>\!0)$-parts only, the $(T\!=\!0)$-parts 
(and $(T\!=\!0)$-parts of 
subgraphs with $\omega \geq 0$, e.g. in the third order SE) are understood 
to have been treated correctly, i.e. by splitting them non-trivially.   
We do not write down this explicitly (cf. also the remark at the end of Sec. 
\ref{subsubsec:tztzcalc}).   

\subsubsection{The Vertex Graph}\label{subsubsec:vertgraph}
$D_{3}^{\mathrm{V}}$ is calculated in appendix \ref{subsec:vertcalc}, 
its singular order as 
well. We will only consider the terms with a non-tilde field $A_{\nu}$ in 
$x_{3}$ (since this will be replaced by a classical external potential in 
our calculations for the IR-behaviour) and $\bar{\Psi}$ or 
$\tilde{\Psi}$ in $x_{1}$, $\Psi$ or $\bar{\tilde{\Psi}}$ in $x_{2}$. The 
terms with permuted arguments give the same result 
in momentum space (as can be shown 
by calculating just as we do for the terms we are concerned with here). Thus 
this gives an additional factor $3!=6$ from the various permutations. But 
from the expansion of the $S$-matrix, (\ref{eq:smatr}), there is another 
global factor, $\frac{1}{3!}=\frac{1}{6}$, which cancels the one above.

Here we give the result for $T_{3}^{\mathrm{V}}$ and some hints for the 
calculation:

We take the first part of $D_{3}^{\mathrm{V}}$ 
(see App. \ref{subsec:vertcalc}) and 
change its form in a suitable way\footnote{The arguments of the products in 
the following formula are $(x_{1}-x_{3}),\; (x_{3}-x_{2}), \;(x_{1}-x_{2})$, 
always in this order.}:
\bel{d3certumform}
\begin{split}
D_{3}^{\mathrm{V}} &= \propSoa{}{-}{}{}\propSoa{}{+}{}{}
          \propDoa{}{\F}{}-
          \propSoa{}{\ret}{}{}\propSoa{}{+}{}{}
           \propDoa{}{+}{}
          -\propSoa{}{-}{}{}\propSoa{}{\av}{}{}
         \propDoa{}{-}{}-\\
          &-\propSoa{}{+}{}{}\propSoa{}{-}{}{}
         \propDoa{}{\F}{}+
          \propSoa{}{\av}{}{}\propSoa{}{-}{}{}
          \propDoa{}{-}{}
          +\propSoa{}{+}{}{}\propSoa{}{\ret}{}{}
          \propDoa{}{+}{}=\\
        =&-\propSoa{}{\av}{}{}\propSoa{}{\ret}{}{}
            \propDoa{}{+}{}
          +\propSoa{}{-}{}{}\propSoa{}{\ret}{}{}
            \propDoa{}{\av}{}
          -\propSoa{}{-}{}{}\propSoa{}{\av}{}{} 
            \propDoa{}{\ret}{}+\\
          &+\propSoa{}{\ret}{}{}\propSoa{}{\av}{}{}
              \propDoa{}{+}{}
          -\propSoa{}{\ret}{}{}\propSoa{}{-}{}{}
            \propDoa{}{\av}{}
          +\propSoa{}{\av}{}{}\propSoa{}{-}{}{}
           \propDoa{}{\ret}{}.
\end{split}
\end{equation}
We take the retarded part of it 
(by trivial splitting since $\omega\leq -1$, see 
App. \ref{subsec:vertcalc}) which gives:
\bel{vertret}
R_{3}^{\mathrm{V}}=
-\propSoa{}{-}{}{}\propSoa{}{\av}{}{}
            \propDoa{}{\ret}{}
          +\propSoa{}{\ret}{}{}\propSoa{}{\av}{}{}
            \propDoa{}{+}{}
          -\propSoa{}{\ret}{}{}\propSoa{}{-}{}{} 
            \propDoa{}{\av}{}.
\end{equation}
Then we have 
\bel{vertt3prop}
\begin{split}
T_{3}=R_{3}-R_{3}'=
&-\propSoa{}{-}{}{}\propSoa{}{\av}{}{}
            \propDoa{}{\ret}{}
          +\propSoa{}{\ret}{}{}\propSoa{}{\av}{}{}
            \propDoa{}{+}{}
          -\propSoa{}{\ret}{}{}\propSoa{}{-}{}{} 
            \propDoa{}{\av}{}\\
&+\propS{}{-}{x_{1}-x_{3}}{}\propSoa{}{+}{}{}
            \propDoa{}{\F}{}
          -\propSoa{}{\ret}{}{}\propSoa{}{+}{}{}
            \propDoa{}{+}{}
          -\propSoa{}{-}{}{}\propSoa{}{\av}{}{} 
            \propDoa{}{-}{}=\\
=&-\propSoa{}{\F}{}{}\propSoa{}{\F}{}{} 
            \propDoa{}{\F}{}.
\end{split}
\end{equation}
To obtain this result we have added 
$S^{-}S^{\av}D^{\F}-S^{-}S^{\av}D^{\F}$ and 
$S^{-}S^{\F}D^{\F}-S^{-}S^{\F}D^{\F}$, skilfully splitted 
$S^{\ret}S^{\F}D^{\F}$ in $S^{\ret}S^{\av}D^{+}$ etc. and used 
the fact that $\supp(S^{\ret}S^{\ret}D^{\av})=\{0\}$. 
Considering all factors we thus get: 
\bel{vertt3}
\begin{split}
T_{3}^{\mathrm{V}}=
&-e^3:\bar{\Psi}(x_{1})\gamma^{\mu}
        \propS{11}{\F}{x_{1}-x_{3}}{}\gamma^{\nu}
        \propS{11}{\F}{x_{3}-x_{2}}{}\propD{11}{\F}{x_{1}-x_{2}}\gamma_{\mu}
        \Psi(x_{2}):A_{\nu}(x_{3})\\
&+e^3:\tilde{\Psi}(x_{1})\gamma^{\mu}
        \propS{21}{\F}{x_{1}-x_{3}}{}\gamma^{\nu}
        \propS{11}{\F}{x_{3}-x_{2}}{}\propD{12}{\F}{x_{1}-x_{2}}\gamma_{\mu}
        \Psi(x_{2}):A_{\nu}(x_{3})\\
&-e^3:\bar{\Psi}(x_{1})\gamma^{\mu}
        \propS{11}{\F}{x_{1}-x_{3}}{}\gamma^{\nu}
        \propS{12}{\F}{x_{3}-x_{2}}{}\propD{12}{\F}{x_{1}-x_{2}}\gamma_{\mu}
        \bar{\tilde{\Psi}}(x_{2}):A_{\nu}(x_{3})\\
&+e^3:\tilde{\Psi}(x_{1})\gamma^{\mu}
        \propS{21}{\F}{x_{1}-x_{3}}{}\gamma^{\nu}
        \propS{12}{\F}{x_{3}-x_{2}}{}\propD{22}{\F}{x_{1}-x_{2}}\gamma_{\mu}
        \bar{\tilde{\Psi}}(x_{2}):A_{\nu}(x_{3}).
\end{split}
\end{equation}

\subsubsection{Third Order Self Energy}\label{subsubsec:3ordSE}
There are two graphs that contribute to the third order SE: `A', with  
non-tilde fields $A_{\nu}$ and $\bar{\Psi}$ in $x_{3}$ and $\Psi$ or 
$\bar{\tilde{\Psi}}$ in $x_{1}$, and a self energy insertion on the incoming 
leg; and `B', with  
non-tilde fields $A_{\nu}$ and $\Psi$ in $x_{3}$ and $\bar{\Psi}$ or 
$\tilde{\Psi}$ in $x_{1}$, and a self energy insertion on the outgoing leg. 
We first consider graph `B'. For some more remarks on 
the choice of the external legs have a look at the 
begin of Sec. \ref{subsubsec:vertgraph}.
The calculation of $D_{3}^{3.\mathrm{ord.SE}}$ and 
$\omega^{3.\mathrm{ord.SE}} = -\infty$ 
is done in appendix \ref{subsec:3ordSEcalc}. Here we just give the result for 
$T_{3}^{3.\mathrm{ord.SE}}$. We split $D_{3}^{3.\mathrm{ord.SE}}$ 
trivially and build $T=R-R'$:   
\bel{t3SEB}
\begin{split}
T_{3}^{3.\mathrm{ord.SE,B}}=e^3:&\bar{\Psi}(x_{3})\bigg(
  -\gamma^{\nu}\propS{11}{\F}{x_{3}-x_{2}}{}\gamma^{\mu}
    \propS{11}{\F}{x_{2}-x_{1}}{}\gamma_{\mu}\propD{11}{\F}{x_{1}-x_{2}}\\
  &-\gamma^{\nu}\propS{12}{\F}{x_{3}-x_{2}}{}\gamma^{\mu}
    \propS{12}{\F}{x_{2}-x_{1}}{}\gamma_{\mu}\propD{12}{\F}{x_{1}-x_{2}}
    \bigg)\Psi(x_{1}):A_{\nu}(x_{3})+\\
+e^3:&\bar{\Psi}(x_{3})\bigg(
  -\gamma^{\nu}\propS{11}{\F}{x_{3}-x_{2}}{}\gamma^{\mu}
    \propS{12}{\F}{x_{2}-x_{1}}{}\gamma_{\mu}\propD{12}{\F}{x_{1}-x_{2}}\\
  &+\gamma^{\nu}\propS{12}{\F}{x_{3}-x_{2}}{}\gamma^{\mu}
    \propS{22}{\F}{x_{2}-x_{1}}{}\gamma_{\mu}\propD{22}{\F}{x_{1}-x_{2}}
    \bigg)\bar{\tilde{\Psi}}(x_{1}):A_{\nu}(x_{3}). 
\end{split}
\end{equation}
Graph A is calculated in the same way:
\bel{t3SEA}
\begin{split}
T_{3}^{3.\mathrm{ord.SE,A}}=e^3:&\bar{\Psi}(x_{1})\bigg(
  -\gamma^{\mu}\propS{11}{\F}{x_{1}-x_{2}}{}\gamma_{\mu}
    \propS{11}{\F}{x_{2}-x_{3}}{}\gamma^{\nu}\propD{11}{\F}{x_{2}-x_{1}}\\
  &-\gamma^{\mu}\propS{12}{\F}{x_{1}-x_{2}}{}\gamma_{\mu}
    \propS{21}{\F}{x_{2}-x_{3}}{}\gamma^{\nu}\propD{12}{\F}{x_{2}-x_{1}}
    \bigg)\Psi(x_{3}):A_{\nu}(x_{3})+\\
+e^3:&\tilde{\Psi}(x_{1})\bigg(
  +\gamma^{\mu}\propS{21}{\F}{x_{1}-x_{2}}{}\gamma_{\mu}
    \propS{11}{\F}{x_{2}-x_{3}}{}\gamma^{\nu}\propD{12}{\F}{x_{2}-x_{1}}\\
  &-\gamma^{\mu}\propS{22}{\F}{x_{1}-x_{2}}{}\gamma_{\mu}
    \propS{21}{\F}{x_{2}-x_{3}}{}\gamma^{\nu}\propD{22}{\F}{x_{2}-x_{1}}
    \bigg)\Psi(x_{3}):A_{\nu}(x_{3}). 
\end{split}
\end{equation}

\subsubsection{Third order Vacuum Polarization}\label{subsubsec:thirdVP}
$D_{3}^{3.\mathrm{ord.VP}}$ and $\omega^{3.\mathrm{ord.VP}}$ are 
calculated as in 
Sec. \ref{subsubsec:3ordSE}. This leads to the following expression for 
$T_{3}^{3.\mathrm{ord.VP}}$: 
\bel{t3vpdrittord} 
\bar{\Psi}(x_{2})\gamma^{\nu}\Psi(x_{2})
\propD{11/12/\textrm{etc}.}{\F}{x_{2}-x_{3}}
  \Pi_{11/21,\mu\nu}(x_{3}-x_{1})A^{\mu}(x_{1}).
\end{equation}
We now insert the correct expressions for the vacuum polarization 
$\Pi$ and for 
$D^{\F}$. Disregarding some factors that are inessential 
with respect to convergence properties, we thus arrive at 
\bel{vp3ordconv}
\begin{split}
\int\!d^4k_{2}d^4k_{3}\hat{g}(k_{2})\hat{g}(k_{3})\bigg(& 
  +\sqrt{(1-f_{-}(p))(1-f_{-}(q))}
  \propDft{11}{\F}{p-q-k_{2}}\hat{\Pi}^{\mu\nu}_{11}(p-q-k_{2}-k_{3})\\
  &+\sqrt{(1-f_{-}(p))(1-f_{-}(q))}
  \propDft{12}{\F}{p-q-k_{2}}\hat{\Pi}^{\mu\nu}_{21}(p-q-k_{2}-k_{3})\\
  &+\sqrt{f_{-}(p)f_{-}(q)}
  \propDft{21}{\F}{p-q-k_{2}}\hat{\Pi}^{\mu\nu}_{11}(p-q-k_{2}-k_{3})\\
  &+\sqrt{f_{-}(p)f_{-}(q)}
  \propDft{22}{\F}{p-q-k_{2}}\hat{\Pi}^{\mu\nu}_{21}(p-q-k_{2}-k_{3})\bigg).
\end{split}
\end{equation}
$\Pi$ is given by $T_{2}^{\mathrm{VP}}$. This leads to\footnote{Here we  
set $Q := p-q-k_{2}-k_{3}$.}
\begin{multline*}
\int\!d^4k_{2}d^4k_{3}\hat{g}(k_{2})\hat{g}(k_{3})\int\!d^4k\underbrace{
(\sim k^2 + \sim k +\textrm{const.})}
_{`\gamma\textrm{-matrix structure}'}\Bigg\{\\
+\bigg(\frac{1}{(2\pi)^2}\frac{-1}{(p-q-k_{2})^2+i0}+\frac{i}{2\pi}\delta(
 (p-q-k_{2})^2)f(p-q-k_{2})\bigg)\cdot\\\cdot
  \bigg(\frac{1}{k^2-m^2+i0}+\frac{i}{2\pi}\delta(k^2-m^2)\bigg[f_{-}(-k)\theta
  (-k_{0})+f_{+}(-k)\theta(k_{0})\bigg]\bigg)\cdot\\\cdot
\bigg(\frac{1}{(Q-k)^2-m^2+i0}+\frac{i}{2\pi}
  \delta((Q-k)^2-m^2)\cdot\\\cdot\bigg[f_{-}(Q-k)\theta
  (Q_{0}-k_{0})+f_{+}(Q-k)\theta
  (-Q_{0}+k_{0})\bigg]\bigg)\sqrt{(1-f_{-}(p))(1-f_{-}(q))}+
\end{multline*}
\begin{multline*}
+\frac{i}{2\pi}\delta((p-q-k_{2})^2)\sqrt{f(p-q-k_{2})(1+f(p-q-k_{2}))}\cdot\\
 \cdot\frac{i}{2\pi}\delta(k^2-m^2)
 \bigg[\theta(-k_{0})\sqrt{f_{-}(-k)(1-f_{-}(-k))}-\theta(k_{0})
 \sqrt{f_{+}(-k)(1-f_{+}(-k))}\;\bigg]\cdot\\\cdot
 \frac{i}{2\pi}\delta((Q-k)^2-m^2) 
 \bigg[\theta(Q_{0}-k_{0})\sqrt{f_{-}(Q-k)
 (1-f_{-}(Q-k))}-\\-\theta(-Q_{0}+k_{0})
 \sqrt{f_{+}(Q-k)(1-f_{+}(Q-k))}\bigg]
 \sqrt{(1-f_{-}(p))(1-f_{-}(q))}+
\end{multline*}
\begin{multline*}
+\frac{i}{2\pi}\delta((p-q-k_{2})^2)\sqrt{f(p-q-k_{2})(1+f(p-q-k_{2}))}\cdot\\
 \cdot\bigg(\frac{1}{k^2-m^2+i0}+\frac{i}{2\pi}\delta(k^2-m^2)
  \bigg[f_{-}(-k)\theta
  (-k_{0})+f_{+}(-k)\theta(k_{0})\bigg]\bigg)\cdot\\\cdot
  \bigg(\frac{1}{(Q-k)^2-m^2+i0}+\frac{i}{2\pi}
  \delta(Q-k)^2-m^2)\cdot\\\cdot\bigg[f_{-}(Q-k)\theta
  (Q_{0}-k_{0})+f_{+}(Q-k)\theta
  (-Q_{0}+k_{0})\bigg]\bigg)\sqrt{f_{-}(p)f_{-}(q)}+
\end{multline*}
\begin{multline}
+\bigg(\frac{1}{(2\pi)^2}\frac{1}{(p-q-k_{2})^2-i0}+\frac{i}{2\pi}\delta(
 (p-q-k_{2})^2)f(p-q-k_{2})\bigg)\cdot\\\cdot
 \frac{i}{2\pi}\delta(k^2-m^2)
 \bigg[\theta(-k_{0})\sqrt{f_{-}(-k)(1-f_{-}(-k))}-\theta(k_{0})
 \sqrt{f_{+}(-k)(1-f_{+}(-k))}\;\bigg]\cdot\\\cdot
 \frac{i}{2\pi}\delta((Q-k)^2-m^2) 
 \bigg[\theta(Q_{0}-k_{0})\sqrt{f_{-}(Q-k)
 (1-f_{-}(Q-k))}-\\-\theta(-Q_{0}+k_{0})
 \sqrt{f_{+}(Q-k)(1-f_{+}(Q-k))}\bigg] 
 \sqrt{f_{-}(p)f_{-}(q)}\bigg\}\;\;.
\end{multline}
Performing the multiplications and investigating the adiabatic 
limit we see that 
it exists and is finite (pay attention to the fact that the whole expression 
as a distribution need not make sense evaluated at specified points $p, q$. 
It has to be smeared out in these variables, therefore $p=q$ causes no 
problems). Thus the third order VP is not involved in the discussion of 
the IR-problem in the cross sections to fourth order (we will see later 
(see after (\ref{eq:vertcrosec4})) that 
a third order graph contributes with twice its real part to the fourth order 
cross section). 

\subsubsection{Graphs with Two Bremsstrahlung-Photons}\label{subsubsec:gwtBS}
This contribution has the following form (here we choose  
special external legs again; cf. the discussion in Sec. 
\ref{subsubsec:vertgraph}):
\bel{2bs}
\begin{split}
\int\!d^4x_{1}d^4x_{2}d^4x_{3}\bar{\Psi}(x_{1})\slas{A}^{\mathrm{ext}}(x_{1})
   \propS{11}{\F}{x_{1}\!-\!x_{2}}{}\slas{A}(x_{2})\cdot\\\cdot
\propS{11}{\F}{x_{2}\!-\!x_{3}}{}
\slas{A}(x_{3})\Psi(x_{3})g(x_{1})g(x_{2})g(x_{3}).
\end{split}
\end{equation}
We form  matrix elements with emitted/absorbed photons (described by 
$\langle\aerz{\mu}{k}\Omega_{\beta}\rvert$ and  
$\langle\aerzt{\mu}{k}\Omega_{\beta}\rvert$, see Sec. \ref{subsubsec:BS}) 
of momentum  
$k$ and $k'$, which are equal to  
\bel{2bsforts}
\begin{split}
\int\!d^4k_{1}d^4k_{2}d^4k_{3}\sqrt{1\!-\!f_{-}(p)}\bar{u}_{s}(p)
  \slas{\hat{A}}^{\mathrm{ext}}
  (p\!-\!q\!-\!k\!-\!k'\!-\!k_{1}\!-\!k_{2}\!-\!k_{3})\cdot\\\cdot
  \propS{11}{\F}{q+k+k'+k_{2}+k_{3}}{}
  \sqrt{1\!+\!f(k)}\frac{\slas{\epsilon}(\vec{k})}
  {\sqrt{2\lvert \vec{k} \rvert}}
  \propS{11}{\F}{q\!+\!k'\!+\!k_{3}}{}\sqrt{1\!+\!f(k')}
   \frac{\slas{\epsilon}(\vec{k'})}
  {\sqrt{2\lvert \vec{k'} \rvert}}\sqrt{1\!-\!f_{-}(q)}\cdot\\\cdot
 u_{\sigma}(q)\hat{g}(k_{1})\hat{g}(k_{2})\hat{g}(k_{3}).
\end{split}
\end{equation}
Here we can have $\sqrt{f(k)}$ instead of $\sqrt{1+f(k)}$, or  
$12$-, $21$- and $22$-propagators instead of $11$-propagators. In addition,  
not both propagators have to depend on $q+k+...$, one or both of them 
can as well depend on $p+k+...$. But all these modifications do not 
change anything in the following considerations on the existence of the 
adiabatic limit in these contributions. In the further calculations we use
\bel{spinor2bs}
\bar{u}\gamma^{\mu}(\slas{q}+m)\slas{\epsilon}(\slas{q}+m)\slas{\epsilon}u 
  \sim \bar{u}\gamma^{\mu}u,
\end{equation}
and the corresponding for $(\slas{p}+m)$ instead of one or both of the factors 
$(\slas{q}+m)$. Here we set $k=k'$ which is valid in the contributions from 
two bremsstrahlungs photons to the cross section to fourth order, cf. Sec. 
\ref{subsubsec:mixstatcanc}. 
 
If we calculate now the cross section and integrate over small\footnote{
Since possible divergencies origin from small values of $k$ and $k'$.} 
$k$ and $k'$, it is 
clear that we have to investigate integrals of the following form 
(here $p^2 = m^2 = q^2$ and `P' 
denotes the `principal value' of the integral):
\begin{multline}
\int\!d^4k_{1}d^4k_{2}\hat{g}(k_{1})\hat{g}(k_{2})\int_{0}^{\omega_{0}}
  \int_{0}^{\omega'_{0}}drdr'r^2{r'}^2\int_{-1}^{1}\int_{-1}^{1}dxdx'\\
  \frac{1}{rr'}\mathrm{P}\frac{1}{(q_{0}r'+q_{0}r+q(k_{1}+k_{2})
-\lvert\vec{q}\rvert
  r'x'-\lvert\vec{q}\rvert rx)(q_{0}r-\lvert\vec{q}\rvert rx
  +q(k_{1}+k_{2}))}.
\end{multline}
This is finite and so are the corresponding integrals with one or two 
$\delta$-distributions from $S^{\F}$. Thus the graphs with two 
Bremsstrahlung-photons need not be considered in order to investigate the 
IR-behaviour in cross sections to fourth order.

\subsection{Thermal Corrections to \boldmath{$\mu_{\mathrm{e}^{-}}$}}
\label{subsec:calcmue}
\subsubsection{Calculation}
In section \ref{subsubsec:vertgraph} we calculated the vertex graph, 
i.e. $T_{3}^{\mathrm{V}}$, eq. (\ref{eq:vertt3}). One part of it 
will be important for the discussion of the IR-problem in the 
cross sections to 
fourth order (see Sec. \ref{subsec:IRlimit}). The other parts will give 
some thermal corrections to the magnetic moment of the electron. In this 
section we will be concerned with the 
calculation of these corrections. Before 
we explain how the vertex contributes to the electron magnetic moment, we 
first investigate its structure more closely to be able to identify the 
before-mentioned parts. 
To do that we consider the vertex in 
momentum space, i.e. its Fourier transform. Since we will be interested in 
electron scattering to determine $\mu_{\mathrm{e}^{-}}$ (see below, after  
(\ref{eq:vertfct})) we study the matrix element 
\bel{matrelemmuel}
S_{fi} = 
 \langle b_{s}(\vec{p},\beta)^{+}\Omega_{\beta}\lvert 
 T_{3}^{\mathrm{Vertex}} \rvert b_{\sigma}(\vec{q},\beta)^{+}\Omega_{\beta}
\rangle
\end{equation}
 and write it down explicitly with Fourier transformed quantities. 
This leads to
\bel{vertmom}
\begin{split}
\int d^4x_{1}d^4x_{2}&d^4x_{3}g(x_{1})g(x_{2})g(x_{3})(-e^3)\frac{1}
{(2\pi)^\frac{3}{2}}\sqrt{1-f_{-}(p)}\bar{u}(\vec{p})e^{ipx_{1}}\cdot\\
 &\cdot\gamma^{\mu}
 \propS{11}{\F}{x_{1}-x_{3}}{}\gamma^{\nu}
 \propS{11}{\F}{x_{3}-x_{2}}{}\gamma_{\mu}
 \propD{11}{\F}{x_{1}-x_{2}}\cdot\\
 &\cdot\frac{1}{(2\pi)^\frac{3}{2}}\sqrt{1-f_{-}(q)}
 u(\vec{q})e^{-iqx_{2}}A_{\nu}^{\mathrm{ext}}+\\
&\hspace{2.0 cm}+\;\textrm{the other three terms of}\;T_{3}^{\mathrm{V}}=\\
=-e^3\int d^4&x_{1,2,3}\frac{1}{(2\pi)^6}\int d^4k_{1,2,3}\hat{g}(k_{1,2,3})
 e^{-ik_{1}x_{1}-ik_{2}x_{2}-ik_{3}x_{3}}\cdot\\
&\cdot\frac{1}{(2\pi)^3}
 \sqrt{(1-f_{-}(p))(1-f_{-}(q))}\bar{u}(p)e^{ipx_{1}-iqx_{2}}\cdot \\
 &\cdot\gamma^{\mu}\frac{1}{(2\pi)^2}\int d^4Q
  \propSft{11}{\F}{Q}{}e^{-iQ(x_{1}-x_{3})}
 \gamma^{\nu}\frac{1}{(2\pi)^2}\int d^4R\propSft{11}{\F}{R}{}
 e^{-iR(x_{3}-x_{2})}\gamma_{\mu}\cdot\\ &\cdot\frac{1}{(2\pi)^2}\int d^4k
 \propDft{11}{\F}{k}e^{-ik(x_{1}-x_{2})}u(q)
 \frac{1}{(2\pi)^2}\int d^4r\hat{A}_{\nu}^{\mathrm{ext}}(r)e^{-irx_{3}}+\\
&\hspace{2.0 cm}+\;\textrm{the other three terms of}\;T_{3}^{\mathrm{V}}.
\end{split}
\end{equation}
Having performed the $x$-integrations we get three $\delta$-distributions, 
each with a factor $(2\pi)^4$. The $k_{3}$-integration is trivial and gives 
a factor $(2\pi)^2$. In the other three terms we proceed in the same way,  
which finally gives   
\bel{vertmomconcr}
\begin{split}
-\frac{e^3}{(2\pi)^3}&\int d^4k_{1}d^4k_{2}\hat{g}(k_{1})\hat{g}(k_{2})
 \bar{u}_{s}(p)\cdot\\\cdot
  \bigg[ &+\sqrt{(1-f_{-}(p))(1-f_{-}(q))}\int d^4k\gamma^{\mu}
 \propSft{11}{\F}{p-k_{1}-k}{}\gamma^{\nu}\propSft{11}{\F}{q+k_{2}-k}{}
 \propDft{11}{\F}{k}\\
&-i\sqrt{f_{-}(p)(1-f_{-}(q))}\int d^4k\gamma^{\mu}
 \propSft{21}{\F}{p-k_{1}-k}{}
 \gamma^{\nu}\propSft{11}{\F}{q+k_{2}-k}{}\propDft{12}{\F}{k}\\
&+(-i)\sqrt{(1-f_{-}(p))f_{-}(q)}\int d^4k\gamma^{\mu}
 \propSft{11}{\F}{p-k_{1}-k}{}
 \gamma^{\nu}\propSft{12}{\F}{q+k_{2}-k}{}\propDft{12}{\F}{k}\\
&-\sqrt{f_{-}(p)f_{-}(q)}\int d^4k\gamma^{\mu}\propSft{21}{\F}{p-k_{1}-k}{}
 \gamma^{\nu}\propSft{12}{\F}{q+k_{2}-k}{}\propDft{22}{\F}{k}\bigg]\cdot\\
 &\hspace{4.0 cm}\cdot\gamma_{\mu}\hat{A}^{ext}_{\nu}(p-q-k_{1}-k_{2})
 u_{\sigma}(q).
\end{split}
\end{equation}
In this expression, the factors $\gamma^{\mu}\propSft{}{\F}{p-k_{1}-k}{}
\gamma^{\nu}\propSft{}{\F}{q+k_{2}-k}{}\gamma_{\mu}$ lead to the following 
`$\gamma$-matrix structure' of the vertex contribution (see the explicit 
representation of the propagators in momentum space, App. 
\ref{subsec:covnotmomspre}.):
\bel{spinstrvert}
\gamma^{\mu}\propSft{}{\F}{p-k_{1}-k}{}
\gamma^{\nu}\propSft{}{\F}{q+k_{2}-k}{}\gamma_{\mu}\sim
\gamma^{\mu}(\slas{p}-\slas{k}_{1}-\slas{k}+m)\gamma^{\nu}
(\slas{q}+\slas{k}_{2}-\slas{k}+m)\gamma_{\mu}.
\end{equation}
Since we are interested in the adiabatic limit of these quantities 
(cf. the discussion in Sec. 
\ref{subsec:IRlimit} and after eq. (\ref{eq:numerverttot})) 
and this can be performed trivially, we can 
omit $k_{i},\; i=1,2$, in 
the numerator (cf. as well the end of Sec. \ref{subsubsec:caussm}) 
and get (using the definition $\slas{p}:= 
p_{\lambda}\gamma^{\lambda}$ and $
\gamma^{\mu}\gamma^{\nu}+\gamma^{\nu}\gamma^{\mu}=2g^{\mu\nu}, \;
\gamma^{\lambda}\gamma^{\mu}\gamma_{\lambda} = -2\gamma^{\mu}$ etc.)
\bel{nennervert}
\begin{split}
\gamma^{\mu}(\slas{p}-\slas{k}+m)\gamma^{\nu}
(\slas{q}-\slas{k}+m)\gamma_{\mu} =\hspace{3.4 cm}\\= 
(-2\slas{q}\gamma^{\nu}\slas{p}+4mq^{\nu}+4mp^{\nu}-2m^2\gamma^{\nu}) + 
(2\slas{k}\gamma^{\nu}\slas{p}+2\slas{q}\gamma^{\nu}\slas{k}-8mk^{\nu})
-2\slas{k}\gamma^{\nu}\slas{k}.
\end{split}
\end{equation}
Because of $\bar{u}_{}(p)$ on the left and $u(q)$ on the right of the 
whole expression (\ref{eq:vertmomconcr}) this gives 
(using $-2\slas{q}\gamma^{\nu}\slas{p} = 
2\slas{p}\gamma^{\nu}\slas{q}-4p^{\nu}\slas{q}-
4q^{\nu}\slas{p}+4pq\gamma^{\nu}$ and the Dirac equation)
\bel{numerverttot}
\bar{u}(p)\Big[4pq\gamma^{\nu}+4(\slas{k}(p^{\nu}+q^{\nu})-mk^{\nu}-(pk+qk)
\gamma^{\nu})-2\slas{k}\gamma^{\nu}\slas{k}\Big]u(q).
\end{equation}
The first of these terms, independent of $k$, is the one we will discuss in 
section \ref{subsec:IRlimit} where we investigate the IR-limit. It is 
the only term of (\ref{eq:numerverttot}) that leads to IR-divergencies. Here 
we will be concerned with the finite part of the vertex, for which the 
adiabatic limit can be performed trivially without problems. The goal of the 
following calculations is to determine the thermal corrections to the magnetic 
moment of the electron, $\mu_{\mathrm{e}^{-}}$. We will not do that in the 
most general setting, we will restrict ourselves to `small' temperatures: 
$k_{{\scriptscriptstyle\mathrm{B}}}T
\ll m_{\mathrm{e}^{-}}$. From the four terms of 
$T_{3}^{\mathrm{V}}$, (\ref{eq:vertt3}), then only the first survives, and 
only the third term of 
(\ref{eq:numerverttot}), proportional to $k^{\lambda}k^{\rho}$, contributes.  
In this approximation, the first term of (\ref{eq:numerverttot}) does not 
contain any finite parts and gets cancelled completely by the contribution 
of the bremsstrahlung (see Sec. \ref{subsec:IRlimit}), 
and the second term (proportional to $k^{\lambda}$) 
vanishes identically as can be seen by symmetry arguments\footnote
{Consider the 
transformation $k\rightarrow -k$ in (\ref{eq:vertmomconcr}) after inserting 
the explicit form of the propagators from App. \ref{subsec:covnotmomspre} 
and considering only the terms $\sim k^{\lambda}$ in the numerator.}.

Thus we have to calculate the following part of the matrix element\footnote{
So this is the temperature dependent part of the first term of 
(\ref{eq:vertmomconcr}) in the approximation  $k_{B}T\ll m_{\mathrm{e}^{-}}$. 
In addition we 
performed the adiabatic limit: we set $k_{i}=0$ in the arguments of 
$T_{\mathrm{V}}^{\mu\nu}$ and $\hat{A}^{\mathrm{ext}}_{\alpha}$, $g=1$ 
and get a factor $(2\pi)^2$ from each $k_{i}$-integration.}
(\ref{eq:vertmomconcr}):
\bel{maelcorr}
\begin{split}
-2\pi e^3\frac{i}{(2\pi)^5}&\bar{u}_{s}(p)(\gamma_{\mu}
\gamma^{\nu}\gamma_{\lambda})u_{\sigma}(q)\hat{A}^{\mathrm{ext}}_{\nu}\cdot\\
&\cdot\int\!d^4k(k^{\mu}k^{\nu})\frac{1}{(-2pk+i0)}\cdot
\frac{1}{(-2qk+i0)}\delta(k^2)
\frac{1}{e^{\beta\lvert ku\rvert}-1}.
\end{split}
\end{equation}
This can be written as
\bel{maelanders}
-\frac{ie^3}{(2\pi)^4}\bar{u}_{s}(p)(\gamma_{\mu}
\gamma^{\nu}\gamma_{\lambda})u_{\sigma}(q)\hat{A}^{\mathrm{ext}}_{\nu}
T_{\mathrm{V}}^{\mu\lambda}(p,q,u),
\end{equation}
with
\bel{muint}
T_{\mathrm{V}}^{\mu\nu}(p,q,u):=-2\int\!d^4k\delta(k^2)
\frac{1}{e^{\beta\lvert ku\rvert}-1}\cdot
\frac{k^{\mu}k^{\nu}}{(-2pk+i0)(-2qk+i0)}. 
\end{equation}
This integral has to be calculated. 
Because of the Lorentz covariance of this expression and the symmetry with 
respect to interchanging $p\leftrightarrow q$ it 
must be of the form 
\bel{ansatzmue}
\begin{split}
T_{\mathrm{V}}^{\mu\nu}(p,q,u)=Ag^{\mu\nu}+&B(p^{\mu}p^{\nu}+q^{\mu}q^{\nu})
+C(p^{\mu}q^{\nu}+
q^{\mu}p^{\nu})+\\
&+D(p^{\mu}u^{\nu}+u^{\mu}p^{\nu}+q^{\mu}u^{\nu}+u^{\mu}q^{\nu})
+G(u^{\mu}u^{\nu}).
\end{split}
\end{equation}
Now we take the trace in this expression or multiply it with $p_{\mu}p_{\nu}, 
\;p_{\mu}q_{\nu},\; p_{\mu}u_{\nu}$ or $u_{\mu}u_{\nu}$, respectively, and 
thus get the following system of five equations:
\bel{gls1}
0=4A+2m^2B+2pqC+2(pu+qu)D+G
\end{equation}
\bel{gls2}
\begin{split}
I_{2}:=
\int\!d^4k\delta(k^2)\frac{1}{(e^{\beta\lvert ku\rvert}-1)}\cdot
\frac{pk}{(-2qk+i0)}=
  m^2A+\Big(m^4+(pq)^2\Big)B+2m^2(pq)C+\\+
  \Big(2m^2(pu)+2(pq)(pu)\Big)D+(pu)^2G
\end{split}
\end{equation}
\bel{gls3}
\begin{split}
I_{3}:=-\frac{1}{2}\int\!d^4k\delta(k^2)&\frac{1}{e^{\beta\lvert ku\rvert}-1}=
  (pq)A+2m^2(pq)B+\Big(m^4+(pq)^2\Big)C+\\&+
  \Big(m^2(qu)+m^2(pu)+(pq)(qu)
  +(pq)(pu)\Big)D+(pu)(qu)G
\end{split}
\end{equation}
\bel{gls4}
\begin{split}
I_{4}:=
\int\!d^4k\delta(k^2)&\frac{1}{(e^{\beta\lvert ku\rvert}-1)}\cdot
\frac{ku}{(-2qk+i0)}=\\
  =&(pu)A+\Big(m^2(pu)+(pq)(qu)\Big)B+\Big(m^2(qu)+(pq)(pu)\Big)C+\\&+
  \Big(m^2+(pu)^2+(pq)+(pu)(qu)\Big)D+(pu)G
\end{split}
\end{equation}
\bel{gls5}
\begin{split}
I_{5}:=-\frac{1}{2}\int\!&d^4k\delta(k^2)\frac{1}{e^{\beta\lvert ku\rvert}-1}
  \cdot\frac{(ku)^2}{(pk-i0)(qk-i0)}=\\
  &=A+\Big((pu)^2+(qu)^2\Big)B+2(pu)(qu)C+\Big(2(pu)+2(qu)\Big)D+G.
\end{split}
\end{equation}
We will solve this linear system later.
 
Now we explain why it is the vertex that determines the magnetic moment. 
First we define $\Lambda^{\alpha}:=T_{\mathrm{V}}^{\mu\nu}\gamma_{\mu}
\gamma^{\alpha}\gamma_{\nu}$. By the definition of the 
$\gamma$-matrices we have then 
\bel{lambdatee}
\Lambda^{\alpha}=2T_{\mathrm{V}}^{\mu\nu}g^{\alpha}_{\ \nu}\gamma_{\mu}-
 T_{\mathrm{V}}^{\mu\nu}\gamma_{\mu}\gamma_{\nu}\gamma^{\alpha}. 
\end{equation} 
With the Gordon decomposition\footnote{Use 
$\sigma^{\mu\nu}:= \frac{i}{2}[\gamma^{\mu},\gamma^{\nu}]$.},
\bel{gordon}
 \bar{u}_{s}(p)(p^{\mu}+q^{\mu})u_{\sigma}(q)=2m\bar{u}_{s}(p)\gamma^{\mu}
 u_{\sigma}(q)-i(p_{\nu}-q_{\nu})\sigma^{\mu\nu}u_{\sigma}(q),
\end{equation}
we get from (\ref{eq:ansatzmue}), after some manipulations and 
taking the $\bar{u}$ on the left and the $u$ on the right end into account,  
\bel{vertfct}
\begin{split}
 \Lambda^{\alpha}(p,q,u)=&\gamma^{\alpha}\Big(-2A+2m^2B+4m^2C-2(pq)C-2(pu)D-
 2(qu)D-G\Big)\\&-2im(B+C)(p_{\mu}-q_{\mu})\sigma^{\alpha\mu}+
 2u^{\alpha}(2mD+\slas{u}G)+2(p^{\alpha}+q^{\alpha})\slas{u}D.
\end{split}
\end{equation}
To extract  
information on $\mu_{\mathrm{e}^{-}}$ from (\ref{eq:vertfct}) 
we proceed as described 
for the case $T\!=\!0$ in \cite{ScharfBuch}, p.236. We consider electron 
scattering on an external static C-number potential at low energies. 
We calculate in the rest frame of the heat bath: $u=(1,\vec{0})$ and 
use the standard representation for spinors with the same 
normalization convention as in \cite{ScharfBuch}:
\bel{spinorstand}
u_{s}(\vec{p})=\sqrt{\frac{E+m}{2E}}
   \begin{pmatrix}\chi_{s}\\\frac{\vec{\sigma}\vec{p}}{E+m}\chi_{s}
   \end{pmatrix}, 
\end{equation}
where $E:=\sqrt{{\vec{p}}^2+m^2}$. 
This is most convenient for the non-relativistic case 
$\lvert\vec{p}\rvert\ll m_{\mathrm{e}^{-}}$ we are interested in. 
Using this approximation and 
\bel{gamma}
\gamma^{0}=
\begin{pmatrix}1&0\\0&-1\end{pmatrix},\; 
\gamma^{j}=
\begin{pmatrix}0&\sigma_{j}\\-\sigma_{j}&0\end{pmatrix},\; 
\sigma^{kl}=\epsilon^{klm}
\begin{pmatrix}\sigma_{m}&0\\0&\sigma_{m}\end{pmatrix}, 
\end{equation}
where $\sigma_{j}$ are the Pauli-matrices, we find to leading order 
in $\frac{\lvert\vec{p}\rvert}{m}$
\bel{chichi}
\begin{split}
\bar{u}_{s}(\vec{p})\gamma^{0}u_{\sigma}(\vec{q})
    &=(\chi_{s}^{+},\chi_{\sigma}),\\
\bar{u}_{s}(\vec{p})\gamma^{j}u_{\sigma}(\vec{q})&=
 \frac{1}{2m}\Big(\chi_{s}^{+},
 \big(\vec{p}+\vec{q}+i\vec{\sigma}\times(\vec{p}-\vec{q})\big)^{j}
  \chi_{\sigma}\Big),\\
\bar{u}_{s}(\vec{p})\sigma^{\mu\nu}(p_{\mu}-q_{\mu})
u_{\sigma}(\vec{q})A_{\nu}&=
-(\chi_{s}^{+},\vec{\sigma}\chi_{\sigma})\Big((\vec{p}-\vec{q})\times
\vec{A}\Big). 
\end{split}
\end{equation}
These expressions will now be used to calculate the matrix 
element describing the beforementioned electron scattering we are 
interested in. In the low temperature approximation this matrix element 
is given by (\ref{eq:matrelemmuel}) or 
(\ref{eq:maelcorr}), respectively, and just this term 
contributes as explained above. To compare with the value of 
$\mu_{\mathrm{e}^{-}}$ without any corrections, we consider the 
first order matrix element, which is given by 
\bel{mumefirstord}
\langle b_{s}(\vec{p},\beta)^{+}\Omega_{\beta}\lvert 
 T_{1} \rvert b_{\sigma}(\vec{q},\beta)^{+}\Omega_{\beta}
\rangle
\end{equation}
in addition, where the $T_{1}$ here consists of the non-tilde part of 
(\ref{eq:teins}) only, 
since $A^{\mathrm{ext}}_{\mu}$ 
has to be a classical potential, i.e. of non-tilde type. 
In momentum space this is given by  
\bel{maelerst}
ie\frac{1}{2\pi}\bar{u}_{s}(p)\gamma^{\alpha}u_{\sigma}(q)
\hat{A}^{\mathrm{ext}}_{\alpha}(p-q).
\end{equation}
Altogether, the first and third order matrix elements 
(\ref{eq:mumefirstord}) and (\ref{eq:maelcorr}) are equal to 
(use (\ref{eq:vertfct}))
\bel{mumumu}
\begin{split}
S_{fi}:=ie&\frac{1}{2\pi}\bar{u}_{s}(p)\cdot\\&\cdot
\Bigg[\gamma^{\alpha}-\frac{e^2}{(2\pi)^3}\gamma^{\alpha}
\Big(-2A+2m^2B+4m^2C-2(pq)C-2(pu)D-
 2(qu)D-G\Big)+\\&+\frac{e^2}{(2\pi)^3}2im(B\!+\!C)(p_{\mu}\!-\!q_{\mu})
 \sigma^{\alpha\mu}\!-\!\frac{e^2}{(2\pi)^3}2u^{\alpha}(2mD\!+\!\slas{u}G)\!
\!-\frac{e^2}{(2\pi)^3}2(p^{\alpha}\!+\!q^{\alpha})
  \slas{u}D   \Bigg]\cdot\\&\cdot
u_{\sigma}(q)\hat{A}^{\mathrm{ext}}_{\alpha}(p-q).
\end{split}
\end{equation}
Now we substitute (\ref{eq:chichi}) in (\ref{eq:mumumu}) and get\footnote{As 
mentioned above: we calculate in the system $u=(1,\vec{0})$.}
\bel{finalmu}
\begin{split}
S_{fi}=\frac{ie}{2\pi}\Bigg[(\chi_{s},\chi_{\sigma})\bigg\{&
\hat{A}_{0}^{\mathrm{ext}}-\frac{1}{2m}(\vec{p}+\vec{q})
\hat{\vec{A}}^{\mathrm{ext}}-\frac{e^2}{(2\pi)^3}W
\hat{A}_{0}^{\mathrm{ext}}+
\frac{e^2}{2m(2\pi)^3}W(\vec{p}+\vec{q})\hat{\vec{A}}^{\mathrm{ext}}-\\-
\frac{2e^2}{(2\pi)^3}&(2mD+G)\hat{A}_{0}^{\mathrm{ext}}
-\frac{2e^2}{(2\pi)^3}D(p^{0}+q^{0})\hat{A}_{0}^{\mathrm{ext}}+
\frac{2e^2}{(2\pi)^3}D(\vec{p}+\vec{q})\hat{\vec{A}}^{\mathrm{ext}}\bigg\}
+\\+
(\chi_{s},\vec{\sigma}\chi_{\sigma})&\Big((\vec{p}-\vec{q})\times
\hat{\vec{A}}^{\mathrm{ext}}\Big)\bigg\{-\frac{i}{2m}
+\frac{ie^2}{2m(2\pi)^3}W-\frac{i2me^2}{(2\pi)^3}(B+C)\bigg\}\Bigg],
\end{split}
\end{equation}
where
\bel{WWWW}
W:=-2A+2m^2B+4m^2C-2(pq)C-2(pu)D-2(qu)D-G.
\end{equation}
The second summand in (\ref{eq:finalmu}) is proportional to the magnetic 
field\footnote{The following formula is the Fourier transform of 
$\vec{B}=\mathrm{rot}\vec{A}$.}
$\hat{\vec{B}}(\vec{p}-\vec{q})=i(\vec{p}-\vec{q})\times
\hat{\vec{A}}(\vec{p}-\vec{q})$, therefore we can identify the magnetic 
moment\footnote{These are the first order term and the thermal correction; 
the QED corrections 
at $T\!=\!0$ are not included here. - 
To distinguish factors that do not belong 
to $\mu_{\mathrm{e}^{-}}$ we use the fact that $\frac{e}{2m}$ is the magnetic 
moment of the electron determined in the framework of relativistic QM. To 
set $p=q$ to identify the magnetic moment stems from a less intuitive 
discussion of expressions like (\ref{eq:finalmu}) as it is done in 
\cite{itzzub}, p.347. There it is shown that the quasi-static limit $p=q$ 
allows the discussion and identification of $\mu_{\mathrm{e}^{-}}$ 
quite similar to the one in the framework of relativistic QM.} 
\bel{magnetmom}
\mu_{\mathrm{e}^{-}}=\frac{e}{2m}\bigg\{1
-\frac{e^2}{(2\pi)^3}W+\frac{4m^2e^2}{(2\pi)^3}(B+C)\bigg\}\Bigg\rvert_{p=q}.
\end{equation}
Using (\ref{eq:gls1}) and (\ref{eq:WWWW}) this gives
\bel{magnetmomeinfach}
\mu_{\mathrm{e}^{-}}=\frac{e}{2m}\bigg\{1
-\frac{e^2}{(2\pi)^3}(2A)\bigg\}\Bigg\rvert_{p=q}.
\end{equation}
Thus we only have to know $A$, the other parameters are irrelevant for 
the magnetic moment in this approximation. Nevertheless we cannot avoid 
solving the whole system (\ref{eq:gls1}) - 
(\ref{eq:gls5}). The solution will be calculated in appendix 
\ref{subsec:mucalcappend}. We only need the solution in the case of $p=q$, 
therefore we investigate the solution of 
the system for $q:=p-\eta$ with $q^2=(p-\eta)^2=m^2$ 
and $\eta_{0}^2-{\vec{\eta}}^2\ll m^2$ and then consider the limit 
$\eta \rightarrow 0$. Pay attention to the facht that due to the above 
definitions we have $\eta_{0}=p_{0}-p_{0}\sqrt{1-(\frac{2\vec{p}\vec{\eta}}{p_{0}^2}-\frac{\vec{\eta}^2}{p_{0}^2})}$. We emphasize that setting $p=q$ 
in the linear system and then solving it is in general not allowed, but 
here it leads to the same result. We will 
discuss that and what could go wrong this way at the end of Sec. 
\ref{subsubsec:discmumu}. 

Now we present the result for the magnetic moment of the electron, including 
the QED-correction for $T\!=\!0$ and the corrections due to $T\!>\!0$, each up to 
third order in $e$ 
(see App. \ref{subsec:mucalcappend}, eq. (\ref{eq:muresult})):
\bel{muresultmaintext}
\mu_{\mathrm{e}^{-}}=\frac{e}{2m}\bigg\{1+\frac{e^2}{8\pi^2}
-\frac{e^2}{36\beta^2m^2}\bigg\}.
\end{equation}
\subsubsection{Discussion of the Result}\label{subsubsec:discmumu}
In this section we want to discuss the result (\ref{eq:muresultmaintext}) for 
$\mu_{\mathrm{e}^{-}}$ a little bit and to compare it with the existing 
results in literature. We are aware of nine papers wherein 
$\mu_{\mathrm{e}^{-}}$ is calculated\footnote{There are other papers that  
give values for $\mu_{\mathrm{e}^{-}}$, e.g. \cite{ElSk} or \cite{AhMa}, 
but for the calculation they refer 
to one of this nine papers. One paper, \cite{ShXuT}, we could not find.}: 
\cite{PlaTa} and \cite{DHR}  - \cite{YabKana}. 
The results of these papers disagree partially; we can find three different 
values, which we report here, in units of `magnetons' $\frac{e}{2m}$, where 
in some of the references temperature dependent values for $e$ and $m$ are 
used. But in the approximation $k_{{\scriptscriptstyle\mathrm{B}}}T
\ll m_{\mathrm{e}^{-}}$, which 
we are considering here this temperature-dependence can be omitted (we refer 
to the cited literature for a more exact treating, here we are just 
interested in giving an idea of what the results in literature are like and 
how they differ from our). Now we give the values from the 
literature\footnote{\cite{PeresSkag} has the wrong 
sign due to some error; it is corrected in \cite{JoPeSk}. And \cite{Bart} 
argues that a QFT calculation should reproduce a correction of 
$-\frac{5e^2}{36\beta^2m^2}$, which was obtained by some other method, an 
effective Hamiltonian method, incorporating some intuitive ideas on 
the nature of the electron and not using covariant QED. He 
then identifies the value of $-\frac{e^2}{18\beta^2m^2}$ as part of it and 
gives some arguments why the calculations that lead to 
the first result in (\ref{eq:muval}) just can account for $\frac{2}{5}$ of 
the desired value.\label{fn:compare}}: 
\bel{muval}
\begin{split}
-\frac{e^2}{18\beta^2m^2}, &\;\;\;
 \textrm{reported by \cite{DHR} - \cite{JoPeSk}, 
 \cite{Bart} and \cite{YabKana}},\\
-\frac{e^2}{36\beta^2m^2}, &\;\;\;\textrm{reported by \cite{PlaTa} and 
\cite{Spanier}},\\ 
-\frac{19e^2}{270\beta^2m^2}, &\;\;\;\textrm{reported by \cite{CoHeY}}.
\end{split}
\end{equation}
So all the results are of the same magnitude with respective 
factors $1, 2, \sim\frac{5}{2}$. We agree with the second result and in 
absolute numbers it reads approximatively (for $T=293\;\mathrm{K}$, $\hbar$ 
and $c$ not set equal to $1$ anymore)
\bel{munumer}
\mu_{\mathrm{e}^{-}}=-0.62\cdot10^{-17}\cdot\frac{e}{2m}\;. 
\end{equation} 
For comparison we give the value of the ($T\!=\!0$)-corrections up to the same 
order: $\frac{e}{2m}\cdot\frac{e^2}{8\pi^2} = 
1.16\cdot10^{-3}\cdot\frac{e}{2m}$ . 

Among all these calculations there is just one 
that is formulated covariantly, \cite{Spanier}, - as is ours. We agree  
with them that all the differing results quoted arise because it is quite 
tricky to identify the magnetic moment of the electron in a non-covariant 
setting.

To caution the reader we emphasize that there could be a  
difference between our result and the one from \cite{Spanier} that could be   
traced back to their setting $p=q$ too early (cf. the remark after 
(\ref{eq:A}): Because the determinant of the linear system is zero for 
$p=q$ it is not allowed to set $p=q$ at the beginning. The procedures of 
taking the limit $q=p-\eta,\;\eta\rightarrow 0$ 
and solving the equation system do 
not commute!). If we set $p=q \leftrightarrow 
\eta=0$ at the very beginning of our calculation, i.e. in (\ref{eq:ggllss}), 
we get five equations. Two of them are identical. We throw away one of them, 
arguing with (\ref{eq:ansatzmue}) that $B$ and $C$ must not be distinguished 
for $p=q$ and therefore just four variables, $A,B,D,G$, have to be determined. 
Solving this new linear system we end up with 
\bel{areduz}
A=\frac{\pi^3}{9\beta^2m^2},
\end{equation}
which fortunately is the same result as we have - but this will not be true 
in general.  

\subsection{The Infrared Problem in thermal QED in Cross Sections to 
 Fourth Order}
\label{subsec:IRlimit}
In this section we investigate the adiabatic limit in the cross 
section for scattering of an electron on an external classical C-number 
potential to fourth order. 
To do that we have to consider the contributions from the vertex, 
the third order self energy and the bremsstrahlung with one BS-photon. There 
are no other second order contributions and the other third order 
graphs that could contribute are finite (see Sec. \ref{subsubsec:thirdVP} and 
\ref{subsubsec:gwtBS}). We calculate the 
cross sections for these processes, which will contain IR-divergent 
parts in the limit $g\rightarrow 1$. Then we investigate if the various 
divergent parts perhaps cancel. This is \emph{not} the case for arbitrary 
temperatures, whereas it is in the approximation 
$k_{{\scriptscriptstyle\mathrm{B}}}T\ll 
m_{\mathrm{e}^{-}}$. We caution the reader that we often write down the 
expressions for the $(T\!=\!0)$-parts as if they had been calculated using 
ordinary Feynman rules. This is not the case, they have to be treated 
correctly according to the causal theory as described in 
\ref{subsubsec:split}. But  
we have chosen to write them down in this simple manner to keep track of them 
without complicating the notation. This is possible since the $T$-independent 
parts can be treated separately (cf. end of Sec. \ref{subsubsec:calcdzwei}) 
and their cancellation was prooved in 
\cite {ScharfBuch}. The same remark applies to some 
terms of the temperature dependent part of the third order SE including 
sub-graphs of singular order $\omega \geq 0$.  

Thus we have to calculate the cross sections for these various 
processes, i.e. 
\bel{crosec4}
\begin{split}
\bigg(\frac{\partial\sigma}{\partial\Omega}\bigg)^{\textrm{div.}}
_{4.\mathrm{ord.}}&=
\lvert S_{fi, 1.\mathrm{ord.}}+S^{\mathrm{SE}}_{fi, 3.\mathrm{ord.}}
+S^{\mathrm{V}}_{fi, 3.\mathrm{ord.}}
\rvert^2\bigg\rvert_{4.\mathrm{ord.}}
+\lvert S^{\mathrm{em}}_{fi, 2.\mathrm{ord.}}\rvert^2
+\lvert S^{\mathrm{abs}}_{fi, 2.\mathrm{ord.}}\rvert^2=\\
  &= S^{\mathrm{em}}_{fi, 2.\mathrm{ord.}}
(S^{\mathrm{em}}_{fi, 2.\mathrm{ord.}})^{*}+
S^{\mathrm{abs}}_{fi, 2.\mathrm{ord.}}
(S^{\mathrm{abs}}_{fi, 2.\mathrm{ord.}})^{*}+\\&\hspace{2.0 cm}+
S_{fi, 1.\mathrm{ord.}}(S^{\mathrm{V}+\mathrm{SE}}_{fi, 3.\mathrm{ord.}})^{*}+
S^{\mathrm{V}+\mathrm{SE}}_{fi, 3.\mathrm{ord.}}(S_{fi, 1.\mathrm{ord.}})^{*}.
\end{split}
\end{equation}
To shorten the notation in the calculations 
we define the following quantities:
\bel{defquant}
\begin{split}
<\theta \sqrt{f_{\pm}(p)}> &:= \bigg(\theta(p_{0})\sqrt{f_{-}(p)(1-f_{-}(p))}-
           \theta(-p_{0})\sqrt{f_{+}(p)(1-f_{+}(p))}\bigg),\\
<\theta f_{\pm}(p)>
  &:=\bigg(\theta(p_{0})f_{-}(p)+\theta(-p_{0})f_{+}(p)\bigg). 
\end{split}
\end{equation}
Since we are interested in the adiabatic limit we set $k_{i}=0$ wherever 
this is already possible (according to the remark after 
(\ref{eq:genadlim3})). 
\subsubsection{Cross Sections of the Involved Processes}
\label{subsubsec:crocrocro}
\textsc{Vertex:} 
We consider the matrix element (\ref{eq:matrelemmuel}) 
\bel{matrelemmuel2}
S_{fi} = 
 \langle b_{s}(\vec{p},\beta)^{+}\Omega_{\beta}\lvert 
 T_{3}^{\mathrm{V}} \rvert b_{\sigma}(\vec{q},\beta)^{+}\Omega_{\beta}
\rangle.
\end{equation}
We proceed as in Sec. \ref{subsec:calcmue}, but now we are interested in the 
first, i.e. $k$-independent part in (\ref{eq:numerverttot}). Since the 
vertex is of third order it contributes to the cross section 
(\ref{eq:crosec4}) with 
\bel{vertcrosec4}
S_{fi, 1.\mathrm{ord.}}(S^{\mathrm{V}}_{fi})^{*}+
S^{\mathrm{V}}_{fi}(S_{fi, 1.\mathrm{ord.}})^{*}.
\end{equation}
From (\ref{eq:numerverttot}) we see that the interesting part of the vertex 
is proportional to $\bar{u}(p)\gamma^{\nu}u(q)$, i.e. proportional to 
$S_{fi, 1.\mathrm{ord.}}$. Using $S^{\mathrm{V}}_{fi} =: 
hS_{fi, 1.\mathrm{ord.}}$ and by inserting the explicit expressions for the  
propagators into (\ref{eq:vertmomconcr}), we thus have  
\bel{vertcroseceinfach}
\begin{split}
\bigg(\frac{\partial\sigma}{\partial\Omega}&\bigg)^{\mathrm{V}}
_{4.\mathrm{ord.}}=\lvert S_{fi, 1.\mathrm{ord.}}\rvert^{2}2\mathrm{Re}(h)=\\
&=\frac{ie}{2\pi}\bar{u}_{s}(p)\hat{\slas{A}}^{\mathrm{ext}}(p-q)
u_{\sigma}(q)\bar{u}_{\sigma}(q)\hat{\slas{A}}^{\mathrm{ext},*}(p-q)
  u_{s}(p)\frac{-ie}{2\pi}
\sqrt{1-f_{-}(p)}\sqrt{1-f_{-}(q)}\cdot\\
\cdot \mathrm{Re}\Bigg[&\frac{ie^2}{(2\pi)^2}\int d^4k_{1}
 d^4k_{2}\hat{g}(k_{1})\hat{g}(k_{2})8(pq)\int d^4k\cdot\\
\cdot\Bigg\{
&+\bigg(\frac{1}{(2\pi)^2}\frac{1}{(p-k_{1}-k)^2-m^2+i0}+\\
&\hspace{4.0 cm}+\frac{i}{2\pi}
 \delta((p-k_{1}-k)^2-m^2)\thermf{p-k_{1}-k}\bigg)\cdot\\
&\hspace{1.0 cm}\cdot\bigg(\frac{1}{(2\pi)^2}\frac{1}{(q+k_{2}-k)^2-m^2+i0}+\\
 &\hspace{4.0 cm}+\frac{i}{2\pi}
 \delta((q+k_{2}-k)^2-m^2)\thermf{q+k_{2}-k}\bigg)\cdot\\
&\hspace{1.0 cm}\cdot\bigg(\frac{1}{(2\pi)^2}
 \frac{-1}{k^2+i0}+\frac{i}{2\pi}\delta(k^2)f(k)\bigg)
 \sqrt{(1-f_{-}(p))(1-f_{-}(q))}-\\
&-i\bigg((-1)\frac{1}{2\pi}\delta((p-k_{1}-k)^2-m^2)\thermroot{p-k_{1}-k}
 \bigg)\cdot\\&\hspace{1.0 cm}\cdot\bigg(\frac{1}{(2\pi)^2}
 \frac{1}{(q+k_{2}-k)^2-m^2+i0}+\\&\hspace{4.0 cm}+\frac{i}{2\pi}
 \delta((q+k_{2}-k)^2-m^2)\thermf{q+k_{2}-k}\bigg)\cdot\\
&\hspace{1.0 cm}\cdot\bigg(\frac{i}{2\pi}\delta(k^2)
 \sqrt{f(k)(1+f(k))}\bigg)\sqrt{f_{-}(p)(1-f_{-}(q))}+\\
&+(-i)\bigg(\frac{1}{(2\pi)^2}\frac{1}{(p-k_{1}-k)^2-m^2+i0}+\\
 &\hspace{4.0 cm}+\frac{i}{2\pi}
 \delta((p-k_{1}-k)^2-m^2)\thermf{p-k_{1}-k}\bigg)\cdot\\
&\hspace{1.0 cm}\cdot\bigg((-1)\frac{1}{2\pi}\delta((q+k_{2}-k)^2-m^2)
\thermroot{q+k_{2}-k}\bigg)\cdot\\&\hspace{1.0 cm}\cdot
 \bigg(\frac{i}{2\pi}\delta(k^2)
 \sqrt{f(k)(1+f(k))}\bigg)\sqrt{(1-f_{-}(p))f_{-}(q)}-\\
&-\bigg((-1)\frac{1}{2\pi}\delta((p-k_{1}-k)^2-m^2)
 \thermroot{p-k_{1}-k}\bigg)\cdot\\&\hspace{1.0 cm}\cdot
 \bigg((-1)\frac{1}{2\pi}\delta((q+k_{2}-k)^2-m^2)\thermroot{q+k_{2}-k}\bigg)
 \cdot\\&\hspace{1.0 cm}\cdot\bigg(\frac{-1}{(2\pi)^2}
 \frac{1}{-k^2+i0}+\frac{i}{2\pi}\delta(k^2)f(k)\bigg)\sqrt{f_{-}(p)f_{-}(q)}
\Bigg\}\Bigg].
\end{split}
\end{equation}
Performing the multiplications this gives the following 14 terms
\begin{equation*}
\begin{split}
\bigg(\frac{\partial\sigma}{\partial\Omega}&\bigg)^{\mathrm{V}}
_{4.\mathrm{ord.}}=
\bigg(\frac{d\sigma}{d\Omega}\bigg)^{1.\mathrm{ord.}}\frac{8(pq)e^2}{(2\pi)^2}
 \mathrm{Re}\Bigg[
 \int d^4k_{1}d^4k_{2}\hat{g}(k_{1})
 \hat{g}(k_{2})\int d^4k\Bigg\{\\
16:\hspace{1.1 cm}
 &+\frac{-i}{(2\pi)^6}\frac{(1-f_{-}(p))(1-f_{-}(q))}{((p-k_{1}-k)^2-m^2+i0)
 ((q+k_{2}-k)^2-m^2+i0)(k^2+i0)}+\\
1:\hspace{1.1 cm}&+\frac{-1}{(2\pi)^5}
 \frac{(1-f_{-}(p))(1-f_{-}(q))}{((p-k_{1}-k)^2-m^2+i0)((q+k_{2}-k)^2-m^2+i0)}
 \delta(k^2)f(k)+\\
2:\hspace{1.1 cm}
&+\frac{1}{(2\pi)^5}\frac{(1-f_{-}(p))(1-f_{-}(q))}{((p-k_{1}-k)^2-m^2+i0)
 (k^2+i0)}\cdot\\&\hspace{2.0 cm}\cdot
  \delta((q+k_{2}-k)^2-m^2)\thermf{q+k_{2}-k}+\\
4:\hspace{1.1 cm}
&+\frac{-i}{(2\pi)^4}\frac{(1-f_{-}(p))(1-f_{-}(q))}{((p-k_{1}-k)^2-m^2+i0)}
 \cdot\\&\hspace{2.0 cm}\cdot  
 \delta((q+k_{2}-k)^2-m^2)\delta(k^2)\thermf{q+k_{2}-k}f(k)+\\
3:\hspace{1.1 cm}
&+\frac{1}{(2\pi)^5}\frac{(1-f_{-}(p))(1-f_{-}(q))}{((q+k_{2}-k)^2-m^2+i0)
 (k^2+i0)}\cdot\\&\hspace{2.0 cm}\cdot
 \delta((p-k_{1}-k)^2-m^2)\thermf{p-k_{1}-k}+\\
5:\hspace{1.1 cm}
&+\frac{-i}{(2\pi)^4}\frac{(1-f_{-}(p))(1-f_{-}(q))}{((q+k_{2}-k)^2-m^2+i0)}
 \cdot\\&\hspace{2.0 cm}\cdot
 \delta((p-k_{1}-k)^2-m^2)\delta(k^2)\thermf{p-k_{1}-k}f(k)+\\
6:\hspace{1.1 cm}
&+\frac{i}{(2\pi)^4}\frac{(1-f_{-}(p))(1-f_{-}(q))}
 {(k^2+i0)}\cdot\\&\hspace{2.0 cm}\cdot\delta((p-k_{1}-k)^2-m^2)
 \thermf{p-k_{1}-k}\cdot\\&\hspace{2.0 cm}\cdot
  \delta((q+k_{2}-k)^2-m^2)\thermf{q+k_{2}-k}+
\end{split}
\end{equation*}
\bel{vertcrse4ausm}
\begin{split}
7:\hspace{1.1 cm}
&+\frac{1}{(2\pi)^3}(1-f_{-}(p))(1-f_{-}(q))\delta((q+k_{2}-k)^2-m^2)
 \thermf{q+k_{2}-k}\cdot\\
  &\hspace{2.0 cm}\cdot\delta((p-k_{1}-k)^2-m^2)
 \delta(k^2)\thermf{p-k_{1}-k}f(k)+\\
12:\hspace{1.1 cm}
&+\frac{-i}{(2\pi)^4}\frac{(1-f_{-}(q))
 \sqrt{f_{-}(p)(1-f_{-}(p))}}{((q+k_{2}-k)^2-m^2+i0)}
 \delta((p-k_{1}-k)^2-m^2)\cdot\\
 &\hspace{2.0 cm}\cdot\thermroot{p-k_{1}-k}\delta(k^2)
 \sqrt{f(k)(1+f(k))}+\\
13:\hspace{1.1 cm}
&+\frac{1}{(2\pi)^3}(1-f_{-}(q))\sqrt{f_{-}(p)(1-f_{-}(p))}
 \delta((p-k_{1}-k)^2-m^2)\cdot\\
 &\hspace{2.0 cm}\cdot\thermroot{p-k_{1}-k}
 \delta((q+k_{2}-k)^2-m^2)\cdot\\
  &\hspace{2.0 cm}\cdot\thermf{q+k_{2}-k}\delta(k^2)\sqrt{f(k)(1+f(k))}+\\
10:\hspace{1.1 cm}
&+\frac{-i}{(2\pi)^4}\frac{(1-f_{-}(p))\sqrt{(1-f_{-}(q))f_{-}(q)}}
{(p-k_{1}-k)^2-m^2+i0}
 \delta((q+k_{2}-k)^2-m^2)\cdot\\
 &\hspace{2.0 cm}\cdot\thermroot{q+k_{2}-k}\delta(k^2)
\sqrt{f(k)(1+f(k))}+\\
11:\hspace{1.1 cm}
&+\frac{1}{(2\pi)^3}(1-f_{-}(p))\sqrt{(1-f_{-}(q))f_{-}(q)}
 \delta((p-k_{1}-k)^2-m^2)\cdot\\
 &\hspace{2.0 cm}\cdot\thermf{p-k_{1}-k}\delta((q+k_{2}-k)^2-m^2)
 \cdot\\&\hspace{2.0 cm}\cdot\thermroot{q+k_{2}-k}
 \delta(k^2)\sqrt{f(k)(1+f(k))}+\\
14:\hspace{1.1 cm}
&+\frac{-i}{(2\pi)^4}\frac{\sqrt{f_{-}(p)(1-f_{-}(p))f_{-}(q)(1-f_{-}(q))}}
 {k^2-i0}\cdot\\
 &\hspace{2.0 cm}\cdot\delta((p-k_{1}-k)^2-m^2)
 \thermroot{p-k_{1}-k}\cdot\\
 &\hspace{2.0 cm}\cdot\delta((q+k_{2}-k)^2-m^2)
 \thermroot{q+k_{2}-k}+\\
15:\hspace{1.1 cm}
&+\frac{1}{(2\pi)^3}\sqrt{f_{-}(p)(1-f_{-}(p))f_{-}(q)(1-f_{-}(q))}
 \cdot\\&\hspace{2.0 cm}\cdot\delta((p-k_{1}-k)^2-m^2)
 \thermroot{p-k_{1}-k}\cdot\\&\hspace{2.0 cm}\cdot\delta((q+k_{2}-k)^2-m^2)
 \thermroot{q+k_{2}-k}\delta(k^2)f(k)
\Bigg\}\Bigg].
\end{split}
\end{equation}
\newpage
\textsc{Self Energy:} We consider the matrix element 
\bel{matrelemseradcorr}
S_{fi} = 
 \langle b_{s}(\vec{p},\beta)^{+}\Omega_{\beta}\lvert 
 T_{3}^{\mathrm{SE}} \rvert b_{\sigma}(\vec{q},\beta)^{+}\Omega_{\beta}
\rangle,
\end{equation}
with $T_{3}^{\mathrm{SE}}$ from (\ref{eq:t3SEB}) and (\ref{eq:t3SEA}). 
Proceeding as we did with the vertex to obtain (\ref{eq:vertmomconcr}) 
we get in the momentum space representation
\bel{SEAimpulsradcorr}
\begin{split}
S_{fi}^{3.\mathrm{ord.SE,A}}=&\frac{e^3}{(2\pi)^3}
     \int d^4k_{1}d^4k_{2}\hat{g}(k_{1})\hat{g}(k_{2})\bar{u}_{s}(p)\cdot\\
\cdot\bigg[&+\sqrt{(1-f_{-}(p))(1-f_{-}(q))}\cdot\\
  &\hspace{1.5 cm}\cdot\int\!d^4k\big[-\gamma^{\mu}
  \propSft{11}{\F}{p+k-k_{1}}{}\gamma_{\mu}\propDft{11}{\F}{k}\big]
  \propSft{11}{\F}{p-k_{1}-k_{2}}{}+\\
&+\sqrt{(1-f_{-}(p))(1-f_{-}(q))}\cdot\\
  &\hspace{1.5 cm}\cdot\int\!d^4k\big[-\gamma^{\mu}
  \propSft{12}{\F}{p+k-k_{1}}{}\gamma_{\mu}\propDft{12}{\F}{k}\big]
  \propSft{21}{\F}{p-k_{1}-k_{2}}{}+\\
&+i\sqrt{f_{-}(p)(1-f_{-}(q))}\cdot\\
  &\hspace{1.5 cm}\cdot\int\!d^4k\big[+\gamma^{\mu}
  \propSft{21}{\F}{p+k-k_{1}}{}\gamma_{\mu}\propDft{12}{\F}{k}\big]
  \propSft{11}{\F}{p-k_{1}-k_{2}}{}+\\
&+i\sqrt{f_{-}(p)(1-f_{-}(q))}\cdot\\
  &\hspace{1.5 cm}\cdot\int\!d^4k\big[-\gamma^{\mu}
  \propSft{22}{\F}{p+k-k_{1}}{}\gamma_{\mu}\propDft{22}{\F}{k}\big]
  \propSft{21}{\F}{p-k_{1}-k_{2}}{}\bigg]\cdot\\
&\hspace{4.0 cm}\cdot\gamma^{\nu}u_{\sigma}(q)
 \hat{A}^{\mathrm{ext}}_{\nu}(p-q-k_{1}-k_{2})
\end{split}
\end{equation}
and
\bel{SEBimpulsradcorr}
\begin{split}
S_{fi}^{3.\mathrm{ord.SE,B}}=&\frac{e^3}{(2\pi)^3}
     \int d^4k_{1}d^4k_{2}\hat{g}(k_{1})\hat{g}(k_{2})\bar{u}_{s}(p)
     \gamma^{\nu}\cdot\\
\cdot\bigg[&+\sqrt{(1-f_{-}(p))(1-f_{-}(q))}\cdot\\
  &\hspace{1.5 cm}\cdot\int\!d^4k\big[-
  \propSft{11}{\F}{q+k_{1}+k_{2}}{}\gamma^{\mu}\propSft{11}{\F}{q+k+k_{1}}{}
  \gamma_{\mu}\propDft{11}{\F}{k}\big]+\\
&+\sqrt{(1-f_{-}(p))(1-f_{-}(q))}\cdot\\
  &\hspace{1.5 cm}\cdot\int\!d^4k\big[-
  \propSft{12}{\F}{q+k_{1}+k_{2}}{}\gamma^{\mu}\propSft{12}{\F}{q+k+k_{1}}{}
  \gamma_{\mu}\propDft{12}{\F}{k}\big]+\\
&+(-i)\sqrt{(1-f_{-}(p))f_{-}(q)}\cdot\\
  &\hspace{1.5 cm}\cdot\int\!d^4k\big[-
 \propSft{11}{\F}{q+k_{1}+k_{2}}{}\gamma^{\mu}\propSft{12}{\F}{q+k+k_{1}}{}
  \gamma_{\mu}\propDft{12}{\F}{k}\big]+\\
&+(-i)\sqrt{(1-f_{-}(p))f_{-}(q)}\cdot\\
  &\hspace{1.5 cm}\cdot\int\!d^4k\big[+
 \propSft{12}{\F}{q+k_{1}+k_{2}}{}\gamma^{\mu}\propSft{22}{\F}{q+k+k_{1}}{}
  \gamma_{\mu}\propDft{22}{\F}{k}\big]\bigg]\cdot\\
&\hspace{4.0 cm}\cdot u_{\sigma}(q)
  \hat{A}^{\mathrm{ext}}_{\nu}(p-q-k_{1}-k_{2}).
\end{split}
\end{equation}
As in the vertex contribution (see after (\ref{eq:vertcrosec4})) 
$S_{fi}^{\mathrm{SE}}$ is proportional to $S_{fi,1.\mathrm{ord}.}$ and thus  
the cross section is given by 
\begin{equation*}
\begin{split}
\bigg(&\frac{\partial\sigma}{\partial\Omega}\bigg)^{\mathrm{SE}}
_{4.\mathrm{ord.}}=\lvert S_{fi, 1.\mathrm{ord.}}\rvert^{2}2\mathrm{Re}
\Big(S_{fi}^{3.\mathrm{ord.SE,A}}+S_{fi}^{3.\mathrm{ord.SE,B}}\Big)=\\
&=\frac{ie}{2\pi}\bar{u}_{s}(p)\hat{\slas{A}}^{\mathrm{ext}}(p-q)
 u_{\sigma}(q)
 \bar{u}_{\sigma}(q)\hat{\slas{A}}^{\mathrm{ext},*}(p-q)
 u_{s}(p)\frac{-ie}{2\pi}
 \sqrt{(1-f_{-}(p))(1-f_{-}(q))}\cdot\\
 &\hspace{2.0 cm}\cdot\mathrm{Re}\bigg[
 8m^2\frac{-ie^2}{(2\pi)^2}\int d^4k_{1}d^4k_{2}\hat{g}(k_{1})
  \hat{g}(k_{2})\int d^4k\cdot\\
\bigg\{
&+\sqrt{(1-f_{-}(p))(1-f_{-}(q))}(-1)\cdot
 \bigg(\frac{1}{(2\pi)^2}\frac{-1}{k^2+i0}+\frac{i}{2\pi}\delta(k^2)f(k)\bigg) 
 \cdot\\&\hspace{0.3 cm}\cdot
 \bigg(\frac{1}{(2\pi)^2}\frac{1}
  {(p+k-k_{1})^2-m^2+i0}+\frac{i}{2\pi}\delta((p+k-k_{1})^2-m^2)
 \thermf{p+k-k_{1}}
  \bigg)\cdot\\
 &\hspace{0.3 cm}\cdot\bigg(\frac{1}{(2\pi)^2}\frac{1}
  {(p-k_{1}-k_{2})^2-m^2+i0}+\\
  &\hspace{4.0 cm}+\frac{i}{2\pi}\delta((p-k_{1}-k_{2})^2-m^2)\!\!
  \thermf{p-k_{1}-k_{2}}
  \bigg)+\\
&+\sqrt{(1-f_{-}(p))(1-f_{-}(q))}(-1)\bigg(\frac{i}{2\pi}\delta(k^2)
  \sqrt{f(k)(1+f(k))}\bigg)\cdot\\&\hspace{0.3 cm}\cdot
  \bigg(\frac{-1}{2\pi}
  \delta((p+k-k_{1})^2-m^2)
  \thermroot{p+k-k_{1}}\bigg)\cdot\\&\hspace{0.3 cm}\cdot
  \bigg(\frac{-1}{2\pi}\delta((p-k_{1}-k_{2})^2-m^2)
  \thermroot{p-k_{1}-k_{2}}\bigg)+\\
&+i\sqrt{f_{-}(p)(1-f_{-}(q))}\bigg(\frac{-1}{2\pi}\delta((p+k-k_{1})^2-m^2)
  \thermroot{p+k-k_{1}}\bigg)\cdot\\&\hspace{0.3 cm}\cdot
  \bigg(\frac{i}{2\pi}\delta(k^2)\sqrt{f(k)(1+f(k))}\bigg)
  \bigg(\frac{1}{(2\pi)^2}\frac{1}
  {(p-k_{1}-k_{2})^2-m^2+i0}+\\
  &\hspace{4.0 cm}+\frac{i}{2\pi}\delta((p-k_{1}-k_{2})^2-m^2)
  \thermf{p-k_{1}-k_{2}}
  \bigg)+\\
&+i\sqrt{f_{-}(p)(1-f_{-}(q))}(-1)
  \bigg(\frac{1}{(2\pi)^2}
  \frac{-1}{-k^2+i0}+\frac{i}{2\pi}\delta(k^2)f(k)\bigg)
  \cdot\\&\hspace{0.3 cm}\cdot\bigg(\frac{1}{(2\pi)^2}\frac{1}
  {(p+k-k_{1})^2-m^2-i0}+\frac{-i}{2\pi}\delta((p+k-k_{1})^2-m^2)
 \thermf{p+k-k_{1}}
   \bigg)\cdot\\
  &\hspace{0.3 cm}\cdot
   \bigg(\frac{-1}{2\pi}\delta((p-k_{1}-k_{2})^2-m^2)
  \thermroot{p-k_{1}-k_{2}}\bigg)+
\end{split}
\end{equation*}
\bel{crosecSE}
\begin{split}
&+\sqrt{(1-f_{-}(p))(1-f_{-}(q))}(-1)
  \bigg(\frac{1}{(2\pi)^2}\frac{-1}{k^2+i0}+\frac{i}{2\pi}\delta(k^2)f(k)\bigg)
  \cdot\\&\hspace{1.0 cm}\cdot\bigg(\frac{1}{(2\pi)^2}\frac{1}
  {(q+k_{1}+k_{2})^2-m^2+i0}+\\
  &\hspace{4.0 cm}\frac{i}{2\pi}\delta((q+k_{1}+k_{2})^2-m^2)
  \thermf{q+k_{1}+k_{2}}
  \bigg)\cdot\\&\hspace{1.0 cm}\cdot
   \bigg(\frac{1}{(2\pi)^2}\frac{1}
  {(q+k+k_{1})^2-m^2+i0}+\\
  &\hspace{4.0 cm}+\frac{i}{2\pi}\delta((q+k+k_{1})^2-m^2)
 \thermf{q+k+k_{1}}
  \bigg)+\\
&+\sqrt{(1-f_{-}(p))(1-f_{-}(q))}(-1)
  \bigg(\frac{i}{2\pi}\delta(k^2)
  \sqrt{f(k)(1+f(k))}\bigg)\cdot\\&\hspace{1.0 cm}\cdot
  \bigg(\frac{-1}{2\pi}
  \delta((q+k_{1}+k_{2})^2-m^2)
  \thermroot{q+k_{1}+k_{2}}\bigg)\cdot\\&\hspace{1.0 cm}\cdot
  \bigg(\frac{-1}{2\pi}\delta((q+k+k_{1})^2-m^2)
  \thermroot{q+k+k_{1}}\bigg)+\\
&+(-i)\sqrt{f_{-}(q)(1-f_{-}(p))}(-1)\bigg(\frac{i}{2\pi}\delta(k^2)
  \sqrt{f(k)(1+f(k))}\bigg)\cdot\\
  &\hspace{1.0 cm}\cdot
  \bigg(\frac{1}{(2\pi)^2}\frac{1}
  {(q+k_{1}+k_{2})^2-m^2+i0}+\\
  &\hspace{4.0 cm}+\frac{i}{2\pi}\delta((q+k_{1}+k_{2})^2-m^2)
 \thermf{q+k_{1}+k_{2}}
  \bigg)\cdot\\&\hspace{1.0 cm}\cdot
  \bigg(\frac{-1}{2\pi}\delta((q+k+k_{1})^2-m^2)
  \thermroot{q+k+k_{1}}\bigg)+\\
&+(-i)\sqrt{f_{-}(q)(1-f_{-}(p))}
  \bigg(\frac{-1}{2\pi}\delta((q+k_{1}+k_{2})^2-m^2)
  \thermroot{q+k_{1}+k_{2}}\bigg)\cdot\\&\hspace{1.0 cm}\cdot
  \bigg(\frac{1}{(2\pi)^2}\frac{-1}{-k^2+i0}+\frac{i}{2\pi}\delta(k^2)f(k)
  \bigg)\bigg(\frac{1}{(2\pi)^2}\frac{1}
  {(q+k+k_{1})^2-m^2-i0}+\\
  &\hspace{4.0 cm}+\frac{-i}{2\pi}\delta((q+k+k_{1})^2-m^2)
 \thermf{q+k+k_{1}}\bigg)
   \bigg\}\bigg].
\end{split}
\end{equation}
After having performed the multiplications, this gives the following 30 terms:
\begin{equation*}
\begin{split}
\bigg(&\frac{\partial\sigma}{\partial\Omega}\bigg)^{\mathrm{SE}}
_{4.\mathrm{ord.}}=
\bigg(\frac{d\sigma}{d\Omega}\bigg)^{1.\mathrm{ord.}}
\frac{8m^2e^2}{(2\pi)^2}\mathrm{Re}\Bigg[\int d^4k_{1}d^4k_{2}\hat{g}(k_{1})
 \hat{g}(k_{2})\int d^4k\bigg\{\\
\mathrm{o}:\hspace{1.1 cm}
 &+\frac{-i}{(2\pi)^6}\frac{(1-f_{-}(p))(1-f_{-}(q))}{((p+k-k_{1})^2-m^2+i0)
(k^2+i0)
 ((p-k_{1}-k_{2})^2-m^2+i0)}+\\
\mathrm{a}:\hspace{1.1 cm}
 &+\frac{1}{(2\pi)^5}\frac{(1-f_{-}(p))(1-f_{-}(q))}
 {((p+k-k_{1})^2-m^2+i0)(k^2+i0)}\cdot\\
 &\hspace{1.5 cm}\cdot\delta((p-k_{1}-k_{2})^2-m^2)\thermf{p-k_{1}-k_{2}}+\\
\mathrm{b}:\hspace{1.1 cm}
 &+\frac{-1}{(2\pi)^5}\frac{(1-f_{-}(p))(1-f_{-}(q))}{((p+k-k_{1})^2-m^2+i0)
 ((p-k_{1}-k_{2})^2-m^2+i0)}\delta(k^2)f(k)+\\
\mathrm{c}:\hspace{1.1 cm}
 &+\frac{-i}{(2\pi)^4}\frac{(1-f_{-}(p))(1-f_{-}(q))}{((p+k-k_{1})^2-m^2+i0)}
 \delta(k^2)f(k)\cdot\\
 &\hspace{1.5 cm}\cdot\delta((p-k_{1}-k_{2})^2-m^2)\thermf{p-k_{1}-k_{2}}+\\
\mathrm{d}:\hspace{1.1 cm}
 &+\frac{1}{(2\pi)^5}\frac{(1-f_{-}(p))(1-f_{-}(q))}{(k^2+i0)
 ((p-k_{1}-k_{2})^2-m^2+i0)}\cdot\\
 &\hspace{1.5 cm}\cdot\delta((p+k-k_{1})^2-m^2)\thermf{p+k-k_{1}}+\\
\mathrm{e}:\hspace{1.1 cm}
 &+\frac{i}{(2\pi)^4}\frac{(1-f_{-}(p))(1-f_{-}(q))}{(k^2+i0)}
 \delta((p+k-k_{1})^2-m^2)\delta((p-k_{1}-k_{2})^2-m^2)\cdot\\
  &\hspace{1.5 cm}\cdot\thermf{p+k-k_{1}}
   \thermf{p-k_{1}-k_{2}}+\\
\mathrm{f}:\hspace{1.1 cm}
 &+\frac{-i}{(2\pi)^4}\frac{(1-f_{-}(p))(1-f_{-}(q))}
{((p-k_{1}-k_{2})^2-m^2+i0)}\delta(k^2)f(k)\cdot\\
 &\hspace{1.5 cm}\cdot\delta((p+k-k_{1})^2-m^2)\thermf{p+k-k_{1}}+\\
\mathrm{g}:\hspace{1.1 cm}
 &+\frac{1}{(2\pi)^3}(1-f_{-}(p))(1-f_{-}(q))\delta(k^2)f(k)\cdot\\
 &\hspace{1.5 cm}\cdot
 \delta((p+k-k_{1})^2-m^2)\delta((p-k_{1}-k_{2})^2-m^2)\cdot\\
   &\hspace{1.5 cm}\cdot\thermf{p+k-k_{1}}
   \thermf{p-k_{1}-k_{2}}+
\end{split}
\end{equation*}
\begin{equation*}
\begin{split}
\mathrm{h}:\hspace{1.1 cm}
 &+\frac{-1}{(2\pi)^3}(1-f_{-}(p))(1-f_{-}(q))\delta(k^2)\sqrt{f(k)(1+f(k))}
 \cdot\\&\hspace{1.5 cm}\cdot\delta((p+k-k_{1})^2-m^2)
 \delta((p-k_{1}-k_{2})^2-m^2)\cdot\\
   &\hspace{1.5 cm}\cdot\thermroot{p+k-k_{1}}
 \thermroot{p-k_{1}-k_{2}}+\\
\mathrm{i}:\hspace{1.1 cm}
 &+\frac{-i}{(2\pi)^4}\frac{(1-f_{-}(q))\sqrt{f_{-}(p)(1-f_{-}(p))}}
 {(p-k_{1}-k_{2})^2-m^2+i0}\delta(k^2)\sqrt{f(k)(1+f(k))}\cdot\\
 &\hspace{1.5 cm}\cdot
 \delta((p+k-k_{1})^2-m^2)
  \thermroot{p+k-k_{1}}+\\
\mathrm{j}:\hspace{1.1 cm}
 &+\frac{1}{(2\pi)^3}(1-f_{-}(q))\sqrt{f_{-}(p)(1-f_{-}(p))}
  \delta(k^2)\sqrt{f(k)(1+f(k))}\cdot\\
 &\hspace{1.5 cm}\cdot
 \delta((p-k_{1}-k_{2})^2-m^2)
 \delta((p+k-k_{1})^2-m^2)\cdot\\
   &\hspace{1.5 cm}\cdot\thermf{p-k_{1}-k_{2}}\thermroot{p+k-k_{1}}+\\
\mathrm{k}:\hspace{1.1 cm}
 &+\frac{-1}{(2\pi)^5}\frac{(1-f_{-}(q))\sqrt{f_{-}(p)(1-f_{-}(p))}}
 {((p+k-k_{1})^2-m^2-i0)(-k^2+i0)}\cdot\\
 &\hspace{1.5 cm}\cdot\delta((p-k_{1}-k_{2})^2-m^2)\thermroot{p-k_{1}-k_{2}}+\\
\mathrm{m}:\hspace{1.1 cm}
 &+\frac{i}{(2\pi)^4}\frac{(1-f_{-}(q))\sqrt{f_{-}(p)(1-f_{-}(p))}}
 {(p+k-k_{1})^2-m^2-i0}\delta(k^2)f(k)\cdot\\
 &\hspace{1.5 cm}\delta((p-k_{1}-k_{2})^2-m^2)
 \thermroot{p-k_{1}-k_{2}}+\\
\mathrm{l}:\hspace{1.1 cm}
 &+\frac{i}{(2\pi)^4}\frac{(1-f_{-}(q))\sqrt{f_{-}(p)(1-f_{-}(p))}}{-k^2+i0}
 \cdot\\&\hspace{1.5 cm}\cdot
 \delta((p+k-k_{1})^2-m^2)\delta((p-k_{1}-k_{2})^2-m^2)\cdot\\
   &\hspace{1.5 cm}\cdot
 \thermf{p+k-k_{1}}\thermroot{p-k_{1}-k_{2}}+\\
\mathrm{n}:\hspace{1.1 cm}
 &+\frac{1}{(2\pi)^3}(1-f_{-}(q))\sqrt{f_{-}(p)(1-f_{-}(p))}\delta(k^2)f(k)
 \cdot\\&\hspace{1.5 cm}\cdot
 \delta((p-k_{1}-k_{2})^2-m^2)\delta((p+k-k_{1})^2-m^2)
 \cdot\\&\hspace{1.5 cm}\cdot
 \thermf{p+k-k_{1}}\thermroot{p-k_{1}-k_{2}}+
\end{split}
\end{equation*}
\begin{equation*}
\begin{split}
\mathrm{o'}:\hspace{1.1 cm}
 &+\frac{-i}{(2\pi)^6}\frac{(1-f_{-}(p))(1-f_{-}(q))}{((q+k+k_{1})^2-m^2+i0)
(k^2+i0)((q+k_{1}+k_{2})^2-m^2+i0)}+\\
\mathrm{a'}:\hspace{1.1 cm}
 &+\frac{1}{(2\pi)^5}\frac{(1-f_{-}(p))(1-f_{-}(q))}
 {((q+k+k_{1})^2-m^2+i0)(k^2+i0)}\cdot\\&\hspace{1.5 cm}\cdot
 \delta((q+k_{1}+k_{2})^2-m^2)\thermf{q+k_{1}+k_{2}}+\\
\mathrm{b'}:\hspace{1.1 cm}
 &+\frac{-1}{(2\pi)^5}\frac{(1-f_{-}(p))(1-f_{-}(q))}{((q+k+k_{1})^2-m^2+i0)
 ((q+k_{1}+k_{2})^2-m^2+i0)}\delta(k^2)f(k)+\\
\mathrm{e'}:\hspace{1.1 cm}
 &+\frac{-i}{(2\pi)^4}\frac{(1-f_{-}(p))(1-f_{-}(q))}{((q+k+k_{1})^2-m^2+i0)}
 \cdot\\&\hspace{1.5 cm}\cdot
 \delta(k^2)f(k)\delta((q+k_{1}+k_{2})^2-m^2)\thermf{q+k_{1}+k_{2}}+\\
\mathrm{c'}:\hspace{1.1 cm}
 &+\frac{1}{(2\pi)^5}\frac{(1-f_{-}(p))(1-f_{-}(q))}{(k^2+i0)
 ((q+k_{1}+k_{2})^2-m^2+i0)}\cdot\\&\hspace{1.5 cm}\cdot
 \delta((q+k+k_{1})^2-m^2)\thermf{q+k+k_{1}}+\\
\mathrm{f'}:\hspace{1.1 cm}
 &+\frac{i}{(2\pi)^4}\frac{(1-f_{-}(p))(1-f_{-}(q))}{(k^2+i0)}
 \delta((q+k+k_{1})^2-m^2)\delta((q+k_{1}+k_{2})^2-m^2)\cdot\\
  &\hspace{1.5 cm}\cdot\thermf{q+k+k_{1}}
   \thermf{q+k_{1}+k_{2}}+\\
\mathrm{d'}:\hspace{1.1 cm}
 &+\frac{-i}{(2\pi)^4}\frac{(1-f_{-}(p))(1-f_{-}(q))}
{((q+k_{1}+k_{2})^2-m^2+i0)}\delta(k^2)f(k)\cdot\\&\hspace{1.5 cm}\cdot
 \delta((q+k+k_{1})^2-m^2)\thermf{q+k+k_{1}}+\\
\mathrm{g'}:\hspace{1.1 cm}
 &+\frac{1}{(2\pi)^3}(1-f_{-}(p))(1-f_{-}(q))\delta(k^2)f(k)
 \cdot\\&\hspace{1.5 cm}\cdot
 \delta((q+k+k_{1})^2-m^2)\delta((q+k_{1}+k_{2})^2-m^2)\cdot\\
   &\hspace{1.5 cm}\cdot\thermf{q+k+k_{1}}
   \thermf{q+k_{1}+k_{2}}+
\end{split}
\end{equation*}
\bel{SEcrosecausmult}
\begin{split}
\mathrm{h'}:\hspace{1.1 cm}
 &+\frac{-1}{(2\pi)^3}(1-f_{-}(p))(1-f_{-}(q))\delta(k^2)\sqrt{f(k)(1+f(k))}
 \cdot\\&\hspace{1.5 cm}\cdot
 \delta((q+k+k_{1})^2-m^2)\delta((q+k_{1}+k_{2})^2-m^2)\cdot\\
  &\hspace{1.5 cm}\cdot\thermroot{q+k+k_{1}}
 \thermroot{q+k_{1}+k_{2}}+\\
\mathrm{i'}:\hspace{1.1 cm}
 &+\frac{-i}{(2\pi)^4}\frac{(1-f_{-}(p))\sqrt{f_{-}(q)(1-f_{-}(q))}}
 {((q+k_{1}+k_{2})^2-m^2+i0)}\delta(k^2)\sqrt{f(k)(1+f(k))}\cdot\\
 &\hspace{1.5 cm}\cdot
 \delta((q+k+k_{1})^2-m^2)
 \thermroot{q+k+k_{1}}\delta(k^2)+\\
\mathrm{j'}:\hspace{1.1 cm}
 &+\frac{1}{(2\pi)^3}(1-f_{-}(p))\sqrt{f_{-}(q)(1-f_{-}(q))}
 \delta(k^2)\sqrt{f(k)(1+f(k))}\cdot\\
 &\hspace{1.5 cm}\cdot
 \delta((q+k_{1}+k_{2})^2-m^2)\delta((q+k+k_{1})^2-m^2)\cdot\\
  &\hspace{1.5 cm}\cdot\thermf{q+k_{1}+k_{2}}\thermroot{q+k+k_{1}}+\\
\mathrm{k'}:\hspace{1.1 cm}
 &+\frac{-1}{(2\pi)^5}\frac{(1-f_{-}(p))\sqrt{f_{-}(q)(1-f_{-}(q))}}
 {((q+k+k_{1})^2-m^2-i0)(-k^2+i0)}\cdot\\&\hspace{1.5 cm}\cdot
 \delta((q+k_{1}+k_{2})^2-m^2)\thermroot{q+k_{1}+k_{2}}+\\
\mathrm{l'}:\hspace{1.1 cm}
 &+\frac{i}{(2\pi)^4}\frac{(1-f_{-}(p))\sqrt{f_{-}(q)(1-f_{-}(q))}}
 {((q+k+k_{1})^2-m^2-i0)}\delta(k^2)f(k)\cdot\\&\hspace{1.5 cm}\cdot
 \delta((q+k_{1}+k_{2})^2-m^2)\thermroot{q+k_{1}+k_{2}}+\\
\mathrm{m'}:\hspace{1.1 cm}
 &+\frac{i}{(2\pi)^4}\frac{(1-f_{-}(p))\sqrt{f_{-}(q)(1-f_{-}(q))}}{-k^2+i0}
 \cdot\\
 &\hspace{1.5 cm}\cdot
 \delta((q+k+k_{1})^2-m^2)\delta((q+k_{1}+k_{2})^2-m^2)\cdot\\
 &\hspace{1.5 cm}\cdot
 \thermf{q+k+k_{1}}\thermroot{q+k_{1}+k_{2}}+\\
\mathrm{n'}:\hspace{1.1 cm}
 &+\frac{1}{(2\pi)^3}(1-f_{-}(p))\sqrt{f_{-}(q)(1-f_{-}(q))}\delta(k^2)f(k)
 \cdot\\&\hspace{1.5 cm}\cdot
 \delta((q+k_{1}+k_{2})^2-m^2)\delta((q+k+k_{1})^2-m^2)
 \cdot\\&\hspace{1.5 cm}\cdot\thermf{q+k+k_{1}}\thermroot{q+k_{1}+k_{2}}
 \bigg\}\Bigg].
\end{split}
\end{equation}
\textsc{Bremsstrahlung:} 
The bremsstrahlung is of second order, therefore (see (\ref{eq:crosec4})) it 
contributes with 
\bel{bscrosec}
S^{\mathrm{BS,em/abs}}_{fi}(S^{\mathrm{BS,em/abs}}_{fi})^{*}
\end{equation}
to the cross section. The matrix elements are known from 
(\ref{eq:matrelmbsexpl}). This gives for the cross section\footnote{ 
The factor $1=\int\!
d^3k\delta(\vec{k}-\vec{k}')$ is inserted in order to have $k'$ independently 
from $k$ in the second factor. We omit the terms proportional to 
$\slas{k_{i}}$ in the numerators since we are interested in the adiabatic 
limit and these terms could at most give finite contributions (as will be 
clear from the divergence properties of the terms proportional to $p\e$ or 
$q\e$. These are investigated in Sec. \ref{subsubsec:poscon}). On similar 
grounds we can omit the terms proportional to $\slas{k}$.}  

\bel{bscrosecexpl}
\begin{split}
\bigg(\frac{\partial\sigma}{\partial\Omega}&\bigg)^{\mathrm{BS,em}}
_{4.\mathrm{ord.}}\hspace{-0.7 cm}(k)\hspace{0.7 cm}=
\frac{4e^4}{(2\pi)^9}\int d^4k_{1}d^4k_{2}\hat{g}(k_{1})\hat{g}^{*}(k_{2})
  \int\!dk_{0}\delta(k^2)\Theta(k_{0})
  \int\!dk_{0}'\delta({k'}^2)\Theta(k_{0}')\\
  \int\!d^3k'&\delta(\vec{k}-\vec{k'})
  \Bigg\{\bar{u}_{s}(p)
  \slas{\hat{A}}^{\mathrm{ext}}(p-q+k)u_{\sigma}(q)\bar{u}_{\sigma}(q)
  \slas{\hat{A}}^{\mathrm{ext},*}(p-q+k')u_{s}(p)\cdot\\
\cdot\bigg[
 &+\frac{(p\e)
  \sqrt{(1-f_{-}(p))(1-f_{-}(q))}\sqrt{1+f(k)}}
    {((p-k_{1}+k)^2-m^2+i0)}+\\&+i2\pi
   (p\e)
    \sqrt{(1-f_{-}(p))(1-f_{-}(q))}\sqrt{1+f(k)}
   \cdot\\&\hspace{3.0 cm}\cdot\delta((p-k_{1}+k)^2-m^2)
   \thermf{p-k_{1}+k}+\\
&+i2\pi(p\e)
   \sqrt{f_{-}(p)(1-f_{-}(q))}\sqrt{f(k)}
   \delta((p-k_{1}+k)^2-m^2)\thermroot{p-k_{1}+k} +\\
&+\frac{(q\epsilon)
  \sqrt{(1-f_{-}(p))(1-f_{-}(q))}
   \sqrt{1+f(k)}}
    {((q+k_{1}-k)^2-m^2+i0)}+\\&+i2\pi(q\e)
    \sqrt{(1-f_{-}(p))(1-f_{-}(q))}\sqrt{1+f(k)}\cdot\\
  &\hspace{3.0 cm}\cdot\delta((q+k_{1}-k)^2-m^2)
   \thermf{q+k_{1}-k}+\\
&+i2\pi(q\epsilon)
   \sqrt{(1-f_{-}(p))f_{-}(q)}\sqrt{f(k)}
   \delta((q+k_{1}-k)^2-m^2)\thermroot{q+k_{1}-k}\bigg]\cdot\\
\cdot\bigg[&+\frac{(p\e)^{*}
  \sqrt{(1-f_{-}(p))(1-f_{-}(q))}\sqrt{1+f(k')}}
    {((p-k_{2}+k')^2-m^2-i0)}-\\&-
   i2\pi(p\epsilon)^{*}
    \sqrt{(1-f_{-}(p))(1-f_{-}(q))}\sqrt{1+f(k')}\cdot\\
 &\hspace{3.0 cm}\cdot\delta((p-k_{2}+k')^2-m^2)
   \thermf{p-k_{2}+k'}-\\
&-i2\pi(p\epsilon)^{*}
   \sqrt{f_{-}(p)(1-f_{-}(q))}\sqrt{f(k')}\cdot\\&\hspace{3.0 cm}\cdot
   \delta((p-k_{2}+k')^2-m^2)\thermroot{p-k_{2}+k'} +\\
&+\frac{(q\epsilon)^{*}
  \sqrt{(1-f_{-}(p))(1-f_{-}(q))}
   \sqrt{1+f(k')}}
    {((q+k_{2}-k')^2-m^2-i0)}-\\
&-i2\pi(q\epsilon)^{*}
    \sqrt{(1-f_{-}(p))(1-f_{-}(q))}\sqrt{1+f(k')}\cdot\\
  &\hspace{3.0 cm}\cdot\delta((q+k_{2}-k')^2-m^2)
   \thermf{q+k_{2}-k'}-\\
&-i2\pi(q\epsilon)^{*}
   \sqrt{(1-f_{-}(p))f_{-}(q)}\sqrt{f(k')}\cdot\\
   &\hspace{3.0 cm}\cdot
   \delta((q+k_{2}-k')^2-m^2)\thermroot{q+k_{2}-k'}\bigg]\Bigg\}.
\end{split} 
\end{equation}

We get the terms 
for the absorption from the ones for the emission by changing  
$k\leftrightarrow -k$ (not in $\Theta(k_{0})$) and $f\leftrightarrow1+f$. 
Performing the $k_{0}$-, 
$k_{0}'$- and $\vec{k}'$-integrations we get a factor 
$\frac{1}{2\lvert\vec{k}\rvert}$, and 
we see that $k'=k$. In order to compare this with 
the contributions from the vertex and the self energy we integrate over small 
photon energies $\lvert \vec{k} \rvert \leq \omega_{0}\ll m_{\mathrm{e}^{-}}$, 
where $\omega_{0}$ 
can be interpreted as some finite energy resolution of the detector. Since 
we are interested in the adiabatic limit we can   
replace $\thermf{p-k_{1}+k}$ by $\thermf{p}$, etc. 
Furthermore, we set $p^2=m^2=q^2$. 
Performing the various multiplications gives the following 2 times 36 
terms\footnote{Here we summed over \emph{four} polarizations of the photons, 
including the unphysical scalar and longitudinal ones (cf. \cite{ScharfBuch} 
p. 245 and eq. (3.12.42)), and we used $\sum_{n=1}^{3}\e_{\mu}^{(n)*}\e_{\nu}
^{(n)}-\e_{\mu}^{(0)*}\e_{\nu}^{(0)}=-g_{\mu\nu}$.}:
\begin{equation*}
\begin{split}
\bigg(\frac{\partial\sigma}{\partial\Omega}&\bigg)^{\mathrm{BS}}
_{4.\mathrm{ord.}}\hspace{-0.5 cm}(k)\hspace{0.5 cm}=
\bigg(\frac{d\sigma}{d\Omega}\bigg)^{1.\mathrm{ord.}}\frac{4e^2}{(2\pi)^7}
 \int d^4k_{1}d^4k_{2}\hat{g}(k_{1})\hat{g}^{*}(k_{2})
\int_{\lvert\vec{k}\rvert\leq\omega_{0}}d^3k\frac{1}
 {2\lvert \vec{k} \rvert} \cdot\Bigg\{\\
\mathrm{A}:\hspace{0.8 cm}
 &+\frac{-m^2(1-f_{-}(p))(1-f_{-}(q))(1+f(k))}{(2p(k-k_{1})+(k-k_{1})^2+i0)
  (2p(k-k_{2})+(k-k_{2})^2-i0)}+\\
 \mathrm{C}:\hspace{0.8 cm}
  &+\frac{-2\pi im^2}{2p(k-k_{2}+(k-k_{2})^2)-i0}(1-f_{-}(p))(1-f_{-}(q))
  (1+f(k))\cdot\\
   &\hspace{2.0 cm}\cdot\delta(2p(k-k_{1})+(k-k_{1})^2)\thermf{p}+\\
\alpha:\hspace{0.8 cm}
  &+\frac{-2\pi im^2}{2p(k-k_{2})+(k-k_{2})^2-i0}
 \sqrt{f_{-}(p)(1-f_{-}(p))}(1-f_{-}(q))
  \sqrt{f(k)(1+f(k))}\cdot\\
   &\hspace{2.0 cm}\cdot\delta(2p(k-k_{1})+(k-k_{1})^2)\thermroot{p}+\\
\mathrm{V}:\hspace{0.8 cm}
  &+\frac{-pq(1-f_{-}(p))(1-f_{-}(q))(1+f(k))}{(2q(k_{1}-k)+(k_{1}-k)^2+i0)
  (2p(k-k_{2})+(k-k_{2})^2-i0)}+\\
 \mathrm{VII}:\hspace{0.8 cm}
  &+\frac{-2\pi ipq}{2p(k-k_{2})+(k-k_{2})^2-i0}(1-f_{-}(p))(1-f_{-}(q))
  (1+f(k))\cdot\\
   &\hspace{2.0 cm}\cdot\delta(2q(k_{1}-k)+(k_{1}-k)^2)\thermf{q}+\\
1\mathrm{-I}:\hspace{0.8 cm}
  &+\frac{-2\pi ipq}{2p(k-k_{2})+(k-k_{2})^2-i0}
 \sqrt{f_{-}(q)(1-f_{-}(q))}(1-f_{-}(p))
  \sqrt{f(k)(1+f(k))}\cdot\\
   &\hspace{2.0 cm}\cdot\delta(2q(k_{1}-k)+(k_{1}-k)^2)\thermroot{q}+\\
\mathrm{B}:\hspace{0.8 cm}
  &+\frac{2\pi im^2}{2p(k-k_{1})+(k-k_{1})^2+i0}
  (1-f_{-}(p))(1-f_{-}(q))(1+f(k))\cdot\\
   &\hspace{2.0 cm}\cdot
  \delta(2p(k-k_{2})+(k-k_{2})^2)\thermf{p}+\\
 \mathrm{D}:\hspace{0.8 cm}
  &+4\pi^2(-m^2)(1-f_{-}(p))(1-f_{-}(q))(1+f(k))\delta(2p(k-k_{1})+(k-k_{1})^2)
  \cdot\\
   &\hspace{2.0 cm}\cdot\delta(2p(k-k_{2})+(k-k_{2})^2)\thermf{p}
  \thermf{p}+\\
\beta:\hspace{0.8 cm}
  &+4\pi^2(-m^2)\sqrt{f_{-}(p)(1-f_{-}(p))}(1-f_{-}(q))\sqrt{f(k)(1+f(k))}
  \cdot\\&\hspace{2.0 cm}\cdot
   \delta(2p(k-k_{1})+(k-k_{1})^2)\delta(2p(k-k_{2})+(k-k_{2})^2)\cdot\\
   &\hspace{2.0 cm}\cdot\thermroot{p}\thermf{p}+
\end{split}
\end{equation*}
\begin{equation*}
\begin{split}
\mathrm{VI}:\hspace{0.6 cm}
  &+\frac{2\pi ipq}{2q(k_{1}-k)+(k_{1}-k)^2+i0}(1-f_{-}(p))(1-f_{-}(q))
  (1+f(k))\cdot\\
   &\hspace{2.0 cm}\cdot
  \delta(2p(k-k_{2})+(k-k_{2})^2)\thermf{p}+\\
 \mathrm{VIII}:\hspace{0.6 cm}
  &+4\pi^2(-pq)(1-f_{-}(p))(1-f_{-}(q))(1+f(k))\delta(2q(k_{1}-k)+(k_{1}-k)^2)
  \cdot\\
   &\hspace{2.0 cm}\cdot
   \delta(2p(k-k_{2})+(k-k_{2})^2)\thermf{q}\thermf{p}+\\
1\mathrm{-II}:\hspace{0.6 cm}
  &+4\pi^2(-pq)\sqrt{f_{-}(q)(1-f_{-}(q))}(1-f_{-}(p))\sqrt{f(k)(1+f(k))}
  \cdot\\&\hspace{2.0 cm}\cdot 
  \delta(2q(k_{1}-k)+(k_{1}-k)^2)\delta(2p(k-k_{2})+(k-k_{2})^2)\cdot\\
   &\hspace{2.0 cm}\cdot\thermroot{q}\thermf{p}+\\
\gamma:\hspace{0.6 cm}
  &+\frac{2\pi im^2}{2p(k-k_{1})+(k-k_{1})^2+i0}
  \sqrt{f_{-}(p)(1-f_{-}(p))}(1-f_{-}(q))
  \sqrt{f(k)(1+f(k))}\cdot\\
   &\hspace{2.0 cm}\cdot\delta(2p(k-k_{2})+(k-k_{2})^2)\thermroot{p}+\\
 \delta:\hspace{0.6 cm}
  &+4\pi^2(-m^2)\sqrt{f_{-}(p)(1-f_{-}(p))}(1-f_{-}(q))\sqrt{f(k)(1+f(k))}
  \cdot\\&\hspace{2.0 cm}\cdot 
  \delta(2p(k-k_{1})+(k-k_{1})^2)\delta(2p(k-k_{2})+(k-k_{2})^2)\cdot\\
   &\hspace{2.0 cm}\cdot\thermf{p}\thermroot{p}+\\
\epsilon:\hspace{0.6 cm}
  &+4\pi^2(-m^2)f_{-}(p)(1-f_{-}(q))f(k)\delta(2p(k-k_{1})+(k-k_{1})^2)\cdot\\
   &\hspace{2.0 cm}\cdot\delta(2p(k-k_{2})+(k-k_{2})^2)
  \thermroot{p}\thermroot{p}+\\
1\mathrm{-III}:\hspace{0.6 cm}
  &+\frac{2\pi ipq}{2q(k_{1}-k)+(k_{1}-k)^2+i0}
  \sqrt{f_{-}(p)(1-f_{-}(p))}(1-f_{-}(q))
  \sqrt{f(k)(1+f(k))}\cdot\\
   &\hspace{2.0 cm}\cdot\delta(2p(k-k_{2})+(k-k_{2})^2)\thermroot{p}+\\
 1\mathrm{-IV}:\hspace{0.6 cm}
  &+4\pi^2(-pq)\sqrt{f_{-}(p)(1-f_{-}(p))}(1-f_{-}(q))\sqrt{f(k)(1+f(k))}
  \cdot\\&\hspace{2.0 cm}\cdot
  \delta(2q(k_{1}-k)+(k_{1}-k)^2)\cdot\delta(2p(k-k_{2})+(k-k_{2})^2)\cdot\\
   &\hspace{2.0 cm}\thermf{q}\thermroot{p}+\\
1\mathrm{-V}:\hspace{0.6 cm}
  &+4\pi^2(-pq)\sqrt{f_{-}(p)(1-f_{-}(p))f_{-}(q)(1-f_{-}(q))}f(k)
  \cdot\\&\hspace{2.0 cm}\cdot
  \delta(2q(k_{1}-k)+(k_{1}-k)^2)\delta(2p(k-k_{2})+(k-k_{2})^2)\cdot\\
   &\hspace{2.0 cm}\cdot\thermroot{q}\thermroot{p}+
\end{split}
\end{equation*}
\begin{equation*}
\begin{split}
\mathrm{I}:\hspace{0.8 cm}
  &+\frac{-pq(1-f_{-}(p))(1-f_{-}(q))(1+f(k))}{(2p(k-k_{1}+(k-k_{1})^2)+i0)
  (2q(k_{2}-k+(k_{2}-k)^2)-i0)}+\\
\mathrm{III}:\hspace{0.6 cm}
 &+\frac{-2\pi ipq}{2q(k_{2}-k)+(k_{2}-k)^2-i0}(1-f_{-}(p))(1-f_{-}(q))
  (1+f(k))\cdot\\
   &\hspace{2.0 cm}\cdot\delta(2p(k-k_{1})+(k-k_{1})^2)\thermf{p}+\\
1\mathrm{-X}:\hspace{0.6 cm}
  &+\frac{-2\pi ipq}{2q(k_{2}-k)+(k_{2}-k)^2-i0}
  \sqrt{f_{-}(p)(1-f_{-}(p))}(1-f_{-}(q))
  \sqrt{f(k)(1+f(k))}\cdot\\
   &\hspace{2.0 cm}\cdot\delta(2p(k-k_{1})+(k-k_{1})^2)\thermroot{p}+\\
\mathrm{E}:\hspace{0.6 cm}
  &+\frac{-m^2(1-f_{-}(p))(1-f_{-}(q))(1+f(k))}{(2q(k_{1}-k)+(k_{1}-k)^2+i0)
  (2q(k_{2}-k)+(k_{2}-k)^2-i0)}+\\
\mathrm{G}:\hspace{0.6 cm}
  &+\frac{-2\pi im^2}{2q(k_{2}-k)+(k_{2}-k)^2-i0}(1-f_{-}(p))(1-f_{-}(q))
  (1+f(k))\cdot\\
   &\hspace{2.0 cm}\cdot\delta(2q(k_{1}-k)+(k_{1}-k)^2)\thermf{q}+\\
\eta:\hspace{0.6 cm}
  &+\frac{-2\pi im^2}{2q(k_{2}-k)+(k_{2}-k)^2-i0}
  \sqrt{f_{-}(q)(1-f_{-}(q))}(1-f_{-}(p))
  \sqrt{f(k)(1+f(k))}\cdot\\
   &\hspace{2.0 cm}\cdot\delta(2q(k_{1}-k)+(k_{1}-k)^2)\thermroot{q}+\\
\mathrm{II}:\hspace{0.6 cm}
  &+\frac{2\pi ipq}{2p(k-k_{1})+(k-k_{1})^2+i0}(1-f_{-}(p))(1-f_{-}(q))(1+f(k))
  \cdot\\
   &\hspace{2.0 cm}\cdot\delta(2q(k_{2}-k)+(k_{2}-k)^2)\thermf{q}+\\
\mathrm{IV}:\hspace{0.6 cm}
  &+4\pi^2(-pq)(1-f_{-}(p))(1-f_{-}(q))(1+f(k))\delta(2p(k-k_{1})+(k-k_{1})^2)
  \cdot\\&\hspace{2.0 cm}\cdot
       \delta(2q(k_{2}-k)+(k_{2}-k)^2)\thermf{p}\thermf{q}+\\
1\mathrm{-VI}:\hspace{0.6 cm}
  &+4\pi^2(-pq)\sqrt{f_{-}(p)(1-f_{-}(p))}(1-f_{-}(q))\sqrt{f(k)(1+f(k))}
  \cdot\\&\hspace{2.0 cm}\cdot
  \delta(2p(k-k_{1})+(k-k_{1})^2)\delta(2q(k_{2}-k)+(k_{2}-k)^2)\cdot\\
   &\hspace{2.0 cm}\cdot\thermroot{p}\thermf{q}+
\end{split}
\end{equation*}
\bel{bscrosecausmultipl}
\begin{split}
\mathrm{F}:\hspace{0.4 cm}
  &+\frac{2\pi im^2}{2q(k_{1}-k)+(k_{1}-k)^2+i0}(1-f_{-}(p))(1-f_{-}(q))
 (1+f(k))\cdot\\
   &\hspace{2.0 cm}\cdot
  \delta(2q(k_{2}-k)+(k_{2}-k)^2)\thermf{q}+\\
\mathrm{H}:\hspace{0.4 cm}
  &+4\pi^2(-m^2)(1-f_{-}(p))(1-f_{-}(q))(1+f(k))\delta(2q(k_{1}-k)+(k_{1}-k)^2)
  \cdot\\
   &\hspace{2.0 cm}\cdot
    \delta(2q(k_{2}-k)+(k_{2}-k)^2)\thermf{q}\thermf{q}+\\
\theta:\hspace{0.4 cm}
  &+4\pi^2(-m^2)\sqrt{f_{-}(q)(1-f_{-}(q))}(1-f_{-}(p))\sqrt{f(k)(1+f(k))}
  \cdot\\&\hspace{2.0 cm}\cdot
   \delta(2q(k_{1}-k)+(k_{1}-k)^2)\delta(2q(k_{2}-k)+(k_{2}-k)^2)\cdot\\
   &\hspace{2.0 cm}\cdot\thermroot{q}\thermf{q}+\\
1\mathrm{-VII}:\hspace{0.4 cm}
  &+\frac{2\pi ipq}{2p(k-k_{1})+(k-k_{1})^2+i0}
  \sqrt{f_{-}(q)(1-f_{-}(q))}(1-f_{-}(p))
  \sqrt{f(k)(1+f(k))}\cdot\\
   &\hspace{2.0 cm}\cdot\delta(2q(k_{2}-k)+(k_{2}-k)^2)\thermroot{q}+\\
1\mathrm{-VIII}:\hspace{0.4 cm}
  &+4\pi^2(-pq)\sqrt{f_{-}(q)(1-f_{-}(q))}(1-f_{-}(p))\sqrt{f(k)(1+f(k))}
  \cdot\\&\hspace{2.0 cm}\cdot
  \delta(2p(k-k_{1})+(k-k_{1})^2)\delta(2q(k_{2}-k)+(k_{2}-k)^2)\cdot\\
   &\hspace{2.0 cm}\cdot\thermf{p}\thermroot{q}+\\
1\mathrm{-IX}:\hspace{0.4 cm}
  &+4\pi^2(-pq)\sqrt{f_{-}(p)(1-f_{-}(p))f_{-}(q)(1-f_{-}(q))}f(k)
  \cdot\\&\hspace{2.0 cm}\cdot
  \delta(2p(k-k_{1})+(k-k_{1})^2)\delta(2q(k_{2}-k)+(k_{2}-k)^2)\cdot\\
   &\hspace{2.0 cm}\cdot\thermroot{p}\thermroot{q}+\\
\iota:\hspace{0.4 cm}
  &+\frac{2\pi im^2}{2q(k_{1}-k)+(k_{1}-k)^2+i0}
 \sqrt{f_{-}(q)(1-f_{-}(q))}(1-f_{-}(p))
  \sqrt{f(k)(1+f(k))}\cdot\\
   &\hspace{2.0 cm}\cdot\delta(2q(k_{2}-k)+(k_{2}-k)^2)\thermroot{q}+\\
\kappa:\hspace{0.4 cm}
  &+4\pi^2(-m^2)\sqrt{f_{-}(q)(1-f_{-}(q))}(1-f_{-}(p))\sqrt{f(k)(1+f(k))}
  \cdot\\&\hspace{2.0 cm}\cdot
 \delta(2q(k_{1}-k)+(k_{1}-k)^2)\delta(2q(k_{2}-k)+(k_{2}-k)^2)\cdot\\
   &\hspace{2.0 cm}\cdot\thermf{q}\thermroot{q}+\\
\lambda:\hspace{0.4 cm}
  &+4\pi^2(-m^2)(1-f_{-}(p))f_{-}(q)f(k)
  \delta(2q(k_{1}-k)+(k_{1}-k)^2)\cdot\\
   &\hspace{2.0 cm}\cdot\delta(2q(k_{2}-k)+(k_{2}-k)^2)
  \thermroot{q}\thermroot{q}\Bigg\}+\\
+&\textrm{dito}(k \leftrightarrow -k, f \leftrightarrow 1+f).
\end{split}
\end{equation} 

\subsubsection{Investigation of Possible Cancellations}
\label{subsubsec:poscon}
Now we compare the cross sections we calculated in the foregoing section. 
We investigate if there is a net cancellation of all the divergent terms or 
not. The result will be that there is no complete 
cancellation, certain contributions 
from the self energy and the bremsstrahlung do not cancel. The vertex on the 
other hand causes no problems. Furthermore, we will see that there is a 
complete cancellation if we have small temperatures 
$k_{{\scriptscriptstyle\mathrm{B}}}T\ll 
m_{\mathrm{e}^{-}}$.

Concretely we will consider the following things: We will 
integrate the BS-contribution 
from $\lvert\vec{k}\rvert=0$ to $\omega_{0}\ll 1$, which accounts for 
soft photons that remain undetected (here $\omega_{0}$ can be interpreted as 
some finite energy resolution of the detector). This also fixes the 
Lorentz system: it is the laboratory system, i.e. the rest system of the bath, 
i.e. $u=(1,\vec{0})$. The $k$-integration in the 
V- and SE-contributions will then be considered for values 
$\lvert\vec{k}\rvert\in[0,\omega_{0}]$ as well,  
since possible divergencies origin 
from $k=0$ only. Thus we can treat $k_{0}=\vert\vec{k}\rvert$ as small and 
omit it where this is possible. Since we are interested in the adiabatic 
limit, we will set $\epsilon = 0 \leftrightarrow k_{i}=0$ where this 
is allowed (cf. the remark after (\ref{eq:genadlim3})). 
Thus we can for example replace $\thermroot{q+k_{1}+k_{2}}$ by 
$\thermroot{q}$, $\thermf{p+k-k_{2}}$ by $\thermf{p}$, etc. In addition 
we choose the test function $g$ to be symmetric and real, which is no big 
restriction since we are interested in the limit $g\rightarrow 1$. Now we list 
how several terms compensate and which terms remain in the end. 
\\
\textsc{Vertex and} $pq$\textsc{-terms of BS:}
We often use the fact that the two BS-terms that are considered together 
with one V-term are the complex conjugate of each other and thus the sum is 
twice the real part.  
\begin{itemize}
\item[-:]$\ve^{1}+\bs^{\mathrm{I}}+\bs^{\mathrm{V}}$ gives the terms 
proportional to $pq$ of the $(T\!=\!0)$-BS  
multiplied with $(1-f_{-}(p))(1-f_{-}(q))$ and compensates $\ve^{16}$ 
(cf. \cite{ScharfBuch}, Sec. 3.11). This can be seen as follows: 
$\ve^{1}$ has an integrand proportional to (since we are interested in the 
real part we consider the principal value part only)
\bel{viint}
\mathrm{P}
 \frac{-f}{(2p(-k_{1}-k)+(-k_{1}-k)^2+i0)(2q(k_{2}-k)+(k_{2}-k)^2+i0)}.
\end{equation}
The BS-contributions $\bs^{\mathrm{I}}+\bs^{\mathrm{V}}$ plus the 
corresponding terms from absorption give after some transformation of 
variables an integrand proportional to 
\bel{bs11}
\mathrm{P}
 \frac{-\Theta(k_{0})-f}
{(2p(k_{1}+k)+(k_{1}+k)^2+i0)(2q(k_{2}-k)+(k_{2}-k)^2+i0)}.
\end{equation}
Now we expand the factors in the denominators (using a formula like  
`$\frac{1}{1+x}\cong 1-x$' adapted to this case) 
and compare the resulting terms. We see 
that the leading divergencies ad up to the ($T\!=\!0$)-part of the bremsstrahlung 
and thus compensate $\ve^{16}$. The next to leading order term is of 
the form 
\bel{nlodivbs}
\mathrm{P}\frac{f(k)(k+k_{1})^2}{(2p(k+k_{1}))^2}\frac{1}{2q(k_{2}-k)}.
\end{equation}
The transformation $k\rightarrow-k,\;\;k_{i}\rightarrow-k_{i}$ then generates 
a change in the global sign and thus the integral is identical zero. The other 
terms are finite and vanish in the adiabatic limit.
\item[-:]$\ve^{4}+\bs^{\mathrm{II}}+\bs^{\mathrm{VII}}$: Proceeding as above 
we get for the vertex an integrand proportional to
\bel{v4gghht}
+\delta(2p(-k-k_{1})+(-k-k_{1})^2)\delta(2q(k_{2}-k)+(k_{2}-k)^2)f(k)
\end{equation}
and for the bremsstrahlung
\bel{bsffdrf}
-\delta(2p(k-k_{1})+(k-k_{1})^2)\delta(2q(k_{2}-k)+(k_{2}-k)^2)
(\Theta(k_{0})+f(k)).
\end{equation}
With the help of $\delta(k^2)$ it can be seen that the terms `$(k_{2}-k)^2$' 
etc. in the other $\delta$-distributions can be neglected. Then it follows 
that the divergent terms (the terms 
proportional to $f(k)$) cancel between BS and V and the $f$-independent 
term from BS is finite.
\item[-:]$\ve^{5}+\bs^{\mathrm{III}}+\bs^{\mathrm{VI}}$: analogous to the 
    previous calculation
\item[-:]$\ve^{7}+\bs^{\mathrm{IV}}+\bs^{\mathrm{VIII}}$: dito 
(Because we consider the real part only).
\item[-:]$\ve^{10}+\bs^{1\mathrm{-I}}+\bs^{1\mathrm{-VII}}$: dito. 
\item[-:]$\ve^{11}+\bs^{1\mathrm{-II}}+\bs^{1\mathrm{-VIII}}$: dito.
\item[-:]$\ve^{12}+\bs^{1\mathrm{-III}}+\bs^{1\mathrm{-X}}$: dito.
\item[-:]$\ve^{13}+\bs^{1\mathrm{-IV}}+\bs^{1\mathrm{-VI}}$: dito.
\item[-:]$\ve^{15}+\bs^{1\mathrm{-V}}+\bs^{1\mathrm{-IX}}$: dito.
\item[-:]$\ve^{14}$: as above since we are interested in the real part only.
\item[-:]$\ve^{6}$: dito.
\item[-:]$\ve^{2}$: The relevant structure is
\bel{v2erer}
\frac{1}{((2p(-k_{1}-k)+(-k_{1}-k)^2)+i0)(k^2+i0)}
\delta(2q(k_{2}-k)+(k_{2}-k)^2).
\end{equation}
We use the $\delta$-distribution to perform the $k_{0}$-integration and 
therefore decompose it according to $\delta(y(x))=\sum_{k,y(x_{k})=0}
\frac{\delta(x-x_{k})}{\lvert y'(x_{k})\rvert}$. We see that this gives 
$\frac{1}{\lvert q_{0}\rvert}[\delta(k_{0}-2q_{0})+\delta(k_{0})]$. Using 
this in the complete expression for $\ve^{2}$ leads to a leading order 
divergency that vanishes because of antisymmetry under the transformation of 
variables $k_{i} \rightarrow -k_{i}$ (cf. (\ref{eq:nlodivbs})). 
The next to leading order term is finite.   
\item[-:]$\ve^{3}$ analogous to $\ve^{2}$.
\end{itemize} 
\textsc{Self energy and} $m$\textsc{-terms of BS:}
We use the same methods as for the vertex; in addition we use the following 
partial fraction decomposition:
\bel{partfrac}
\begin{split}
\frac{1}{(2p(k-k_{1})+(k-k_{1})^2+i0)}\cdot
\frac{1}{(2p(k-k_{2})+(k-k_{2})^2-i0)}\overset{!}{=}\\\overset{!}{=}
\frac{A}{(2p(k-k_{1})+(k-k_{1})^2+i0)}+\frac{B}{(2p(k-k_{2})+(k-k_{2})^2-i0)},
\end{split}
\end{equation}
where we demand $A\overset{!}{=}-B$,  which gives 
\bel{aaaaa}
A=\frac{1}{2p(k_{1}-k_{2})+(k_{1}-k_{2})(2k-k_{1}-k_{2})-i0}.
\end{equation}
\begin{itemize}
\item[-:]$\se^{\mathrm{b}}+\bs^{\mathrm{A}}$ is discussed as the terms 
$\ve^{1}+...$, in addition using (\ref{eq:partfrac}). The leading order 
gives the terms proportional to $m^2$ of the $(T\!=\!0)$-BS  
multiplied with $(1-f_{-}(p))(1-f_{-}(q))$ and compensates $\se^{\mathrm{o}}$ 
(cf. \cite{ScharfBuch}, Sec. 3.11). The other terms are finite or can be 
shown  
to vanish because of asymmetry under certain transformations of variables (cf. 
$\ve^{1}+...$). 
\item[-:]$\se^{\mathrm{b}'}+\bs^{\mathrm{E}}+\se^{\mathrm{o}'}$: 
compensation in an analogous way.
\item[-:]$\se^{\mathrm{c}}+\se^{\mathrm{f}}$: the relevant term is 
\bel{bbffggh}
-4\delta(2p(k-k_{1})+(k-k_{1})^2)\delta(2p(-k_{1}-k_{2})+(-k_{1}-k_{2})^2)
\delta(k^2)f,\;\textrm{for}
\end{equation}
$\bs^{\mathrm{B}}+\bs^{\mathrm{C}}$ it is
\bel{bbffiuy}
2\delta(2p(k-k_{1})+(k-k_{1})^2)\delta(2p(k-k_{2})+(k-k_{2})^2)
(\Theta(k_{0})+f).
\end{equation}
Using $\delta(k^2)$ we see that the terms $\sim(k-k_{1})^2$ etc. in the 
argument of the other $\delta$-distributions are negligible. Thus there is 
\emph{no} cancellation between the contributions proportional to $f$ 
because of the factors `4' and `2', respectively, in 
front of these expressions. The term proportional to $\Theta(k_{0})$ is 
finite. This factor `2' which inhibits a cancellation shows up here since we 
cannot use the afore-mentioned partial fraction decomposition which precisely 
led to the compensation in the case of $\se^{\mathrm{b}}+...$.
\item[-:]$\se^{\mathrm{d}'}+\se^{\mathrm{e}'}+\bs^{\mathrm{F}}
    +\bs^{\mathrm{G}}$: is investigated analogously.
\item[-:]$\bs^{\mathrm{D}}+\se^{\mathrm{g}}$: dito.
\item[-:]$\bs^{\mathrm{H}}+\se^{\mathrm{g}'}$: dito.
\item[-:]$\se^{\mathrm{i}}+\se^{\mathrm{m}}+\bs^{\alpha}
   +\bs^{\gamma}$: dito.
\item[-:]$\se^{\mathrm{i}'}+\se^{\mathrm{l}'}+\bs^{\eta}+\bs^{\iota}$: dito.
\item[-:]$\se^{\mathrm{j}}+\se^{\mathrm{n}}+\bs^{\beta}+\bs^{\delta}$: dito.
\item[-:]$\se^{\mathrm{j}'}+\se^{\mathrm{n}'}+\bs^{\kappa}+\bs^{\theta}$: 
   dito.
\item[-:]$\se^{\mathrm{e}}, \se^{\mathrm{f}'}, \se^{\mathrm{l}}, 
  \se^{\mathrm{m}'}$: These terms have the same structure as the finite 
contribution to $\se^{\mathrm{c}}+...\;$ . Thus they are 
finite as well. 
\item[-:]$\se^{\mathrm{a}}, \se^{\mathrm{a}'}, \se^{\mathrm{k}}, 
  \se^{\mathrm{k}'}, \se^{\mathrm{d}}, \se^{\mathrm{c}'}$: 
Here we have the case of a subgraph with singular order $\omega\geq 0$. 
The self energy part has to be splitted using the causal method. The result 
is given in \cite{ScharfBuch}, eq. (3.7.41). If we have a close look at 
that for the momentum $p-\e k_{1}$ with $\e \rightarrow 0$, we see that 
the terms that could diverge vanish because they are antisymmetric under the 
transformation of the integration variable $k_{1} \rightarrow -k_{1}$ 
(cf. $\ve^{1}...$).  
\item[-:]$\se^{\mathrm{h}}+\bs^{\epsilon}$: These terms contain 
three $\delta$-distributions that are treated as in $\ve^4+...\;$ . More 
important, however, are the factors that multiply them. For the self energy 
it is $2\sqrt{f(1+f)}(1-f_{-}(p))(1-f_{-}(q))$, whereas for the 
bremsstrahlung it is $f(1-f_{-}(q))f_{-}(p)$ (the term proportional to 
$\Theta(k_{0})$ is finite). Thus there is \emph{no} cancellation because of 
the factor `2' \emph{and} the different structure in $f_{\pm}$.   
\item[-:]$\se^{\mathrm{h}'}+\bs^{\lambda}$: analogously.
\end{itemize}
We want to add some remarks on the crucial factor `two' in the calculations 
above: each vertex contribution comes with two corresponding BS-contributions, 
thus resulting in a net cancellation. For each self energy contribution, 
on the other hand, there is just one BS-contribution. For 
$\se^{\mathrm{b}}+\bs^{\mathrm{A}}+\se^{\mathrm{o}}$ the partial fraction 
decomposition (\ref{eq:partfrac}) leads to an additional factor two 
in $\bs^{\mathrm{A}}$, thus leading to a net cancellation. The same 
considerations apply to 
$\se^{\mathrm{b'}}+\bs^{\mathrm{E}}+\se^{\mathrm{o'}}$. However, 
there is no partial fraction decomposition (\ref{eq:partfrac}) involved 
in the ($T>0$)-contributions since one or two of these 
factors are replaced by a $\delta$-distribution.
Thus there is no additional factor two for these terms and in consequence no 
cancellation either.
 
\subsubsection{Investigation of Possible Cancellations with Mixed States}
\label{subsubsec:mixstatcanc}
In this section we want to analyse possible cancellations using more general 
asymptotic states. It will turn out that not even the use of general mixed 
states or coherent superpositions 
leads to a cancellation of the IR-divergencies. 

First, 
we consider incoherent mixtures of states with one electron, on electron and 
one photon, ordinary and `tilde', one electron and two photons, ordinary and 
`tilde'. Higher numbers of photons do not have to be considered since we are 
interested in the cross section to fourth order only. From more than two 
photons some are merely spectators and not involved in the processes.  

We will not use a density matrix formulation for the problem. Thus we will 
have to calculate the cross sections corresponding to the various parts of 
the initial and final state, take the weightened mean over the initial 
contributions and the sum over the final ones. We write  
for the initial state (we have suppressed to write down the vacuum $\Omega$ 
on which the emission operators act) 
\bel{initialstate}
\begin{split}
\lvert\Psi_{\mathrm{initial}}\rangle := &+\int_{\mathrm{MS}} 
  \!\!d^4q\lvert b^{+}(q)\rangle\langle b^{+}\cdots\rvert c_{0}(q)+\\
&+\sum_{i}\int_{\mathrm{MS}} \!\!d^4q\int_{\leq \omega_{0}}\!\!d^3 k_{1}
  \lvert b^{+}(q)a^{+}_{i}(k_{1})\rangle\langle\cdots\rvert
   c_{1}(q,k_{1})_{i}+\\
&+\sum_{i}\int_{\mathrm{MS}} \!\!d^4q\int_{\leq \omega_{0}}\!\!d^3 k_{1}
  \lvert b^{+}(q)\tilde{a}^{+}_{i}(k_{1})\rangle\langle\cdots\rvert
   \tilde{c}_{1}(q,k_{1})_{i}+\\
&+\sum_{i,j}\int_{\mathrm{MS}} \!\!d^4q\int_{\leq \omega_{0}}\!\!d^3 k_{1}
  \int_{\leq \omega_{0}}\!\!d^3 k_{2}
  \lvert b^{+}(q)a^{+}_{i}(k_{1})a^{+}_{j}(k_{2})\rangle\langle\cdots\rvert 
  c_{2}(q,k_{1},k_{2})_{ij}+\\
&+\sum_{i,j}\int_{\mathrm{MS}} \!\!d^4q\int_{\leq \omega_{0}}\!\!d^3 k_{1}
  \int_{\leq \omega_{0}}\!\!d^3 k_{2}
  \lvert b^{+}(q)a^{+}_{i}(k_{1})\tilde{a}^{+}_{j}(k_{2})\rangle 
  \langle\cdots\rvert\tilde{c}_{2}(q,k_{1},k_{2})_{ij}+\\
&+\sum_{i,j}\int_{\mathrm{MS}} \!\!d^4q\int_{\leq \omega_{0}}\!\!d^3 k_{1}
  \int_{\leq \omega_{0}}\!\!d^3 k_{2}
  \lvert b^{+}(q)\tilde{a}^{+}_{i}(k_{1})\tilde{a}^{+}_{j}(k_{2})\rangle 
  \langle\cdots\rvert\tilde{\tilde{c}}_{2}(q,k_{1},k_{2})_{ij}.
\end{split} 
\end{equation} 
The $q$-integration is over the mass shell (MS), the 
$k_{1,2}$-integration over $\lvert\vec{k}_{1,2}\rvert \leq \omega_{0}$ and the 
sum runs over four polarizations\footnote{We decided for shortness and 
completeness to indicate the four involved photon polarizations in this way 
directly in the emission operators.} 
$i,j=1,...,4$. The $\beta$-dependence in the various thermal emission 
operators is suppressed in the notation. Since 
the $c_{i}$'s etc. represent probability densities we have $c_{0}, c_{1},...,
\tilde{\tilde{c}}_{2} \geq 0$ and the normalization condition
\bel{normalc}
\begin{split}
\int_{MS}d^4q\Big[ &c_{0}(q)
  +\sum_{i}\int_{\leq \omega_{0}}d^3k_{1}c_{1}(q,k_{1})_{i}
  +\sum_{i}\int_{\leq \omega_{0}}d^3k_{1}\tilde{c}_{1}(q,k_{1})_{i}+\\
  &+\sum_{i,j}\int_{\leq \omega_{0}}d^3k_{1}\int_{\leq \omega_{0}}d^3k_{2}
   c_{2}(q,k_{1},k_{2})_{ij}
  +\sum_{i,j}\int_{\leq \omega_{0}}d^3k_{1}\int_{\leq \omega_{0}}d^3k_{2}
   \tilde{c}_{2}(q,k_{1},k_{2})_{ij}+\\
  &\hspace{3cm}
   +\sum_{i,j}\int_{\leq \omega_{0}}d^3k_{1}\int_{\leq \omega_{0}}d^3k_{2}
   \tilde{\tilde{c}}_{2}(q,k_{1},k_{2})_{ij}\Big] = 1.
\end{split}
\end{equation}
Furtheron $c_{2}$ and $\tilde{\tilde{c}}_{2}$ are symmetric in the arguments 
$k_{1}$ and $k_{2}$ and the indices $i, j$. 

The $S$-matrix terms involved are the same as in the sections 
\ref{subsubsec:crocrocro} and \ref{subsubsec:poscon}. We 
group them according to the number of bremsstrahlung photons (BSP) involved: 
zero BSP : the first order contribution and third order vertex and self energy 
\footnote{The third order vacuum polarisation could at most give next to 
leading order in divergencies (cf. Sec. \ref{subsubsec:thirdVP}).}. 
One BSP : the second order BS 
contributions. Two BSP: the three third order terms with two BS photons. 

Performing the sum and integration over the possible final states we get 36 
contributions to the cross section, each involving a weighting factor 
$c_{0},\;c_{1},...$ and integrals over the electron and photon momenta. 
They are listed in appendix \ref{sec:canc}.

Now we investigate the cross section given by (\ref{eq:canc}). 
We first consider the case of sharp momenta for the incoming 
and outgoing electron, i.e. there is no integration over these momenta. 
Then\footnote{In the following we write $\lvert 0\rvert$, $\lvert 1 \rvert$  
and $\lvert 2 \rvert$ in the matrix elements for the zero, one and two 
BSP-contribution to the $S$-matrix.} 
the coefficient of $\lvert\langle b^+\rvert 0\lvert b^+\rangle\rvert^2$ 
equals 1 by the normalization condition (\ref{eq:normalc}).

Using the $\delta$-distributions, the fact that $f\approx 1+f \approx 
\frac{1}{\beta\lvert ku\rvert}$ for the 
leading contributions to the IR-divergencies and the structure of $A_{\mu}$ 
applied to $\lvert a^{+}\rangle$, etc. 
we see that the contributions proportional 
to the product of zero and two BSP factors are of the following form:

\bel{nullzwei}
\int d^3k_{1}\langle 0\rangle\langle b^+a^+(k_{1})\rvert 2\lvert b^+a^+(k_{1})
 \rangle\cdot[\textrm{ terms involving }c_{1},...].
\end{equation}
    
Since the $c$'s obey the normalization condition and are positive, no 
integral of the form $\int dk c(k)$ can show IR-divergencies. Furthermore, 
the products $\langle 0\rangle\langle 2\rangle$ are finite as well (see Sec. 
\ref{subsubsec:gwtBS}), thus the whole contribution involving these products 
of zero and two BSP contributions is finite. 
 
Therefore, we only have to consider the zero BSP contributions, 
which are the same 
as in the calculation with pure states as already mentioned and the one 
BSP contributions we will examine now. 
 
Using the relations between $\langle b^+a^+\rvert 1\lvert b^+\rangle$, 
$\langle b^+\tilde{a}^+\rvert 1\lvert b^+\rangle$, 
$\langle b^+\rvert 1\lvert b^+a^+\rangle$ and 
$\langle b^+\rvert 1\lvert b^+\tilde{a}^+\rangle$  
($k\leftrightarrow-k$ and $f\leftrightarrow 1+f$ replacements are involved)  
and the fact that $f\approx 1+f \approx 
\frac{1}{\beta\lvert ku\rvert}$ for the 
leading contributions to the IR-divergencies
we can write the leading one BPS contribution as follows:

\bel{onebps}
\begin{split}
1\textrm{BPS} =& \\=\sum_{i}\int d^3k_{1}\bigg\{
 &+\lvert\langle b^+a^{+}_{i}(k_{1})\rvert 1 \lvert b^+\rangle\rvert^2\Big[
  c_{0}(q)+\sum_{m}\int d^3l_{1}c_{1}(q,l_{1})_{m}+
  \sum_{m}\int d^3l_{1}\tilde{c}_{1}(q,l_{1})_{m}+\\
 &\hspace{1.0 cm}+\tilde{c}_{1}(q,k_{1})_{i}+
  \sum_{n}\int d^3l_{2}\tilde{c}_{2}(q,k_{1},l_{2})_{in}+
  \sum_{n}\int d^3l_{2}\tilde{\tilde{c}}_{2}(q,k_{1},l_{2})_{in}\Big]+\\
 &+\lvert\langle b^+\tilde{a}^{+}_{i}(k_{1})\rvert 1 \lvert b^+
  \rangle\rvert^2\Big[
  c_{0}(q)+\sum_{m}\int d^3l_{1}c_{1}(q,l_{1})_{m}+
  \sum_{m}\int d^3l_{1}\tilde{c}_{1}(q,l_{1})_{m}+\\
 &\hspace{1.0 cm}+c_{1}(q,k_{1})_{i}+
  \sum_{n}\int d^3l_{2}c_{2}(q,k_{1},l_{2})_{in}+
  \sum_{n}\int d^3l_{2}\tilde{c}_{2}(q,k_{1},l_{2})_{in}\Big]\bigg\}
\end{split}
\end{equation}

We look at the $p^2, q^2$- and $pq$-contributions in the cross section factors 
of this expression separately as we did in Sec. \ref{subsubsec:poscon}.  
Since the zero BPS contribution are the same as before, we see that each  
$pq$-contribution should finally give the same value as before, thus 
cancelling the vertex contributions. The $p^2, q^2$-contributions, however,  
should undergo a change which amounts to a factor 2 for the ($T>0$)-parts, 
accounts especially for the SE-graphs $\se^{\mathrm{h, h'}}$ 
($\mathrm{BS}^{\lambda},\mathrm{BS}^{\epsilon}$ have to change in this way) 
and does not spoil the cancellation of the ($T=0$)-parts 
(i.e. $\mathrm{BS}^{\mathrm{A}},\mathrm{BS}^{\mathrm{E}}$ should give the same 
as before); cf. Sec. \ref{subsubsec:poscon}. 
 
To investigate this further we write the one BSP cross section in the 
following form (cf. eq. (\ref{eq:bscrosecexpl})). Unimportant global 
factors are omitted and the `$+$' in the argument of $F$ indicates the 
$\Theta(+k_{0})$ involved):
\bel{crosepppqqq}
\begin{split}
\sum_{i}\int d^3k\lvert\langle b^{+}(p)a^{+}_{i}(k)\rvert &1
 \lvert b^{+}(q)\rangle\rvert^2=\\
=&\sum_{i}\int d^4k_{1}\int d^4k_{2}\hat{g}(k_{1})\hat{g}^{*}(k_{2})\int d^3k
  \cdot\\
 &\cdot\Big[F(p,q,k_{1},k,\Theta(k))_{i}+F(q,p,-k_{1},-k,\Theta(k))_{i}\Big]
  \cdot\\
 &\cdot\Big[F(p,q,k_{2},k,\Theta(k))_{i}^*+F(q,p,-k_{2},-k,\Theta(k))_{i}^*
  \Big]
\end{split}
\end{equation} 
The analogous term with $\tilde{a}^+$ instead of $a^+$ reads the same 
after having replaced $k$ with $ -k$ (not in the $\Theta$-function involved). 
We have from (\ref{eq:onebps})

\bel{1bpsstruct}
\begin{split}
1\textrm{BPS} =&+\sum_{i}
 \int d^4k_{1}\int d^4k_{2}\hat{g}(k_{1})\hat{g}^{*}(k_{2})
 \int d^3k\bigg\{\\
&\hspace{-1.0 cm}
  +\Big(F(p,q,k_{1},k,+)_{i}F(q,p,-k_{2},-k,+)^*_{i}
  +F(q,p,-k_{1},-k,+)_{i}F(p,q,k_{2},k,+)^*_{i}\Big)\cdot\\
  &\hspace{0.0 cm}\cdot\Big[Q(q,\omega_{0})+Q'(q,\omega_{0},k)_{i}\Big]+\\
 &\hspace{-1.0 cm}
  +\Big(F(p,q,k_{1},-k,+)_{i}F(q,p,-k_{2},k,+)^*_{i}
  +F(q,p,-k_{1},k,+)_{i}F(p,q,k_{2},-k,+)^*_{i}\Big)\cdot\\
   &\hspace{0.0 cm}\cdot\Big[Q(q,\omega_{0})
   +\tilde{Q}'(q,\omega_{0},k)_{i}\Big]\bigg\}+\\
&+\sum_{i}
 \int d^4k_{1}\int d^4k_{2}\hat{g}(k_{1})\hat{g}^{*}(k_{2})
 \int d^3k\bigg\{\\
&\hspace{-1.0 cm}+\Big(
  F(p,q,k_{1},k,+)_{i}F(p,q,k_{2},k,+)^*_{i}
  +F(q,p,-k_{1},-k,+)_{i}F(q,p,-k_{2},-k,+)^*_{i}
  \Big)\cdot\\
  &\hspace{0.0 cm}\cdot\Big[Q(q,\omega_{0})+Q'(q,\omega_{0},k)_{i}\Big]+\\
 &\hspace{-1.0 cm}+\Big(
  F(p,q,k_{1},-k,+)_{i}F(p,q,k_{2},-k,+)^*_{i}
  +F(q,p,-k_{1},k,+)_{i}F(q,p,-k_{2},k,+)^*_{i}
  \Big)\cdot\\&\hspace{0.0 cm}\cdot\Big[Q(q,\omega_{0})
   +\tilde{Q}'(q,\omega_{0},k)_{i}\Big]\bigg\}.
\end{split}
\end{equation}
The $Q$'s abbreviate the terms involving the $c$'s. Consider the leading 
contributions to the IR-divergencies in $GG^*$, i.e. the terms `$(k-k_{i})^2$' 
can be omitted. Then each `$pq$'-contribution equals the positive or negative 
of the corresponding `$p^2,q^2$'-contribution for $p\rightarrow q$. Using the 
continuity of $G$ in $p$ and $q$ and $G(p,p,...)\neq 0$ we 
see\footnote{For $p=q$, perform the transformation 
$k_{1,2} \longrightarrow -k_{1,2}$ in 
the second and fourth summand. Then we see that we can write 
``$(GG^{*}+GG^{*})[2Q+Q'+\tilde{Q}']$'' 
for the sum of the first two and last two 
summands, respectively.} that it is 
impossible to find $Q$'s (and hence $c$'s) that meet the afore-mentioned 
conditions (cf. after eq. (\ref{eq:onebps})) and lead to a cancellation of all 
IR-divergencies. 

The same consideration applies to the case of sharply peaked momenta centered 
around $q$ and $p$ of the in- and outgoing electrons, respectively, even if 
we allow small deviations from $p^2=m^2=q^2$. The case 
of arbitrary coherent superposition of states can be investigated using 
similar arguments and does not lead to a cancellation as well.    
 
\subsubsection{Discussion of the Results}\label{subsubsec:disccan}
In this section we make several remarks on the results of the last one and 
discuss the results established in literature in relation to ours. In 
section \ref{sec:conclus} we will present some ideas how the problems that 
arise here in the IR-behaviour possibly could be cured. 
\begin{itemize}
\item[-] First we want to emphasize that there is \emph{no} cancellation of 
IR-divergencies in QED at finite temperature in the cross sections to fourth 
order if we do not restrict 
ourselves to small temperatures (see the second point). The vertex 
contributions to the divergency cancel with some part of the bremsstrahlung, 
whereas the self energy contributions do not. Especially the terms 
`$\se^{\mathrm{h}}$' and `$\se^{\mathrm{h}'}$' 
cause problems since they are of a structure that does not arise 
in the bremsstrahlung (in `$\se^{\mathrm{h}}$' and `$\se^{\mathrm{h}'}$' 
we have 
non-tilde external legs leading to a factor 
$(1-f_{-}(p))(1-f_{-}(q))$ and an inner tilde vertex leading to 
$12$-propagators. The BS terms either have a factor 
$(1-f_{-}(p))(1-f_{-}(q))$ and no $12$-propagator or a $12$-propagator and 
the wrong factor involving $f_{-}$). 
The other SE-terms that do not compensate differ 
by a factor two from corresponding BS-terms. Thus we disagree with the 
existing literature that claims IR-finiteness for all temperatures 
(\cite{indum1}, \cite{ahmedsaleem}, etc.) and have doubts if proofs of 
IR-finiteness to all orders (\cite{indum1}) could rigorously be performed. 
We will discuss the shortcomings in the literature in point three. In 
addition to the external states considered usually we have investigated 
mixed states as well. We find the same result: there is no cancellation of 
IR-divergencies.    
\item[-] Secondly we see that the restriction to small temperatures 
$k_{{\scriptscriptstyle\mathrm{B}}}T
\ll m_{\mathrm{e}^{-}}$ finally leads to IR-finite values. 
In this case only $\ve^{1},\;\ve^{16},\;\se^{\mathrm{b}},\;\se^{\mathrm{o}},\;
\se^{\mathrm{b'}},\;\se^{\mathrm{o'}}$, 
$\bs^{\mathrm{I}},\;\bs^{\mathrm{V}},\;\bs^{\mathrm{A}}$ and 
$\bs^{\mathrm{E}}$ contribute. In contrast to the literature we 
have this cancellation in a very straightforward way without the necessity 
to introduce temperature dependent spinors or a thermal Dirac equation as 
it is introduced in (\cite{DH}) and used by many other authors. 
Thus we see that the consideration of the Fermi-Dirac-distribution 
$f_{\pm}$, which in itself is IR-finite causes more than just minor irrelevant 
changes with respect to the low temperature case (as claimed in 
\cite{indum1}, \cite{ahmedsaleem}): 
the finiteness present in the latter gets completely lost.   

Pay attention to the fact that in this approximation the contributions with 
$\lvert\vec{k}\rvert > \omega_{0}$ from  
$\ve^{1},\;\bs^{\mathrm{I}}$ and $\bs^{\mathrm{V}}$ cancel exactly. 
That is why  
there is no contribution to the magnetic moment from this part of the vertex.  

\item[-] In this third point we discuss several papers and the reasons for 
which they end up with finite results and why they are wrong or do not 
contradict our results. 

A first group of papers (\cite{DHR}, \cite{JoPeSk}, \cite{DH} and 
\cite{jaga} - \cite{varma}) 
presents calculations in the low temperature 
approximation $k_{{\scriptscriptstyle\mathrm{B}}}T\ll m_{\mathrm{e}^{-}}$ 
only. As mentioned above we agree with the IR-finiteness found 
by them in this approximation; but as our calculations show, one cannot 
extract from this any results concerning arbitrary temperatures, 
and to use this finiteness as 
a hint for the existence of IR-cancellations to all orders at $T\!>\!0$ is 
too optimistic.

A second group of papers (\cite{tyr} - \cite{wel2}) 
is concerned with arbitrary temperatures but neglects  
the thermal modification of the fermion propagators via $f_{\pm}$ on 
more or less clear reasons. Thus the fact that they find a complete 
cancellation of IR-divergencies does 
not contradict our results since we have this as well if we omit the terms 
with factors $f_{\pm}$. 

The papers (\cite{ahmedsaleem}, \cite{PlaTa} and \cite{eim}) 
leave out the doubling of fields in their 
cancellation and thus especially do not have to face the contributions 
$\se^{\mathrm{h}}, \se^{\mathrm{h}'}$, which cause the major problems in 
our calculations. Especially in \cite{ahmedsaleem} there are some terms 
missing: without reason, they claim to have no modification of the 
bremsstrahlung by the 
Fermi-Dirac-distribution and in their contribution to the vertex and the self 
energy they only have the terms $\ve^{2}, \ve^{3}, \se^{\mathrm{a}}, 
\se^{\mathrm{d}}, \se^{\mathrm{a}'}$ and $\se^{\mathrm{c}'}$, which do not 
cause any problem anyway. 

In the paper \cite{indum1} there are some shortcomings in the 
considerations after their equation (32): In their general investigation of 
the n-th order they consider $S\slas{k}S$ for each vertex and set $g=1$ from 
the very beginning. But in the third 
order we are interested in, one has to investigate $\bar{u}\slas{k}S
\slas{k}S\gamma^{\nu}u\cdot b$ and the limit $g\rightarrow 1$ has to be 
performed at the very end. The net effect of the former treatment 
with respect to 
the latter is again a the omission of contributions as  
$\se^{\mathrm{h}}, \se^{\mathrm{h}'}$. 

The paper \cite{keil} also neglects the doubling of fields. 
  
Besides that there are some papers that consider other problems than we did, 
\cite{phi1} and  \cite{phi2}, to mention just two, 
which consider $\varphi^3$ and $\varphi^4$ 
theories, respectively, which have quite a different structure from QED. 
\item[-] Furthermore it is clear from the discussion that there are a lot of 
finite contributions from these various terms considered above. As in the 
($T\!=\!0$)-case they possibly depend on the test function g (see 
\cite{ScharfBuch}, Sec. 3.12). We do not consider this here any further since 
it is of no importance as long as the whole theory stays divergent in the 
adiabatic limit as it does now. 
\end{itemize}
\section{Conclusion}\label{sec:conclus}
\setcounter{equation}{0}
In this work we investigated QED at finite temperatures in perturbation 
theory. Using the causal 
approach to QFT we are not confronted with any UV-divergency and have a 
method at hand to rigorously discuss the IR-behaviour of the theory by 
investigating the adiabatic limit. 

We have calculated several second and third order graphs and 
derived corrections to $m_{\mathrm{e}^{-}}$ and 
$\mu_{\mathrm{e}^{-}}$ due to the effects of the temperature. Our results 
agree with the ones in the literature. 

The main result, however, is the fact that the theory is not IR-convergent. 
Already in the cross sections to fourth order there is no complete 
cancellation of the 
various IR-divergent terms in the case of arbitrary temperatures. This 
result remains valid even if we set the chemical potential $\mu$ equal zero 
with the consequence that $f_{-}=f_{+}$ or if we use more general 
mixed states or coherent superpositions as initial and final ones. For small 
temperatures the cancellation can be established in a straightforward 
way without the necessity to introduce thermal spinors. 

Now we give some ideas, which perhaps could 
be pursued to construct an IR-finite theory for QED at $T\!>\!0$.  

A first idea would be to use the normalization freedom of the causal approach
due to the distribution-splitting with singular order $\omega \geq 0$ 
to get rid 
of the divergent terms in the adiabatic limit. But as a close inspection of 
these terms readily shows, this is not possible since these terms do not 
have the structure of a polynomial - not even in the adiabatic limit.

Another idea is to change the decomposition of the field 
operators (\ref{eq:bosfeldop}), (\ref{eq:fermfeldop}) 
into positive and negative frequency 
parts or to define the thermal absorption and emission operators 
in a different 
way. But it seems quite impossible to do that in a reasonable way without 
violating the causality of the support of the $D_{2}$-distribution. 
Since this property is crucial for the causal construction of the $S$-matrix, 
the implementation of this approach seems impossible. 
  
The third - and perhaps most promising - idea is to change the involved 
scattering states in a suitable way. On grounds of the observation that 
the divergent vertex contributions cancel and only the self energy (and the 
corresponding bremsstrahlung terms) cause difficulties we could try to absorb 
these SE-contributions in the fermionic states by summation (proceeding as 
for the thermal mass corrections, see Sec. \ref{subsubsec:thermMass}). 
This reasoning leads to a modified Dirac equation and temperature dependent 
spinors as it was done in \cite{DH} for the 
low temperature limit. Difficulties surely will arise through the \emph{two} 
types of vertices that should be considered in a complete summation scheme and 
through the mathematical subtleties that origin in the structure of  
a thermal Dirac equation.\\  
Compare to the case of charged particles in the ($T=0$)-theory. To have 
a unique finite adiabatic limit the unphysical degrees of freedom of the 
photon have to be considered in the bremsstrahlung 
(\cite{ScharfBuch}, Sec. 3.12.). This can be seen as some manifestation that 
the physical scattering states are not free electrons but have always a 
Coulomb-field attached to them. The presence of a thermal bath perhaps 
influences somehow the effects of these unphysical degrees of freedom.

Another point that is not yet clear is the process of `thermalization': 
intuitively it should be the case that a particle immersed in a heath-bath 
looses momentum by repeated scattering processes; this is not the case in 
our framework. But to take intuition as a guide is not always the best thing 
to do. 

Some other ideas we did not investigate at all are the following:  
Implementation of the temperature in a different way, i.e. not as a scalar 
(cf. footnote \ref{fn:tscal}),  
different treatment of the fermionic 
part of the theory, i.e. not through the implementation of the 
Fermi-Dirac-distribution in the theory in a way so similar to the photons or 
description of the theory using infraparticles (see e.g \cite{hen1} and 
references therein) to mention just three. 

Finally, we mention that in the opinion of some authors  
(e.g. \cite{landsman}) a consistent 
perturbation theory for a QFT including temperature is not possible at all or 
at least not in such a direct way as known from the ($T\!=\!0$)-case. 
We do not 
adopt this view, but the physical content and foundations of the 
theory have to be investigated and thought over very carefully in order to 
set up a sensible thermal QED. Thus it could be that the theory is not 
good for arbitrary temperatures because the fermionic contribution from the 
heat-bath (virtual pair generation) is not incorporated correctly by treating 
it more or less analogously to the photon-part.  
   
\section*{Acknowledgment}
\addcontentsline{toc}{section}{Acknowledgment}
I want to thank Professor Scharf for the continuous support during this work, 
his explanations, the discussions, the freedom he left me pursuing it and 
his patience with my handwriting. Further  
I want to thank Nicola Grillo for clearing many points, answering even  
my silliest questions and the fun it is to be with him in the office. 
Ivo Schorn I thank for all the things he made me see clear when 
I was a student. 
These were the `professional' acknowledgments. The `private' ones I will 
express personally - except the point with the handwriting: I thank   
everybody who knows me for her and his patience with it... 

\begin{appendix}

\addcontentsline{toc}{section}{\numberline{}APPENDIX}

\section*{APPENDIX}
\section{Contractions, Propagators etc.}\label{sec:contrprop}
\setcounter{equation}{0}
\subsection{Contractions, Causal Functions and Feynman Propagators at 
\boldmath{$T>0$} in 
\boldmath{$x$}-space}
In the following let $A^{-}, \Psi^{-}$ etc., and  $A^{+}, \Psi^{+}$ etc., 
resp. denote the parts of $A, \Psi$ etc. (see (\ref{eq:bosfeldop}), 
(\ref{eq:fermfeldop})) with absorption and emission operators, respectively.\\
\subsubsection*{Contractions:}
\begin{align*}
[A_{\mu}^{(-)}(x),A_{\nu}^{(+)}(y)] &= \frac{-g_{\mu\nu}}{(2\pi)^{3}}\int
\frac{d^{3} k}{2 \lvert \vec{k} \rvert} \Bigg( (1+f(k))e^{-ik(x-y)} + 
f(k)e^{ik(x-y)} \Bigg) =:\\
&=: g_{\mu\nu}i \; {^{\beta}}\!D_{11}^{(+)}(x-y)\\
[A_{\mu}^{(-)}(x),\tilde{A}_{\nu}^{(+)}(y)] &= 
\frac{-g_{\mu\nu}}{(2\pi)^{3}}\int
\frac{d^{3} k}{2 \lvert \vec{k} \rvert} \Bigg( \sqrt{f(k)(1+f(k))}
(e^{-ik(x-y)} + e^{ik(x-y)}) \Bigg) =:\\ 
&=:g_{\mu\nu}i \; {^{\beta}}\!D_{12}^{(+)}(x-y)\\
[\tilde{A}_{\mu}^{(-)}(x),A_{\nu}^{(+)}(y)] &= 
\frac{-g_{\mu\nu}}{(2\pi)^{3}}\int
\frac{d^{3} k}{2 \lvert \vec{k} \rvert} \Bigg( \sqrt{f(k)(1+f(k))}
(e^{-ik(x-y)} + e^{ik(x-y)}) \Bigg) =:\\ 
&=:g_{\mu\nu}i \; {^{\beta}}\!D_{21}^{(+)}(x-y)\\
[\tilde{A}_{\mu}^{(-)}(x),\tilde{A}_{\nu}^{(+)}(y)] &= 
\frac{-g_{\mu\nu}}{(2\pi)^{3}}\int
\frac{d^{3} k}{2 \lvert \vec{k} \rvert} \Bigg( f(k)e^{-ik(x-y)} + 
(1+f(k))e^{ik(x-y)} \Bigg) =:\\
&=:g_{\mu\nu}i \; {^{\beta}}\!D_{22}^{(+)}(x-y)\\
[A_{\mu}^{(+)}(x),A_{\nu}^{(-)}(y)] &=: g_{\mu\nu}i\propD{11}{(-)}{x-y} =
-g_{\mu\nu}i\propD{11}{(+)}{y-x}\\
[\tilde{A}_{\mu}^{(+)}(x),A_{\nu}^{(-)}(y)] &=: g_{\mu\nu}i\propD{12}{(-)}{x-y}
= -g_{\mu\nu}i\propD{12}{(+)}{y-x}\\
[A_{\mu}^{(+)}(x),\tilde{A}_{\nu}^{(-)}(y)] &=: g_{\mu\nu}i\propD{21}{(-)}{x-y}
= -g_{\mu\nu}i\propD{21}{(+)}{y-x}\\
[\tilde{A}_{\mu}^{(+)}(x),\tilde{A}_{\nu}^{(-)}(y)] &=: 
g_{\mu\nu}i\propD{22}{(-)}{x-y} = -g_{\mu\nu}i\propD{22}{(+)}{y-x}
\end{align*}
\begin{align*} 
\{\Psi_{a}^{(-)}(x),\bar{\Psi}_{b}^{(+)}(y) \} &=     
    \frac{1}{(2\pi)^{3}}\int\frac{d^{3}p}{2\sqrt{{\vec{p}}^{2} + m^2}}
    \Bigg( (1-f_{-}(p))(\slas{p}+m)_{ab}e^{-ip(x-y)} +\\
   &\hspace{4.0 cm}+ 
    f_{+}(p)(\slas{p}-m)_{ab}e^{ip(x-y)}\Bigg)=: \\
    &=: \frac{1}{i}\propS{11}{(+)}{x-y}{ab}\\  
\{\Psi_{a}^{(-)}(x),\tilde{\Psi}_{b}^{(+)}(y) \} &=
    \frac{1}{(2\pi)^{3}}\int\frac{d^{3}p}{2\sqrt{{\vec{p}}^{2} + m^2}}
    \Bigg( i\sqrt{f_{-}(p)(1-f_{-}(p))}(\slas{p}+m)_{ab}e^{-ip(x-y)} +\\ 
    &\hspace{4.0 cm}
    +i\sqrt{f_{+}(p)(1-f_{+}(p))}(\slas{p}-m)_{ab}e^{ip(x-y)}\Bigg)=:\\ 
    &=:i\propS{12}{(+)}{x-y}{ab}\\  
\{\bar{\tilde{\Psi}}_{a}^{(-)}(x),\bar{\Psi}_{b}^{(+)}(y) \} &=
    \frac{1}{(2\pi)^{3}}\int\frac{d^{3}p}{2\sqrt{{\vec{p}}^{2} + m^2}}
    \Bigg( -i\sqrt{f_{-}(p)(1-f_{-}(p))}(\slas{p}+m)_{ab}e^{-ip(x-y)} -\\ 
     &\hspace{4.0 cm}
     -i\sqrt{f_{+}(p)(1-f_{+}(p))}(\slas{p}-m)_{ab}e^{ip(x-y)}\Bigg)=:\\ 
    &=:-i\propS{21}{(+)}{x-y}{ab}\\
\{\bar{\tilde{\Psi}}_{a}^{(-)}(x),\tilde{\Psi}_{b}^{(+)}(y) \} &=
    \frac{1}{(2\pi)^{3}}\int\frac{d^{3}p}{2\sqrt{{\vec{p}}^{2} + m^2}}
    \Bigg( f_{-}(p)(\slas{p}+m)_{ab}e^{-ip(x-y)} +\\
     &\hspace{4.0 cm}+ 
    (1-f_{+}(p))(\slas{p}-m)_{ab}e^{ip(x-y)}\Bigg)=: \\ 
    &=:\frac{1}{i}\propS{22}{(+)}{x-y}{ab}
\end{align*}
\begin{align*}
\{\bar{\Psi}_{a}^{(-)}(x),\Psi_{b}^{(+)}(y) \} &=
    \frac{1}{(2\pi)^{3}}\int\frac{d^{3}p}{2\sqrt{{\vec{p}}^{2} + m^2}}
    \Bigg( f_{-}(p)(\slas{p}+m)_{ba}e^{ip(x-y)} +\\
     &\hspace{3.8 cm}+ 
    (1-f_{+}(p))(\slas{p}-m)_{ba}e^{-ip(x-y)}\Bigg) =:\\ 
    &=:\frac{1}{i}\propS{11}{(-)}{y-x}{ba}\\ 
\{\tilde{\Psi}_{a}^{(-)}(x),\Psi_{b}^{(+)}(y) \} &=
    \frac{1}{(2\pi)^{3}}\int\frac{d^{3}p}{2\sqrt{{\vec{p}}^{2} + m^2}}
    \Bigg( -i\sqrt{f_{-}(p)(1-f_{-}(p))}(\slas{p}+m)_{ba}e^{ip(x-y)} - \\
     &\hspace{3.8 cm}
    -i\sqrt{f_{+}(p)(1-f_{+}(p))}(\slas{p}-m)_{ba}e^{-ip(x-y)}\Bigg) =:\\ 
    &=:i\propS{12}{(-)}{y-x}{ba}\\
\{\bar{\Psi}_{a}^{(-)}(x),\bar{\tilde{\Psi}}_{b}^{(+)}(y) \} &=
    \frac{1}{(2\pi)^{3}}\int\frac{d^{3}p}{2\sqrt{{\vec{p}}^{2} + m^2}}
    \Bigg( i\sqrt{f_{-}(p)(1-f_{-}(p))}(\slas{p}+m)_{ba}e^{ip(x-y)} +\\ 
     &\hspace{3.8 cm}
    +i\sqrt{f_{+}(p)(1-f_{+}(p))}(\slas{p}-m)_{ba}e^{-ip(x-y)}\Bigg)=:\\  
    &=:-i\propS{21}{(-)}{y-x}{ba}\\
\{\tilde{\Psi}_{a}^{(-)}(x),\bar{\tilde{\Psi}}_{b}^{(+)}(y) \} &=
    \frac{1}{(2\pi)^{3}}\int\frac{d^{3}p}{2\sqrt{{\vec{p}}^{2} + m^2}}  
    \Bigg( (1-f_{-}(p))(\slas{p}+m)_{ba}e^{ip(x-y)} +\\
     &\hspace{3.8 cm}+ 
    f_{+}(p)(\slas{p}-m)_{ba}e^{-ip(x-y)}\Bigg) =:\\ 
    &=:\frac{1}{i}\propS{22}{(-)}{y-x}{ba}
\end{align*}
\subsubsection*{Causal Functions:}
$\underline{\boldsymbol{D = D^{(-)} + D^{(+)}}}$:
\begin{align*}
\propD{11}{}{x-y} &= 
    \frac{i}{(2\pi)^{3}}\int\frac{d^{3}k}{2 \lvert \vec{k} \rvert}
    \Bigg( e^{-ik(x-y)} - e^{ik(x-y)}\Bigg) = \; {^{T=0}}\!D(x-y) \\
\propD{12}{}{x-y} &= 0 = \propD{21}{}{x-y}\\
\propD{22}{}{x-y} &= -\; {^{T=0}}\!D(x-y)
\end{align*}
$\underline{\boldsymbol{S = S^{(-)} + S^{(+)}}}$:
\begin{align*}
\propS{11}{}{x-y}{ab} &= 
    \frac{i}{(2\pi)^{3}}\int\frac{d^{3}p}{2\sqrt{{\vec{p}}^{2} + m^2}}
    \Bigg( (\slas{p}+m)_{ab}e^{-ip(x-y)} + 
    (\slas{p}-m)_{ab}e^{ip(x-y)}\Bigg) =\\&= {^{T=0}}\!S(x-y)_{ab} \\
\propS{12}{}{x-y}{ab} &= 0 = \propS{21}{}{x-y}{ab}\\
\propS{22}{}{x-y}{ab} &= \; {^{T=0}}\!S(x-y)_{ab}    
\end{align*}
\subsubsection*{Feynman Propagators:}
$\underline{\boldsymbol{D^{\F} = D^{\ret} - D^{(-)} = D^{\av} + D^{(+)}}}$:
\begin{align*}
\propD{11}{\F}{x-y} &= 
    \frac{i}{(2\pi)^{3}}\int\frac{d^{3}k}{2 \lvert \vec{k} \rvert}
    \Bigg( e^{-ik(y-x)} + f(k)(e^{-ik(y-x)} + e^{ik(y-x)}) \Bigg)
    \!+ {^{T=0}}\!D^{\ret}(x-y)=\\ 
    &= \; {^{T=0}}\!D^{\F}(x-y)
    + \frac{i}{(2\pi)^{3}}\int\frac{d^{3}k}{2 \lvert \vec{k} \rvert}
    \Bigg( f(k)(e^{-ik(x-y)} + e^{ik(x-y)}) \Bigg)\\
\propD{12}{\F}{x-y} &= -\propD{12}{(-)}{x-y}\\
\propD{21}{\F}{x-y} &= -\propD{21}{(-)}{x-y}\\
\propD{22}{\F}{x-y} &= 
    \frac{i}{(2\pi)^{3}}\int\frac{d^{3}k}{2 \lvert \vec{k} \rvert}
    \Bigg( e^{ik(y-x)} + f(k)(e^{-ik(y-x)} + e^{ik(y-x)}) \Bigg) 
    - \; {^{T=0}}\!D^{\ret}(x-y)=\\
    &= -(\; {^{T=0}}\!D^{\F}(x-y))^{*}
    + \frac{i}{(2\pi)^{3}}\int\frac{d^{3}k}{2 \lvert \vec{k} \rvert}
    \Bigg( f(k)(e^{-ik(x-y)} + e^{ik(x-y)}) \Bigg)
\end{align*}
$\underline{\boldsymbol{S^{\F} = S^{(-)} - S^{\ret} = -S^{(+)} - S^{\av}}}$:
\begin{align*}
\propS{11}{\F}{x-y}{ab} &= 
    \frac{i}{(2\pi)^{3}}\int\frac{d^{3}p}{2\sqrt{{\vec{p}}^{2} + m^2}}
    \Bigg( (\slas{p} -m)_{ab}e^{ip(x-y)} +\\ 
    &\hspace{-1.5 cm}+f_{-}(p)(\slas{p}+m)_{ab}e^{-ip(x-y)} - 
    f_{+}(p)(\slas{p}-m)_{ab}e^{ip(x-y)}\Bigg) 
    -\; {^{T=0}}\!S^{\ret}(x-y)_{ab} \\
    &= \; {^{T=0}}\!S^{\F}(x-y)_{ab} +\\ 
    &\hspace{-1.5 cm}
       +\frac{i}{(2\pi)^{3}}\int\frac{d^{3}p}{2\sqrt{{\vec{p}}^{2} + m^2}}
    \Bigg(f_{-}(p)(\slas{p}+m)_{ab}e^{-ip(x-y)} - 
    f_{+}(p)(\slas{p}-m)_{ab}e^{ip(x-y)}\Bigg)\\
\propS{12}{\F}{x-y}{ab} &= \propS{12}{(-)}{x-y}{ab}\\
\propS{21}{\F}{x-y}{ab} &= \propS{21}{(-)}{x-y}{ab}\\
\propS{22}{\F}{x-y}{ab} &= 
    \frac{i}{(2\pi)^{3}}\int\frac{d^{3}p}{2\sqrt{{\vec{p}}^{2} + m^2}}
    \Bigg( (\slas{p} +m)_{ab}e^{-ip(x-y)} +\\ 
    &\hspace{-1.5 cm}+f_{+}(p)(\slas{p}-m)_{ab}e^{ip(x-y)} - 
    f_{-}(p)(\slas{p}+m)_{ab}e^{-ip(x-y)}\Bigg) 
    -\; {^{T=0}}\!S^{\ret}(x-y)_{ab} \\
    &=  (\; {^{T=0}}\!S^{\F}(y-x)_{ab})^{*} +\\ 
    &\hspace{-1.5 cm}
       +\frac{i}{(2\pi)^{3}}\int\frac{d^{3}p}{2\sqrt{{\vec{p}}^{2} + m^2}}
    \Bigg(f_{+}(p)(\slas{p}-m)_{ab}e^{ip(x-y)} - 
    f_{-}(p)(\slas{p}+m)_{ab}e^{-ip(x-y)}\Bigg)
\end{align*}
\subsection{Relations}\label{subsec:rel}
\begin{gather*}
\propD{11}{(-)}{z} = -\propD{11}{(+)}{-z},  \qquad \propD{12}{(-)}{z} = 
       -\propD{12}{(+)}{-z}\\
\propD{22}{(-)}{z} = -\propD{22}{(+)}{-z},  \qquad \propD{22}{(+)}{z} =
       \propD{11}{(+)}{-z}\\
\propD{21}{(+)}{z} = \propD{12}{(+)}{z} = \propD{12}{(+)}{-z} 
\end{gather*}
\begin{gather*}
\propD{11}{\F}{z} = \propD{11}{\F}{-z}, \qquad \propD{22}{\F}{z} =
       \propD{22}{\F}{-z}\\ 
\propD{12}{\F}{z} = \propD{12}{+}{z} = \propD{21}{\F}{z} 
\end{gather*}
\begin{gather*}
\; {^{T>0}}\!D^{\F}_{11}(z) = \; {^{T>0}}\!D^{\F}_{22}(z) =
       \; {^{T>0}}\!D^{(+)}_{11}(z) 
       = \; {^{T>0}}\!D^{(+)}_{11}(-z) 
\end{gather*}
\begin{gather*}
\propS{11}{(-)}{z}{} = \propS{22}{(+)}{z}{}, \qquad \propS{22}{(-)}{z}{} =
       \propS{11}{(+)}{z}{}\\
\propS{12}{(+)}{z}{} = \propS{21}{(+)}{z}{},  \qquad \propS{12}{(-)}{z}{} =
       \propS{21}{(-)}{z}{}\\
\propS{12}{(-)}{z}{} = -\propS{12}{(+)}{z}{} 
\end{gather*}
\begin{gather*}
  \propS{12}{\F}{z}{} =
       \propS{12}{(-)}{z}{}, \ 
        \propS{21}{\F}{z}{} = \propS{12}{\F}{z}{} 
\end{gather*}
\begin{gather*}
\; {^{T>0}}\!S^{\F}_{11}(z) = -\; {^{T>0}}\!S^{\F}_{22}(z) =
       -\; {^{T>0}}\!S^{(+)}_{11}(z) 
       = \; {^{T>0}}\!S^{(-)}_{11}(z) 
\end{gather*}

\subsection{Covariant Notation and Momentum Space Representation}
\label{subsec:covnotmomspre}
\subsubsection*{Covariant Notation}
\begin{align*}
\propD{11}{(+)}{x-y} &= \frac{i}{(2\pi)^{3}}\int\!d^{4}k \delta(k^2)
      e^{-ik(x-y)}(\theta(k_{0}) + f(k))\\
\propD{12}{(+)}{x-y} &= \frac{i}{(2\pi)^{3}}\int\!d^{4}k \delta(k^2)
      e^{-ik(x-y)}\sqrt{f(k)(1+f(k))}\\
\propD{11}{\F}{x-y} &= \; {^{T=0}}\!D^{\F}(x-y)
      + \frac{i}{(2\pi)^{3}}\int\!d^{4}k \delta(k^2)
      e^{-ik(x-y)}f(k)\\
\end{align*}
\begin{align*}
\propS{11}{(+)}{x-y}{} &= \frac{i}{(2\pi)^{3}} \int\!d^4p \delta(p^2 -m^2)
     \theta(p_{0})\cdot\\
     &\hspace{2.0 cm}\cdot\bigg( (1-f_{-}(p))(\slas{p}+m)e^{-ip(x-y)} +
     f_{+}(p)(\slas{p}-m)e^{ip(x-y)} \bigg) =\\
     &= \frac{i}{(2\pi)^{3}} \int\!d^4p \delta(p^2 -m^2)(\slas{p}+m)
     e^{-ip(x-y)}\cdot\\
     &\hspace{2.0 cm}\cdot\bigg( \theta(p_{0}) -f_{-}(p)\theta(p_{0})
     -f_{+}(p)\theta(-p_{0}) \bigg)\\
\propS{12}{(+)}{x-y}{} &= \frac{1}{(2\pi)^{3}} \int\!d^4p 
     \delta(p^2 \!-\!m^2)(\slas{p}+m)e^{-ip(x-y)}\cdot\\
     &\hspace{2.0 cm}\cdot\bigg(\! \theta(p_{0})
     \sqrt{f_{-}(p)(1\!-\!f_{-}(p))}
     \!-\! \theta(-p_{0})\sqrt{f_{+}(p)(1\!-\!f_{+}(p))}\! \bigg)\\
\propS{22}{(+)}{x-y}{} &= \frac{i}{(2\pi)^{3}} \int\!d^4p 
     \delta(p^2 -m^2)(\slas{p}+m)e^{-ip(x-y)}\cdot\\
     &\hspace{2.0 cm}\cdot
     \bigg(-\theta(-p_{0}) +f_{-}(p)\theta(p_{0}) +f_{+}(p)\theta(-p_{0})
     \bigg)\\
\propS{11}{\F}{x-y}{} &= \; {^{T=0}}\!S^{\F}(x-y)
     + \frac{i}{(2\pi)^{3}} \int\!d^4p \delta(p^2 -m^2)(\slas{p}+m)
     e^{-ip(x-y)}\cdot\\
     &\hspace{2.0 cm}\cdot
      \bigg( f_{-}(p)\theta(p_{0}) +f_{+}(p)\theta(-p_{0})\bigg)\\
\end{align*} 
\subsubsection*{Momentum Space Representation}
\begin{align*}
\propDft{11}{(+)}{k} &= \frac{i}{2\pi}\delta(k^2)(\theta(k_{0}) + f(k))\\
\propDft{12}{(+)}{k} &= \frac{i}{2\pi}\delta(k^2)\sqrt{f(k)(1+f(k))}\\
\propDft{11}{\F}{k} &= \frac{1}{(2\pi)^2}\frac{-1}{k^2 + i0} 
     + \frac{i}{2\pi}\delta(k^2)f(k)=\\ 
     &= \; {^{T=0}}\!\widehat{D}^{\F}(k) + \frac{i}{2\pi}\delta(k^2)f(k)\\
\propDft{22}{\F}{k} &= \frac{-1}{(2\pi)^2}\frac{1}{-k^2 + i0} 
     + \frac{i}{2\pi}\delta(k^2)f(k)=\\
     &= -\; {^{T=0}}\!\widehat{D}^{\F}(k)^{*} + \frac{i}{2\pi}\delta(k^2)f(k)
\end{align*}
\begin{align*}
\propSft{11}{(+)}{p}{} &= \frac{i}{2\pi}\delta(p^2-m^2)(\slas{p} +m)
   \bigg( \theta(p_{0}) -f_{-}(p)\theta(p_{0})-f_{+}(p)\theta(-p_{0}) \bigg)\\
\propSft{12}{(+)}{p}{} &= \frac{1}{2\pi}\delta(p^2-m^2)(\slas{p} +m)
     \bigg( \theta(p_{0})\sqrt{f_{-}(p)(1-f_{-}(p))}
     - \theta(-p_{0})\sqrt{f_{+}(p)(1-f_{+}(p))} \bigg)\\
\propSft{22}{(+)}{p}{} &= \frac{i}{2\pi}\delta(p^2-m^2)(\slas{p} +m)
   \bigg(-\theta(-p_{0}) +f_{-}(p)\theta(p_{0})+f_{+}(p)\theta(-p_{0}) \bigg)\\
\propSft{11}{\F}{p}{} &= \frac{1}{(2\pi)^2}\frac{(\slas{p} +m)}{p^2-m^2+i0} +
     \frac{i}{2\pi}\delta(p^2-m^2)(\slas{p}+m)
     \bigg(f_{-}(p)\theta(p_{0})+f_{+}(p)\theta(-p_{0}) \bigg)=\\
   &= \; {^{T=0}}\!\widehat{S}^{\F}(p)+
   \frac{i}{2\pi}\delta(p^2-m^2)(\slas{p}+m)
     \bigg(f_{-}(p)\theta(p_{0})+f_{+}(p)\theta(-p_{0}) \bigg)\\
\propSft{22}{\F}{p}{} &= \big[\propSft{11}{\F}{p}{}\big]^{*}
\end{align*}   

\section{Three `Intuitive' Observations in TFD}\label{sec:motiv}
\setcounter{equation}{0}
In this appendix we present some unrigorous considerations, which may help 
to develop some intuition for TFD:
\\

\textsc{Particle Number Operator:}
Calculate the expectation value of the ordinary ($T\!=\!0$)-particle number 
operator for scalar bosons in the thermal vacuum:
\bel{pno}
\begin{split}
&\hspace{1.8 cm}
\langle\Omega_{\beta}\lvert a(\vec{k})^{+}a(\vec{k})\rvert\Omega_{\beta}
\rangle \overset{(\ref{eq:bogolbos})}{=}\\
=\Bigl\langle\Omega_{\beta}\Bigl\lvert \Bigl(\sqrt{1+f}\;&\aerz{}{k}
-\sqrt{f}\;\avert{}{k}\Bigr)
\Bigl(\sqrt{1+f}\;\aver{}{k}-\sqrt{f}\;\aerzt{}{k}\Bigr) \Bigr\rvert
\Omega_{\beta}\Bigr\rangle =\\
&= \Bigl\langle\Omega_{\beta}\Bigl\lvert\Bigl(-\sqrt{f}\;\avert{}{k}\Bigr)
\Bigl(-\sqrt{f}\;\aerzt{}{k}\Bigr)\Bigr\rvert\Omega_{\beta}\Bigr\rangle=f(k).
\end{split}
\end{equation}
Here we made use of $\langle \Omega_{\beta}\lvert\aerz{}{k} =0$ and 
$\aver{}{k}\rvert
\Omega_{\beta}
\rangle =0$. So the thermal vacuum seems to mimic a black body heath 
bad as we expect it to do.
\\

\textsc{Electron in a Heath Bath:}
Consider the following calculation (for the formulae with $e^{-iG}$ we refer 
to \cite{ojima}):
\bel{termel}
\begin{split}
\berz{}{p}\rvert\Omega_{\beta}\rangle = e^{-iG}b(\vec{p})^{+}e^{iG}
\rvert\Omega_{\beta}\rangle = e^{-iG}b(\vec{p})^{+}e^{iG}e^{-iG}\rvert0
\rangle=\\
= e^{-iG}b(\vec{p})^{+}\rvert0\rangle=b(\vec{p})^{+}e^{-iG}\rvert0\rangle=
b(\vec{p})^{+} \rvert\Omega_{\beta}\rangle.
\end{split}
\end{equation}
So it could be said that $\berz{}{p}$ generates an electron with momentum 
$\vec{p}$ in the thermal setting. But this should not be taken as a sound  
physical interpretation of the formulae.
\\

A third observation: set $T\!=\!0$ in the formulae of subsection 
\ref{subsubsec:causthermth}; then we get two copies of ordinary QED at $T\!=\!0$. 

\section{Second Order Calculations} \label{sec:secord}
\setcounter{equation}{0}
\subsection{The \boldmath{$D_{2}$}-Distribution}\label{subsec:dzwei}
\bel{DzweiVorstufeEins}
\begin{split}
&D_{2}(x,y) = -e^2\gamma^{\mu}_{ab}\gamma^{\nu}_{cd}\Bigg\{\\
+&\Bigg(\wl\bar{\Psi}_{a}(x)\Psi_{b}(x)\bar{\Psi}_{c}(y)\Psi_{d}(y)\wre +
   \big\{ \bar{\Psi}_{a}^{(-)}(x),\Psi_{d}^{(+)}(y) \big\}
         \wl\Psi_{b}(x)\bar{\Psi}_{c}(y)\wre +\\
  & + \big\{\Psi_{b}^{(-)}(x),\bar{\Psi}_{c}^{(+)}(y) \big\}
         \wl\bar{\Psi}_{a}(x)\Psi_{d}(y)\wre +\\ 
 &+\big\{ \bar{\Psi}_{a}^{(-)}(x),\Psi_{d}^{(+)}(y) \big\}
     \big\{\Psi_{b}^{(-)}(x),\bar{\Psi}_{c}^{(+)}(y) \big\} \Bigg)
   \bigg(\wl A_{\mu}(x)A_{\nu}(y)\wre + \big[A_{\mu}^{(-)}(x),
        A_{\nu}^{(+)}(y) \big]\bigg)-\\
-&\Bigg(\wl\bar{\Psi}_{a}(x)\Psi_{b}(x)
       \tilde{\Psi}_{c}(y)\bar{\tilde{\Psi}}_{d}(y)\wre +
   \big\{ \bar{\Psi}_{a}^{(-)}(x),\bar{\tilde{\Psi}}_{d}^{(+)}(y) \big\}
         \wl\Psi_{b}(x)\tilde{\Psi}_{c}(y)\wre +\\ 
  &+ \big\{\Psi_{b}^{(-)}(x),\tilde{\Psi}_{c}^{(+)}(y) \big\}
         \wl\bar{\Psi}_{a}(x)\bar{\tilde{\Psi}}_{d}(y)\wre +\\ 
 &+\big\{ \bar{\Psi}_{a}^{(-)}(x),\bar{\tilde{\Psi}}_{d}^{(+)}(y) \big\}
     \big\{\Psi_{b}^{(-)}(x),\tilde{\Psi}_{c}^{(+)}(y) \big\} \Bigg)
   \bigg(\wl A_{\mu}(x)\tilde{A}_{\nu}(y)\wre + \big[A_{\mu}^{(-)}(x),
        \tilde{A}_{\nu}^{(+)}(y) \big]\bigg)-\\
-&\Bigg(\wl\tilde{\Psi}_{a}(x)\bar{\tilde{\Psi}}_{b}(x)
  \bar{\Psi}_{c}(y)\Psi_{d}(y)\wre +
   \big\{ \tilde{\Psi}_{a}^{(-)}(x),\Psi_{d}^{(+)}(y) \big\}
         \wl\bar{\tilde{\Psi}}_{b}(x)\bar{\Psi}_{c}(y)\wre +\\
 & + \big\{\bar{\tilde{\Psi}}_{b}^{(-)}(x),\bar{\Psi}_{c}^{(+)}(y) \big\}
         \wl\tilde{\Psi}_{a}(x)\Psi_{d}(y)\wre +\\
 &+\big\{ \tilde{\Psi}_{a}^{(-)}(x),\Psi_{d}^{(+)}(y) \big\}
     \big\{\bar{\tilde{\Psi}}_{b}^{(-)}(x),\bar{\Psi}_{c}^{(+)}(y) \big\}\Bigg)
   \bigg(\wl\tilde{A}_{\mu}(x)A_{\nu}(y)\wre + \big[\tilde{A}_{\mu}^{(-)}(x),
        A_{\nu}^{(+)}(y) \big]\bigg)+\\
+&\Bigg(\wl\tilde{\Psi}_{a}(x)\bar{\tilde{\Psi}}_{b}(x)
   \tilde{\Psi}_{c}(y)\bar{\tilde{\Psi}}_{d}(y)\wre +
   \big\{ \tilde{\Psi}_{a}^{(-)}(x),\bar{\tilde{\Psi}}_{d}^{(+)}(y) \big\}
         \wl\bar{\tilde{\Psi}}_{b}(x)\tilde{\Psi}_{c}(y)\wre +\\
 & + \big\{\bar{\tilde{\Psi}}_{b}^{(-)}(x),\tilde{\Psi}_{c}^{(+)}(y) \big\}
         \wl\tilde{\Psi}_{a}(x)\bar{\tilde{\Psi}}_{d}(y)\wre +\\ 
  &+\big\{ \tilde{\Psi}_{a}^{(-)}(x),\bar{\tilde{\Psi}}_{d}^{(+)}(y) \big\}
     \big\{\bar{\tilde{\Psi}}_{b}^{(-)}(x),\tilde{\Psi}_{c}^{(+)}(y)\big\}
  \Bigg)\bigg(\wl\tilde{A}_{\mu}(x)\tilde{A}_{\nu}(y)\wre +
   \big[\tilde{A}_{\mu}^{(-)}(x),\tilde{A}_{\nu}^{(+)}(y) \big]\bigg)\Bigg\}\\
&\\
&\underbrace{-\Bigg( -e^2\gamma^{\mu}_{ab}\gamma^{\nu}_{cd}\bigg\{
 \textrm{dito with} \: x \leftrightarrow y \bigg\}\Bigg)}_{+R'_{2}} \qquad =
\end{split}
\end{equation}
\bel{DzweiVorstufeZwei}
\begin{split}
 = -e^2\gamma^{\mu}_{ab}&\gamma^{\nu}_{cd}\Bigg\{\\
+\Bigg(&\wl\bar{\Psi}_{a}(x)\Psi_{b}(x)\bar{\Psi}_{c}(y)\Psi_{d}(y)\wre +
   \frac{1}{i}\propS{11}{(-)}{y-x}{da}
         \wl\Psi_{b}(x)\bar{\Psi}_{c}(y)\wre +\\
  & + \frac{1}{i}\propS{11}{(+)}{x-y}{bc}
         \wl\bar{\Psi}_{a}(x)\Psi_{d}(y)\wre +\\ 
 &+\frac{1}{i}\propS{11}{(-)}{y-x}{da}
     \frac{1}{i}\propS{11}{(+)}{x-y}{bc} \Bigg)
   \bigg(\wl A_{\mu}(x)A_{\nu}(y)\wre + g_{\mu\nu}i\propD{11}{(+)}{x-y} 
       \bigg)-\\
-\Bigg(&\wl\bar{\Psi}_{a}(x)\Psi_{b}(x)
       \tilde{\Psi}_{c}(y)\bar{\tilde{\Psi}}_{d}(y)\wre +
   (-i)\propS{21}{(-)}{y-x}{da}
         \wl\Psi_{b}(x)\tilde{\Psi}_{c}(y)\wre +\\ 
  &+ i\propS{12}{(+)}{x-y}{bc}
         \wl\bar{\Psi}_{a}(x)\bar{\tilde{\Psi}}_{d}(y)\wre +\\ 
 &+(-i)\propS{21}{(-)}{y-x}{da}
      i\propS{12}{(+)}{x-y}{bc}\Bigg)
   \bigg(\wl A_{\mu}(x)\tilde{A}_{\nu}(y)\wre + 
      g_{\mu\nu}i\propD{12}{(+)}{x-y}\bigg)-\\
-\Bigg(&\wl\tilde{\Psi}_{a}(x)\bar{\tilde{\Psi}}_{b}(x)
  \bar{\Psi}_{c}(y)\Psi_{d}(y)\wre +
   i\propS{12}{(-)}{y-x}{da}
         \wl\bar{\tilde{\Psi}}_{b}(x)\bar{\Psi}_{c}(y)\wre +\\
 & + (-i)\propS{21}{(+)}{x-y}{bc}
         \wl\tilde{\Psi}_{a}(x)\Psi_{d}(y)\wre +\\
 &+i\propS{12}{(-)}{y-x}{da}(-i)\propS{21}{(+)}{x-y}{bc}\Bigg)
   \bigg(\wl\tilde{A}_{\mu}(x)A_{\nu}(y)\wre + 
    g_{\mu\nu}i\propD{12}{(+)}{x-y}\bigg)+\\
+\Bigg(&\wl\tilde{\Psi}_{a}(x)\bar{\tilde{\Psi}}_{b}(x)
   \tilde{\Psi}_{c}(y)\bar{\tilde{\Psi}}_{d}(y)\wre+
   \frac{1}{i}\propS{22}{(-)}{y-x}{da}
         \wl\bar{\tilde{\Psi}}_{b}(x)\tilde{\Psi}_{c}(y)\wre +\\
 & + \frac{1}{i}\propS{22}{(+)}{x-y}{bc}
         \wl\tilde{\Psi}_{a}(x)\bar{\tilde{\Psi}}_{d}(y)\wre+\\ 
  &+\frac{1}{i}\propS{22}{(-)}{y-x}{da}\frac{1}{i}\propS{22}{(+)}{x-y}{bc}
  \Bigg)\bigg(\wl\tilde{A}_{\mu}(x)\tilde{A}_{\nu}(y)\wre +
   g_{\mu\nu}i\propD{22}{(+)}{x-y}\bigg)\Bigg\}\\
&\\
&\hspace{-1.0 cm}
 \underbrace{-\Bigg( -e^2\gamma^{\mu}_{ab}\gamma^{\nu}_{cd}\bigg\{
 \textrm{dito with} \: x \leftrightarrow y \;\;(\textrm{{\footnotesize and
: $\;a\leftrightarrow c, b\leftrightarrow d, 
\mu \leftrightarrow \nu$.}})\bigg\}\Bigg)}_{+R'_{2}}.
\end{split}
\end{equation}
The four summands of the $A_{2}'$-part of this expression shall be denoted 
by $A^{\alpha}, B^{\alpha}, C^{\alpha}, D^{\alpha}$, the respective four 
summands in the first factor of 
$A^{\alpha},...,D^{\alpha}$ by $1, 2, 3, 4$, the respective two in the 
second factor of 
$A^{\alpha},...,D^{\alpha}$ by $1,2$. For the $R_{2}'$-part we choose the same 
notation with a superscript ${\;}^{\beta}$ instead of ${\;}^{\alpha}$. 
After carrying out the multiplications in this expression we then can refer to 
the several parts by $A_{11}^{\alpha}, D_{32}^{\beta}$ etc. By $A_{kl},...,
D_{kl}$ we will denote $A_{kl}^{\alpha}+A_{kl}^{\beta},...,D_{kl}^{\alpha}+
D_{kl}^{\beta}$.

After having done this calculation, 
we arrive at the following expression for $D_{2}$:
\begin{equation*}
\begin{split}
{^{\beta}}\!D_{2}(x,y) = & `A+B+C+D'=\\
A_{11}^{\alpha}:\hspace{1.2 cm}
  -&e^2\gamma_{ab}^{\mu}\gamma_{cd}^{\nu}\wl\bar{\Psi}_{a}(x)\Psi_{b}(x)
     \bar{\Psi}_{c}(y)\Psi_{d}(y)\wre\wl A_{\mu}(x)A_{\nu}(y)\wre  \\
    A_{11}^{\beta}:\hspace{1.2 cm}&\qquad-
     (-e^2\gamma_{cd}^{\nu}\gamma_{ab}^{\mu}\wl \bar{\Psi}_{c}(y)\Psi_{d}(y)
     \bar{\Psi}_{a}(x)\Psi_{b}(x)\wre\wl A_{\nu}(y)A_{\mu}(x)\wre )\\
A_{12}^{\alpha}:\hspace{1.2 cm}
-&e^2\gamma_{ab}^{\mu}\gamma_{cd}^{\nu}\wl \bar{\Psi}_{a}(x)\Psi_{b}(x)
     \bar{\Psi}_{c}(y)\Psi_{d}(y)\wre g_{\mu\nu}i\propD{11}{(+)}{x-y}\\
    A_{12}^{\beta}:\hspace{1.2 cm}&\qquad-
     (-e^2\gamma_{cd}^{\nu}\gamma_{ab}^{\mu}\wl \bar{\Psi}_{c}(y)\Psi_{d}(y)
     \bar{\Psi}_{a}(x)\Psi_{b}(x)\wre  g_{\mu\nu}i\propD{11}{(+)}{y-x})\\
A_{21}^{\alpha}:\hspace{1.2 cm}
-&e^2\gamma_{ab}^{\mu}\gamma_{cd}^{\nu}\wl \Psi_{b}(x)\bar{\Psi}_{c}(y)\wre 
     \frac{1}{i}\propS{11}{(-)}{y-x}{da}\wl A_{\mu}(x)A_{\nu}(y)\wre  \\
      A_{21}^{\beta}:\hspace{1.2 cm}
   &\qquad-(-e^2\gamma_{cd}^{\nu}\gamma_{ab}^{\mu}\wl 
    \Psi_{d}(y)\bar{\Psi}_{a}(x)
     \wre\frac{1}{i}\propS{11}{(-)}{x-y}{bc}\wl A_{\nu}(y)A_{\mu}(x)\wre )\\
A_{22}^{\alpha}:\hspace{1.2 cm}
-&e^2\gamma_{ab}^{\mu}\gamma_{cd}^{\nu}\wl \bar{\Psi}_{a}(x)\Psi_{d}(y)\wre 
     \frac{1}{i}\propS{11}{(+)}{x-y}{bc}\wl A_{\mu}(x)A_{\nu}(y)\wre  \\
   A_{22}^{\beta}:\hspace{1.2 cm}&\qquad-
     (-e^2\gamma_{cd}^{\nu}\gamma_{ab}^{\mu}\wl \bar{\Psi}_{c}(y)\Psi_{b}(x)
   \wre\frac{1}{i}\propS{11}{(+)}{y-x}{da}\wl A_{\nu}(y)A_{\mu}(x)\wre )\\
A_{31}^{\alpha}:\hspace{1.2 cm}
-&e^2\gamma_{ab}^{\mu}\gamma_{cd}^{\nu}\wl \Psi_{b}(x)\bar{\Psi}_{c}(y)\wre 
     \frac{1}{i}\propS{11}{(-)}{y-x}{da}g_{\mu\nu}i\propD{11}{(+)}{x-y} 
          \\A_{31}^{\beta}:\hspace{1.2 cm}&\qquad-
     (-e^2\gamma_{cd}^{\nu}\gamma_{ab}^{\mu}\wl \Psi_{d}(y)\bar{\Psi}_{a}(x)
    \wre\frac{1}{i}\propS{11}{(-)}{x-y}{bc}g_{\mu\nu}i\propD{11}{(+)}{y-x})\\
A_{32}^{\alpha}:\hspace{1.2 cm}
-&e^2\gamma_{ab}^{\mu}\gamma_{cd}^{\nu}\wl \bar{\Psi}_{a}(x)\Psi_{d}(y)\wre 
     \frac{1}{i}\propS{11}{(+)}{x-y}{bc}g_{\mu\nu}i\propD{11}{(+)}{x-y} 
    \\A_{32}^{\beta}:\hspace{1.2 cm}&\qquad-
     (-e^2\gamma_{cd}^{\nu}\gamma_{ab}^{\mu}\wl \bar{\Psi}_{c}(y)\Psi_{b}(x)
    \wre\frac{1}{i}\propS{11}{(+)}{y-x}{da}g_{\mu\nu}i\propD{11}{(+)}{y-x})\\
A_{41}^{\alpha}:\hspace{1.2 cm}
-&e^2\gamma_{ab}^{\mu}\gamma_{cd}^{\nu}\frac{1}{i}\propS{11}{(-)}{y-x}{da}
    \frac{1}{i}\propS{11}{(+)}{x-y}{bc}\wl A_{\mu}(x)A_{\nu}(y)\wre  \\
   A_{41}^{\beta}:\hspace{1.2 cm}&\qquad-
     (-e^2\gamma_{cd}^{\nu}\gamma_{ab}^{\mu}\frac{1}{i}\propS{11}{(-)}{x-y}{bc}
     \frac{1}{i}\propS{11}{(+)}{y-x}{da}\wl A_{\nu}(y)A_{\mu}(x)\wre )\\
A_{42}^{\alpha}:\hspace{1.2 cm}
-&e^2\gamma_{ab}^{\mu}\gamma_{cd}^{\nu}\frac{1}{i}\propS{11}{(-)}{y-x}{da}
     \frac{1}{i}\propS{11}{(+)}{x-y}{bc}g_{\mu\nu}i\propD{11}{(+)}{x-y} 
      \\A_{42}^{\beta}:\hspace{1.2 cm}&\qquad-
     (-e^2\gamma_{cd}^{\nu}\gamma_{ab}^{\mu}\frac{1}{i}\propS{11}{(-)}{x-y}{bc}
     \frac{1}{i}\propS{11}{(+)}{y-x}{da}g_{\mu\nu}i\propD{11}{(+)}{y-x})+
\end{split}
\end{equation*}
\begin{equation*}
\begin{split}
B_{11}^{\alpha}:\hspace{1.2 cm}
+&e^2\gamma_{ab}^{\mu}\gamma_{cd}^{\nu}\wl \bar{\Psi}_{a}(x)\Psi_{b}(x)
     \tilde{\Psi}_{c}(y)\bar{\tilde{\Psi}}_{d}(y)\wre 
     \wl A_{\mu}(x)\tilde{A}_{\nu}(y)\wre \\
   B_{11}^{\beta}:\hspace{1.2 cm}&\qquad+
     (-e^2\gamma_{cd}^{\nu}\gamma_{ab}^{\mu}\wl \bar{\Psi}_{c}(y)\Psi_{d}(y)
     \tilde{\Psi}_{a}(x)\bar{\tilde{\Psi}}_{b}(x)\wre 
     \wl A_{\nu}(y)\tilde{A}_{\mu}(x)\wre )\\
B_{12}^{\alpha}:\hspace{1.2 cm}
+&e^2\gamma_{ab}^{\mu}\gamma_{cd}^{\nu}\wl \bar{\Psi}_{a}(x)\Psi_{b}(x)
     \tilde{\Psi}_{c}(y)\bar{\tilde{\Psi}}_{d}(y)\wre 
     g_{\mu\nu}i\propD{12}{(+)}{x-y}\\
   B_{12}^{\beta}:\hspace{1.2 cm}&\qquad+
     (-e^2\gamma_{cd}^{\nu}\gamma_{ab}^{\mu}\wl \bar{\Psi}_{c}(y)\Psi_{d}(y)
     \tilde{\Psi}_{a}(x)\bar{\tilde{\Psi}}_{b}(x)\wre 
     g_{\mu\nu}i\propD{12}{(+)}{y-x})\\
B_{21}^{\alpha}:\hspace{1.2 cm}
+&e^2\gamma_{ab}^{\mu}\gamma_{cd}^{\nu}\wl \Psi_{b}(x)\tilde{\Psi}_{c}(y)\wre 
     (-i)\propS{21}{(-)}{y-x}{da}
      \wl A_{\mu}(x)\tilde{A}_{\nu}(y)\wre  \\
   b_{21}^{\beta}:\hspace{1.2 cm}&\qquad+
     (-e^2\gamma_{cd}^{\nu}\gamma_{ab}^{\mu}\wl \Psi_{d}(y)
     \tilde{\Psi}_{a}(x)\wre 
     (-i)\propS{21}{(-)}{x-y}{bc}\wl A_{\nu}(y)\tilde{A}_{\mu}(x)\wre )\\
B_{22}^{\alpha}:\hspace{1.2 cm}
+&e^2\gamma_{ab}^{\mu}\gamma_{cd}^{\nu}
       \wl \bar{\Psi}_{a}(x)\bar{\tilde{\Psi}}_{d}(y)\wre 
     i\propS{12}{(+)}{x-y}{bc}
      \wl A_{\mu}(x)\tilde{A}_{\nu}(y)\wre  \\
     B_{22}^{\beta}:\hspace{1.2 cm} &\qquad+
     (-e^2\gamma_{cd}^{\nu}\gamma_{ab}^{\mu}
     \wl \bar{\Psi}_{c}(y)\bar{\tilde{\Psi}}_{b}(x)\wre 
     i\propS{12}{(+)}{y-x}{da}
     \wl A_{\nu}(y)\tilde{A}_{\mu}(x)\wre )\\
B_{31}^{\alpha}:\hspace{1.2 cm}
+&e^2\gamma_{ab}^{\mu}\gamma_{cd}^{\nu}\wl \Psi_{b}(x)\tilde{\Psi}_{c}(y)\wre 
     (-i)\propS{21}{(-)}{y-x}{da}g_{\mu\nu}i\propD{12}{(+)}{x-y} 
          \\B_{31}^{\beta}:\hspace{1.2 cm}&\qquad+
     (-e^2\gamma_{cd}^{\nu}\gamma_{ab}^{\mu}\wl \Psi_{d}(y)\tilde{\Psi}_{a}(x)
     \wre (-i)\propS{21}{(-)}{x-y}{bc}g_{\mu\nu}i\propD{12}{(+)}{y-x})\\
B_{32}^{\alpha}:\hspace{1.2 cm}
+&e^2\gamma_{ab}^{\mu}\gamma_{cd}^{\nu}
     \wl \bar{\Psi}_{a}(x)\bar{\tilde{\Psi}}_{d}(y)\wre 
     i\propS{12}{(+)}{x-y}{bc}g_{\mu\nu}i\propD{12}{(+)}{x-y}  
    \\B_{32}^{\beta}:\hspace{1.2 cm}&\qquad+
     (-e^2\gamma_{cd}^{\nu}\gamma_{ab}^{\mu}
     \wl \bar{\Psi}_{c}(y)\bar{\tilde{\Psi}}_{b}(x)\wre 
     i\propS{12}{(+)}{y-x}{da}g_{\mu\nu}i\propD{12}{(+)}{y-x})\\
B_{41}^{\alpha}:\hspace{1.2 cm}
+&e^2\gamma_{ab}^{\mu}\gamma_{cd}^{\nu}(-i)\propS{21}{(-)}{y-x}{da}
     i\propS{12}{(+)}{x-y}{bc}\wl A_{\mu}(x)\tilde{A}_{\nu}(y)\wre  \\
   B_{41}^{\beta}:\hspace{1.2 cm}&\qquad+
     (-e^2\gamma_{cd}^{\nu}\gamma_{ab}^{\mu}(-i)\propS{21}{(-)}{x-y}{bc}
     i\propS{12}{(+)}{y-x}{da}\wl A_{\nu}(y)\tilde{A}_{\mu}(x)\wre )\\
B_{42}^{\alpha}:\hspace{1.2 cm}
+&e^2\gamma_{ab}^{\mu}\gamma_{cd}^{\nu}(-i)\propS{21}{(-)}{y-x}{da}
     i\propS{12}{(+)}{x-y}{bc}g_{\mu\nu}i\propD{12}{(+)}{x-y} 
      \\B_{42}^{\beta}:\hspace{1.2 cm}&\qquad+
     (-e^2\gamma_{cd}^{\nu}\gamma_{ab}^{\mu}(-i)\propS{21}{(-)}{x-y}{bc}
     i\propS{12}{(+)}{y-x}{da}g_{\mu\nu}i\propD{12}{(+)}{y-x})\\
C_{11}^{\alpha}:\hspace{1.2 cm}
+&e^2\gamma_{ab}^{\mu}\gamma_{cd}^{\nu}
     \wl \tilde{\Psi}_{a}(x)\bar{\tilde{\Psi}}_{b}(x)
     \bar{\Psi}_{c}(y)\Psi_{d}(y)\wre \wl \tilde{A}_{\mu}(x)A_{\nu}(y)\wre  \\
  C_{11}^{\beta}:\hspace{1.2 cm}&\qquad-(e^2\gamma_{cd}^{\nu}\gamma_{ab}^{\mu}
      \wl \tilde{\Psi}_{c}(y)\bar{\tilde{\Psi}}_{d}(y)
     \bar{\Psi}_{a}(x)\Psi_{b}(x)\wre \wl \tilde{A}_{\nu}(y)A_{\mu}(x)\wre )\\
C_{12}^{\alpha}:\hspace{1.2 cm}
+&e^2\gamma_{ab}^{\mu}\gamma_{cd}^{\nu}
    \wl \tilde{\Psi}_{a}(x)\bar{\tilde{\Psi}}_{b}(x)
     \bar{\Psi}_{c}(y)\Psi_{d}(y)\wre g_{\mu\nu}i\propD{12}{(+)}{x-y}\\
    C_{12}^{\beta}:\hspace{1.2 cm}&\qquad-
     (e^2\gamma_{cd}^{\nu}\gamma_{ab}^{\mu}
     \wl \tilde{\Psi}_{c}(y)\bar{\tilde{\Psi}}_{d}(y)
     \bar{\Psi}_{a}(x)\Psi_{b}(x)\wre g_{\mu\nu}i\propD{12}{(+)}{y-x})\\
C_{21}^{\alpha}:\hspace{1.2 cm}
+&e^2\gamma_{ab}^{\mu}\gamma_{cd}^{\nu}
     \wl \bar{\tilde{\Psi}}_{b}(x)\bar{\Psi}_{c}(y)\wre 
     i\propS{12}{(-)}{y-x}{da}\wl \tilde{A}_{\mu}(x)A_{\nu}(y)\wre  \\
  C_{21}^{\beta}:\hspace{1.2 cm}&\qquad-
     (e^2\gamma_{cd}^{\nu}\gamma_{ab}^{\mu}
     \wl \bar{\tilde{\Psi}}_{d}(y)\bar{\Psi}_{a}(x)\wre 
     i\propS{12}{(-)}{x-y}{bc}\wl \tilde{A}_{\nu}(y)A_{\mu}(x)\wre )\\
C_{22}^{\alpha}:\hspace{1.2 cm}
+&e^2\gamma_{ab}^{\mu}\gamma_{cd}^{\nu}
      \wl \tilde{\Psi}_{a}(x)\Psi_{d}(y)\wre 
     (-i)\propS{21}{(+)}{x-y}{bc}\wl \tilde{A}_{\mu}(x)A_{\nu}(y)\wre  \\
  C_{22}^{\beta}:\hspace{1.2 cm}&\qquad-
     (e^2\gamma_{cd}^{\nu}\gamma_{ab}^{\mu}\wl 
     \tilde{\Psi}_{c}(y)\Psi_{b}(x)\wre 
     (-i)\propS{21}{(+)}{y-x}{da}\wl \tilde{A}_{\nu}(y)A_{\mu}(x)\wre )\\
C_{31}^{\alpha}:\hspace{1.2 cm}
+&e^2\gamma_{ab}^{\mu}\gamma_{cd}^{\nu}
      \wl \bar{\tilde{\Psi}}_{b}(x)\bar{\Psi}_{c}(y)\wre 
     i\propS{12}{(-)}{y-x}{da}g_{\mu\nu}i\propD{12}{(+)}{x-y} 
          \\C_{31}^{\beta}:\hspace{1.2 cm}&\qquad-
     (e^2\gamma_{cd}^{\nu}\gamma_{ab}^{\mu}
     \wl \bar{\tilde{\Psi}}_{d}(y)\bar{\Psi}_{a}(x)\wre 
     i\propS{12}{(-)}{x-y}{bc}g_{\mu\nu}i\propD{12}{(+)}{y-x})\\
C_{32}^{\alpha}:\hspace{1.2 cm}
+&e^2\gamma_{ab}^{\mu}\gamma_{cd}^{\nu}
     \wl \tilde{\Psi}_{a}(x)\Psi_{d}(y)\wre 
     (-i)\propS{21}{(+)}{x-y}{bc}g_{\mu\nu}i\propD{12}{(+)}{x-y} 
    \\C_{32}^{\beta}:\hspace{1.2 cm}&\qquad-
     (e^2\gamma_{cd}^{\nu}\gamma_{ab}^{\mu}\wl \tilde{\Psi}_{c}(y)\Psi_{b}(x)
    \wre (-i)\propS{21}{(+)}{y-x}{da}g_{\mu\nu}i\propD{12}{(+)}{y-x})\\
C_{41}^{\alpha}:\hspace{1.2 cm}
+&e^2\gamma_{ab}^{\mu}\gamma_{cd}^{\nu}i\propS{12}{(-)}{y-x}{da}
     (-i)\propS{21}{(+)}{x-y}{bc}\wl \tilde{A}_{\mu}(x)A_{\nu}(y)\wre  \\
     C_{41}^{\beta}:\hspace{1.2 cm}&
        \qquad-(e^2\gamma_{cd}^{\nu}\gamma_{ab}^{\mu}i\propS{12}{(-)}{x-y}{bc}
     (-i)\propS{21}{(+)}{y-x}{da}\wl \tilde{A}_{\nu}(y)A_{\mu}(x)\wre )\\
C_{42}^{\alpha}:\hspace{1.2 cm}
+&e^2\gamma_{ab}^{\mu}\gamma_{cd}^{\nu}i\propS{12}{(-)}{y-x}{da}
     (-i)\propS{21}{(+)}{x-y}{bc}g_{\mu\nu}i\propD{12}{(+)}{x-y} 
      \\C_{42}^{\beta}:\hspace{1.2 cm}&\qquad-
     (e^2\gamma_{cd}^{\nu}\gamma_{ab}^{\mu}i\propS{12}{(-)}{x-y}{bc}
     (-i)\propS{21}{(+)}{y-x}{da}g_{\mu\nu}i\propD{12}{(+)}{y-x}) -
\end{split}
\end{equation*}
\bel{Dzwei}
\begin{split}
D_{11}^{\alpha}:\hspace{1.2 cm}
-&e^2\gamma_{ab}^{\mu}\gamma_{cd}^{\nu}
     \wl \tilde{\Psi}_{a}(x)\bar{\tilde{\Psi}}_{b}(x)
     \tilde{\Psi}_{c}(y)\bar{\tilde{\Psi}}_{d}(y)\wre 
     \wl \tilde{A}_{\mu}(x)\tilde{A}_{\nu}(y)\wre \\
    D_{11}^{\beta}:\hspace{1.2 cm}&\qquad-
     (-e^2\gamma_{cd}^{\nu}\gamma_{ab}^{\mu}
     \wl \tilde{\Psi}_{c}(y)\bar{\tilde{\Psi}}_{d}(y)
     \tilde{\Psi}_{a}(x)\bar{\tilde{\Psi}}_{b}(x)\wre 
     \wl \tilde{A}_{\nu}(y)\tilde{A}_{\mu}(x)\wre )\\
D_{12}^{\alpha}:\hspace{1.2 cm}
-&e^2\gamma_{ab}^{\mu}\gamma_{cd}^{\nu}
     \wl \tilde{\Psi}_{a}(x)\bar{\tilde{\Psi}}_{b}(x)
     \tilde{\Psi}_{c}(y)\bar{\tilde{\Psi}}_{d}(y)\wre 
     g_{\mu\nu}i\propD{22}{(+)}{x-y}\\
    D_{12}^{\beta}:\hspace{1.2 cm}&\qquad-
     (-e^2\gamma_{cd}^{\nu}\gamma_{ab}^{\mu}
     \wl \tilde{\Psi}_{c}(y)\bar{\tilde{\Psi}}_{d}(y)
     \tilde{\Psi}_{a}(x)\bar{\tilde{\Psi}}_{b}(x)\wre 
     g_{\mu\nu}i\propD{22}{(+)}{y-x})\\
D_{21}^{\alpha}:\hspace{1.2 cm}
-&e^2\gamma_{ab}^{\mu}\gamma_{cd}^{\nu}
    \wl \bar{\tilde{\Psi}}_{b}(x)\tilde{\Psi}_{c}(y)\wre 
     \frac{1}{i}\propS{22}{(-)}{y-x}{da}
      \wl \tilde{A}_{\mu}(x)\tilde{A}_{\nu}(y)\wre  \\
    D_{21}^{\beta}:\hspace{1.2 cm}&\qquad-
     (-e^2\gamma_{cd}^{\nu}\gamma_{ab}^{\mu}
    \wl \bar{\tilde{\Psi}}_{d}(y)\tilde{\Psi}_{a}(x)\wre 
     \frac{1}{i}\propS{22}{(-)}{x-y}{bc}
     \wl \tilde{A}_{\nu}(y)\tilde{A}_{\mu}(x)\wre )\\
D_{22}^{\alpha}:\hspace{1.2 cm}
-&e^2\gamma_{ab}^{\mu}\gamma_{cd}^{\nu}
       \wl \tilde{\Psi}_{a}(x)\bar{\tilde{\Psi}}_{d}(y)\wre 
     \frac{1}{i}\propS{22}{(+)}{x-y}{bc}
      \wl \tilde{A}_{\mu}(x)\tilde{A}_{\nu}(y)\wre  \\
    D_{22}^{\beta}:\hspace{1.2 cm}&\qquad-
     (-e^2\gamma_{cd}^{\nu}\gamma_{ab}^{\mu}
     \wl \tilde{\Psi}_{c}(y)\bar{\tilde{\Psi}}_{b}(x)\wre 
     \frac{1}{i}\propS{22}{(+)}{y-x}{da}
     \wl \tilde{A}_{\nu}(y)\tilde{A}_{\mu}(x)\wre )\\
D_{31}^{\alpha}:\hspace{1.2 cm}
-&e^2\gamma_{ab}^{\mu}\gamma_{cd}^{\nu}
     \wl \bar{\tilde{\Psi}}_{b}(x)\tilde{\Psi}_{c}(y)\wre 
     \frac{1}{i}\propS{22}{(-)}{y-x}{da}g_{\mu\nu}i\propD{22}{(+)}{x-y} 
          \\D_{31}^{\beta}:\hspace{1.2 cm}&\qquad-
     (-e^2\gamma_{cd}^{\nu}\gamma_{ab}^{\mu}
     \wl \bar{\tilde{\Psi}}_{d}(y)\tilde{\Psi}_{a}(x)\wre 
     \frac{1}{i}\propS{22}{(-)}{x-y}{bc}g_{\mu\nu}i\propD{22}{(+)}{y-x})\\
D_{32}^{\alpha}:\hspace{1.2 cm}
-&e^2\gamma_{ab}^{\mu}\gamma_{cd}^{\nu}
       \wl \tilde{\Psi}_{a}(x)\bar{\tilde{\Psi}}_{d}(y)\wre 
     \frac{1}{i}\propS{22}{(+)}{x-y}{bc}
      g_{\mu\nu}i\propD{22}{(+)}{x-y} \\
    D_{32}^{\beta}:\hspace{1.2 cm}&\qquad-
     (-e^2\gamma_{cd}^{\nu}\gamma_{ab}^{\mu}
     \wl \tilde{\Psi}_{c}(y)\bar{\tilde{\Psi}}_{b}(x)\wre 
     \frac{1}{i}\propS{22}{(+)}{y-x}{da}
     g_{\mu\nu}i\propD{22}{(+)}{y-x})\\
D_{41}^{\alpha}:\hspace{1.2 cm}
-&e^2\gamma_{ab}^{\mu}\gamma_{cd}^{\nu}
     \frac{1}{i}\propS{22}{(-)}{y-x}{da}
     \frac{1}{i}\propS{22}{(+)}{x-y}{bc}
     \wl \tilde{A}_{\mu}(x)\tilde{A}_{\nu}(y)\wre ) 
      \\D_{41}^{\beta}:\hspace{1.2 cm}&\qquad-
     (-e^2\gamma_{cd}^{\nu}\gamma_{ab}^{\mu}
     \frac{1}{i}\propS{22}{(-)}{x-y}{bc}
     \frac{1}{i}\propS{22}{(+)}{y-x}{da}
     \wl \tilde{A}_{\nu}(y)\tilde{A}_{\mu}(x)\wre )\\
D_{42}^{\alpha}:\hspace{1.2 cm}
-&e^2\gamma_{ab}^{\mu}\gamma_{cd}^{\nu}
     \frac{1}{i}\propS{22}{(-)}{y-x}{da}
     \frac{1}{i}\propS{22}{(+)}{x-y}{bc}g_{\mu\nu}i\propD{22}{(+)}{x-y} 
      \\D_{42}^{\beta}:\hspace{1.2 cm}&\qquad-
     (-e^2\gamma_{cd}^{\nu}\gamma_{ab}^{\mu}
     \frac{1}{i}\propS{22}{(-)}{x-y}{bc}
     \frac{1}{i}\propS{22}{(+)}{y-x}{da}g_{\mu\nu}i\propD{22}{(+)}{y-x}). 
\end{split}
\end{equation}
\subsection{Proof of the Causal Support of \boldmath{$D_{2}$}}
\label{subsec:caussupp}
Pay attention to  $\wl\Psi_{d}\bar{\Psi}_{a}\wre
  =-\wl\bar{\Psi}_{a}\Psi_{d}\wre $.
\begin{align*}
A_{11} :& 0 : \textrm{causal}\\
A_{12} :& \propD{11}{(+)}{x-y} - \propD{11}{(+)}{y-x} = \propD{11}{(+)}{x-y} 
          + \propD{11}{(-)}{x-y} =\\
        & \hspace{1.5 cm}= \propD{11}{}{x-y} = \; {^{T=0}}\!D(x-y) : 
  \textrm{causal}\\
A_{21}+A_{31} :&1.+2.\textrm{term}: \propS{11}{(-)}{y-x}{da} + 
      \propS{11}{(+)}{y-x}{da} =\\&\hspace{1.5 cm}= 
         \propS{11}{}{y-x}{da} = {^{T=0}}\!S_{da}(y-x) : \textrm{causal}\\
   &2.+1.\textrm{term}: \propS{11}{(+)}{x-y}{bc} + \propS{11}{(-)}{x-y}{bc} :
       \textrm{analogously} \longrightarrow \textrm{causal}\\
A_{22}+A_{32} :&1.+2.\textrm{term}: 
                \propS{11}{(-)}{y-x}{da}\propD{11}{(+)}{x-y}+
                  \propS{11}{(+)}{y-x}{da}\propD{11}{(+)}{y-x} = \\
             &\hspace{1.5 cm}= \propS{11}{(-)}{y-x}{da}\propD{11}{(+)}{x-y}
                 + \propS{11}{(+)}{y-x}{da}\propD{11}{(+)}{x-y}+ \\
               &\hspace{2.0 cm}+ \propS{11}{(+)}{y-x}{da}\propD{11}{(+)}{y-x} 
                 - \propS{11}{(+)}{y-x}{da}\propD{11}{(+)}{x-y}=\\
             &\hspace{1.5 cm}= \propS{11}{}{y-x}{da}\propD{11}{(+)}{x-y} -\\
              &\hspace{2.0 cm}- \propS{11}{(+)}{y-x}{da}
    \underbrace{(-\propD{11}{(+)}{y-x}+\propD{11}{(+)}{x-y})}
              _{=D(x-y)} : \textrm{causal}\\
        &2.+1.\textrm{term}: \propS{11}{(+)}{x-y}{bc}\propD{11}{(+)}{x-y}+
           \propS{11}{(-)}{x-y}{bc}\propD{11}{(+)}{y-x} :\\&
            \hspace{1.5 cm}
    \textrm{analogously} \longrightarrow \textrm{causal}\\
A_{41} :&\propS{11}{(-)}{y-x}{da}\propS{11}{(+)}{x-y}{bc} -
         \propS{11}{(-)}{x-y}{bc}\propS{11}{(+)}{y-x}{da} =\\
         &\hspace{1.5 cm}=\propS{11}{(-)}{y-x}{da}\propS{11}{(+)}{x-y}{bc} +
         \propS{11}{(-)}{y-x}{da}\propS{11}{(-)}{x-y}{bc}- \\
         &\hspace{2.0 cm}-\!\propS{11}{(-)}{x-y}{bc}\propS{11}{(+)}{y-x}{da}\!
         -\!\!
         \propS{11}{(-)}{y-x}{da}\propS{11}{(-)}{x-y}{bc}\!= \\
       &\hspace{1.5 cm}=\propS{11}{(-)}{y-x}{da}\propS{11}{}{x-y}{bc} -\\
      &\hspace{2.0 cm}-
         \propS{11}{(-)}{x-y}{bc}[
                    \propS{11}{(+)}{y-x}{da} + \propS{11}{(-)}{y-x}{da}]=\\
       &\hspace{1.5 cm}=\propS{11}{(-)}{y-x}{da}\propS{11}{}{x-y}{bc} -\\
      &\hspace{2.0 cm}-
         \propS{11}{(-)}{x-y}{bc}\propS{11}{}{y-x}{da} : \textrm{causal}\\
A_{42}:&\propS{11}{(-)}{y-x}{da}\propS{11}{(+)}{x-y}{bc}\propD{11}{(+)}{x-y}-\\
      &\hspace{2.0 cm}
      -\propS{11}{(-)}{x-y}{bc}\propS{11}{(+)}{y-x}{da}\propD{11}{(+)}{y-x}=\\
    &\hspace{1.5 cm}
     =\propS{11}{(-)}{y-x}{da}\propS{11}{(+)}{x-y}{bc}\propD{11}{(+)}{x-y}+\\
    &\hspace{2.0 cm}
     +\propS{11}{(-)}{y-x}{da}\propS{11}{(-)}{x-y}{bc}\propD{11}{(+)}{x-y}-\\
     &\hspace{2.0 cm}
     -\propS{11}{(-)}{x-y}{bc}\propS{11}{(+)}{y-x}{da}\propD{11}{(+)}{y-x}-\\
    &\hspace{2.0 cm}
       -\propS{11}{(-)}{y-x}{da}\propS{11}{(-)}{x-y}{bc}\propD{11}{(+)}{x-y}\\
    &\hspace{1.5 cm}
     \textrm{the first two terms are proportional to} \propS{11}{}{x-y}{bc},\\
    &\hspace{1.5 cm}
        \textrm{the other terms are treated as the first one in} \;
     A_{22}+A_{32}\;\longrightarrow \textrm{causal} 
\end{align*}
\begin{align*}
B_{11}+ C_{11} :& 0 : \textrm{causal}\\
B_{12}+C_{12}:&1.+2.\textrm{term}:\propD{12}{(+)}{x-y}-\propD{12}{(+)}{y-x}=0\\
            &2.+1.\textrm{term}:-\propD{12}{(+)}{y-x}+\propD{12}{(+)}{x-y}=0\\
B_{21}+C_{31}:&1.+2.\textrm{term}:-\propS{21}{(-)}{y-x}{da}-
                               \propS{21}{(+)}{y-x}{da}=0\\
             &2.+1.\textrm{term}:\propS{21}{(-)}{x-y}{bc}+
                                \propS{21}{(+)}{x-y}{bc}=0\\
B_{31}+C_{21}:&1.+2.\textrm{term}: \propS{12}{(+)}{x-y}{bc}+
                                \propS{12}{(-)}{x-y}{bc}=0\\
             &2.+1.\textrm{term}:-\propS{12}{(+)}{y-x}{da}
                                -\propS{12}{(-)}{y-x}{da}=0\\
B_{22}+C_{32}:&1.+2.\textrm{term}
              :-\propS{21}{(-)}{y-x}{da}\propD{12}{(+)}{x-y}-
                     \propS{21}{(+)}{y-x}{da}\propD{12}{(+)}{y-x}=\\
             &\hspace{1.2 cm}=-\propD{12}{(+)}{x-y}\propS{21}{}{y-x}{da} =0\\
             &2.+1.\textrm{term}:\propS{21}{(-)}{x-y}{bc}\propD{12}{(+)}{y-x}+
                     \propS{21}{(+)}{x-y}{bc}\propD{12}{(+)}{x-y}=\\
             &\hspace{1.2 cm}=\propD{12}{(+)}{x-y}\propS{21}{}{x-y}{bc} =0\\
B_{32}+C_{22}:&1.+2.\textrm{term}:\propS{12}{(+)}{x-y}{bc}\propD{12}{(+)}{x-y}+
                     \propS{12}{(-)}{x-y}{bc}\propD{12}{(+)}{y-x}=\\
             &\hspace{1.2 cm}=\propD{12}{(+)}{y-x}\propS{12}{}{x-y}{bc} =0\\
             &2.+1.\textrm{term}:-\propS{12}{(+)}{y-x}{da}\propD{12}{(+)}{y-x}
                     -\propS{12}{(-)}{y-x}{da}\propD{12}{(+)}{x-y}=\\
             &\hspace{1.2 cm}=-\propD{12}{(+)}{x-y}\propS{12}{}{y-x}{da} =0\\
B_{41}+C_{41}:&1.+2.\textrm{term}:
           -\propS{21}{(-)}{y-x}{da}\propS{12}{(+)}{x-y}{bc}+\\
              &\hspace{2.0 cm}
      +\propS{21}{(+)}{y-x}{da}\propS{12}{(-)}{x-y}{bc}=\\
             &\hspace{1.2 cm}
           =-\propS{12}{(+)}{x-y}{bc}\propS{21}{}{y-x}{da}=0\\
             &2.+1.\textrm{term}:
           \propS{21}{(-)}{x-y}{bc}\propS{12}{(+)}{y-x}{da}
                    - \propS{21}{(+)}{x-y}{bc}\propS{12}{(-)}{y-x}{da}=\\
             &\hspace{1.2 cm}
         =-\propS{21}{(+)}{x-y}{bc}\propS{12}{}{y-x}{da}=0\\
B_{42}+C_{42}:&1.+2.\textrm{term}:
              -\propS{21}{(-)}{y-x}{da}\propS{12}{(+)}{x-y}{bc}
                                 \propD{12}{(+)}{x-y}+\\ 
                     &\hspace{2.0 cm}
         +\propS{21}{(+)}{y-x}{da}\propS{12}{(-)}{x-y}{bc}
                                 \propD{12}{(+)}{y-x}=\\
             &\hspace{1.2 cm}=\propS{21}{(-)}{y-x}{da}\propS{12}{(+)}{x-y}{bc}[
                      -\propD{12}{(+)}{x-y}+\propD{12}{(+)}{y-x}]=0\\
             &2.+1.\textrm{term}:
          \propS{21}{(-)}{x-y}{bc}\propS{12}{(+)}{y-x}{da}
                                 \propD{12}{(+)}{y-x}-\\ 
                     &\hspace{2.0 cm}
        -\propS{21}{(+)}{x-y}{bc}\propS{12}{(-)}{y-x}{da}
                                 \propD{12}{(+)}{x-y}=\\
             &\hspace{1.2 cm}=\propS{21}{(+)}{x-y}{bc}\propS{12}{(-)}{y-x}{da}[
                      \propD{12}{(+)}{y-x}-\propD{12}{(+)}{x-y}]=0
\end{align*}
\begin{align*}
D_{11} :& 0 : \textrm{causal}\\
D_{12} :& \propD{22}{(+)}{x-y} - \propD{22}{(+)}{y-x} = \propD{11}{(+)}{y-x} 
          + \propD{11}{(-)}{y-x} =\\
        & \hspace{0.9 cm}
    = \propD{11}{}{y-x} = \; {^{T=0}}\!D(y-x) : \textrm{causal}\\
D_{21}+D_{31} :&1.+2.\textrm{term}:
         -\propS{22}{(-)}{y-x}{da}-\propS{22}{(+)}{y-x}{da} =\\
        &\hspace{0.9 cm}= 
         -\propS{22}{}{y-x}{da} = -\; {^{T=0}}\!S_{da}(y-x) : \textrm{causal}\\
       &2.+1.\textrm{term}: 
            \propS{22}{(-)}{x-y}{bc} + \propS{22}{(+)}{x-y}{bc} :
       \textrm{analogously} \longrightarrow \textrm{causal}\\
D_{22}+D_{32} :&1.+2.\textrm{term}:
                   -\propS{22}{(-)}{y-x}{da}\propD{22}{(+)}{x-y}-
                  \propS{22}{(+)}{y-x}{da}\propD{22}{(+)}{y-x} = \\
             &\hspace{0.9 cm}=-\propS{22}{(-)}{y-x}{da}\propD{22}{(+)}{x-y}
                 - \propS{22}{(+)}{y-x}{da}\propD{22}{(+)}{x-y}- \\
               &\hspace{1.7 cm}- \propS{22}{(+)}{y-x}{da}\propD{22}{(+)}{y-x} 
                 + \propS{22}{(+)}{y-x}{da}\propD{22}{(+)}{x-y}=\\
             &\hspace{0.9 cm}= -\propS{22}{}{y-x}{da}\propD{22}{(+)}{x-y} +\\
            &\hspace{1.7 cm}+ 
                \propS{22}{(+)}{y-x}{da}
    \underbrace{[-\propD{22}{(+)}{y-x}+\propD{22}{(+)}{x-y}]}
              _{=\;{^{T=0}}\!D(y-x)} : \textrm{causal}\\
        &2.+1.\textrm{term}: \propS{22}{(-)}{x-y}{bc}\propD{22}{(+)}{y-x}+\\
           &\hspace{1.7 cm}+
           \propS{22}{(+)}{x-y}{bc}\propD{22}{(+)}{x-y} :
            \textrm{analogously} \longrightarrow \textrm{causal}\\
D_{41} :&-\propS{22}{(-)}{y-x}{da}\propS{22}{(+)}{x-y}{bc} +
         \propS{22}{(+)}{y-x}{da}\propS{22}{(-)}{x-y}{bc} =\\
         &\hspace{0.9 cm}=-\propS{22}{(-)}{y-x}{da}\propS{22}{(+)}{x-y}{bc} -
         \propS{22}{(+)}{y-x}{da}\propS{22}{(+)}{x-y}{bc} +\\
         &\hspace{1.7 cm}+\propS{22}{(+)}{y-x}{da}\propS{22}{(-)}{x-y}{bc} +\\
         &\hspace{2.0 cm}+
         \propS{22}{(+)}{y-x}{da}\propS{22}{(+)}{x-y}{bc} =\\
       &\hspace{0.9 cm}=-\propS{22}{}{y-x}{da}\propS{22}{(+)}{x-y}{bc} +\\
        &\hspace{1.7 cm}+
         \propS{22}{(+)}{y-x}{da}[
                    \propS{22}{(-)}{x-y}{bc} + \propS{22}{(+)}{x-y}{bc}]=\\
       &\hspace{0.9 cm}=-\propS{22}{}{y-x}{da}\propS{22}{(+)}{x-y}{bc} +\\
       &\hspace{2.0 cm}+
         \propS{22}{(+)}{y-x}{da}\propS{22}{}{x-y}{bc} : \textrm{causal}\\
D_{42} :&-\propS{22}{(-)}{y-x}{da}\propS{22}{(+)}{x-y}{bc} 
           \propD{22}{(+)}{x-y}+\\
         &\hspace{1.7 cm}+
      \propS{22}{(+)}{y-x}{da}\propS{22}{(-)}{x-y}{bc}\propD{22}{(+)}{y-x}=\\
    &\hspace{0.9 cm}
    =-\propS{22}{(-)}{y-x}{da}\propS{22}{(+)}{x-y}{bc}\propD{22}{(+)}{x-y}-\\
    &\hspace{1.7 cm}     
    -\propS{22}{(+)}{y-x}{da}\propS{22}{(+)}{x-y}{bc}\propD{22}{(+)}{x-y}+\\
     &\hspace{1.7 cm}
    +\propS{22}{(+)}{y-x}{da}\propS{22}{(-)}{x-y}{bc}\propD{22}{(+)}{y-x}+\\
     &\hspace{1.7 cm}+ 
     \propS{22}{(+)}{y-x}{da}\propS{22}{(+)}{x-y}{bc}\propD{22}{(+)}{x-y}\\
    &\hspace{0.9 cm}
      \textrm{the first two terms are proportional to} \propS{22}{}{y-x}{da},\\
    &\hspace{0.9 cm}
      \textrm{the other terms are treated as the first one in} \;\;
    D_{22}+D_{32}\;\; \longrightarrow \textrm{causal}
\end{align*}
Thus we proved that $D_{2}$ has causal support, and the requirements 
for the inductive causal construction are met. In addition we can see that the 
sum of the $B$- and $C$-contributions to $D_{2}$ equals zero. 

\subsection{Some Calculations for \boldmath{$T_{2}$}}\label{subsec:tzwei}

\subsubsection*{1. Calculation of \boldmath{$\omega^{\mathrm{VP}}$}}
The ($T>0$)-parts of the vacuum polarization have $\omega \leq -1$, 
thus they can be split trivially. This will be shown in the following. 
\bel{omegavp1}
\begin{split}
A_{41}^{\alpha}+A_{41}^{\beta}:\hspace{1.0 cm}&\bigg[\propS{11}{(-)}{y-x}{}
           \gamma^{\mu}\propS{11}{(+)}{x-y}{}-
           \propS{11}{(+)}{y-x}{}\gamma^{\mu}
             \propS{11}{(-)}{x-y}{}\bigg]^{\ret}=\\
             =&\bigg[\propS{11}{(-)}{y-x}{}
           \gamma^{\mu}\propS{11}{}{x-y}{}-
           \propS{11}{}{y-x}{}\gamma^{\mu}
                 \propS{11}{(-)}{x-y}{}\bigg]^{\ret}=\\
             =&\bigg[\propStgl{11}{(-)}{y-x}{}
           \gamma^{\mu}\propStgl{11}{}{x-y}{}-\\
              &\hspace{2.3 cm}-
           \propStgl{11}{}{y-x}{}\gamma^{\mu}
                 \propStgl{11}{(-)}{x-y}{}\bigg]^{\ret}+\\
            &\hspace{1.3 cm}+\bigg[\propStgr{11}{(-)}{y-x}{}
           \gamma^{\mu}\propStgl{11}{}{x-y}{}-\\
           &\hspace{2.3 cm}-
           \propStgl{11}{}{y-x}{}\gamma^{\mu}
                 \propStgr{11}{(-)}{x-y}{}\bigg]^{\ret}+\\
           =&\big[(T=0)\textrm{-part}]^{\ret} + \bigg[\propStgr{11}{(-)}{y-x}{}
           \gamma^{\mu}\propStgl{11}{}{x-y}{}\bigg]^{\ret} +\\
           &\hspace{2.0 cm}+ \bigg[-\propStgl{11}{}{y-x}{}\gamma^{\mu}
                 \propStgr{11}{(-)}{x-y}{}\bigg]^{\ret}. 
\end{split}
\end{equation}
We consider the singular order of the $(T>0)$-parts in momentum space. 
According to (\ref{eq:quasiasimp}) we thus  
have to investigate the following expression (we suppress some unimportant 
factors):
$$\lim_{\delta\rightarrow 0}\rho(\delta)\langle\hat{\Pi}(\frac{p}{\delta}),
  \check{\phi}(p)\rangle\overset{!}{=}\langle\hat{\Pi}_{0}(p),\check{\phi}(p)
  \rangle,$$
with
\begin{equation*}
\begin{split}
\hat{\Pi}(p) := &\int\!d^4q \propSfttgr{11}{(-)}{q}{}\gamma^{\mu}
  \propSfttgl{11}{}{q-p}{}\sim\\
&\sim\int\!d^4q tr\bigg[\gamma^{\nu}(\slas{q}+m)\gamma^{\mu}
     (\slas{q}-\slas{p}+m)\bigg]\delta(q^2-m^2)\cdot\\
   &\hspace{1.5 cm}\cdot
  \big[f_{-}(q)\theta(q_{0})+f_{+}(q)\theta(-q_{0})\big]
  \delta((q-p)^2-m^2)\sign(q_{0}-p_{0}).
\end{split}
\end{equation*}
Now we calculate the $q_{0}$-integral using 
$$\delta(q^2-m^2)=\frac{1}{\sqrt{\vec{q}^2+m^2}}\bigg(\delta(q_{0}-
  \sqrt{\vec{q}^2+m^2})+\delta(q_{0}+\sqrt{\vec{q}^2+m^2})\bigg),$$
which gives us (we choose polar coordinates, $x:= \cos(\theta)$)
\begin{align*}
\lim_{\delta\rightarrow 0}\rho(\delta)&\int\!d^4p\check{\phi}(p)
  \int_{0}^{\infty}dr r^2 
  \int_{-1}^{1}dx\int_{0}^{2\pi}d\phi\bigg\{\\
&\big[4g^{\nu\mu}\frac{pq}{\delta}+
  4\frac{p^{\nu}}{\delta}q^{\mu}-4q^{\nu}\frac{p^{\mu}}{\delta}\big]
  \frac{1}{\sqrt{r^2+m^2}}\frac{1}{e^{\beta\lvert qu\rvert-\beta\mu}+1}\cdot\\
  &\hspace{2.5 cm}\cdot
  \sign(\sqrt{r^2+m^2}-\frac{p_{0}}{\delta})\frac{1}{2}
  \delta(-\frac{p_{0}}{\delta}\sqrt{r^2+m^2}
  +\frac{\lvert\vec{p}\rvert}{\delta} rx+\frac{p^2}{2\delta^2})+\\
&+\big[4g^{\nu\mu}\frac{pq}{\delta}+
  4\frac{p^{\nu}}{\delta}q^{\mu}-4q^{\nu}\frac{p^{\mu}}{\delta}\big]
  \frac{1}{\sqrt{r^2+m^2}}\frac{1}{e^{\beta\lvert qu\rvert+\beta\mu}+1}\cdot\\
  &\hspace{2.5 cm}\cdot
  \sign(-\sqrt{r^2+m^2}-\frac{p_{0}}{\delta})\frac{1}{2}
  \delta(\frac{p_{0}}{\delta}\sqrt{r^2+m^2}
  +\frac{\lvert\vec{p}\rvert}{\delta} rx+\frac{p^2}{2\delta^2})\bigg\},
\end{align*}
where we have $q_{0}=+\sqrt{r^2+m^2}$ in the first summand and  
$q_{0}=-\sqrt{r^2+m^2}$ in the second.
With the remaining $\delta$-distribution we calculate the $x$-integral:
\begin{align*}
&=\lim_{\delta\rightarrow 0}\rho(\delta)\int\!d^4p\int_{0}^{\infty}dr
 \int_{0}^{2\pi}d\phi\frac{\check{\phi}(p)}
 {\lvert\vec{p}\rvert}\frac{\delta}{\sqrt{r^2+m^2}}\cdot\\
 &\hspace{4.0 cm}\cdot
 \big[4g^{\nu\mu}\frac{pq}{\delta}+
  4\frac{p^{\nu}}{q^{\mu}\delta}-4\frac{q^{\nu}p^{\mu}}{\delta}\big]
  \sign(\sqrt{r^2+m^2}-\frac{p_{0}}{\delta})\cdot\\&\cdot
 \Big(e^{\beta\bigl\lvert\sqrt{r^2+m^2}u_{0}-(u_{1}\sin{\phi}
  +u_{2}\cos{\phi}) 
  \frac{1}{\lvert\vec{p}\rvert}\sqrt{\!-\!p^2r^2\!-\!m^2p_{0}^2\!+\!
  2\frac{p^2}{2\delta}p_{0}\sqrt{r^2+m^2}\!-\!\frac{p^4}{4\delta^2}}-
  \frac{u_{3}}{\lvert \vec{p}\rvert}(p_{0}\sqrt{r^2+m^2}\!
  -\!\frac{p^2}{2\delta})
  \bigr\rvert-\beta\mu}\!+\!1\Big)^{\!\!-1}\\
&+\textrm{analogous terms}.
\end{align*}
In the first summand we have $x= 
\frac{p_{0}\sqrt{r^2+m^2}}{\lvert\vec{p}\rvert r}
 -\frac{p^2}{\lvert\vec{p}\rvert r2\delta}$. Because of $-1 \leq  
x\leq 1$, this equation requires $r$ to fulfil some additional 
condition. This claims $r \sim \frac{1}{\delta}$ for 
$\delta \rightarrow 0$, and the following considerations are thus valid also 
for $u=(1,0)$. After investigation of the integrand we can verify that all 
assumptions of the theorem of Lebesgue are met and it thus applies\footnote{
Some problems possibly could be caused by 
$p \in \{p\in\mathbb{M}\;\vert\; p^2=0 \}$. But since this is a 
set of measure zero with respect to the measure $d^{4}p$ this is 
not the case.}. Thus we can perform the limit in the integrand and it 
follows that the whole expression converges to zero,  
\emph{for all}  $\rho(\delta)\sim\delta^{-m}, m \in 
\mathbb{N}$ (because $\lim_{\delta\rightarrow 0}(\frac{1}{\delta^m}
 \frac{1}{e^{\frac{\lvert \vec{p} \rvert}{\delta}}+1})=0 \;\forall \vec{p} \neq
 \vec{0}$)! Thus we have $\omega^{\mathrm{VP}}=-\infty$, and the quasi 
asymptotics equals zero. 

\subsubsection*{2. On the Vacuum of Thermal QED in 2. Order}
We have to show that the vacuum expectation value of the vacuum graph is 
finite in the adiabatic limit. Thus we consider the Fourier transform of 
$\int\!d^4xd^4yT_{2}^{\mathrm{VG}}(x,y)g(x)g(y)$ (use $T^{\mathrm{VG}} \sim 
SSD$, cf. (\ref{eq:vgt2})), which gives  
$$\frac{1}{\epsilon^4}\int\!d^4p\hat{T}^{\mathrm{VG}}_{2}
  (\epsilon p)\hat{g}_{0}(p)\hat{g}_{0}(-p).$$
For the $B+C$-terms this equals\footnote{Here $\sqrt{1/2 \pm}:= 
\sqrt{f_{\pm}(q_{1/2})(1-f_{\pm}(q_{1/2}))}$.}
\begin{multline*}
\frac{1}{\epsilon^4}\int\!d^4p\hat{g}_{0}(p)\hat{g}_{0}(-p)\int\!d^4q_{1}d^4q_{2}
 \tr\bigg[\gamma^{\mu}(\slas{q}_{1}+m)\gamma_{\mu}(\slas{q}_{2}+m)\bigg]
 \delta(q_{1}^2-m^2)\delta(q_{2}^2-m^2)\cdot\\ \cdot
 \big[\theta(q_{1,0})\sqrt{1-}-
 \theta(-q_{1,0})\sqrt{1+}\big]\big[\theta(q_{2,0})\sqrt{2-}-
 \theta(-q_{2,0})\sqrt{2+}\big]\delta((q_{1}-q_{2}-\epsilon p)^2)\sqrt{f(1+f)}.
\end{multline*}
Evaluation of the trace gives $8(2m^2-q_{1}q_{2})$, and with 
$\delta(q_{i}^2-m^2)$ the $q_{1,0}$- and $q_{2,0}$-integrals are calculated. 
then we insert $1=\int\!d^4k\delta(q_{1}-q_{2}-\epsilon p-k)$ 
and evaluate the  
$k_{0}$-integral with $\delta(k^2)$ and the $d^3q_{2}$-integral using  
$\delta(\vec{q}_{1}-\vec{q}_{2}-\epsilon\vec{p}-\vec{k})$. Thus we arrive at 
terms of the following form:
\begin{multline*}
\frac{1}{\epsilon^4}\int\!d^4p\hat{g}_{0}(p)\hat{g}_{0}(-p)\int\!d^3k
 \int\!d^3q_{1}\cdot\\\cdot\big[2m^2-\sqrt{\lvert \vec{q}_{1}\rvert^2+m^2}
 \sqrt{(\vec{q}_{1}-\epsilon\vec{p}-\vec{k})^2+m^2}+\vec{q}_{1}^2-
 \epsilon\vec{q}_{1}\vec{p}-\vec{q}_{1}\vec{k}\big]\cdot\\ \cdot
 \frac{1}{\sqrt{\lvert \vec{q}_{1}\rvert^2+m^2}
  \sqrt{(\vec{q}_{1}-\epsilon\vec{p}-\vec{k})^2+m^2}}\sqrt{1-}\sqrt{2-}
 \sqrt{f(k)(1+f(k))}\cdot\\ \cdot
 \delta(\sqrt{\lvert \vec{q}_{1}\rvert^2+m^2}-
  \sqrt{(\vec{q}_{1}-\epsilon\vec{p}-\vec{k})^2+m^2}-\epsilon p_{0} 
 -\lvert \vec{k} \rvert).
\end{multline*}
Here the limit $\epsilon \rightarrow 0$ can be performed without problems, 
the whole expression (\emph{without} the factor $\frac{1}{\epsilon^4}$) 
converges to a $p$-independent constant, 
which can be cancellated by a suitable 
normalization term\footnote{See the discussion in the main text after 
eq. (\ref{eq:vgt2}).}. Thus we established the existence of the 
adiabatic limit 
of the vacuum expectation value for the $B+C$-part of the vacuum graph. 

As for the $A+D$-term:
\begin{equation*}
\begin{split}
\int\!d^4q_{1}d^4q_{2}&\tr[....]\delta(q_{1}^2-m^2)\delta(q_{2}^2-m^2)\cdot\\
 &\cdot\bigg[
  \theta(q_{1,0})-f_{-}(q_{1})\theta(q_{1,0})-
  f_{+}(q_{1})\theta(-q_{1,0})\bigg]\cdot \\ &\cdot
  \bigg[-\theta(-q_{2,0})+f_{-}(q_{2})\theta(q_{2,0})+
  f_{+}(q_{2})\theta(-q_{2,0})\bigg]\cdot\\
 &\cdot\delta((q_{1}-q_{2}-p)^2)\bigg[
  \theta(q_{1,0}-q_{2,0}-p_{0})+f(q_{1}-q_{2}-p)\bigg].
\end{split}
\end{equation*} 
Its $(T\!=\!0)$-part is discussed in \cite{ScharfBuch}, p.263 (it equals 
$d(p)$ as it is defined there) and does not cause any problems because of 
the factor $\theta(p^2-4m^2)$. The temperature dependent part has the same 
`$\delta$-structure' as the term discussed above and thus converges in the 
adiabatic limit to a constant, 
which can be cancelled by suitable normalization 
as well.

\subsubsection*{3. Calculation of \boldmath{$\omega^{\mathrm{SE}}$}}
The ($T\!>\!0$)-parts of the self energy have a singular order of $\omega = 
-\infty$. This will be shown in the following.
\begin{align*}
A_{22}^{\alpha}+A_{32}^{\beta}:\hspace{0.5 cm}&\bigg[ 
                  \propS{11}{(-)}{y-x}{}\propD{11}{(+)}{x-y}+
                   \propS{11}{(+)}{y-x}{}\propD{11}{(+)}{y-x}\bigg]^{\ret} =\\
                &=\bigg[\propS{11}{}{y-x}{}\propD{11}{(+)}{x-y}+
                   \propS{11}{(+)}{y-x}{}\propD{11}{}{y-x}\bigg]^{\ret}=\\
                &=\bigg[\propStgl{11}{}{y-x}{}\propDtgl{11}{(+)}{x-y}+
                   \propStgl{11}{(+)}{y-x}{}\propDtgl{11}{}{y-x}\bigg]^{\ret}
                     +\\
                &+\bigg[\propStgl{11}{}{y-x}{}\propDtgr{11}{(+)}{x-y}+
                   \propStgr{11}{(+)}{y-x}{}\propDtgl{11}{}{y-x}\bigg]^{\ret}
                    =\\
                &=\bigg[(T=0)\textrm{-part}\bigg]^{\ret} + 
          \bigg[\propStgr{11}{(+)}{y-x}{}\propDtgl{11}{}{y-x}\bigg]^{\ret}+\\
         &+ \bigg[\propStgl{11}{}{y-x}{}\propDtgr{11}{(+)}{x-y}\bigg]^{\ret}. 
\end{align*}
We consider the singular order of the temperature dependent parts in momentum 
space, according to (\ref{eq:quasiasimp}) we thus have to calculate the 
following limit (we disregard some unessential factors): 
$$\lim_{\delta\rightarrow 0}\rho(\delta)\langle\hat{\Sigma}(\frac{p}{\delta}),
  \check{\phi}(p)\rangle\overset{!}{=}\langle\hat{\Sigma}_{0}(p),
  \check{\phi}(p)\rangle,$$
with
\begin{align*}
\hat{\Sigma}(p):=&\int\!d^4q\propSfttgr{11}{(+)}{p-q}{}\propDfttgl{11}{}{q}
  \sim\\
\sim&\int\!d^4q\delta((p-q)^2-m^2)\gamma^{\nu}(\slas{p}-\slas{q}+m)
         \gamma_{\nu}\cdot\\
       &\hspace{1.5 cm}\cdot\big[
         -f_{-}(p-q)\theta(p_{0}-q_{0})-f_{+}(p-q)\theta(q_{0}-p_{0})\big]
         \delta(q^2)\sign(q_{0}).
\end{align*}
We calculate the $q_{0}$-integral using $\delta(q^2)$, which gives terms 
of the following form: 
$$\lim_{\delta \rightarrow 0}\rho(\delta)\int\!d^4p\check{\phi}(p)
 \int\!d^3q\frac{1}{\lvert \vec{q}\rvert}(\slas{q}-\frac{\slas{p}}{\delta}+2m)
 f_{-}(\frac{p}{\delta}-q)
 \theta(\frac{p_{0}}{\delta}-q_{0})\delta(\frac{p^2}{\delta^2}-
 2\frac{p_{0}}{\delta}\lvert\vec{q}\rvert+2\frac{\lvert\vec{p}\rvert}{\delta}
 \lvert\vec{q}\rvert x-m^2).$$
With the remaining $\delta$-distribution we calculate the $x$-integral and 
get (we use $\delta(\frac{p^2}{\delta^2}-
 2\frac{p_{0}}{\delta}\lvert\vec{q}\rvert+2\frac{\lvert\vec{p}\rvert}{\delta}
 \lvert\vec{q}\rvert x-m^2)
 =\frac{\delta}{2\lvert\vec{p}\rvert\lvert\vec{q}\rvert}
 \delta(x-\frac{-\frac{p^2}{\delta}+2p_{0}\lvert\vec{q}\rvert+m^2\delta}
  {2\lvert\vec{p}\rvert\lvert\vec{q}\rvert})$):
\begin{multline*}
=\lim_{\delta \rightarrow 0}\rho(\delta)\int\!d^4p\int_{0}^{\infty}dr
 \int_{0}^{2\pi}d\phi\check{\phi}(p)
  \frac{(\slas{q}-\frac{\slas{p}}{\delta}+2m)\delta}
  {2\lvert\vec{p}\rvert}\theta(\frac{p_{0}}{\delta}-r)\cdot\\\cdot
\Big(e^{\beta\lvert ru_{0}\!-\!(u_{1}\sin{\phi}\!+\!
  u_{2}\cos{\phi})\frac{1}{2\pi}
  \sqrt{\!-\!4p^2r^2+2p^2m^2\!-\!\delta^2m^4\!+\!2\frac{p_{0}p^2}{\delta}r
  \!-\!4p_{0}\delta m^2r}\!-\!u_{3}
  \frac{1}{2\lvert\vec{p}\rvert}(\!-\!
  \frac{p^2}{\delta}\!+\!m^2\delta\!+\!2p_{0}r)\rvert
  \!-\!\beta\mu}\!+\!1\Big)^{\!-\!1}.
\end{multline*}
Proceeding as above for the vacuum polarization we finally get 
$\omega^{\mathrm{SE}}= -\infty$.
\subsubsection*{4. Some Calculations for the Thermal Mass}
$C_{0}$: We choose $u=(1,\vec{0}),\;\;\;\vec{p}\;\|\;k_{3}\textrm{-axis}$. The 
$k_{0}$-integration is performed using the $\delta$-distribution. Then we 
write the integral in polar coordinates, the $\phi$-integration is trivial, 
giving a factor $2\pi$, and  
the $x$-integration is straightforward and yields an integrand proportional to 
\begin{equation*}
\begin{split}
\int_{-1}^{1}dx\bigg[\frac{m}{p^2-m^2-2p_{0}r+2\lvert\vec{p}\rvert rx+i0}
 +\frac{m}{p^2-m^2+2p_{0}r+2\lvert\vec{p} \rvert rx +i0}\bigg] =\\= 
\frac{1}{2\lvert\vec{p}\rvert r}\ln\bigg(\frac{p^2-m^2-2p_{0}r+
   2\lvert\vec{p}\rvert r +i0}
 {p^2-m^2-2p_{0}r-2\lvert\vec{p}\rvert r+i0}\bigg)+\frac{1}
  {2\lvert\vec{p}\rvert r}
 \ln\bigg(\frac{p^2-m^2+2p_{0}r+2\lvert\vec{p}\rvert r+i0}
 {p^2-m^2+2p_{0}r-2\lvert\vec{p}\rvert r+i0}\bigg).
\end{split}
\end{equation*}
This is well defined for all values of $r$ and $x$ as can be seen readily. 
If we now have $p^2=m^2$ it can be verified using $\mathrm{ln}(a+ib) = 
\mathrm{ln}(\sqrt{a^2+b^2}) + i\mathrm{arg}(a+ib)$ that the real and the 
imaginary parts are zero each. 

$C_{p,u}$: Because the parts proportional to $(pu)$ and $p^2$, respectively, 
in the integrals for $C_{p}, C_{u}$ in (\ref{eq:thermmassbestimmgl}) 
have the same structure as $C_{0}$, which was defined by the corresponding 
integral with $m^2$ in the numerator, these parts can be omitted for 
$p^2=m^2$. Solving the remaining integrals as above and additionally using 
$\int_{0}^{\infty} dr \frac{r}{e^{\beta r}-1} = \frac{\pi^2}{6\beta^2}$,  
equation (\ref{eq:resultthermmass}) can be verified.
 
\section{Third Order Calculations} \label{sec:thirdord}
\setcounter{equation}{0}
For $D_{3}$ the general calculation scheme is as follows: we write down the 
various involved products of $T_{k}$'s according to the causal approach. Then 
we perform the contractions which lead to contributions with external legs we 
are interested in and consider only these terms.  

\subsection{Calculations for the Vertex Graph}\label{subsec:vertcalc}
\subsubsection*{1. Calculation of \boldmath{$D_{3}^{\mathrm{V}}$}}
\begin{equation*}
\begin{split}
R_{3}'(x_{1},x_{2},x_{3}) =& \sum_{P_{2}}T(Y,x_{3})\overset{\frown}{T}(X)=\\
    &=\underbrace{T_{2}(x_{1},x_{3})\overset{\frown}{T}_{1}(x_{2})}_{R_{31}'}+
      \underbrace{T_{2}(x_{2},x_{3})\overset{\frown}{T}_{1}(x_{1})}_{R_{32}'}+
    \underbrace{T_{1}(x_{3})\overset{\frown}{T}_{2}(x_{1},x_{2})}_{R_{33}'},\\
 A_{3}'(x_{1},x_{2},x_{3}) =& \sum_{P_{2}}\overset{\frown}{T}(X)T(Y,x_{3})=\\ 
    &=\underbrace{\overset{\frown}{T}_{1}(x_{2})T_{2}(x_{1},x_{3})}_{A_{31}'}+
      \underbrace{\overset{\frown}{T}_{1}(x_{1})T_{2}(x_{2},x_{3})}_{A_{32}'}+
      \underbrace{\overset{\frown}{T}_{2}(x_{1},x_{2})T_{1}(x_{3})}_{A_{33}'}.
\end{split}
\end{equation*}
This is calculated using the following formulae (see (\ref{eq:invt})):
\begin{equation*}
\overset{\frown}{T}_{1}(x)=-T_{1}(x),\;\;
\overset{\frown}{T}_{2}(x_{1},x_{2}) = 
-T_{2}(x_{1},x_{2}) + T_{1}(x_{1})T_{1}(x_{2}) +T_{1}(x_{2})T_{1}(x_{1}). 
\end{equation*}
The three terms of the second equation define the three parts of $R_{33}'$:
$R_{33,1}'+R_{33,2}'+R_{33,3}'$, analogously for $A_{33}'$. 
We take here just the terms with external fields we are interested in 
(cf. Sec. \ref{subsubsec:vertgraph}): 
In $x_{3}$ there shall be an `ordinary field' 
$A_{\mu}^{\mathrm{ext}}(x_{3})$, no `tilde field'; in $x_{1}$ shall be 
$\bar{\Psi}(x_{1})$ or $\tilde{\Psi}(x_{1})$, in $x_{2}\;\; 
\Psi(x_{2})$ or $\bar{\tilde{\Psi}}(x_{2})$. 
\begin{equation*}
\begin{split}
R_{31}'(x_{1},x_{2},x_{3})=&-\underbrace{T_{2}(x_{1},x_{3})}_{=T_{2}^
                 {\mathrm{Compton}}}T_{1}(x_{2})=\\
=&e^3:\bar{\Psi}(x_{1})\gamma^{\mu}\propS{11}{\F}{x_{1}-x_{3}}{}(-1)
                \gamma^{\nu}
        \Psi(x_{3})::A_{\mu}(x_{1})A_{\nu}(x_{3}):\cdot\\ 
        &\hspace{1.5 cm}\cdot\bigg\{
       :\bar{\Psi}(x_{2})\gamma^{\alpha}\Psi(x_{2}):A_{\alpha}(x_{2})-
        :\tilde{\Psi}(x_{2})\gamma^{\alpha}\bar{\tilde{\Psi}}(x_{2}):
        \tilde{A}_{\alpha}(x_{2})\bigg\}\\
&-e^3:\tilde{\Psi}(x_{1})\gamma^{\mu}\propS{21}{\F}{x_{1}-x_{3}}{}
        (-1)\gamma^{\nu}
        \Psi(x_{3})::\tilde{A}_{\mu}(x_{1})A_{\nu}(x_{3}):\cdot\\
        &\hspace{1.5 cm}\cdot\bigg\{
       :\bar{\Psi}(x_{2})\gamma^{\alpha}\Psi(x_{2}):A_{\alpha}(x_{2})-
        :\tilde{\Psi}(x_{2})\gamma^{\alpha}\bar{\tilde{\Psi}}(x_{2}):
        \tilde{A}_{\alpha}(x_{2})\bigg\}.
\end{split}
\end{equation*}
After performing the contractions, we have 
\begin{equation*}
\begin{split}
R_{31}'=&-e^3:\bar{\Psi}(x_{1})\gamma^{\mu}\propS{11}{\F}{x_{1}-x_{3}}{}
  \gamma^{\nu}\frac{1}{i}\propS{11}{(+)}{x_{3}-x_{2}}{}\gamma_{\mu}\Psi(x_{2}):
  i\propD{11}{(+)}{x_{1}-x_{2}}A_{\nu}(x_{3})\\
  &+e^3:\bar{\Psi}(x_{1})\gamma^{\mu}\propS{11}{\F}{x_{1}-x_{3}}{}
  \gamma^{\nu}i\propS{12}{(+)}{x_{3}-x_{2}}{}\gamma_{\mu}
  \bar{\tilde{\Psi}}(x_{2}):i\propD{12}{(+)}{x_{1}-x_{2}}A_{\nu}(x_{3})\\
&+e^3:\tilde{\Psi}(x_{1})\gamma^{\mu}\propS{21}{\F}{x_{1}-x_{3}}{}
  \gamma^{\nu}\frac{1}{i}\propS{11}{(+)}{x_{3}-x_{2}}{}\gamma_{\mu}\Psi(x_{2}):
  i\propD{21}{(+)}{x_{1}-x_{2}}A_{\nu}(x_{3})\\
  &-e^3:\tilde{\Psi}(x_{1})\gamma^{\mu}\propS{21}{\F}{x_{1}-x_{3}}{}
  \gamma^{\nu}i\propS{12}{(+)}{x_{3}-x_{2}}{}\gamma_{\mu}
  \bar{\tilde{\Psi}}(x_{2}):i\propD{22}{(+)}{x_{1}-x_{2}}A_{\nu}(x_{3}).
\end{split}
\end{equation*}
Now to $R_{32}'$:
\begin{equation*}
\begin{split}
R_{32}'(x_{1},x_{2},x_{3})=&-\underbrace{T_{2}(x_{2},x_{3})}_{=T_{2}^
                 {\mathrm{Compton}}}T_{1}(x_{1})=\\
&=-e^3:\bar{\Psi}(x_{3})\gamma^{\nu}\propS{11}{\F}{x_{3}-x_{2}}{}\gamma^{\mu}
        \Psi(x_{2})::A_{\mu}(x_{2})A_{\nu}(x_{3}):\cdot\\
        &\cdot\bigg\{
       :\bar{\Psi}(x_{1})\gamma^{\alpha}\Psi(x_{1}):A_{\alpha}(x_{1})-
        :\tilde{\Psi}(x_{1})\gamma^{\alpha}\bar{\tilde{\Psi}}(x_{1}):
        \tilde{A}_{\alpha}(x_{1})\bigg\}\\
&-e^3:\bar{\Psi}(x_{3})\gamma^{\nu}\propS{12}{\F}{x_{3}-x_{2}}{}
        \gamma^{\mu}
        \bar{\tilde{\Psi}}(x_{2}):
        :\tilde{A}_{\mu}(x_{2})A_{\nu}(x_{3}):\cdot\\
       &\cdot\bigg\{
       :\bar{\Psi}(x_{1})\gamma^{\alpha}\Psi(x_{1}):A_{\alpha}(x_{1})-
        :\tilde{\Psi}(x_{1})\gamma^{\alpha}\bar{\tilde{\Psi}}(x_{1}):
        \tilde{A}_{\alpha}(x_{1})\bigg\}.
\end{split}
\end{equation*}
After performing the contractions (pay attention to  
$\wl\Psi(x_{2})....\bar{\Psi}(x_{1})\wre\rightarrow -\wl\bar{\Psi}(x_{1})....
\Psi(x_{2})\wre$\;!), this gives 
\begin{equation*}
\begin{split}
R_{32}'&=\\&+e^3:\bar{\Psi}(x_{1})\gamma^{\mu}
  \frac{1}{i}\propS{11}{(-)}{x_{1}-x_{3}}{}
  \gamma^{\nu}\propS{11}{\F}{x_{3}-x_{2}}{}\gamma_{\mu}\Psi(x_{2}):
  i\propD{11}{(+)}{x_{2}-x_{1}}A_{\nu}(x_{3})\\
  &-e^3:\tilde{\Psi}(x_{1})\gamma^{\mu}(-i)\propS{21}{(-)}{x_{1}-x_{3}}{}
  \gamma^{\nu}\propS{11}{\F}{x_{3}-x_{2}}{}\gamma_{\mu}
  \Psi(x_{2}):i\propD{12}{(+)}{x_{2}-x_{1}}A_{\nu}(x_{3})\\
&+e^3:\bar{\Psi}(x_{1})\gamma^{\mu}\frac{1}{i}\propS{11}{(-)}{x_{1}-x_{3}}{}
  \gamma^{\nu}\propS{12}{\F}{x_{3}-x_{2}}{}\gamma_{\mu}
  \bar{\tilde{\Psi}}(x_{2}):
  i\propD{21}{(+)}{x_{2}-x_{1}}A_{\nu}(x_{3})\\
  &-e^3:\tilde{\Psi}(x_{1})\gamma^{\mu}(-i)
  \propS{21}{(-)}{x_{1}-x_{3}}{}
  \gamma^{\nu}\propS{12}{\F}{x_{3}-x_{2}}{}\gamma_{\mu}
  \bar{\tilde{\Psi}}(x_{2}):i\propD{22}{(+)}{x_{2}-x_{1}}A_{\nu}(x_{3}).
\end{split}
\end{equation*}
Now we turn to $R_{33}'$: 
\begin{equation*}
\begin{split}
R_{33,1}'=&-T_{1}(x_{3})\underbrace{T_{2}(x_{1},x_{2})}_
            {=T_{2}^{\textrm{M\o ller}}}=\\
=&-e^3\bigg\{:\bar{\Psi}(x_{3})\gamma^{\alpha}\Psi(x_{3}):A_{\alpha}(x_{3})-
        :\tilde{\Psi}(x_{3})\gamma^{\alpha}\bar{\tilde{\Psi}}(x_{3}):
        \tilde{A}_{\alpha}(x_{3})\bigg\}\cdot\\ 
 &\hspace{3.0 cm}\cdot:\bar{\Psi}(x_{1})\gamma^{\alpha}
        \Psi(x_{1})\bar{\Psi}(x_{2})\gamma_{\alpha}\Psi(x_{2}):
        \propD{11}{\F}{x_{1}-x_{2}}\\
&+e^3\bigg\{:\bar{\Psi}(x_{3})\gamma^{\alpha}\Psi(x_{3}):A_{\alpha}(x_{3})-
        :\tilde{\Psi}(x_{3})\gamma^{\alpha}\bar{\tilde{\Psi}}(x_{3}):
        \tilde{A}_{\alpha}(x_{3})\bigg\}\cdot \\ 
    &\hspace{3.0 cm}\cdot:\bar{\Psi}(x_{1})\gamma^{\alpha}
        \Psi(x_{1})\tilde{\Psi}(x_{2})\gamma_{\alpha}
        \bar{\tilde{\Psi}}(x_{2}):
        \propD{12}{\F}{x_{2}-x_{1}}\\
&+e^3\bigg\{:\bar{\Psi}(x_{3})\gamma^{\alpha}\Psi(x_{3}):A_{\alpha}(x_{3})-
        :\tilde{\Psi}(x_{3})\gamma^{\alpha}\bar{\tilde{\Psi}}(x_{3}):
        \tilde{A}_{\alpha}(x_{3})\bigg\}\cdot\\
   &\hspace{3.0 cm}\cdot:\tilde{\Psi}(x_{1})\gamma^{\alpha}
        \bar{\tilde{\Psi}}(x_{1})\bar{\Psi}(x_{2})\gamma_{\alpha}\Psi(x_{2}):
        \propD{12}{\F}{x_{2}-x_{1}}\\
&-e^3\bigg\{:\bar{\Psi}(x_{3})\gamma^{\alpha}\Psi(x_{3}):A_{\alpha}(x_{3})-
        :\tilde{\Psi}(x_{3})\gamma^{\alpha}\bar{\tilde{\Psi}}(x_{3}):
        \tilde{A}_{\alpha}(x_{3})\bigg\}\cdot\\
    &\hspace{3.0 cm}\cdot:\tilde{\Psi}(x_{1})\gamma^{\alpha}
        \bar{\tilde{\Psi}}(x_{1})\tilde{\Psi}(x_{2})\gamma_{\alpha}
        \bar{\tilde{\Psi}}(x_{2}):\propD{22}{\F}{x_{1}-x_{2}}.
\end{split}
\end{equation*}
Considering only the interesting terms here as well we get
\begin{align*}
R&_{33,1}'=\\&+e^3:\bar{\Psi}(x_{1})\gamma^{\mu}\frac{1}{i}
     \propS{11}{(-)}{x_{1}-x_{3}}{}\gamma^{\nu}\frac{1}{i}
     \propS{11}{(+)}{x_{3}-x_{2}}{}\gamma_{\mu}\Psi(x_{2}):
     \propD{11}{\F}{x_{1}-x_{2}}A_{\nu}(x_{3})\\
   &-e^3:\bar{\Psi}(x_{1})\gamma^{\mu}\frac{1}{i}
     \propS{11}{(-)}{x_{1}-x_{3}}{}\gamma^{\nu}i
     \propS{12}{(+)}{x_{3}-x_{2}}{}\gamma_{\mu}\bar{\tilde{\Psi}}(x_{2}):
     \propD{12}{\F}{x_{2}-x_{1}}A_{\nu}(x_{3})\\
   &-e^3:\tilde{\Psi}(x_{1})\gamma^{\mu}(-i)
     \propS{21}{(-)}{x_{1}-x_{3}}{}\gamma^{\nu}\frac{1}{i}
     \propS{11}{(+)}{x_{3}-x_{1}}{}\gamma_{\mu}\Psi(x_{2}):
     \propD{12}{\F}{x_{2}-x_{1}}A_{\nu}(x_{3})\\
   &+e^3:\tilde{\Psi}(x_{1})\gamma^{\mu}(-i)
     \propS{21}{(-)}{x_{1}-x_{3}}{}\gamma^{\nu}i
     \propS{12}{(+)}{x_{3}-x_{2}}{}\gamma_{\mu}\bar{\tilde{\Psi}}(x_{2}):
     \propD{22}{\F}{x_{1}-x_{2}}A_{\nu}(x_{3}).
\end{align*}
Now to the next term:
\begin{align*}
R_{33,2}'=T_{1}(x_{3})T_{1}(x_{1})T_{1}(x_{2})=
&ie\bigg\{:\bar{\Psi}(x_{3})\gamma^{\alpha}\Psi(x_{3}):A_{\alpha}(x_{3})-
        :\tilde{\Psi}(x_{3})\gamma^{\alpha}\bar{\tilde{\Psi}}(x_{3}):
        \tilde{A}_{\alpha}(x_{3})\bigg\}\cdot\\
\cdot&ie\bigg\{:\bar{\Psi}(x_{1})\gamma^{\alpha}\Psi(x_{1}):
        A_{\alpha}(x_{1})-
        :\tilde{\Psi}(x_{1})\gamma^{\alpha}\bar{\tilde{\Psi}}(x_{1}):
        \tilde{A}_{\alpha}(x_{1})\bigg\}\cdot\\
\cdot&ie\bigg\{:\bar{\Psi}(x_{2})\gamma^{\alpha}\Psi(x_{2}):
        A_{\alpha}(x_{2})-
        :\tilde{\Psi}(x_{2})\gamma^{\alpha}\bar{\tilde{\Psi}}(x_{2}):
        \tilde{A}_{\alpha}(x_{2})\bigg\}\;\;.
\end{align*}
Contracting as above and disregarding uninteresting external legs we have 
\begin{align*}
R&_{33,2}'=\\&+ie^3:\bar{\Psi}(x_{1})\gamma^{\mu}\frac{1}{i}
     \propS{11}{(-)}{x_{1}-x_{3}}{}\gamma^{\nu}\frac{1}{i}
     \propS{11}{(+)}{x_{3}-x_{2}}{}\gamma_{\mu}\Psi(x_{2}):
     i\propD{11}{(+)}{x_{1}-x_{2}}A_{\nu}(x_{3})\\
   &-ie^3:\bar{\Psi}(x_{1})\gamma^{\mu}\frac{1}{i}
     \propS{11}{(-)}{x_{1}-x_{3}}{}\gamma^{\nu}i
     \propS{12}{(+)}{x_{3}-x_{2}}{}\gamma_{\mu}\bar{\tilde{\Psi}}(x_{2}):
     i\propD{12}{(+)}{x_{2}-x_{1}}A_{\nu}(x_{3})\\
   &-ie^3:\tilde{\Psi}(x_{1})\gamma^{\mu}(-i)
     \propS{21}{(-)}{x_{1}-x_{3}}{}\gamma^{\nu}\frac{1}{i}
     \propS{11}{(+)}{x_{3}-x_{2}}{}\gamma_{\mu}\Psi(x_{2}):
     i\propD{12}{(+)}{x_{2}-x_{1}}A_{\nu}(x_{3})\\
   &+ie^3:\tilde{\Psi}(x_{1})\gamma^{\mu}(-i)
     \propS{21}{(-)}{x_{1}-x_{3}}{}\gamma^{\nu}i
     \propS{12}{(+)}{x_{3}-x_{2}}{}\gamma_{\mu}\bar{\tilde{\Psi}}(x_{2}):
     i\propD{22}{(+)}{x_{1}-x_{2}}A_{\nu}(x_{3}).
\end{align*}
For the last term of $R_{33}'$ we have 
\begin{align*}
R_{33,3}'=T_{1}(x_{3})T_{1}(x_{2})T_{1}(x_{1})=
&ie\bigg\{:\bar{\Psi}(x_{3})\gamma^{\alpha}\Psi(x_{3}):A_{\alpha}(x_{3})-
        :\tilde{\Psi}(x_{3})\gamma^{\alpha}\bar{\tilde{\Psi}}(x_{3}):
        \tilde{A}_{\alpha}(x_{3})\bigg\}\cdot\\
\cdot&ie\bigg\{:\bar{\Psi}(x_{2})\gamma^{\alpha}\Psi(x_{2}):
        A_{\alpha}(x_{2})-
        :\tilde{\Psi}(x_{2})\gamma^{\alpha}\bar{\tilde{\Psi}}(x_{2}):
        \tilde{A}_{\alpha}(x_{2})\bigg\}\cdot\\
\cdot&ie\bigg\{:\bar{\Psi}(x_{1})\gamma^{\alpha}\Psi(x_{1}):
        A_{\alpha}(x_{1})-
        :\tilde{\Psi}(x_{1})\gamma^{\alpha}\bar{\tilde{\Psi}}(x_{1}):
        \tilde{A}_{\alpha}(x_{1})\bigg\}\;\;.
\end{align*}
Again contracting and considering interesting external legs only 
(Pay again attention to $\wl\Psi(x_{2})....
\bar{\Psi}(x_{1})\wre \rightarrow -\wl\bar{\Psi}(x_{1})....\Psi(x_{2})\wre$!) 
this leads to 
\begin{align*}
R&_{33,3}'=\\&+ie^3:\bar{\Psi}(x_{1})\gamma^{\mu}\frac{1}{i}
     \propS{11}{(-)}{x_{1}-x_{3}}{}\gamma^{\nu}\frac{1}{i}
     \propS{11}{(+)}{x_{3}-x_{2}}{}\gamma_{\mu}\Psi(x_{2}):
     i\propD{11}{(+)}{x_{2}-x_{1}}A_{\nu}(x_{3})\\
   &-ie^3:\tilde{\Psi}(x_{1})\gamma^{\mu}(-i)
     \propS{21}{(-)}{x_{1}-x_{3}}{}\gamma^{\nu}\frac{1}{i}
     \propS{11}{(+)}{x_{3}-x_{2}}{}\gamma_{\mu}\Psi(x_{2}):
     i\propD{12}{(+)}{x_{2}-x_{1}}A_{\nu}(x_{3})\\
   &-ie^3:\bar{\Psi}(x_{1})\gamma^{\mu}\frac{1}{i}
     \propS{11}{(-)}{x_{1}-x_{3}}{}\gamma^{\nu}i
     \propS{12}{(+)}{x_{3}-x_{2}}{}\gamma_{\mu}\bar{\tilde{\Psi}}(x_{2}):
     i\propD{21}{(+)}{x_{2}-x_{1}}A_{\nu}(x_{3})\\
   &+ie^3:\tilde{\Psi}(x_{1})\gamma^{\mu}(-i)
     \propS{21}{(-)}{x_{1}-x_{3}}{}\gamma^{\nu}i
     \propS{12}{(+)}{x_{3}-x_{2}}{}\gamma_{\mu}\bar{\tilde{\Psi}}(x_{2}):
     i\propD{22}{(+)}{x_{2}-x_{1}}A_{\nu}(x_{3}).
\end{align*}
Now we calculate $A_{3}'$ in quite the same way:
\begin{equation*}
\begin{split}
A_{31}'(x_{1},x_{2},x_{3})=&-T_{1}(x_{2})
   \underbrace{T_{2}(x_{1},x_{3})}_{=T_{2}^{\mathrm{Compton}}}=\\
=&+e^3\bigg\{
       :\bar{\Psi}(x_{2})\gamma^{\alpha}\Psi(x_{2}):A_{\alpha}(x_{2})-
        :\tilde{\Psi}(x_{2})\gamma^{\alpha}\bar{\tilde{\Psi}}(x_{2}):
        \tilde{A}_{\alpha}(x_{2})\bigg\}\cdot\\
     &\hspace{1.5 cm}\cdot
   :\bar{\Psi}(x_{1})\gamma^{\mu}\propS{11}{\F}{x_{1}-x_{3}}{}(-1)\gamma^{\nu}
        \Psi(x_{3})::A_{\mu}(x_{1})A_{\nu}(x_{3}):-\\
&-e^3\bigg\{
       :\bar{\Psi}(x_{2})\gamma^{\alpha}\Psi(x_{2}):A_{\alpha}(x_{2})-
        :\tilde{\Psi}(x_{2})\gamma^{\alpha}\bar{\tilde{\Psi}}(x_{2}):
        \tilde{A}_{\alpha}(x_{2})\bigg\}\cdot\\
    &\hspace{1.5 cm}\cdot
    :\tilde{\Psi}(x_{1})\gamma^{\mu}\propS{21}{\F}{x_{1}-x_{3}}{}
        (-1)\gamma^{\nu}
        \Psi(x_{3})::\tilde{A}_{\mu}(x_{1})A_{\nu}(x_{3}):.
\end{split}
\end{equation*}
Doing the contractions this results in 
\begin{align*}
A&_{31}'=\\&+e^3:\bar{\Psi}(x_{1})\gamma^{\mu}\propS{11}{\F}{x_{1}-x_{3}}{}
  \gamma^{\nu}\frac{1}{i}\propS{11}{(-)}{x_{3}-x_{2}}{}\gamma_{\mu}\Psi(x_{2}):
  i\propD{11}{(+)}{x_{2}-x_{1}}A_{\nu}(x_{3})\\
  &-e^3:\bar{\Psi}(x_{1})\gamma^{\mu}\propS{11}{\F}{x_{1}-x_{3}}{}
  \gamma^{\nu}i\propS{12}{(-)}{x_{3}-x_{2}}{}\gamma_{\mu}
  \bar{\tilde{\Psi}}(x_{2}):i\propD{21}{(+)}{x_{2}-x_{1}}A_{\nu}(x_{3})\\
&-e^3:\tilde{\Psi}(x_{1})\gamma^{\mu}\propS{12}{\F}{x_{1}-x_{3}}{}
  \gamma^{\nu}\frac{1}{i}\propS{11}{(-)}{x_{3}-x_{2}}{}\gamma_{\mu}\Psi(x_{2}):
  i\propD{12}{(+)}{x_{2}-x_{1}}A_{\nu}(x_{3})\\
  &+e^3:\tilde{\Psi}(x_{1})\gamma^{\mu}\propS{12}{\F}{x_{1}-x_{3}}{}
  \gamma^{\nu}i\propS{12}{(-)}{x_{3}-x_{2}}{}\gamma_{\mu}
  \bar{\tilde{\Psi}}(x_{2}):i\propD{22}{(+)}{x_{2}-x_{1}}A_{\nu}(x_{3}).
\end{align*}
The next term is 
\begin{equation*}
\begin{split}
A_{32}'(x_{1},x_{2},x_{3})=&-T_{1}(x_{1})
   \underbrace{T_{2}(x_{2},x_{3})}_{=T_{2}^{\mathrm{Compton}}}=\\
=&-e^3\bigg\{
       :\bar{\Psi}(x_{1})\gamma^{\alpha}\Psi(x_{1}):A_{\alpha}(x_{1})-
        :\tilde{\Psi}(x_{1})\gamma^{\alpha}\bar{\tilde{\Psi}}(x_{1}):
        \tilde{A}_{\alpha}(x_{1})\bigg\}\cdot \\
    &\hspace{1.5 cm}\cdot
     :\bar{\Psi}(x_{3})\gamma^{\nu}\propS{11}{\F}{x_{3}-x_{2}}{}\gamma^{\mu}
        \Psi(x_{2})::A_{\mu}(x_{2})A_{\nu}(x_{3}):\\
&-e^3\bigg\{
       :\bar{\Psi}(x_{1})\gamma^{\alpha}\Psi(x_{1}):A_{\alpha}(x_{1})-
        :\tilde{\Psi}(x_{1})\gamma^{\alpha}\bar{\tilde{\Psi}}(x_{1}):
        \tilde{A}_{\alpha}(x_{1})\bigg\}\cdot\\
    &\hspace{1.5 cm}\cdot
     \bar{\Psi}(x_{3})\gamma^{\nu}\propS{12}{\F}{x_{3}-x_{2}}{}
        \gamma^{\mu}
        \bar{\tilde{\Psi}}(x_{2})::\tilde{A}_{\mu}(x_{2})A_{\nu}(x_{3}):\;,
\end{split}
\end{equation*}
which gives after contracting: 
\begin{align*}
A&_{32}'=\\&-e^3:\bar{\Psi}(x_{1})\gamma^{\mu}\frac{1}{i}
  \propS{11}{(+)}{x_{1}-x_{3}}{}
  \gamma^{\nu}\propS{11}{\F}{x_{3}-x_{2}}{}\gamma_{\mu}\Psi(x_{2}):
  i\propD{11}{(+)}{x_{1}-x_{2}}A_{\nu}(x_{3})\\
  &+e^3:\tilde{\Psi}(x_{1})\gamma^{\mu}(-i)\propS{21}{(+)}{x_{1}-x_{3}}{}
  \gamma^{\nu}\propS{11}{\F}{x_{3}-x_{2}}{}\gamma_{\mu}
  \Psi(x_{2}):i\propD{21}{(+)}{x_{1}-x_{2}}A_{\nu}(x_{3})\\
&-e^3:\bar{\Psi}(x_{1})\gamma^{\mu}\frac{1}{i}\propS{11}{(+)}{x_{1}-x_{3}}{}
  \gamma^{\nu}\propS{12}{\F}{x_{3}-x_{2}}{}\gamma_{\mu}
  \bar{\tilde{\Psi}}(x_{2}):
  i\propD{12}{(+)}{x_{1}-x_{2}}A_{\nu}(x_{3})\\
  &+e^3:\tilde{\Psi}(x_{1})\gamma^{\mu}(-i)
  \propS{21}{(+)}{x_{1}-x_{3}}{}
  \gamma^{\nu}\propS{12}{\F}{x_{3}-x_{2}}{}\gamma_{\mu}
  \bar{\tilde{\Psi}}(x_{2}):i\propD{22}{(+)}{x_{1}-x_{2}}A_{\nu}(x_{3}).
\end{align*}
Now to $A_{33}'$: 
\begin{equation*}
\begin{split}
A_{33,1}'=&-\underbrace{T_{2}(x_{1},x_{2})}_
            {=T_{2}^{\textrm{M\o ller}}}T_{1}(x_{3})=\\
&=-e^3:\bar{\Psi}(x_{1})\gamma^{\alpha}
        \Psi(x_{1})\bar{\Psi}(x_{2})\gamma_{\alpha}\Psi(x_{2}):
        \propD{11}{\F}{x_{1}-x_{2}}\cdot\\&\hspace{3.0 cm}\cdot
      \bigg\{:\bar{\Psi}(x_{3})\gamma^{\alpha}\Psi(x_{3}):A_{\alpha}(x_{3})-
        :\tilde{\Psi}(x_{3})\gamma^{\alpha}\bar{\tilde{\Psi}}(x_{3}):
        \tilde{A}_{\alpha}(x_{3})\bigg\}\\
&+e^3:\bar{\Psi}(x_{1})\gamma^{\alpha}
        \Psi(x_{1})\tilde{\Psi}(x_{2})\gamma_{\alpha}
        \bar{\tilde{\Psi}}(x_{2}):
        \propD{12}{\F}{x_{2}-x_{1}}\cdot\\&\hspace{3.0 cm}\cdot
      \bigg\{:\bar{\Psi}(x_{3})\gamma^{\alpha}\Psi(x_{3}):A_{\alpha}(x_{3})-
        :\tilde{\Psi}(x_{3})\gamma^{\alpha}\bar{\tilde{\Psi}}(x_{3}):
        \tilde{A}_{\alpha}(x_{3})\bigg\}\\
&+e^3:\tilde{\Psi}(x_{1})\gamma^{\alpha}
        \bar{\tilde{\Psi}}(x_{1})\bar{\Psi}(x_{2})\gamma_{\alpha}\Psi(x_{2}):
        \propD{12}{\F}{x_{2}-x_{1}}\cdot\\&\hspace{3.0 cm}\cdot
      \bigg\{:\bar{\Psi}(x_{3})\gamma^{\alpha}\Psi(x_{3}):A_{\alpha}(x_{3})-
        :\tilde{\Psi}(x_{3})\gamma^{\alpha}\bar{\tilde{\Psi}}(x_{3}):
        \tilde{A}_{\alpha}(x_{3})\bigg\}\\
&-e^3:\tilde{\Psi}(x_{1})\gamma^{\alpha}
        \bar{\tilde{\Psi}}(x_{1})\tilde{\Psi}(x_{2})\gamma_{\alpha}
        \bar{\tilde{\Psi}}(x_{2}):\propD{22}{\F}{x_{1}-x_{2}}
        \cdot\\&\hspace{3.0 cm}\cdot
      \bigg\{:\bar{\Psi}(x_{3})\gamma^{\alpha}\Psi(x_{3}):A_{\alpha}(x_{3})-
        :\tilde{\Psi}(x_{3})\gamma^{\alpha}\bar{\tilde{\Psi}}(x_{3}):
        \tilde{A}_{\alpha}(x_{3})\bigg\}.
\end{split}
\end{equation*}
Disregarding terms with uninteresting external legs and performing the 
contractions we arrive at
\begin{align*}
A&_{33,1}'=\\&+e^3:\bar{\Psi}(x_{1})\gamma^{\mu}\frac{1}{i}
     \propS{11}{(+)}{x_{1}-x_{3}}{}\gamma^{\nu}\frac{1}{i}
     \propS{11}{(-)}{x_{3}-x_{2}}{}\gamma_{\mu}\Psi(x_{2}):
     \propD{11}{\F}{x_{1}-x_{2}}A_{\nu}(x_{3})\\
   &-e^3:\bar{\Psi}(x_{1})\gamma^{\mu}\frac{1}{i}
     \propS{11}{(+)}{x_{1}-x_{3}}{}\gamma^{\nu}i
     \propS{12}{(-)}{x_{3}-x_{2}}{}\gamma_{\mu}\bar{\tilde{\Psi}}(x_{2}):
     \propD{12}{\F}{x_{2}-x_{1}}A_{\nu}(x_{3})\\
   &-e^3:\tilde{\Psi}(x_{1})\gamma^{\mu}(-i)
     \propS{21}{(+)}{x_{1}-x_{3}}{}\gamma^{\nu}\frac{1}{i}
     \propS{11}{(-)}{x_{3}-x_{2}}{}\gamma_{\mu}\Psi(x_{2}):
     \propD{12}{\F}{x_{2}-x_{1}}A_{\nu}(x_{3})\\
   &+e^3:\tilde{\Psi}(x_{1})\gamma^{\mu}(-i)
     \propS{21}{(+)}{x_{1}-x_{3}}{}\gamma^{\nu}i
     \propS{12}{(-)}{x_{3}-x_{2}}{}\gamma_{\mu}\bar{\tilde{\Psi}}(x_{2}):
     \propD{22}{\F}{x_{1}-x_{2}}A_{\nu}(x_{3}).
\end{align*}
The next term is 
\begin{align*}
A_{33,2}'=T_{1}(x_{1})T_{1}(x_{2})T_{1}(x_{3})=
&ie\bigg\{:\bar{\Psi}(x_{1})\gamma^{\alpha}\Psi(x_{1}):A_{\alpha}(x_{1})-
        :\tilde{\Psi}(x_{1})\gamma^{\alpha}\bar{\tilde{\Psi}}(x_{1}):
        \tilde{A}_{\alpha}(x_{1})\bigg\}\cdot\\
\cdot&ie\bigg\{:\bar{\Psi}(x_{2})\gamma^{\alpha}\Psi(x_{2}):
        A_{\alpha}(x_{2})-
        :\tilde{\Psi}(x_{2})\gamma^{\alpha}\bar{\tilde{\Psi}}(x_{2}):
        \tilde{A}_{\alpha}(x_{2})\bigg\}\cdot\\
\cdot&ie\bigg\{:\bar{\Psi}(x_{3})\gamma^{\alpha}\Psi(x_{3}):
        A_{\alpha}(x_{3})-
        :\tilde{\Psi}(x_{3})\gamma^{\alpha}\bar{\tilde{\Psi}}(x_{3}):
        \tilde{A}_{\alpha}(x_{3})\bigg\}\;\;.
\end{align*}
Once more we perform the procedure of contracting and disregarding 
uninteresting terms with the result 
\begin{align*}
A&_{33,2}'=\\&+ie^3:\bar{\Psi}(x_{1})\gamma^{\mu}\frac{1}{i}
     \propS{11}{(+)}{x_{1}-x_{3}}{}\gamma^{\nu}\frac{1}{i}
     \propS{11}{(-)}{x_{3}-x_{2}}{}\gamma_{\mu}\Psi(x_{2}):
     i\propD{11}{(+)}{x_{1}-x_{2}}A_{\nu}(x_{3})\\
   &-ie^3:\bar{\Psi}(x_{1})\gamma^{\mu}\frac{1}{i}
     \propS{11}{(+)}{x_{1}-x_{3}}{}\gamma^{\nu}i
     \propS{12}{(-)}{x_{3}-x_{2}}{}\gamma_{\mu}\bar{\tilde{\Psi}}(x_{2}):
     i\propD{12}{(+)}{x_{2}-x_{1}}A_{\nu}(x_{3})\\
   &-ie^3:\tilde{\Psi}(x_{1})\gamma^{\mu}(-i)
     \propS{21}{(+)}{x_{1}-x_{3}}{}\gamma^{\nu}\frac{1}{i}
     \propS{11}{(-)}{x_{3}-x_{2}}{}\gamma_{\mu}\Psi(x_{2}):
     i\propD{12}{(+)}{x_{2}-x_{1}}A_{\nu}(x_{3})\\
   &+ie^3:\tilde{\Psi}(x_{1})\gamma^{\mu}(-i)
     \propS{21}{(+)}{x_{1}-x_{3}}{}\gamma^{\nu}i
     \propS{12}{(-)}{x_{3}-x_{2}}{}\gamma_{\mu}\bar{\tilde{\Psi}}(x_{2}):
     i\propD{22}{(+)}{x_{1}-x_{2}}A_{\nu}(x_{3}).
\end{align*}
Finally 
\begin{align*}
A_{33,3}'=T_{1}(x_{2})T_{1}(x_{1})T_{1}(x_{3})=
&ie\bigg\{:\bar{\Psi}(x_{2})\gamma^{\alpha}\Psi(x_{2}):A_{\alpha}(x_{2})-
        :\tilde{\Psi}(x_{2})\gamma^{\alpha}\bar{\tilde{\Psi}}(x_{2}):
        \tilde{A}_{\alpha}(x_{2})\bigg\}\cdot\\
\cdot&ie\bigg\{:\bar{\Psi}(x_{1})\gamma^{\alpha}\Psi(x_{1}):
        A_{\alpha}(x_{1})-
        :\tilde{\Psi}(x_{1})\gamma^{\alpha}\bar{\tilde{\Psi}}(x_{1}):
        \tilde{A}_{\alpha}(x_{1})\bigg\}\cdot\\
\cdot&ie\bigg\{:\bar{\Psi}(x_{3})\gamma^{\alpha}\Psi(x_{3}):
        A_{\alpha}(x_{3})-
        :\tilde{\Psi}(x_{3})\gamma^{\alpha}\bar{\tilde{\Psi}}(x_{3}):
        \tilde{A}_{\alpha}(x_{3})\bigg\}\;\;.
\end{align*}
We calculate as above ( 
$\wl\Psi(x_{2})....\bar{\Psi}(x_{1})\wre \rightarrow 
-\wl\bar{\Psi}(x_{1})....\Psi(x_{2})\wre$!) This gives 
\begin{align*}
A&_{33,3}'=\\&+ie^3:\bar{\Psi}(x_{1})\gamma^{\mu}\frac{1}{i}
     \propS{11}{(+)}{x_{1}-x_{3}}{}\gamma^{\nu}\frac{1}{i}
     \propS{11}{(-)}{x_{3}-x_{2}}{}\gamma_{\mu}\Psi(x_{2}):
     i\propD{11}{(+)}{x_{2}-x_{1}}A_{\nu}(x_{3})\\
   &-ie^3:\tilde{\Psi}(x_{1})\gamma^{\mu}(-i)
     \propS{21}{(+)}{x_{1}-x_{3}}{}\gamma^{\nu}\frac{1}{i}
     \propS{11}{(-)}{x_{3}-x_{2}}{}\gamma_{\mu}\Psi(x_{2}):
     i\propD{12}{(+)}{x_{2}-x_{1}}A_{\nu}(x_{3})\\
   &-ie^3:\bar{\Psi}(x_{1})\gamma^{\mu}\frac{1}{i}
     \propS{11}{(+)}{x_{1}-x_{3}}{}\gamma^{\nu}i
     \propS{12}{(-)}{x_{3}-x_{2}}{}\gamma_{\mu}\bar{\tilde{\Psi}}(x_{2}):
     i\propD{21}{(+)}{x_{2}-x_{1}}A_{\nu}(x_{3})\\
   &+ie^3:\tilde{\Psi}(x_{1})\gamma^{\mu}(-i)
     \propS{21}{(+)}{x_{1}-x_{3}}{}\gamma^{\nu}i
     \propS{12}{(-)}{x_{3}-x_{2}}{}\gamma_{\mu}\bar{\tilde{\Psi}}(x_{2}):
     i\propD{22}{(+)}{x_{2}-x_{1}}A_{\nu}(x_{3}).
\end{align*}
Now we calculate $D = R'- A'$ and order it according to the external legs.  
We thus arrive at the following expression (the arguments of `$SSD$' are 
$(x_{1}-x_{3}),\;(x_{3}-x_{2}),\;(x_{1}-x_{2})$, always in this order)
\begin{align*}
D_{3}^{\mathrm{V}}=&\;
+e^3:\bar{\Psi}(x_{1})\gamma^{\mu}\!\!\cdot \begin{bmatrix}
        -\pnoaS{11}{(-)}{}{}\gamma^{\nu}
        \pnoaS{11}{(+)}{}{}\pnoaD{11}{\F}{}\;+\\
\;+\pnoaS{11}{\ret}{}{}\gamma^{\nu}
        \pnoaS{11}{(+)}{}{}\pnoaD{11}{(+)}{}\;+\\
\;+\pnoaS{11}{(-)}{}{}\gamma^{\nu}
        \pnoaS{11}{\av}{}{}\pnoaD{11}{(-)}{}\;+\\
\;+\pnoaS{11}{(+)}{}{}\gamma^{\nu}
        \pnoaS{11}{(-)}{}{}\pnoaD{11}{\F}{}\;-\\
\;-\pnoaS{11}{\av}{}{}\gamma^{\nu}
        \pnoaS{11}{(-)}{}{}\pnoaD{11}{(-)}{}\;-\\
\;-\pnoaS{11}{(+)}{}{}\gamma^{\nu}
        \pnoaS{11}{\ret}{}{}\pnoaD{11}{(+)}{}\;
     \end{bmatrix}\cdot\gamma_{\mu}\Psi(x_{2}):A_{\nu}(x_{3})\;+\\
&\;+e^3:\tilde{\Psi}(x_{1})\gamma^{\mu}\!\!\cdot\begin{bmatrix}
        +\pnoaS{21}{(-)}{}{}\gamma^{\nu}
        \pnoaS{11}{(+)}{}{}\pnoaD{12}{\F}{}\;-\\
\;-\pnoaS{21}{\ret}{}{}\gamma^{\nu}
        \pnoaS{11}{(+)}{}{}\pnoaD{12}{(+)}{}\;-\\
\;-\pnoaS{21}{(-)}{}{}\gamma^{\nu}
        \pnoaS{11}{\av}{}{}\pnoaD{12}{(-)}{}\;-\\
\;-\pnoaS{21}{(+)}{}{}\gamma^{\nu}
        \pnoaS{11}{(-)}{}{}\pnoaD{12}{\F}{}\;+\\
\;+\pnoaS{21}{\av}{}{}\gamma^{\nu}
        \pnoaS{11}{(-)}{}{}\pnoaD{12}{(-)}{}\;+\\
\;+\pnoaS{21}{(+)}{}{}\gamma^{\nu}
        \pnoaS{11}{\ret}{}{}\pnoaD{12}{(+)}{}\;
     \end{bmatrix}\cdot\gamma_{\mu}\Psi(x_{2}):A_{\nu}(x_{3})\;+\\
&\;+e^3:\bar{\Psi}(x_{1})\gamma^{\mu}\!\!\cdot\begin{bmatrix}
        -\pnoaS{11}{(-)}{}{}\gamma^{\nu}
        \pnoaS{12}{(+)}{}{}\pnoaD{12}{\F}{}\;+\\
\;+\pnoaS{11}{\ret}{}{}\gamma^{\nu}
        \pnoaS{12}{(+)}{}{}\pnoaD{12}{(+)}{}\;+\\
\;+\pnoaS{11}{(-)}{}{}\gamma^{\nu}
        \pnoaS{12}{\av}{}{}\pnoaD{12}{(-)}{}\;+\\
\;+\pnoaS{11}{(+)}{}{}\gamma^{\nu}
        \pnoaS{12}{(-)}{}{}\pnoaD{12}{\F}{}\;-\\
\;-\pnoaS{11}{\av}{}{}\gamma^{\nu}
        \pnoaS{12}{(-)}{}{}\pnoaD{12}{(-)}{}\;-\\
\;-\pnoaS{11}{(+)}{}{}\gamma^{\nu}
        \pnoaS{12}{\ret}{}{}\pnoaD{12}{(+)}{}\;
      \end{bmatrix}\cdot\gamma_{\mu}\bar{\tilde{\Psi}}(x_{2}):
       A_{\nu}(x_{3})\;+\\
&\;+e^3:\tilde{\Psi}(x_{1})\gamma^{\mu}\!\!\cdot\begin{bmatrix}
        +\pnoaS{21}{(-)}{}{}\gamma^{\nu}
        \pnoaS{12}{(+)}{}{}\pnoaD{22}{\F}{}\;-\\
\;-\pnoaS{21}{\ret}{}{}\gamma^{\nu}
        \pnoaS{12}{(+)}{}{}\pnoaD{22}{(+)}{}\;-\\
\;-\pnoaS{21}{(-)}{}{}\gamma^{\nu}
        \pnoaS{12}{\av}{}{}\pnoaD{22}{(-)}{}\;-\\
\;-\pnoaS{21}{(+)}{}{}\gamma^{\nu}
        \pnoaS{12}{(-)}{}{}\pnoaD{22}{\F}{}\;+\\
\;+\pnoaS{21}{\av}{}{}\gamma^{\nu}
        \pnoaS{12}{(-)}{}{}\pnoaD{22}{(-)}{}\;+\\
\;+\pnoaS{21}{(+)}{}{}\gamma^{\nu}
        \pnoaS{12}{\ret}{}{}\pnoaD{22}{(+)}{}\;
\end{bmatrix}\cdot\gamma_{\mu}\bar{\tilde{\Psi}}(x_{2}):A_{\nu}(x_{3})\;\;\;. 
\end{align*}
\subsubsection*{2. Calculation of \boldmath{$\omega^{\mathrm{V}}$}}
We have to investigate (see (\ref{eq:quasiasimp}))
\begin{equation*} 
\lim_{\delta\rightarrow 0}\rho(\delta)\langle\hat{\Lambda}(\frac{p}{\delta},
\frac{q}{\delta}),\check{\phi}(p,q)\rangle \overset{!}{=}
\langle\hat{\Lambda}_{0}(p,q),\check{\phi}(p,q)\rangle
\end{equation*}
with  
\begin{equation*}
 \hat{\Lambda}(p,q):=\int\!d^4k\gamma^{\mu}\propSft{}{\av}{p-k}{}\gamma^{\nu}
 \propSft{}{\ret}{q-k}{}\gamma_{\mu}\propDft{}{(+)}{k}+...,
\end{equation*}
where the ``$...$'' stands for terms without essential differences in their 
structure with respect to the determination of the singular order. 
We choose the Lorentz-system wherein $u=(1,\vec{0})$ and use 
$\delta(k^2)$ to perform the $k$-integration. This gives integrals of the 
following form:   
\begin{multline*}
\lim_{\delta\rightarrow 0}\rho(\delta)\int\!d^4pd^4q\check{\phi}(p,q)
\int\!d^3k\frac{1}{2\lvert \vec{k} \rvert}\big[
\underbrace{\sim\frac{1}{\delta^2}+...}_{\gamma\textrm{- and slash-factors}}
\big]\cdot\\\cdot\delta(\frac{p^2}{\delta^2}-\frac{2pk}{\delta}-m^2)
\delta(\frac{q^2}{\delta^2}-\frac{2qk}{\delta}-m^2)
 \theta(-\frac{p_{0}}{\delta}+k_{0})\theta(\frac{q_{0}}{\delta}-k_{0})
\frac{1}{e^{\beta\lvert\vec{k}\rvert}-1}.
\end{multline*}
We calculate the $\lvert\vec{k}\rvert$-integral with one of the 
$\delta$-distributions and the $x$-integral with the other. Proceeding 
as in the determination of $\omega^{\mathrm{VP}}$ 
(see App. \ref{subsec:tzwei}) we 
derive $\omega^{\mathrm{V}} = -\infty$.

\subsection{Calculations for the Third Order Self Energy}
   \label{subsec:3ordSEcalc}
\subsubsection*{1. Calculation of \boldmath{$D_{3}^{3.\mathrm{ord.SE}}$}}
The procedure is essentially the same as in the corresponding calculation for 
the vertex. Nevertheless, we write it down here in detail to give a 
reference of the various signs, factors $i$, etc., which are crucial for 
the discussion of a possible cancellation of the IR-divergencies.  
There are two graphs that have to be calculated, denoted by `A' and `B', 
respectively, (see Sec. \ref{subsubsec:3ordSE}). As in the calculation for 
the vertex we have: 
\begin{equation*}
\begin{split}
R_{3}'(x_{1},x_{2},x_{3}) = \sum_{P_{2}}&T(Y,x_{3})\tilde{T}(X) = 
      -\underbrace{T_{2}(x_{1},x_{3})T_{1}(x_{2})}_{R_{31}'}-
      \underbrace{T_{2}(x_{2},x_{3})T_{1}(x_{1})}_{R_{32}'}-\\
      &-\underbrace{T_{1}(x_{3})T_{2}(x_{1},x_{2})}_{R_{33,1}'}
        +\underbrace{T_{1}(x_{3})T_{1}(x_{1})T_{1}(x_{2})}_{R_{33,2}'}
        +\underbrace{T_{1}(x_{3})T_{1}(x_{2})T_{1}(x_{1})}_{R_{33,3}'},\\
A_{3}'(x_{1},x_{2},x_{3}) = \sum_{P_{2}}&\tilde{T}(X)T(Y,x_{3})= 
      -\underbrace{T_{1}(x_{2})T_{2}(x_{1},x_{3})}_{A_{31}'}
      -\underbrace{T_{2}(x_{1})T_{2}(x_{2},x_{3})}_{A_{32}'}-\\
      &-\underbrace{T_{2}(x_{1},x_{2})T_{1}(x_{3})}_{A_{33,1}'}+
        \underbrace{T_{1}(x_{1})T_{1}(x_{2})T_{1}(x_{3})}_{A_{33,2}'}+
        \underbrace{T_{1}(x_{2})T_{1}(x_{1})T_{1}(x_{3})}_{A_{33,3}'}.
\end{split}
\end{equation*}
First we consider graph B (disregarding terms with external legs different  
from $\bar{\Psi}(x_{3}), A_{\nu}(x_{3})$ and $\Psi(x_{1})$ or 
$\bar{\tilde{\Psi}}(x_{1})$):
\begin{equation*}
\begin{split}
R_{31}'=&-\underbrace{T_{2}(x_{1},x_{3})}_
      {\mathrm{trivial\; graph}}T_{1}(x_{2})=\\
&+ie^3:\bar{\Psi}(x_{3})\gamma^{\nu}\Psi(x_{3})\bar{\Psi}(x_{1})\gamma^{\mu}
   \Psi(x_{1})::A_{\nu}(x_{3})A_{\mu}(x_{1}):\cdot\\&\hspace{3.0 cm}\cdot
   \bigg(:\bar{\Psi}(x_{2})\gamma^{\alpha}\Psi(x_{2}):A_{\alpha}(x_{2})-
   :\tilde{\Psi}(x_{2})\gamma^{\alpha}\bar{\tilde{\Psi}}(x_{2}):
    \tilde{A}_{\alpha}(x_{2})\bigg)\\
&-ie^3:\bar{\Psi}(x_{3})\gamma^{\nu}\Psi(x_{3})\tilde{\Psi}(x_{1})\gamma^{\mu}
   \bar{\tilde{\Psi}}(x_{1})::A_{\nu}(x_{3})\tilde{A}_{\mu}(x_{1}):
   \cdot\\&\hspace{3.0 cm}\cdot  
   \bigg(:\bar{\Psi}(x_{2})\gamma^{\alpha}\Psi(x_{2}):A_{\alpha}(x_{2})-
   :\tilde{\Psi}(x_{2})\gamma^{\alpha}\bar{\tilde{\Psi}}(x_{2}):
    \tilde{A}_{\alpha}(x_{2})\bigg).
\end{split}
\end{equation*}
After having performed the contractions providing us with the external 
legs we are interested in, we have 
\begin{align*}
R&_{31}' =\\&-ie^3:\bar{\Psi}(x_{3})\gamma^{\nu}\frac{1}{i}
   \propS{11}{(+)}{x_{3}-x_{2}}{}\gamma_{\mu}\frac{1}{i}
   \propS{11}{(-)}{x_{2}-x_{1}}{}\gamma^{\mu}\Psi(x_{1}):A_{\nu}(x_{3})i
   \propD{11}{(+)}{x_{1}-x_{2}}\\
&+ie^3:\bar{\Psi}(x_{3})\gamma^{\nu}i
   \propS{12}{(+)}{x_{3}-x_{2}}{}\gamma_{\mu}(-i)
   \propS{21}{(-)}{x_{2}-x_{1}}{}\gamma^{\mu}\Psi(x_{1}):A_{\nu}(x_{3})i
   \propD{12}{(+)}{x_{1}-x_{2}}\\
&+ie^3:\bar{\Psi}(x_{3})\gamma^{\nu}\frac{1}{i}
   \propS{11}{(+)}{x_{3}-x_{2}}{}\gamma_{\mu}i
   \propS{12}{(-)}{x_{2}-x_{1}}{}\gamma^{\mu}
   \bar{\tilde{\Psi}}(x_{1}):A_{\nu}(x_{3})i
   \propD{21}{(+)}{x_{1}-x_{2}}\\
&-ie^3:\bar{\Psi}(x_{3})\gamma^{\nu}i
   \propS{12}{(+)}{x_{3}-x_{2}}{}\gamma_{\mu}\frac{1}{i}
   \propS{22}{(-)}{x_{2}-x_{1}}{}\gamma^{\mu}
   \bar{\tilde{\Psi}}(x_{1}):A_{\nu}(x_{3})i
   \propD{22}{(+)}{x_{1}-x_{2}}.
\end{align*}
The next term, 
$R_{32}'=-\underbrace{T_{2}(x_{2},x_{3})}_
      {\textrm{Compton}}T_{1}(x_{1})$, is calculated as 
$R_{31}^{\mathrm{Vertex}}$ (see App. \ref{subsec:vertcalc}):
\begin{equation*}
\begin{split}
R_{32}'=&-e^3:\bar{\Psi}(x_{3})\gamma^{\nu}\propS{11}{\F}{x_{3}-x_{2}}{}
     \gamma^{\mu}
        \Psi(x_{2})::A_{\mu}(x_{2})A_{\nu}(x_{3}):\cdot\\&\hspace{3.0 cm}\cdot
        \bigg\{
       :\bar{\Psi}(x_{1})\gamma^{\alpha}\Psi(x_{1}:A_{\alpha}(x_{1})-
        :\tilde{\Psi}(x_{1})\gamma^{\alpha}\bar{\tilde{\Psi}}(x_{1}:
        \tilde{A}_{\alpha}(x_{1})\bigg\}\\
&-e^3:\bar{\Psi}(x_{3})\gamma^{\nu}\propS{12}{\F}{x_{3}-x_{2}}{}\gamma^{\mu}
        \bar{\tilde{\Psi}}(x_{2})::\tilde{A}_{\mu}(x_{2})A_{\nu}(x_{3}):
       \cdot\\&\hspace{3.0 cm}\cdot\bigg\{
       :\bar{\Psi}(x_{1})\gamma^{\alpha}\Psi(x_{1}:A_{\alpha}(x_{1})-
        :\tilde{\Psi}(x_{1})\gamma^{\alpha}\bar{\tilde{\Psi}}(x_{1}:
        \tilde{A}_{\alpha}(x_{1})\bigg\}\;\;.
\end{split}
\end{equation*}
Performing the relevant contractions gives
\begin{align*}
R&_{32}'=\\&-e^3:\bar{\Psi}(x_{3})\gamma^{\nu}\propS{11}{\F}{x_{3}-x_{2}}{}
  \gamma^{\mu}\frac{1}{i}\propS{11}{(+)}{x_{2}-x_{1}}{}\gamma_{\mu}\Psi(x_{1}):
  A_{\nu}(x_{3})i\propD{11}{(+)}{x_{2}-x_{1}}\\
&+e^3:\bar{\Psi}(x_{3})\gamma^{\nu}\propS{11}{\F}{x_{3}-x_{2}}{}
  \gamma^{\mu}i\propS{12}{(+)}{x_{2}-x_{1}}{}\gamma_{\mu}
  \bar{\tilde{\Psi}}(x_{1}):
  A_{\nu}(x_{3})i\propD{12}{(+)}{x_{2}-x_{1}}\\
&-e^3:\bar{\Psi}(x_{3})\gamma^{\nu}\propS{12}{\F}{x_{3}-x_{2}}{}
  \gamma^{\mu}(-i)\propS{21}{(+)}{x_{2}-x_{1}}{}\gamma_{\mu}\Psi(x_{1}):
  A_{\nu}(x_{3})i\propD{21}{(+)}{x_{2}-x_{1}}\\
&+e^3:\bar{\Psi}(x_{3})\gamma^{\nu}\propS{12}{\F}{x_{3}-x_{2}}{}
  \gamma^{\mu}\frac{1}{i}\propS{22}{(+)}{x_{2}-x_{1}}{}\gamma_{\mu}
  \bar{\tilde{\Psi}}(x_{1}):
  A_{\nu}(x_{3})i\propD{22}{(+)}{x_{2}-x_{1}}.
\end{align*}
The next term is 
\begin{equation*}
\begin{split}
R_{33,1}'=&-T_{1}(x_{3})\underbrace{T_{2}(x_{1},x_{2})}_{SE}=\\
&-ie^3\bigg(:\bar{\Psi}(x_{3})\gamma^{\alpha}\Psi(x_{3}):A_{\alpha}(x_{3})-
   :\tilde{\Psi}(x_{3})\gamma^{\alpha}\bar{\tilde{\Psi}}(x_{3}):
    \tilde{A}_{\alpha}(x_{3})\bigg)\cdot\\&\hspace{3.0 cm}\cdot
    :\bar{\Psi}(x_{2})\gamma^{\nu}
 \propS{11}{\F}{x_{2}-x_{1}}{}\propD{11}{\F}{x_{2}-x_{1}}
 \gamma_{\nu}\Psi(x_{1}):\\
&+ie^3\bigg(:\bar{\Psi}(x_{3})\gamma^{\alpha}\Psi(x_{3}):A_{\alpha}(x_{3})-
   :\tilde{\Psi}(x_{3})\gamma^{\alpha}\bar{\tilde{\Psi}}(x_{3}):
    \tilde{A}_{\alpha}(x_{3})\bigg)\cdot\\&\hspace{3.0 cm}\cdot
   :\tilde{\Psi}(x_{2})\gamma^{\nu}
 \propS{21}{\F}{x_{2}-x_{1}}{}\propD{12}{\F}{x_{2}-x_{1}}
 \gamma_{\nu}\Psi(x_{1}):\\
&-ie^3\bigg(:\bar{\Psi}(x_{3})\gamma^{\alpha}\Psi(x_{3}):A_{\alpha}(x_{3})-
   :\tilde{\Psi}(x_{3})\gamma^{\alpha}\bar{\tilde{\Psi}}(x_{3}):
    \tilde{A}_{\alpha}(x_{3})\bigg)\cdot\\&\hspace{3.0 cm}\cdot
   :\bar{\Psi}(x_{2})\gamma^{\nu}(-1)
 \propS{12}{\F}{x_{2}-x_{1}}{}\propD{12}{\F}{x_{2}-x_{1}}\gamma_{\nu}
  \bar{\tilde{\Psi}}(x_{1}):\\
&-ie^3\bigg(:\bar{\Psi}(x_{3})\gamma^{\alpha}\Psi(x_{3}):A_{\alpha}(x_{3})-
   :\tilde{\Psi}(x_{3})\gamma^{\alpha}\bar{\tilde{\Psi}}(x_{3}):
    \tilde{A}_{\alpha}(x_{3})\bigg)\cdot\\&\hspace{3.0 cm}\cdot
   :\tilde{\Psi}(x_{2})\gamma^{\nu}
 \propS{22}{\F}{x_{2}-x_{1}}{}\propD{22}{\F}{x_{2}-x_{1}}\gamma_{\nu}
  \bar{\tilde{\Psi}}(x_{1}):\;\;,
\end{split}
\end{equation*}
which after performing the relevant contractions leads to 
\begin{align*}
R&_{33,1}'=\\&-ie^3:\bar{\Psi}(x_{3})\gamma^{\nu}\frac{1}{i}
  \propS{11}{(+)}{x_{3}-x_{2}}{}\gamma^{\mu}\propS{11}{\F}{x_{2}-x_{1}}{}
  \gamma_{\mu}\Psi(x_{1}):A_{\nu}(x_{3})\propD{11}{\F}{x_{2}-x_{1}}\\
&+ie^3:\bar{\Psi}(x_{3})\gamma^{\nu}i
  \propS{12}{(+)}{x_{3}-x_{2}}{}\gamma^{\mu}\propS{21}{\F}{x_{2}-x_{1}}{}
  \gamma_{\mu}\Psi(x_{1}):A_{\nu}(x_{3})\propD{12}{\F}{x_{2}-x_{1}}\\
&-ie^3:\bar{\Psi}(x_{3})\gamma^{\nu}\frac{1}{i}
  \propS{11}{(+)}{x_{3}-x_{2}}{}\gamma^{\mu}\propS{12}{\F}{x_{2}-x_{1}}{}
  \gamma_{\mu}\bar{\tilde{\Psi}}(x_{1}):A_{\nu}(x_{3})
  \propD{12}{\F}{x_{2}-x_{1}}\\
&-ie^3:\bar{\Psi}(x_{3})\gamma^{\nu}i
  \propS{12}{(+)}{x_{3}-x_{2}}{}\gamma^{\mu}\propS{22}{\F}{x_{2}-x_{1}}{}
  \gamma_{\mu}
  \bar{\tilde{\Psi}}(x_{1}):A_{\nu}(x_{3})\propD{22}{\F}{x_{2}-x_{1}}.
\end{align*}
Now we come to 
\begin{equation*}
\begin{split}
R_{33,2}'=T_{1}(x_{3})T_{1}(x_{1})T_{1}(x_{2})=
&ie\bigg\{:\bar{\Psi}(x_{3})\gamma^{\alpha}\Psi(x_{3}):A_{\alpha}(x_{3})-
        :\tilde{\Psi}(x_{3})\gamma^{\alpha}\bar{\tilde{\Psi}}(x_{3}):
        \tilde{A}_{\alpha}(x_{3})\bigg\}\cdot\\
\cdot&ie\bigg\{:\bar{\Psi}(x_{1})\gamma^{\alpha}\Psi(x_{1}):
        A_{\alpha}(x_{1})-
        :\tilde{\Psi}(x_{1})\gamma^{\alpha}\bar{\tilde{\Psi}}(x_{1}):
        \tilde{A}_{\alpha}(x_{1})\bigg\}\cdot\\
\cdot&ie\bigg\{:\bar{\Psi}(x_{2})\gamma^{\alpha}\Psi(x_{2}):
        A_{\alpha}(x_{2})-
        :\tilde{\Psi}(x_{2})\gamma^{\alpha}\bar{\tilde{\Psi}}(x_{2}):
        \tilde{A}_{\alpha}(x_{2})\bigg\}\;\;,
\end{split}
\end{equation*}
which gives after having done the contractions 
\begin{align*}
R&_{33,2}'=\\&+ie^3:\bar{\Psi}(x_{3})\gamma^{\nu}\frac{1}{i}
  \propS{11}{(+)}{x_{3}-x_{2}}{}\gamma^{\mu}
  \frac{1}{i}\propS{11}{(-)}{x_{2}-x_{1}}{}
  \gamma_{\mu}\Psi(x_{1}):A_{\nu}(x_{3})i\propD{11}{(+)}{x_{1}-x_{2}}\\
&-ie^3:\bar{\Psi}(x_{3})\gamma^{\nu}i
  \propS{12}{(+)}{x_{3}-x_{2}}{}\gamma^{\mu}(-i)\propS{21}{(-)}{x_{2}-x_{1}}{}
  \gamma_{\mu}\Psi(x_{1}):A_{\nu}(x_{3})i\propD{12}{(+)}{x_{1}-x_{2}}\\
&-ie^3:\bar{\Psi}(x_{3})\gamma^{\nu}\frac{1}{i}
  \propS{11}{(+)}{x_{3}-x_{2}}{}\gamma^{\mu}i\propS{12}{(-)}{x_{2}-x_{1}}{}
  \gamma_{\mu}\bar{\tilde{\Psi}}(x_{1}):A_{\nu}(x_{3})
  i\propD{21}{(+)}{x_{1}-x_{2}}\\
&+ie^3:\bar{\Psi}(x_{3})\gamma^{\nu}i
  \propS{12}{(+)}{x_{3}-x_{2}}{}\gamma^{\mu}
  \frac{1}{i}\propS{22}{(-)}{x_{2}-x_{1}}{}
  \gamma_{\mu}
  \bar{\tilde{\Psi}}(x_{1}):A_{\nu}(x_{3})i\propD{22}{(+)}{x_{1}-x_{2}}. 
\end{align*}
$R_{33,3}'$ is calculated analogously, the result is the same as for 
$R_{33,2}'$ but with $x_{1}$ and $x_{2}$ exchanged - \emph{before} 
performing the contractions! After having done them, we have (disregarding 
terms with uninteresting external legs):
\begin{align*}
R&_{33,3}'=\\&-ie^3:\bar{\Psi}(x_{3})\gamma^{\nu}\frac{1}{i}
  \propS{11}{(+)}{x_{3}-x_{2}}{}\gamma^{\mu}
  \frac{1}{i}\propS{11}{(+)}{x_{2}-x_{1}}{}
  \gamma_{\mu}\Psi(x_{1}):A_{\nu}(x_{3})i\propD{11}{(+)}{x_{2}-x_{1}}\\
&+ie^3:\bar{\Psi}(x_{3})\gamma^{\nu}\frac{1}{i}
  \propS{11}{(+)}{x_{3}-x_{2}}{}\gamma^{\mu}i\propS{12}{(+)}{x_{2}-x_{1}}{}
  \gamma_{\mu}\bar{\tilde{\Psi}}(x_{1}):A_{\nu}(x_{3})
  i\propD{12}{(+)}{x_{2}-x_{1}}\\
&+ie^3:\bar{\Psi}(x_{3})\gamma^{\nu}i
  \propS{12}{(+)}{x_{3}-x_{2}}{}\gamma^{\mu}(-i)\propS{21}{(+)}{x_{2}-x_{1}}{}
  \gamma_{\mu}\Psi(x_{1}):A_{\nu}(x_{3})
  i\propD{21}{(+)}{x_{2}-x_{1}}\\
&-ie^3:\bar{\Psi}(x_{3})\gamma^{\nu}i
  \propS{12}{(+)}{x_{3}-x_{2}}{}\gamma^{\mu}
  \frac{1}{i}\propS{22}{(+)}{x_{2}-x_{1}}{}
  \gamma_{\mu}
  \bar{\tilde{\Psi}}(x_{1}):A_{\nu}(x_{3})i\propD{22}{(+)}{x_{2}-x_{1}}. 
\end{align*}
Now we come to $A_{3}'$:\\
$A_{31}'=-T_{1}(x_{2})T_{2}(x_{1},x_{3})$ is calculated just as $R_{31}'$, 
but with the 
order of the factors reversed. With the relevant contractions this gives 
\begin{align*}
A&_{31}' =\\&-ie^3:\bar{\Psi}(x_{3})\gamma^{\nu}\frac{1}{i}
   \propS{11}{(-)}{x_{3}-x_{2}}{}\gamma_{\mu}\frac{1}{i}
   \propS{11}{(+)}{x_{2}-x_{1}}{}\gamma^{\mu}\Psi(x_{1}):A_{\nu}(x_{3})i
   \propD{11}{(+)}{x_{2}-x_{1}}\\
&+ie^3:\bar{\Psi}(x_{3})\gamma^{\nu}i
   \propS{12}{(-)}{x_{3}-x_{2}}{}\gamma_{\mu}(-i)
   \propS{21}{(+)}{x_{2}-x_{1}}{}\gamma^{\mu}\Psi(x_{1}):A_{\nu}(x_{3})i
   \propD{21}{(+)}{x_{2}-x_{1}}\\
&+ie^3:\bar{\Psi}(x_{3})\gamma^{\nu}\frac{1}{i}
   \propS{11}{(-)}{x_{3}-x_{2}}{}\gamma_{\mu}i
   \propS{12}{(+)}{x_{2}-x_{1}}{}\gamma^{\mu}
   \bar{\tilde{\Psi}}(x_{1}):A_{\nu}(x_{3})i
   \propD{12}{(+)}{x_{2}-x_{1}}\\
&-ie^3:\bar{\Psi}(x_{3})\gamma^{\nu}i
   \propS{12}{(-)}{x_{3}-x_{2}}{}\gamma_{\mu}\frac{1}{i}
   \propS{22}{(+)}{x_{2}-x_{1}}{}\gamma^{\mu}
   \bar{\tilde{\Psi}}(x_{1}):A_{\nu}(x_{3})i
   \propD{22}{(+)}{x_{2}-x_{1}}.
\end{align*}
The next term is 
$A_{32}'=-T_{1}(x_{1})\underbrace{T_{2}(x_{2},x_{3})}_{\textrm{Compton}}$. 
Performing the relevant contractions this leads to 
(pay attention to the fact that 
$\wl\Psi...\bar{\Psi}\wre \rightarrow -\wl\bar{\Psi}...\Psi\wre$):
\begin{align*}
A&_{32}' =\\&+e^3:\bar{\Psi}(x_{3})\gamma^{\nu}
   \propS{11}{\F}{x_{3}-x_{2}}{}\gamma_{\mu}\frac{1}{i}
   \propS{11}{(-)}{x_{2}-x_{1}}{}\gamma^{\mu}\Psi(x_{1}):A_{\nu}(x_{3})i
   \propD{11}{(+)}{x_{1}-x_{2}}\\
&-e^3:\bar{\Psi}(x_{3})\gamma^{\nu}
   \propS{11}{\F}{x_{3}-x_{2}}{}\gamma_{\mu}i
   \propS{12}{(-)}{x_{2}-x_{1}}{}\gamma^{\mu}
   \bar{\tilde{\Psi}}(x_{1}):A_{\nu}(x_{3})i
   \propD{21}{(+)}{x_{1}-x_{2}}\\
&+e^3:\bar{\Psi}(x_{3})\gamma^{\nu}
   \propS{12}{\F}{x_{3}-x_{2}}{}\gamma_{\mu}(-i)
   \propS{21}{(-)}{x_{2}-x_{1}}{}\gamma^{\mu}
   \Psi(x_{1}):A_{\nu}(x_{3})i
   \propD{12}{(+)}{x_{1}-x_{2}}\\
&-e^3:\bar{\Psi}(x_{3})\gamma^{\nu}
   \propS{12}{\F}{x_{3}-x_{2}}{}\gamma_{\mu}\frac{1}{i}
   \propS{22}{(-)}{x_{2}-x_{1}}{}\gamma^{\mu}
   \bar{\tilde{\Psi}}(x_{1}):A_{\nu}(x_{3})i
   \propD{22}{(+)}{x_{1}-x_{2}}.
\end{align*}
Now we turn to \\
$A_{33,1}'=-\underbrace{T_{2}(x_{2},x_{3})}_{\textrm{SE}}T_{1}(x_{3})$, 
which gives with the relevant contractions  
\begin{align*}
A&_{33,1}' =\\&+ie^3:\bar{\Psi}(x_{3})\gamma^{\nu}\frac{1}{i}
   \propS{11}{(-)}{x_{3}-x_{2}}{}\gamma_{\mu}
   \propS{11}{\F}{x_{2}-x_{1}}{}\gamma^{\mu}\Psi(x_{1}):A_{\nu}(x_{3})
   \propD{11}{\F}{x_{2}-x_{1}}\\
&-ie^3:\bar{\Psi}(x_{3})\gamma^{\nu}i
   \propS{12}{(-)}{x_{3}-x_{2}}{}\gamma_{\mu}
   \propS{21}{\F}{x_{2}-x_{1}}{}\gamma^{\mu}\Psi(x_{1}):A_{\nu}(x_{3})
   \propD{12}{\F}{x_{2}-x_{1}}\\
&+ie^3:\bar{\Psi}(x_{3})\gamma^{\nu}\frac{1}{i}
   \propS{11}{(-)}{x_{3}-x_{2}}{}\gamma_{\mu}
   \propS{12}{\F}{x_{2}-x_{1}}{}\gamma^{\mu}
   \bar{\tilde{\Psi}}(x_{1}):A_{\nu}(x_{3})
   \propD{12}{\F}{x_{2}-x_{1}}\\
&+ie^3:\bar{\Psi}(x_{3})\gamma^{\nu}i
   \propS{12}{(-)}{x_{3}-x_{2}}{}\gamma_{\mu}
   \propS{22}{\F}{x_{2}-x_{1}}{}\gamma^{\mu}
   \bar{\tilde{\Psi}}(x_{1}):A_{\nu}(x_{3})
   \propD{22}{\F}{x_{2}-x_{1}}.
\end{align*}
The next term is \\
$A_{33,2}'=T_{1}(x_{1})T_{1}(x_{2})T_{1}(x_{3})\Rightarrow$ 
(after having done the relevant contractions)
\begin{align*}
A&_{33,2}' =\\&-ie^3:\bar{\Psi}(x_{3})\gamma^{\nu}\frac{1}{i}
   \propS{11}{(-)}{x_{3}-x_{2}}{}\gamma_{\mu}\frac{1}{i}
   \propS{11}{(-)}{x_{2}-x_{1}}{}\gamma^{\mu}\Psi(x_{1}):A_{\nu}(x_{3})
   i\propD{11}{(+)}{x_{1}-x_{2}}\\
&+ie^3:\bar{\Psi}(x_{3})\gamma^{\nu}i
   \propS{12}{(-)}{x_{3}-x_{2}}{}\gamma_{\mu}(-i)
   \propS{21}{(-)}{x_{2}-x_{1}}{}\gamma^{\mu}\Psi(x_{1}):A_{\nu}(x_{3})
   i\propD{12}{(+)}{x_{1}-x_{2}}\\
&+ie^3:\bar{\Psi}(x_{3})\gamma^{\nu}\frac{1}{i}
   \propS{11}{(-)}{x_{3}-x_{2}}{}\gamma_{\mu}i
   \propS{12}{(-)}{x_{2}-x_{1}}{}\gamma^{\mu}
   \bar{\tilde{\Psi}}(x_{1}):A_{\nu}(x_{3})
   i\propD{21}{(+)}{x_{1}-x_{2}}\\
&-ie^3:\bar{\Psi}(x_{3})\gamma^{\nu}i
   \propS{12}{(-)}{x_{3}-x_{2}}{}\gamma_{\mu}\frac{1}{i}
   \propS{22}{(-)}{x_{2}-x_{1}}{}\gamma^{\mu}
   \bar{\tilde{\Psi}}(x_{1}):A_{\nu}(x_{3})
   i\propD{22}{(+)}{x_{1}-x_{2}}.
\end{align*}
Finally we have to calculate\\
$A_{33,3}'=T_{1}(x_{2})T_{1}(x_{1})T_{1}(x_{3})\Rightarrow$ 
(proceeding as above)
\begin{align*}
A&_{33,3}' =\\&+ie^3:\bar{\Psi}(x_{3})\gamma^{\nu}\frac{1}{i}
   \propS{11}{(-)}{x_{3}-x_{2}}{}\gamma_{\mu}\frac{1}{i}
   \propS{11}{(+)}{x_{2}-x_{1}}{}\gamma^{\mu}\Psi(x_{1}):A_{\nu}(x_{3})
   i\propD{11}{(+)}{x_{2}-x_{1}}\\
&-ie^3:\bar{\Psi}(x_{3})\gamma^{\nu}\frac{1}{i}
   \propS{11}{(-)}{x_{3}-x_{2}}{}\gamma_{\mu}i
   \propS{12}{(+)}{x_{2}-x_{1}}{}\gamma^{\mu}
   \bar{\tilde{\Psi}}(x_{1}):A_{\nu}(x_{3})
   i\propD{12}{(+)}{x_{2}-x_{1}}\\
&-ie^3:\bar{\Psi}(x_{3})\gamma^{\nu}i
   \propS{12}{(-)}{x_{3}-x_{2}}{}\gamma_{\mu}(-i)
   \propS{21}{(+)}{x_{2}-x_{1}}{}\gamma^{\mu}\Psi(x_{1}):A_{\nu}(x_{3})
   i\propD{21}{(+)}{x_{2}-x_{1}}\\
&+ie^3:\bar{\Psi}(x_{3})\gamma^{\nu}i
   \propS{12}{(-)}{x_{3}-x_{2}}{}\gamma_{\mu}\frac{1}{i}
   \propS{22}{(+)}{x_{2}-x_{1}}{}\gamma^{\mu}
   \bar{\tilde{\Psi}}(x_{1}):A_{\nu}(x_{3})
   i\propD{22}{(+)}{x_{2}-x_{1}}.
\end{align*}
Thus we know $A'$ and $R'$ from the third order self energy graph B. 
This gives the following result for $D = R'-A'$ of graph B (the arguments of
`$SS$' are $(x_{3}-x_{2}),\;(x_{2}-x_{1})$, always in this order): 
\begin{align*}
D_{3}^{3.\mathrm{ord.SE,B}}=e^3:\bar{\Psi}(x_{3})\cdot
\begin{bmatrix}
-\gamma^{\nu}
  \pnoaS{11}{(+)}{}{}\gamma^{\mu}\pnoaS{11}{(-)}{}{}
  \gamma_{\mu}\propD{11}{(+)}{x_{1}-x_{2}}-\\
-\gamma^{\nu}
  \pnoaS{12}{(+)}{}{}\gamma^{\mu}\pnoaS{21}{(-)}{}{}
  \gamma_{\mu}\propD{12}{(+)}{x_{1}-x_{2}}-\\
-\gamma^{\nu}
  \pnoaS{11}{\F}{}{}\gamma^{\mu}\pnoaS{11}{(+)}{}{}
  \gamma_{\mu}\propD{11}{(+)}{x_{2}-x_{1}}-\\
-\gamma^{\nu}
  \pnoaS{12}{\F}{}{}\gamma^{\mu}\pnoaS{21}{(+)}{}{}
  \gamma_{\mu}\propD{21}{(+)}{x_{2}-x_{1}}-\\
-\gamma^{\nu}
  \pnoaS{11}{(+)}{}{}\gamma^{\mu}\pnoaS{11}{\F}{}{}
  \gamma_{\mu}\propD{11}{\F}{x_{2}-x_{1}}-\\
-\gamma^{\nu}
  \pnoaS{12}{(+)}{}{}\gamma^{\mu}\pnoaS{21}{\F}{}{}
  \gamma_{\mu}\propD{12}{\F}{x_{2}-x_{1}}+\\
+\gamma^{\nu}
  \pnoaS{11}{(+)}{}{}\gamma^{\mu}\pnoaS{11}{(-)}{}{}
  \gamma_{\mu}\propD{11}{(+)}{x_{1}-x_{2}}+\\
+\gamma^{\nu}
  \pnoaS{12}{(+)}{}{}\gamma^{\mu}\pnoaS{21}{(-)}{}{}
  \gamma_{\mu}\propD{12}{(+)}{x_{1}-x_{2}}-\\
-\gamma^{\nu}
  \pnoaS{11}{(+)}{}{}\gamma^{\mu}\pnoaS{11}{(+)}{}{}
  \gamma_{\mu}\propD{11}{(+)}{x_{2}-x_{1}}-\\
-\gamma^{\nu}
  \pnoaS{12}{(+)}{}{}\gamma^{\mu}\pnoaS{21}{(+)}{}{}
  \gamma_{\mu}\propD{21}{(+)}{x_{2}-x_{1}}+\\
+\gamma^{\nu}
  \pnoaS{11}{(-)}{}{}\gamma^{\mu}\pnoaS{11}{(+)}{}{}
  \gamma_{\mu}\propD{11}{(+)}{x_{2}-x_{1}}+\\
+\gamma^{\nu}
  \pnoaS{12}{(-)}{}{}\gamma^{\mu}\pnoaS{21}{(+)}{}{}
  \gamma_{\mu}\propD{21}{(+)}{x_{2}-x_{1}}-\\
-\gamma^{\nu}
  \pnoaS{11}{\F}{}{}\gamma^{\mu}\pnoaS{11}{(-)}{}{}
  \gamma_{\mu}\propD{11}{(+)}{x_{1}-x_{2}}-\\
-\gamma^{\nu}
  \pnoaS{12}{\F}{}{}\gamma^{\mu}\pnoaS{21}{(-)}{}{}
  \gamma_{\mu}\propD{12}{(+)}{x_{1}-x_{2}}-\\
-\gamma^{\nu}
  \pnoaS{11}{(-)}{}{}\gamma^{\mu}\pnoaS{11}{\F}{}{}
  \gamma_{\mu}\propD{11}{\F}{x_{2}-x_{1}}-\\
-\gamma^{\nu}
  \pnoaS{12}{(-)}{}{}\gamma^{\mu}\pnoaS{21}{\F}{}{}
 \gamma_{\mu}\propD{12}{\F}{x_{2}-x_{1}}+\\
+\gamma^{\nu}
  \pnoaS{11}{(-)}{}{}\gamma^{\mu}\pnoaS{11}{(-)}{}{}
  \gamma_{\mu}\propD{11}{(+)}{x_{1}-x_{2}}+\\
+\gamma^{\nu}
  \pnoaS{12}{(-)}{}{}\gamma^{\mu}\pnoaS{21}{(-)}{}{}
  \gamma_{\mu}\propD{12}{(+)}{x_{1}-x_{2}}-\\
-\gamma^{\nu}
  \pnoaS{11}{(-)}{}{}\gamma^{\mu}\pnoaS{11}{(+)}{}{}
  \gamma_{\mu}\propD{11}{(+)}{x_{2}-x_{1}}-\\
-\gamma^{\nu}
  \pnoaS{12}{(-)}{}{}\gamma^{\mu}\pnoaS{21}{(+)}{}{}
  \gamma_{\mu}\propD{21}{(+)}{x_{2}-x_{1}}
\end{bmatrix}\cdot\Psi(x_{1}):A_{\nu}(x_{3})+
\end{align*}
\begin{align*}
+e^3:\bar{\Psi}(x_{3})\cdot
\begin{bmatrix}
-\gamma^{\nu}
  \pnoaS{11}{(+)}{}{}\gamma^{\mu}\pnoaS{12}{(-)}{}{}
  \gamma_{\mu}\propD{21}{(+)}{x_{1}-x_{2}}+\\
+\gamma^{\nu}
  \pnoaS{12}{(+)}{}{}\gamma^{\mu}\pnoaS{22}{(-)}{}{}
  \gamma_{\mu}\propD{22}{(+)}{x_{1}-x_{2}}-\\
-\gamma^{\nu}
  \pnoaS{11}{\F}{}{}\gamma^{\mu}\pnoaS{12}{(+)}{}{}
  \gamma_{\mu}\propD{12}{(+)}{x_{2}-x_{1}}+\\
+\gamma^{\nu}
  \pnoaS{12}{\F}{}{}\gamma^{\mu}\pnoaS{22}{(+)}{}{}
  \gamma_{\mu}\propD{22}{(+)}{x_{2}-x_{1}}-\\
-\gamma^{\nu}
  \pnoaS{11}{(+)}{}{}\gamma^{\mu}\pnoaS{12}{\F}{}{}
  \gamma_{\mu}\propD{12}{\F}{x_{2}-x_{1}}+\\
+\gamma^{\nu}
  \pnoaS{12}{(+)}{}{}\gamma^{\mu}\pnoaS{22}{\F}{}{}
  \gamma_{\mu}\propD{22}{\F}{x_{2}-x_{1}}+\\
+\gamma^{\nu}
  \pnoaS{11}{(+)}{}{}\gamma^{\mu}\pnoaS{12}{(-)}{}{}
  \gamma_{\mu}\propD{21}{(+)}{x_{1}-x_{2}}-\\
-\gamma^{\nu}
  \pnoaS{12}{(+)}{}{}\gamma^{\mu}\pnoaS{22}{(-)}{}{}
  \gamma_{\mu}\propD{22}{(+)}{x_{1}-x_{2}}-\\
-\gamma^{\nu}
  \pnoaS{11}{(+)}{}{}\gamma^{\mu}\pnoaS{12}{(+)}{}{}
  \gamma_{\mu}\propD{12}{(+)}{x_{2}-x_{1}}+\\
+\gamma^{\nu}
  \pnoaS{12}{(+)}{}{}\gamma^{\mu}\pnoaS{22}{(+)}{}{}
  \gamma_{\mu}\propD{22}{(+)}{x_{2}-x_{1}}+\\
+\gamma^{\nu}
  \pnoaS{11}{(-)}{}{}\gamma^{\mu}\pnoaS{12}{(+)}{}{}
  \gamma_{\mu}\propD{12}{(+)}{x_{2}-x_{1}}-\\
-\gamma^{\nu}
  \pnoaS{12}{(-)}{}{}\gamma^{\mu}\pnoaS{22}{(+)}{}{}
  \gamma_{\mu}\propD{22}{(+)}{x_{2}-x_{1}}-\\
-\gamma^{\nu}
  \pnoaS{11}{\F}{}{}\gamma^{\mu}\pnoaS{12}{(-)}{}{}
  \gamma_{\mu}\propD{21}{(+)}{x_{1}-x_{2}}+\\
+\gamma^{\nu}
  \pnoaS{12}{\F}{}{}\gamma^{\mu}\pnoaS{22}{(-)}{}{}
  \gamma_{\mu}\propD{22}{(+)}{x_{1}-x_{2}}-\\
-\gamma^{\nu}
  \pnoaS{11}{(-)}{}{}\gamma^{\mu}\pnoaS{12}{\F}{}{}
  \gamma_{\mu}\propD{12}{\F}{x_{2}-x_{1}}+\\
+\gamma^{\nu}
  \pnoaS{12}{(-)}{}{}\gamma^{\mu}\pnoaS{22}{\F}{}{}
  \gamma_{\mu}\propD{22}{\F}{x_{2}-x_{1}}+\\
+\gamma^{\nu}
  \pnoaS{11}{(-)}{}{}\gamma^{\mu}\pnoaS{12}{(-)}{}{}
 \gamma_{\mu}\propD{21}{(+)}{x_{1}-x_{2}}-\\
-\gamma^{\nu}
 \pnoaS{12}{(-)}{}{}\gamma^{\mu}\pnoaS{22}{(-)}{}{}
  \gamma_{\mu}\propD{22}{(+)}{x_{1}-x_{2}}-\\
-\gamma^{\nu}
  \pnoaS{11}{(-)}{}{}\gamma^{\mu}\pnoaS{12}{(+)}{}{}
  \gamma_{\mu}\propD{12}{(+)}{x_{2}-x_{1}}+\\
+\gamma^{\nu}
  \pnoaS{12}{(-)}{}{}\gamma^{\mu}\pnoaS{22}{(+)}{}{}
  \gamma_{\mu}\propD{22}{(+)}{x_{2}-x_{1}}
\end{bmatrix}\cdot\bar{\tilde{\Psi}}(x_{1}):A_{\nu}(x_{3}).
\end{align*}
Now we simplify this expression; in the first part, the ten terms with 
a type 1-vertex in $x_{2}$ give after suitable modification (consider the 
support properties of the quantities involved)
\begin{align*}
D_{3,\bar{\Psi}\Psi}^{3.\mathrm{ord.SE,B;}x_{2}:\mathrm{type 1}}=&
-\propS{11}{\av}{x_{3}-x_{2}}{}\gamma^{\mu}\propS{11}{}{x_{2}-x_{1}}{}
  \gamma_{\mu}\propD{11}{(-)}{x_{1}-x_{2}}\\
&+\propS{11}{\ret}{x_{3}-x_{2}}{}\gamma^{\mu}\propS{11}{(-)}{x_{2}-x_{1}}{}
  \gamma_{\mu}\propD{11}{}{x_{1}-x_{2}}\\
&-\propS{11}{}{x_{3}-x_{2}}{}\gamma^{\mu}\propS{11}{\F}{x_{2}-x_{1}}{}
  \gamma_{\mu}\propD{11}{\ret}{x_{1}-x_{2}}\\
&-\propS{11}{}{x_{3}-x_{2}}{}\gamma^{\mu}\propS{11}{\ret}{x_{2}-x_{1}}{}
  \gamma_{\mu}\propD{11}{(-)}{x_{1}-x_{2}}=\\
=&+\propS{11}{\av}{x_{3}-x_{2}}{}\gamma^{\mu}\propS{11}{\av}{x_{2}-x_{1}}{}
  \gamma_{\mu}\propD{11}{(-)}{x_{1}-x_{2}}\\
&-\propS{11}{\ret}{x_{3}-x_{2}}{}\gamma^{\mu}\propS{11}{(-)}{x_{2}-x_{1}}{}
  \gamma_{\mu}\propD{11}{\av}{x_{1}-x_{2}}\\
&+\propS{11}{\av}{x_{3}-x_{2}}{}\gamma^{\mu}\propS{11}{(-)}{x_{2}-x_{1}}{}
  \gamma_{\mu}\propD{11}{\ret}{x_{1}-x_{2}}\\
&-\propS{11}{\ret}{x_{3}-x_{2}}{}\gamma^{\mu}\propS{11}{\ret}{x_{2}-x_{1}}{}
  \gamma_{\mu}\propD{11}{(-)}{x_{1}-x_{2}}.
\end{align*}
The terms with type 2 in $x_{2}$ give
\begin{equation*} 
D_{3,\bar{\Psi}\Psi}^
    {3.\mathrm{ord.SE,B;}x_{2}:\mathrm{type 2}}=0,
\end{equation*}
because both $D_{12}$ and $S_{12}$ are equal to zero. Likewise we have 
\begin{equation*}
D_{3,\bar{\Psi}\bar{\tilde{\Psi}}}^
    {3.\mathrm{ord.SE,B;}x_{2}:\mathrm{type 1}}=0 \;\;\textrm{and}\;\; 
D_{3,\bar{\Psi}\bar{\tilde{\Psi}}}^
    {3.\mathrm{ord.SE,B;}x_{2}:\mathrm{type 2}}=0.
\end{equation*}
$D_{3}$ for graph A is calculated in the same way. 
\subsubsection*{2. Singular Order}
The structure of the third order SE graphs is 
$S(x_{3}-x_{2})S(x_{2}-x_{1})D(x_{1}-x_{2})$. This gives after Fourier 
transformation  
$\hat{S}(q)\int\!d^4k\hat{S}(q+k)\hat{D}(k)$, which means that only  
$\omega^{\mathrm{2.ord.SE}}$ is relevant, because the first factor 
($S$) has singular order $\omega \leq -1$ and the second one is exactly the 
second order SE. Since the temperature dependent parts of the second order SE 
have singular order $\omega \leq -1$ the temperature dependent parts of the 
third order SE can be split trivially. 

\subsection{Calculations for the Thermal Corrections to 
\boldmath{$\mu_{\mathrm{e}^{\mathbf{-}}}$}}
\label{subsec:mucalcappend}
In this appendix we present the calculations to solve the linear system 
(\ref{eq:gls1}) - (\ref{eq:gls5}), which reads in matrix form
\bel{ggllss}
\begin{split}
&\hspace{4.0 cm}\begin{pmatrix}0\\I_{2}\\I_{3}\\I_{4}\\I_{5}\end{pmatrix}=\\
&=\!\!\begin{pmatrix}4&2m^2&2(pq)&2(pu+qu)&1\\
m^2&m^4+(pq)^2&2m^2(pq)&2(pu)(m^2+pq)&(pu)^2\\
pq&2m^2(pq)&m^4+(pq)^2&(m^2+pq)(pu+qu)&(pu)(qu)\\
pu&m^2(pu)\!+\!(pq)(qu)&m^2(qu)\!+\!(pq)(pu)&m^2\!+\!(pu)^2\!+\!(pq)\!+\!
(pu)(qu)&pu\\
1&(pu)^2+(qu)^2&2(pu)(qu)&2(pu+qu)&1\end{pmatrix}\!\cdot\\
&\hspace{8.0 cm}\cdot\begin{pmatrix}A\\B\\C\\D\\G\end{pmatrix}. 
\end{split}
\end{equation}
Before solving that, we calculate the four integrals $I_{2},...,I_{5}$, as 
defined in (\ref{eq:gls1}) - (\ref{eq:gls5}).
\bel{int2}
I_{2}=\int\!d^4k\delta(k^2)\frac{1}{(e^{\beta\lvert ku\rvert}-1)}\cdot\frac{pk}
{(-2qk+i0)}.
\end{equation}
We choose the Lorenz system with $u=(1,\vec{0})$ and 
$q=(q_{0},0,0,\lvert\vec{q}\rvert)$. Then we perform 
the $k_{0}$-integration with 
the $\delta$-distribution and choose polar coordinates for the spatial 
integration ($x:=\cos(\theta)$). This gives 
\bel{int2a}
\begin{split}
I_{2}=
\int_{0}^{\infty}\!dr\frac{r}{2}\int_{-1}^{1}\!dx\int_{0}^{2\pi}\!&d\varphi
\frac{1}{e^{\beta\lvert ku\rvert}-1}\cdot\\
\cdot\Bigg[&\frac{p_{0}r-p_{1}r\sin(\varphi)
\sqrt{1-x^2}-p_{2}r\cos(\varphi)\sqrt{1-x^2}-p_{3}rx}{-2q_{0}r+2\lvert\vec{q}
\rvert rx+i0}+\\&+\frac{-p_{0}r-p_{1}r\sin(\varphi)
\sqrt{1-x^2}-p_{2}r\cos(\varphi)\sqrt{1-x^2}-p_{3}rx}{2q_{0}r+2\lvert\vec{q}
\rvert rx +i0}\Bigg].
\end{split}
\end{equation}
Now we perform the $\varphi$-integration, which leads to a solvable 
$r$-integral (use $\int_{0}^{\infty}\!dr\frac{r}{e^{\beta r}-1}=
\frac{\pi^2}{6\beta^2}$).Thus we have 
\bel{int2b}
I_{2}=
\frac{\pi^3}{12\beta^2}\int_{-1}^{1}\!dx\Bigg[\frac{p_{0}-p_{3}x}{-q_{0}+
\lvert\vec{q}\rvert x+i0}+\frac{-p_{0}-p_{3}x}{q_{0}+
\lvert\vec{q}\rvert x+i0}\Bigg].
\end{equation} 
In the second term we transform $x\rightarrow-x$ and see that in the sum with 
the first term the imaginary part vanishes identically and the real part gives 
the same contribution two times. Now the $x$-integration can be performed, 
and we
have after transforming back to a general Lorentz 
system\footnote{\label{fn:backtrafo}Here we give some 
relations necessary to perform this transformation from the special 
system with $u=(1,\vec{0}), q=(q_{0},0,0,\lvert\vec{q}\rvert)$ back to a 
general one: 
$p_{0}\rightarrow pu,\;\;\lvert\vec{p}\rvert\rightarrow\sqrt{(pu)^2-m^2},\;\;
q_{0}\rightarrow qu,\;\;\lvert\vec{q}\rvert\rightarrow\sqrt{(qu)^2-m^2},\;\;
p_{3}\rightarrow \frac{(pu)(qu)-pq}{\sqrt{(qu)^2-m^2}}$. These relations can 
be verified straightforward.}
\bel{int2c}
\begin{split}
I_{2}=-\frac{\pi^3}{6\beta^2}&\frac{1}{\sqrt{(qu)^2-m^2}}\cdot\\
&\cdot\Bigg(\frac{
-2pq+2(pu)(qu)}{\sqrt{(qu)^2-m^2}}+\bigg[\frac{(pu)m^2-(pq)(qu)}
{(qu)^2-m^2}\bigg]\mathrm{ln}\bigg\lvert\frac{qu-\sqrt{(qu)^2-m^2}}
{qu+\sqrt{(qu)^2-m^2}}\bigg\rvert\Bigg).
\end{split}
\end{equation}
The next integral is easier to calculate (use again $\int_{0}^{\infty}\!d
r\frac{r}{e^{\beta r}-1}=\frac{\pi^2}{6\beta^2}$):
\bel{int3}
I_{3}=-\frac{1}{2}\int\!d^4k\delta(k^2)\frac{1}{e^{\beta\lvert ku\rvert}-1}=
-\int_{0}^{\infty}\!dr\frac{r}{2}\frac{1}{e^{\beta r}-1}\cdot4\pi
=-2\pi\cdot\frac{\pi^2}{6\beta^2}=-\frac{\pi^3}{3\beta^2}.
\end{equation}
Now to $I_{4}$:
\bel{int4}
I_{4}=\int\!d^4k\delta(k^2)\frac{1}{(e^{\beta\lvert ku\rvert}-1)}\cdot\frac{ku}
{(-2qk+i0)}.
\end{equation}
We proceed as in the calculation of $I_{2}$ above and finally get
\bel{int4a}
 I_{4}=\frac{\pi^3}{6\beta^2}\frac{1}{\sqrt{(qu)^2-m^2}}
  \mathrm{ln}\bigg\lvert\frac{qu-\sqrt{(qu)^2-m^2}} 
 {qu+\sqrt{(qu)^2-m^2}}\bigg\rvert.
\end{equation}
There remains one integral to calculate:
\bel{int5}
I_{5}=-\frac{1}{2}\int\!d^4k\delta(k^2)\frac{1}{(e^{\beta\lvert ku\rvert}-1)}
 \cdot\frac{(ku)^2}{(pk-i0)(qk-i0)}.
\end{equation}
Again we choose polar coordinates. 
The $k_{0}$-integration is done with $\delta(k^2)$, and the $r$-integration 
decouples from the other variables and can be performed (using formulae as 
for the $r$-integration in $I_{2}$). Then we transform 
$x\rightarrow-x, \varphi\rightarrow\varphi-\pi$ in the term with $k_{0}=-r$, 
which leads to an integral $\int_{-\pi}^{\pi}$. We transform its part 
$\int_{-\pi}^{0}$ with $\varphi\rightarrow \varphi+2\pi$ to 
$\int_{\pi}^{2\pi}$. Now we can see that the term with $k_{0}=-r$ equals the 
complex conjugate of the term with $k_{0}=r$ and thus the sum is two times 
its real part. We get
\bel{int5a}
I_{5}=-\frac{1}{2}\frac{\pi^2}{6\beta^2}
    \int_{-1}^{1}\!dx\int_{0}^{2\pi}d\varphi
 \frac{1}{(p_{0}-\lvert\vec{p}\rvert x)}\cdot\frac{1}{(q_{0}-q_{1}\sin(\varphi)
\sqrt{1-x^2}-q_{2}\cos(\varphi)\sqrt{1-x^2}-q_{3}x)}.
\end{equation}
The $\varphi$-integration now gives\footnote{It has to be verified that the 
possible poles are integrable. This is done best by cutting them from the 
integration interval and then extend the result to this border values of 
the interval. In addition, using $q^2=m^2>0, 
\lvert x\rvert\leq 1$, etc. it can be shown that the radicand in the 
resulting formulae is always positive.}
\bel{varphi}
\int_{0}^{2\pi}\!d\varphi\frac{1}{q_{0}-q_{1}\sin(\varphi)
\sqrt{1-x^2}-q_{2}\cos(\varphi)\sqrt{1-x^2}-q_{3}x}=\frac{2\pi}{
\sqrt{\lvert \vec{q} \rvert^2x^2-2q_{0}q_{3}x+(m^2+q_{3}^2)}}.
\end{equation}  
Thus we have
\bel{int5b}
I_{5}=-\frac{\pi^3}{6\beta^2}\int_{-1}^{1}\!dx\frac{1}
{(p_{0}-\lvert\vec{p}\rvert x)\sqrt{\lvert \vec{q} \rvert^2x^2
-2q_{0}q_{3}x+(m^2+q_{3}^2)}}.
\end{equation}
Having performed this integration (which can be done quite straightforward), 
we transform the result back into a general Lorentz frame\footnote{
See footnote \ref{fn:backtrafo}.} and thus have
\bel{int5c}
\begin{split}
I_{5}=&\frac{-\pi^3}{12\beta^2\sqrt{(pq)^2-m^4}}\Bigg[\mathrm{ln}\Bigg(
\frac{\sqrt{(pq)^2-m^4}\Big(qu-\frac{(pu)(qu)-pq}{\sqrt{(pu)^2-m^2}}\Big)+
\cdot\cdot\cdot}
{\sqrt{(pq)^2-m^4}\Big(qu-\frac{(pu)(qu)-pq}{\sqrt{(pu)^2-m^2}}\Big)-
\cdot\cdot\cdot}\\
&\hspace{2.0 cm}\frac{\cdot\cdot\cdot+
\frac{1}{\sqrt{(pu)^2-m^2}}\big(m^2(pu)^2-m^4+(pq)^2-(pq)(pu)(qu)\big)-
\cdot\cdot\cdot}
{\cdot\cdot\cdot-
\frac{1}{\sqrt{(pu)^2-m^2}}\big(m^2(pu)^2-m^4+(pq)^2-(pq)(pu)(qu)\big)+
\cdot\cdot\cdot}\\
&\hspace{2.0 cm}
\frac{\cdot\cdot\cdot-m^2(pu)+(pq)(qu)}{\cdot\cdot\cdot+m^2(pu)-(pq)(qu)}
\Bigg)-\\
&-\mathrm{ln}\Bigg(
\frac{\sqrt{(pq)^2-m^4}\Big(qu+\frac{(pu)(qu)-pq}{\sqrt{(pu)^2-m^2}}\Big)+
\cdot\cdot\cdot}
{\sqrt{(pq)^2-m^4}\Big(qu+\frac{(pu)(qu)-pq}{\sqrt{(pu)^2-m^2}}\Big)-
\cdot\cdot\cdot}\\
&\hspace{2.0 cm}\frac{\cdot\cdot\cdot+
\frac{1}{\sqrt{(pu)^2-m^2}}\big(m^2(pu)^2-m^4+(pq)^2-(pq)(pu)(qu)\big)+
\cdot\cdot\cdot}
{\cdot\cdot\cdot-
\frac{1}{\sqrt{(pu)^2-m^2}}\big(m^2(pu)^2-m^4+(pq)^2-(pq)(pu)(qu)\big)-
\cdot\cdot\cdot}\\
&\hspace{2.0 cm}
\frac{\cdot\cdot\cdot+m^2(pu)-(pq)(qu)}{\cdot\cdot\cdot-m^2(pu)+(pq)(qu)}
\Bigg)\Bigg].
\end{split}
\end{equation}
Now we want to solve the linear system (\ref{eq:ggllss}). We did that 
using Maple and get the general solution for $A,...,G$. Then we set 
$q:=p-\eta$ with $q^2=(p-\eta)^2=m^2$ and let it be simplified and 
factorised. Thus we get (since we will 
not need the solutions for $B,C,D$ and $G$ here, we just give the one for $A$):
\bel{A}
\begin{split}
A = \Big(&-8(pu)^2I_{3}m^2-4m^4I_{2}+4m^4I_{3}-4(pu)^4I_{2}+4(pu)^4I_{3}
+8(pu)^2m^2I_{2}-\\&-8(pu)^3I_{4}(p\eta)-2(pu)^2I_{2}(p\eta)
-6(pu)^3I_{3}(\eta u)+6I_{3}(pu)^2(p\eta )
+2m^2I_{2}(p\eta )+\\&+6I_{3}m^2(pu)(\eta u)-4m^2(pu)^2I_{4}(\eta u)
-10m^2(pu)I_{2}(\eta u)
+8(pu)I_{4}m^2(p\eta )+\\&+4(pu)^2m^2I_{5}(p\eta )+4I_{4}m^4(\eta u)
-4I_{5}m^4(p\eta )-6I_{3}m^2(p\eta )+10(pu)^3I_{2}(\eta u)+\\
&+2(pu)^2I_{3}(\eta u)^2-2(pu)^2I_{5}(p\eta )^2+2m^2I_{2}(\eta u)^2
+4I_{5}m^2(p\eta )^2-\\&-8(pu)^2I_{2}(\eta u)^2-4(pu)m^2I_{5}(p\eta )(\eta u)
+4m^2(pu)I_{4}(\eta u)^2-6I_{4}m^2(p\eta )(\eta u)+\\
&+10(pu)^2I_{4}(\eta u)(p\eta )-4(pu)I_{3}(p\eta )(\eta u)
+4(pu)I_{2}(p\eta )(\eta u)+2I_{3}(p\eta )^2-\\&-4(pu)I_{4}(p\eta )^2
-I_{2}(\eta u)^2(p\eta )+2I_{4}(p\eta )^2(\eta u)+\\&+2(pu)I_{2}(\eta u)^3
+2(pu)I_{5}(p\eta )^2(\eta u)-4(pu)I_{4}(\eta u)^2(p\eta )
-I_{5}(p\eta )^3\Big)/\\
/\Big(&+4m^4(p\eta )-8m^2(p\eta )(pu)^2+4(pu)^4(p\eta )
+2m^4(\eta u)^2-\\&-2(pu)^2m^2(\eta u)^2-4m^2(p\eta )^2+4(p\eta )^2(pu)^2
+2(pu)m^2(\eta u)^3-\\&-m^2(p\eta )
(\eta u)^2-2(pu)^2(\eta u)^2(p\eta )+(p\eta )^3\Big).
\end{split}
\end{equation}
Having a close look at the solution we see what is the problem of setting 
$p=q \leftrightarrow \eta = 0$ too early. The determinant of the 
linear system and the denominator in the solutions, resp.,  
is zero for this values. Therefore we must not set $\eta =
0$ but expand the various quantities involved in `powers of $\eta$'. Then we 
will see that all possibly divergent terms cancel and we are left with  
finite values for $\eta \rightarrow 0$. This expansion is rather 
delicate and we have to pay attention not to throw away any term too early or 
even to forget some at all. A crucial thing with the linear system 
considered is the fact that it remains defined even if we set $\eta = 0$ 
at the very beginning; we then can reduce the system to four equations and 
solve it. And we are in the lucky case that we only need to know $A$ for 
which we get the same result this way. But this is not true in general and 
here it is also wrong for the other variables $B$, etc.

To do these things we introduce some notation: set $\vec{\eta}=:\epsilon 
|vec{\eta}_{e}$ with $\epsilon \in \mathbb{R},\;\epsilon \ll 1,\;\;
\vec{\eta}_{e} \in \mathbb{R},\;\vec{\eta}_{e}^2=1,$ 
so that the limit in $\mathbb{M}$, $\eta\rightarrow 0$, can be 
discussed as an ordinary limit in $\mathbb{R}$, $\epsilon \rightarrow 0$. 
Pay attention to the fact that $\eta_{0}$ depends on $\vec{\eta}$ through 
$q^2=(p-\eta)^2=m^2$. We have $\eta_{0}=p_{0}-p_{0}\sqrt{1-(\frac{2\vec{p}
\vec{\eta}}{p_{0}^2}-\frac{\vec{\eta}^2}{p_{0}^2})}$. Thus $\eta_{0}$ is 
not simply proportional to $\epsilon$! This has to be investigated for 
$\vec{q}\neq 0$ and $\vec{q}=0$ separately. But considering the first case it 
will be clear that the second follows by simply setting $\vec{q}=0$. 

Since the more direct but in general 
wrong method of setting $p=q$ at the beginning 
leads to the same result we will not present here a detailed derivation of the 
calculations. We just give the general direction: All quantities involved in 
the numerator and denominator of $A$ have to be expanded in $\epsilon$ 
up to a suitable order. Especially the integrals $I_{2}, ...,I_{5}$ have to 
be expanded. This is a tedious but strightforward task. We emphasize 
especially the following: The structure of $A$ is

\bel{genstruct}
\begin{split}
A=&\frac{\wt{L}_{0}+\wt{L}_{1}\e+\wt{L}_{2}\e^2+\wt{L}_{3}\e^3+\cdots}
{\wt{l}_{2}\e^2+\wt{l}_{3}\e^3+\wt{l}_{4}\e^4}=\\
&=\frac{1}{\wt{l}_{2}\e^2}\cdot\frac{1}{(1+\frac{\wt{l}_{3}}{\wt{l}_{2}}\e
+\frac{\wt{l}_{4}}{\wt{l}_{2}}\e^2)+\cdots}
(\wt{L}_{0}+\wt{L}_{1}\e+\wt{L}_{2}\e^2+\wt{L}_{3}\e^3+\cdots).
\end{split}
\end{equation}

Now we expand the numerator, identify which term of $A$ has to be known up to 
which order and calculate it straightforward. This leads to 

\bel{muresult}
\mu_{\mathrm{e}^{-}}=\frac{e}{2m}\bigg\{1+\frac{e^2}{8\pi^2}
-\frac{e^2}{(2\pi)^3}
\bigg(\frac{2\pi^3}{9\beta^2m^2}\bigg)\bigg\}.
\end{equation}

\section{The Cross Section with General Mixed States}\label{sec:canc}
\setcounter{equation}{0}
The cross section involving the initial state (\ref{eq:initialstate}) and 
the corresponding mixture for the final state reads as follows (`0',`1' and 
`2' in the matrix elements stand for contributions with zero, one or two 
bremsstrahlungs photons involved, the sums run from 1 to 4 and the 
$k_{1},...$-integrations over $k_{1},...\leq \omega_{0}$. `MS' indicates 
the mass shell and `$\mathrm{MS}_{\mathrm{I}}$' a small part of the 
mass shell centered around 
$p$. $b^{+}$ in the final state is always understood with the momentum $p$, 
in the initial one with momentum $q$.):

\begin{align*}
\int_{\mathrm{MS}_{\mathrm{I}}}d^4p\int_{\mathrm{MS}}d^4q&\bigg\{\\
&+\big\lvert\langle b^+\rvert 0+1+2\lvert b^+\rangle\big\rvert^2c_{0}(q)+\\
&+\sum_{i}\int d^3k_{1}
 \big\lvert\langle b^+\rvert 0+1+2\lvert b^+a_{i}^{+}(k_{1})\rangle\big\rvert^2
 c_{1}(q,k_{1})_{i}+\\
&+\sum_{i}\int d^3k_{1}
 \big\lvert\langle b^+\rvert 0+1+2\lvert b^+\tilde{a}_{i}^{+}(k_{1})
 \rangle\big\rvert^2\tilde{c}_{1}(q,k_{1})_{i}+\\
&+\sum_{i,j}\int d^3k_{1}\int d^3k_{2}
 \big\lvert\langle b^+\rvert 0+1+2\lvert 
 b^+a_{i}^{+}(k_{1})a_{j}^{+}(k_{2})\rangle\big\rvert^2
 c_{2}(q,k_{1},k_{2})_{ij}+\\
&+\sum_{i,j}\int d^3k_{1}\int d^3k_{2}
 \big\lvert\langle b^+\rvert 0+1+2\lvert 
 b^+a_{i}^{+}(k_{1})\tilde{a}_{j}^{+}(k_{2})\rangle\big\rvert^2
 \tilde{c}_{2}(q,k_{1},k_{2})_{ij}+\\
&+\sum_{i,j}\int d^3k_{1}\int d^3k_{2}
 \big\lvert\langle b^+\rvert 0+1+2\lvert 
 b^+\tilde{a}_{i}^{+}(k_{1})\tilde{a}_{j}^{+}(k_{2})\rangle\big\rvert^2
 \tilde{\tilde{c}}_{2}(q,k_{1},k_{2})_{ij}+\\
\end{align*}
\begin{align*}
&+\sum_{m}\int d^3l_{1}\big\lvert\langle b^+a^{+}_{m}(l_{1})   
  \rvert 0+1+2\lvert b^+\rangle\big\rvert^2c_{0}(q)+\\
&+\sum_{m}\sum_{i}\int d^3l_{1}\int d^3k_{1}
 \big\lvert\langle b^+a^{+}_{m}(l_{1})
 \rvert 0+1+2\lvert b^+a_{i}^{+}(k_{1})\rangle\big\rvert^2
 c_{1}(q,k_{1})_{i}+\\
&+\sum_{m}\sum_{i}\int d^3l_{1}\int d^3k_{1}
 \big\lvert\langle b^+a^{+}_{m}(l_{1})
 \rvert 0+1+2\lvert b^+\tilde{a}_{i}^{+}(k_{1})
 \rangle\big\rvert^2\tilde{c}_{1}(q,k_{1})_{i}+\\
&+\sum_{m}\sum_{i,j}\int d^3l_{1}\int d^3k_{1}\int d^3k_{2}
 \big\lvert\langle b^+a^{+}_{m}(l_{1})\rvert 0+1+2\lvert 
 b^+a_{i}^{+}(k_{1})a_{j}^{+}(k_{2})\rangle\big\rvert^2
 c_{2}(q,k_{1},k_{2})_{ij}+\\
&+\sum_{m}\sum_{i,j}\int d^3l_{1}\int d^3k_{1}\int d^3k_{2}
 \big\lvert\langle b^+a^{+}_{m}(l_{1})\rvert 0+1+2\lvert 
 b^+a_{i}^{+}(k_{1})\tilde{a}_{j}^{+}(k_{2})\rangle\big\rvert^2
 \tilde{c}_{2}(q,k_{1},k_{2})_{ij}+\\
&+\sum_{m}\sum_{i,j}\int d^3l_{1}\int d^3k_{1}\int d^3k_{2}
 \big\lvert\langle b^+a^{+}_{m}(l_{1})\rvert 0+1+2\lvert 
 b^+\tilde{a}_{i}^{+}(k_{1})\tilde{a}_{j}^{+}(k_{2})\rangle\big\rvert^2
 \tilde{\tilde{c}}_{2}(q,k_{1},k_{2})_{ij}+\\
\end{align*}
\begin{align*}
&+\sum_{m}\int d^3l_{1}\big\lvert\langle b^+\tilde{a}^{+}_{m}(l_{1})   
  \rvert 0+1+2\lvert b^+\rangle\big\rvert^2c_{0}(q)+\\
&+\sum_{m}\sum_{i}\int d^3l_{1}\int d^3k_{1}
 \big\lvert\langle b^+\tilde{a}^{+}_{m}(l_{1})
 \rvert 0+1+2\lvert b^+a_{i}^{+}(k_{1})\rangle\big\rvert^2
 c_{1}(q,k_{1})_{i}+\\
&+\sum_{m}\sum_{i}\int d^3l_{1}\int d^3k_{1}
 \big\lvert\langle b^+\tilde{a}^{+}_{m}(l_{1})
 \rvert 0+1+2\lvert b^+\tilde{a}_{i}^{+}(k_{1})
 \rangle\big\rvert^2\tilde{c}_{1}(q,k_{1})_{i}+\\
&+\sum_{m}\sum_{i,j}\int d^3l_{1}\int d^3k_{1}\int d^3k_{2}
 \big\lvert\langle b^+\tilde{a}^{+}_{m}(l_{1})\rvert 0+1+2\lvert 
 b^+a_{i}^{+}(k_{1})a_{j}^{+}(k_{2})\rangle\big\rvert^2
 c_{2}(q,k_{1},k_{2})_{ij}+\\
&+\sum_{m}\sum_{i,j}\int d^3l_{1}\int d^3k_{1}\int d^3k_{2}
 \big\lvert\langle b^+\tilde{a}^{+}_{m}(l_{1})\rvert 0+1+2\lvert 
 b^+a_{i}^{+}(k_{1})\tilde{a}_{j}^{+}(k_{2})\rangle\big\rvert^2
 \tilde{c}_{2}(q,k_{1},k_{2})_{ij}+\\
&+\sum_{m}\sum_{i,j}\int d^3l_{1}\int d^3k_{1}\int d^3k_{2}
 \big\lvert\langle b^+\tilde{a}^{+}_{m}(l_{1})\rvert 0+1+2\lvert 
 b^+\tilde{a}_{i}^{+}(k_{1})\tilde{a}_{j}^{+}(k_{2})\rangle\big\rvert^2
 \tilde{\tilde{c}}_{2}(q,k_{1},k_{2})_{ij}+\\
\end{align*}
\begin{align*} 
&+\sum_{m,n}\int d^3l_{1}\int d^3l_{2}\big\lvert\langle b^+a^{+}_{m}(l_{1})   
  a^{+}_{n}(l_{2})\rvert 0+1+2\lvert b^+\rangle\big\rvert^2c_{0}(q)+\\
&+\sum_{m,n}\sum_{i}\int d^3l_{1}\int d^3l_{2}\int d^3k_{1}
 \big\lvert\langle b^+a^{+}_{m}(l_{1})a^{+}_{n}(l_{2})
 \rvert 0+1+2\lvert b^+a_{i}^{+}(k_{1})\rangle\big\rvert^2
 c_{1}(q,k_{1})_{i}+\\
&+\sum_{m,n}\sum_{i}\int d^3l_{1}\int d^3l_{2}\int d^3k_{1}
 \big\lvert\langle b^+a^{+}_{m}(l_{1}a^{+}_{n}(l_{2}))
 \rvert 0+1+2\lvert b^+\tilde{a}_{i}^{+}(k_{1})
 \rangle\big\rvert^2\tilde{c}_{1}(q,k_{1})_{i}+\\
&+\sum_{m,n}\sum_{i,j}\int d^3l_{1}\int d^3l_{2}\int d^3k_{1}\int d^3k_{2}\\
&\hspace{2.0 cm}
 \big\lvert\langle b^+a^{+}_{m}(l_{1})a^{+}_{n}(l_{2})\rvert 0+1+2\lvert 
 b^+a_{i}^{+}(k_{1})a_{j}^{+}(k_{2})\rangle\big\rvert^2
 c_{2}(q,k_{1},k_{2})_{ij}+\\
&+\sum_{m,n}\sum_{i,j}\int d^3l_{1}\int d^3l_{2}\int d^3k_{1}\int d^3k_{2}\\
&\hspace{2.0 cm}
 \big\lvert\langle b^+a^{+}_{m}(l_{1})a^{+}_{n}(l_{2})\rvert 0+1+2\lvert 
 b^+a_{i}^{+}(k_{1})\tilde{a}_{j}^{+}(k_{2})\rangle\big\rvert^2
 \tilde{c}_{2}(q,k_{1},k_{2})_{ij}+\\
&+\sum_{m,n}\sum_{i,j}\int d^3l_{1}\int d^3l_{2}\int d^3k_{1}\int d^3k_{2}\\
&\hspace{2.0 cm}
 \big\lvert\langle b^+a^{+}_{m}(l_{1})a^{+}_{n}(l_{2})\rvert 0+1+2\lvert 
 b^+\tilde{a}_{i}^{+}(k_{1})\tilde{a}_{j}^{+}(k_{2})\rangle\big\rvert^2
 \tilde{\tilde{c}}_{2}(q,k_{1},k_{2})_{ij}+\\
\end{align*}
\begin{align*}
&+\sum_{m,n}\int d^3l_{1}\int d^3l_{2}\big\lvert\langle b^+a^{+}_{m}(l_{1})   
  \tilde{a}^{+}_{n}(l_{2})\rvert 0+1+2\lvert b^+\rangle\big\rvert^2c_{0}(q)+\\
&+\sum_{m,n}\sum_{i}\int d^3l_{1}\int d^3l_{2}\int d^3k_{1}
 \big\lvert\langle b^+a^{+}_{m}(l_{1})\tilde{a}^{+}_{n}(l_{2})
 \rvert 0+1+2\lvert b^+a_{i}^{+}(k_{1})\rangle\big\rvert^2
 c_{1}(q,k_{1})_{i}+\\
&+\sum_{m,n}\sum_{i}\int d^3l_{1}\int d^3l_{2}\int d^3k_{1}
 \big\lvert\langle b^+a^{+}_{m}(l_{1}\tilde{a}^{+}_{n}(l_{2}))
 \rvert 0+1+2\lvert b^+\tilde{a}_{i}^{+}(k_{1})
 \rangle\big\rvert^2\tilde{c}_{1}(q,k_{1})_{i}+\\
&+\sum_{m,n}\sum_{i,j}\int d^3l_{1}\int d^3l_{2}\int d^3k_{1}\int d^3k_{2}\\
&\hspace{2.0 cm}\big\lvert\langle b^+a^{+}_{m}(l_{1})\tilde{a}^{+}_{n}(l_{2})
 \rvert 0+1+2\lvert 
 b^+a_{i}^{+}(k_{1})a_{j}^{+}(k_{2})\rangle\big\rvert^2
 c_{2}(q,k_{1},k_{2})_{ij}+\\
&+\sum_{m,n}\sum_{i,j}\int d^3l_{1}\int d^3l_{2}\int d^3k_{1}\int d^3k_{2}\\
&\hspace{2.0 cm}\big\lvert\langle b^+a^{+}_{m}(l_{1})\tilde{a}^{+}_{n}(l_{2})
 \rvert 0+1+2\lvert 
 b^+a_{i}^{+}(k_{1})\tilde{a}_{j}^{+}(k_{2})\rangle\big\rvert^2
 \tilde{c}_{2}(q,k_{1},k_{2})_{ij}+\\
&+\sum_{m,n}\sum_{i,j}\int d^3l_{1}\int d^3l_{2}\int d^3k_{1}\int d^3k_{2}\\
&\hspace{2.0 cm}\big\lvert\langle b^+a^{+}_{m}(l_{1})\tilde{a}^{+}_{n}(l_{2})
 \rvert 0+1+2\lvert 
 b^+\tilde{a}_{i}^{+}(k_{1})\tilde{a}_{j}^{+}(k_{2})\rangle\big\rvert^2
 \tilde{\tilde{c}}_{2}(q,k_{1},k_{2})_{ij}+\\
\end{align*}
\bel{canceinsalles}
\begin{split}
&+\sum_{m,n}\int d^3l_{1}\int d^3l_{2}\big\lvert\langle b^+
 \tilde{a}^{+}_{m}(l_{1})   
  \tilde{a}^{+}_{n}(l_{2})\rvert 0+1+2\lvert b^+\rangle\big\rvert^2c_{0}(q)+\\
&+\sum_{m,n}\sum_{i}\int d^3l_{1}\int d^3l_{2}\int d^3k_{1}
 \big\lvert\langle b^+\tilde{a}^{+}_{m}(l_{1})\tilde{a}^{+}_{n}(l_{2})
 \rvert 0+1+2\lvert b^+a_{i}^{+}(k_{1})\rangle\big\rvert^2
 c_{1}(q,k_{1})_{i}+\\
&+\sum_{m,n}\sum_{i}\int d^3l_{1}\int d^3l_{2}\int d^3k_{1}
 \big\lvert\langle b^+\tilde{a}^{+}_{m}(l_{1}\tilde{a}^{+}_{n}(l_{2}))
 \rvert 0+1+2\lvert b^+\tilde{a}_{i}^{+}(k_{1})
 \rangle\big\rvert^2\tilde{c}_{1}(q,k_{1})_{i}+\\
&+\sum_{m,n}\sum_{i,j}\int d^3l_{1}\int d^3l_{2}\int d^3k_{1}\int d^3k_{2}\\
&\hspace{2.0 cm}
 \big\lvert\langle b^+\tilde{a}^{+}_{m}(l_{1})\tilde{a}^{+}_{n}(l_{2})
 \rvert 0+1+2\lvert 
 b^+a_{i}^{+}(k_{1})a_{j}^{+}(k_{2})\rangle\big\rvert^2
 c_{2}(q,k_{1},k_{2})_{ij}+\\
&+\sum_{m,n}\sum_{i,j}\int d^3l_{1}\int d^3l_{2}\int d^3k_{1}\int d^3k_{2}\\
&\hspace{2.0 cm}
 \big\lvert\langle b^+\tilde{a}^{+}_{m}(l_{1})\tilde{a}^{+}_{n}(l_{2})
 \rvert 0+1+2\lvert 
 b^+a_{i}^{+}(k_{1})\tilde{a}_{j}^{+}(k_{2})\rangle\big\rvert^2
 \tilde{c}_{2}(q,k_{1},k_{2})_{ij}+\\
&+\sum_{m,n}\sum_{i,j}\int d^3l_{1}\int d^3l_{2}\int d^3k_{1}\int d^3k_{2}\\
&\hspace{2.0 cm}
 \big\lvert\langle b^+\tilde{a}^{+}_{m}(l_{1})\tilde{a}^{+}_{n}(l_{2})
 \rvert 0+1+2\lvert 
 b^+\tilde{a}_{i}^{+}(k_{1})\tilde{a}_{j}^{+}(k_{2})\rangle\big\rvert^2
 \tilde{\tilde{c}}_{2}(q,k_{1},k_{2})_{ij}\bigg\}.
\end{split}
\end{equation}
Now we collect the contributions that are not zero and of order not higher 
than four. Thus we are left with:
\begin{align*}
\int_{\mathrm{MS}_{\mathrm{I}}}d^4p\int_{\mathrm{MS}}d^4q&\bigg\{\\
+\lvert\langle b^+\rvert0\lvert b^+\rangle\rvert^2\Big[
 &c_{0}(q)+\sum_{i}\int d^3k_{1}c_{1}(q,k_{1})_{i}
 +\sum_{i}\int d^3k_{1}\tilde{c}_{1}(q,k_{1})_{i}+\\
&+\sum_{i,j}\int d^3k_{1}\int d^3k_{2}c_{2}(q,k_{1},k_{2})_{ij}
 +\sum_{i,j}\int d^3k_{1}\int d^3k_{2}\tilde{c}_{2}(q,k_{1},k_{2})_{ij}+\\
&+\sum_{i,j}\int d^3k_{1}\int d^3k_{2}\tilde{\tilde{c}}_{2}(q,k_{1},k_{2})_{ij}
 \Big]+
\end{align*}
\begin{align*}
+\sum_{i}\int d^3k_{1}\Big[&+\lvert\langle b^+\rvert 1
  \lvert b^+a^{+}_{i}(k_{1})\rangle\rvert^2 c_{1}(q,k_{1})_{i}+
  \lvert\langle b^+\rvert 1\lvert b^+\tilde{a}^{+}_{i}(k_{1})
  \rangle\rvert^2\tilde{c}_{1}(q,k_{1})_{i}+\\
&+\lvert\langle b^+a^{+}_{i}(k_{1})\lvert 1\rvert b^+\rangle\rvert^2c_{0}(q)+
 \sum_{j}\int d^3k_{2} \lvert\langle b^+\rvert 1
  \lvert b^+a^{+}_{i}(k_{1})\rangle\rvert^2 c_{2}(q,k_{1},k_{2})_{ij}+\\
&+\sum_{j}\int d^3k_{2} \lvert\langle b^+\rvert 1
  \lvert b^+\tilde{a}^{+}_{i}(k_{1})\rangle\rvert^2
  \tilde{c}_{2}(q,k_{1},k_{2})_{ij}
 +\lvert\langle b^+\tilde{a}^{+}_{i}(k_{1})\lvert 1\rvert b^{+}
  \rangle\rvert^2c_{0}(q)+\\
&+\sum_{j}\int d^3k_{2} \lvert\langle b^+\rvert 1
  \lvert b^+a^{+}_{i}(k_{1})\rangle\rvert^2\tilde{c}_{2}(q,k_{1},k_{2})_{ij}+\\
&+\sum_{j}\int d^3k_{2} \lvert\langle b^+\rvert 1
  \lvert b^+\tilde{a}^{+}_{i}(k_{1})\rangle\rvert^2 
  \tilde{\tilde{c}}_{2}(q,k_{1},k_{2})_{ij}+\\
&+\sum_{j}\int d^3k_{2}\lvert\langle b^+a^{+}_{i}(k_{1})\rvert 1 \lvert b^+
  \rangle\rvert^2 c_{1}(q,k_{2})_{j}+\\
&+\sum_{j}\int d^3k_{2}\lvert\langle b^+\tilde{a}^{+}_{i}(k_{1})
  \rvert 1 \lvert b^+\rangle\rvert^2 c_{1}(q,k_{2})_{j}+\\
&+\sum_{j}\int d^3k_{2}\lvert\langle b^+a^{+}_{i}(k_{1})\rvert 1 \lvert b^+
  \rangle\rvert^2 \tilde{c}_{1}(q,k_{2})_{j}+\\
&+\sum_{j}\int d^3k_{2}\lvert\langle b^+\tilde{a}^{+}_{i}(k_{1})
  \rvert 1 \lvert b^+\rangle\rvert^2 \tilde{c}_{1}(q,k_{2})_{j}\Big]+  
\end{align*}  
\bel{canc}
\begin{split} 
&+\sum_{m}\sum_{i}\int d^3l_{1}\int d^3k_{1} 2\mathrm{Re}\Big[
 \langle b^+\rvert 0\lvert b^+\rangle\delta_{mi}\delta(l_{1}-k_{1})
 \langle b^+a^{+}_{m}(l_{1})\rvert 2\lvert b^+a^{+}_{i}(k_{1})\rangle^*
 c_{1}(q,k_{1})_{i}\Big]+\\
&+\sum_{m}\sum_{i}\int d^3l_{1}\int d^3k_{1} 2\mathrm{Re}\Big[
 \langle b^+\rvert 0\lvert b^+\rangle\delta_{mi}\delta(l_{1}-k_{1})
 \langle b^+\tilde{a}^{+}_{m}(l_{1})\rvert 2
 \lvert b^+\tilde{a}^{+}_{i}(k_{1})\rangle^*
 \tilde{c}_{1}(q,k_{1})_{i}\Big]+\\
&+\sum_{m,n}\sum_{i,j}\int d^3l_{1}\int d^3l_{2}\int d^3k_{1}\int d^3k_{2}
  \cdot\\ 
&\hspace{2.0 cm}\cdot 2\mathrm{Re}\Big[
 \langle b^+\rvert 0\lvert b^+\rangle\delta_{mi}\delta_{nj}\delta(l_{1}-k_{1})
 \delta(l_{2}-k_{2})\cdot\\&\hspace{4.0 cm}\cdot 
 \langle b^+a^{+}_{m}(l_{1})\rvert 2\lvert b^+a^{+}_{i}(k_{1})\rangle^*
 \delta(l_{2}-k_{2})\delta_{nj}c_{2}(q,k_{1},k_{2})_{ij}\Big]+\\
&+\sum_{m,n}\sum_{i,j}\int d^3l_{1}\int d^3l_{2}\int d^3k_{1}\int d^3k_{2} 
 2\mathrm{Re}\Big[
 \langle b^+\rvert 0\lvert b^+\rangle\delta_{mi}\delta_{nj}\delta(l_{1}-k_{1})
 \delta(l_{2}-k_{2})\cdot\\
&\hspace{2.0 cm}\cdot\Big\{ 
 \langle b^+a^{+}_{m}(l_{1})\rvert 2\lvert b^+a^{+}_{i}(k_{1})\rangle^*
 \frac{1}{2}\delta(l_{2}-k_{2})\delta_{nj}+\\&\hspace{3.0 cm}+
 \langle b^+\tilde{a}^{+}_{n}(l_{2})\rvert 2\lvert 
  b^+\tilde{a}^{+}_{j}(k_{2})\rangle^*
 \frac{1}{2}\delta(l_{1}-k_{1})\delta_{mi}\Big\}
 \tilde{c}_{2}(q,k_{1},k_{2})_{ij}\Big]+\\
&+\sum_{m,n}\sum_{i,j}\int d^3l_{1}\int d^3l_{2}\int d^3k_{1}\int d^3k_{2} 
   \cdot\\ 
  &\hspace{2.0 cm}\cdot2\mathrm{Re}\Big[
 \langle b^+\rvert 0\lvert b^+\rangle\delta_{mi}\delta_{nj}\delta(l_{1}-k_{1})
 \delta(l_{2}-k_{2})\cdot\\&\hspace{4.0 cm}\cdot 
 \langle b^+\tilde{a}^{+}_{m}(l_{1})\rvert 2\lvert 
 b^+\tilde{a}^{+}_{i}(k_{1})\rangle^*
 \delta(l_{2}-k_{2})\delta_{nj}
 \tilde{\tilde{c}}_{2}(q,k_{1},k_{2})_{ij}\Big]\bigg\}.
\end{split}
\end{equation} 
The products of identical $\delta$-distributions showing up in some terms of 
this formula indicate that we should have done this calculation using wave 
packets instead of sharp momentum states in the description of the asymptotic 
states. Since the products of zero- and two-BS contributions play no 
r\^ole in our discussion since they are finite (cf. Sec. 
{\ref{subsubsec:mixstatcanc}}), we will 
not do this here explicitly, we just state that 
the analogous calculation involving wave packets leads to the same result 
with only one of the identical $\delta$'s in each term. 
Thus we throw away one of them in the expression above.
\end{appendix}

\end{document}